\documentclass[b5paper,twoside,openright]{report}
\pdfoutput=1

\usepackage{moreverb}								% List settings
\usepackage{textcomp}								% Fonts, symbols etc.
\usepackage{lmodern}								% Latin modern font
\usepackage{helvet}									% Enables font switching
\usepackage[T1]{fontenc}							% Output settings
\usepackage[english]{babel}							% Language settings
\usepackage[utf8]{inputenc}							% Input settings
\usepackage{amsmath}								% Mathematical expressions (American mathematical society)
\usepackage{amssymb}								% Mathematical symbols (American mathematical society)
\usepackage{graphicx}								% Figures
\usepackage{subfig}									% Enables subfigures
\numberwithin{equation}{chapter}					% Numbering order for equations
\numberwithin{figure}{chapter}						% Numbering order for figures
\numberwithin{table}{chapter}						% Numbering order for tables
\usepackage{listings}								% Enables source code listings
\usepackage{chemfig}								% Chemical structures		
\usepackage{eso-pic}								% Create cover page background

\usepackage{bm} %boldmath
\usepackage{setspace} %for space setting
\usepackage[labelfont=bf, textfont=small,
			justification=justified,
			singlelinecheck=false]{caption} 		

\usepackage{titlesec}		
\titleformat{\chapter}[display]
  {\Huge\bfseries\filcenter}
  {{\fontsize{20pt}{1em}\vspace{-6.2ex}\selectfont \textnormal{\bf{Chapter \thechapter}}}}{1ex}{}[]

\usepackage{fancyhdr}								
\def\layout{2}	% Choose 1 for one-sided or 2 for two-sided layout
\ifnum\layout=2	% Two-sided
	\fancypagestyle{plain}{			% Redefine the plain page style
	\fancyhf{} % clear all header and footer fields
	
	\fancyfoot[LE,RO]{\thepage}}	
\else			% One-sided  	
\fi

\usepackage{epigraph}
\usepackage[numbers]{natbib}
\usepackage{enumitem}% http://ctan.org/pkg/enumerate
\usepackage[hyphens]{url}
\usepackage{hyperref}
\hypersetup{
    colorlinks=true,
    urlcolor=blue,
    citecolor=gray,
    linkcolor=red, %darkgray, %change this to 
    breaklinks=true
}
\usepackage[top=3cm, bottom=3cm, left=3cm, right=3cm]{geometry} % Page margin lengths	
			
\usepackage{makeidx} 
\makeindex 

\begin{document} 
\setcounter{page}{-1}
%-----% COVER PAGE, TITLE PAGE AND IMPRINT PAGE
\pagenumbering{roman}			% Roman numbering (starting with i (one)) until first main c
% !TeX root = ../main.tex

% COVER PAGE	
% TITLE PAGE
\newpage
\thispagestyle{empty}
\begin{center}
\text{ }\\
\vspace{0.3cm}
\setlength{\baselineskip}{30pt}
{\Huge \textbf{One-Dimensional Quantum Systems}\\[0.4cm]\par % Title
  \LARGE\textit{From Few To Many Particles}}
\vspace{0.5cm}

	\begin{figure}[h!]
	\centering
	\includegraphics[width=0.7\columnwidth]{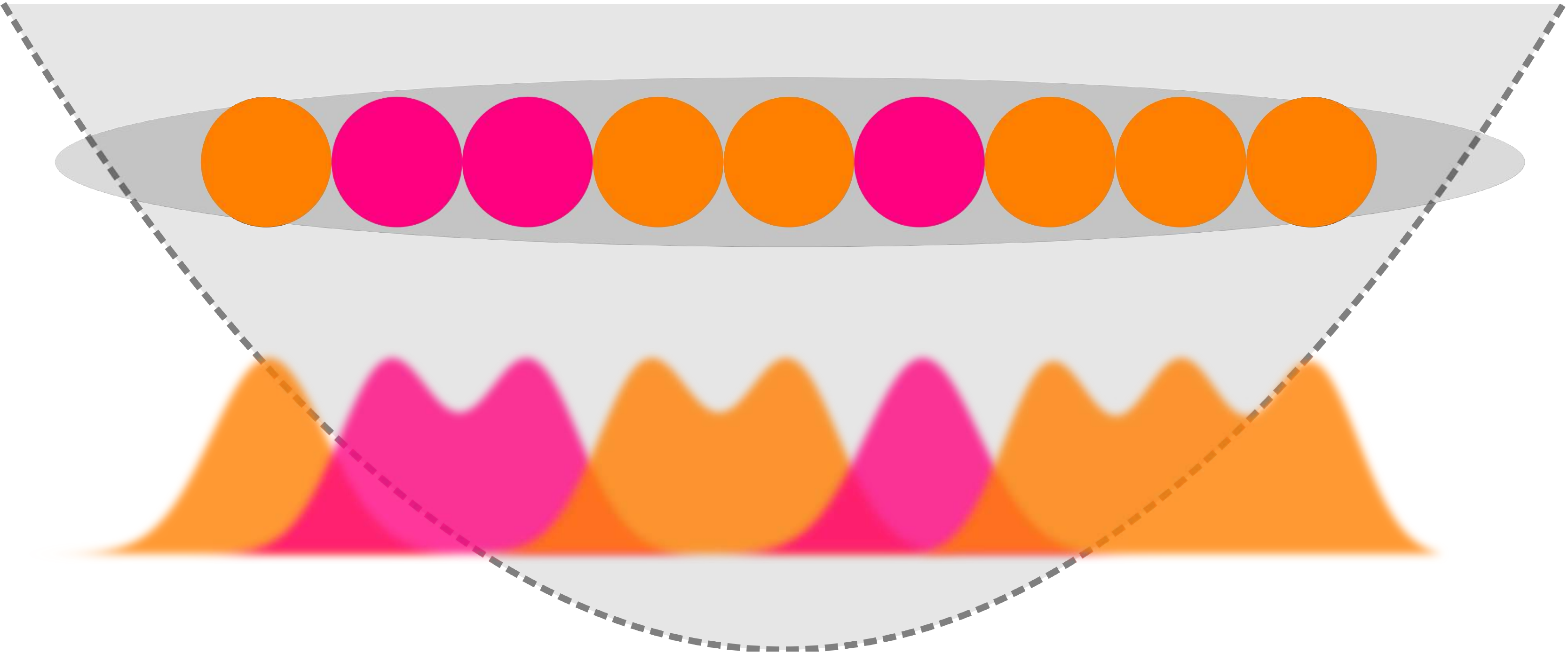}
	\end{figure}	\vspace{5mm}
	
\setlength{\baselineskip}{10pt}
\vspace{0.5cm}
{\large\textsc{Amin~Salami~Dehkharghani}}\par
% \small\textit{amin88phd@gmail.com\\}\par
\vspace{1cm}
\textsc{Dissertation for the degree of}\par
\textsc{Doctor of Philosophy}\par
\vspace{0.5cm}
\textsc{2nd Edition}
\vspace{0.5cm}

	\begin{figure}[h!]
	\centering
	\includegraphics[width=0.2\columnwidth]{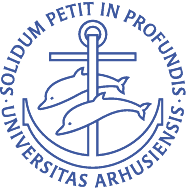}
	\includegraphics[width=0.2\columnwidth]{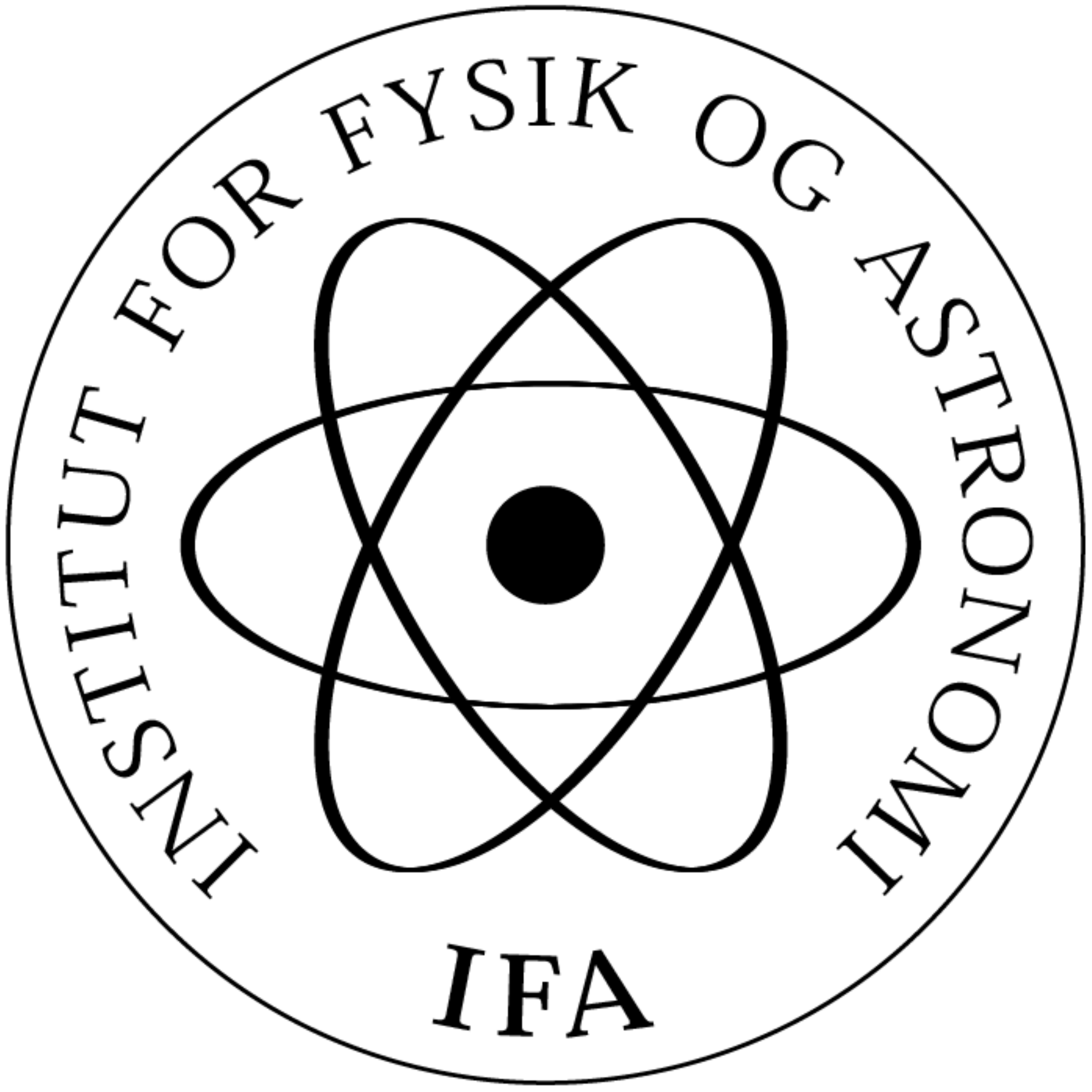}
	\end{figure}	\vspace{5mm}

%\textsc{Supervisor: Nikolaj~T.~Zinner}\par
\vspace{0.0cm}

{
\textsc{Department of Physics and Astronomy}\par
\textsc{Graduate School of Science and Technology}\par
\textsc{Aarhus University, Denmark}\par
\textsc{July 2017}
}
%\enlargethispage{2\onelineskip}
\end{center}

\newpage				% Create empty back of side
\thispagestyle{empty}
\mbox{}

% IMPRINT PAGE (BACK OF TITLE PAGE)
\thispagestyle{empty}
\vfill

\noindent
 \copyright~Copyright, Amin Salami Dehkharghani \\
1st Edition, July 2017\\
2nd Edition, December 2017\\
\href{http://pure.au.dk/portal/da/persons/id(7a929f10-7551-4ac8-ab87-08744b58ab7e).html}{Phys.au.dk $\rightarrow$ Amin Salami Dehkharghani}\\
\href{mailto:amin@phys.au.dk}{E-mail $\rightarrow$  amin@phys.au.dk}\\
\href{https://scholar.google.dk/citations?user=G03M2xoAAAAJ&hl=da&oi=sra}{Google Scholar  $\rightarrow$  Amin Salami Dehkharghani}\\
\href{https://www.linkedin.com/in/aminsalamidehkharghani/}{LinkedIn $\rightarrow$  Amin Salami Dehkharghani}\\[0.5cm]

%\noindent
%Supervisor: Nikolaj T. Zinner\\

\noindent
Dissertation for the degree of Doctor of Philosophy\\ \\
Few Body Group\\
Department of Physics and Astronomy\\
Aarhus University\\
Ny Munkegade 120
DK-8000 Aarhus C\\
Denmark\\
E-mail: phys@au.dk\\
Tel.: +45 8715 0000\\

\noindent
Typeset in \LaTeX, \\
Figures in Python, Gimp, Inkscape and Adobe Photoshop,\\
Printed by AUTRYK, Aarhus University\\

\newpage

%\cleardoublepage
\vfill
\thispagestyle{empty}
\vspace*{2cm}
\noindent
\section*{Preface}
This dissertation has been submitted to the Faculty of Science and Technology at Aarhus University, Denmark, in partial fulfillment of the requirements for the PhD degree in physics. The work and the published articles presented here have been performed in the period from August 2013 to July 2017 under the supervision of Nikolaj T. Zinner from Aarhus University.\\

\noindent
{\bf 2nd Edition}\\
After defending my thesis successfully in September 22nd, 2017, and based on the recommendation I got from many people, I have decided to put my thesis on arXiv so interested students, colleagues and friends can have access to it. Throughout the thesis I have tried to explain one-dimensional topics in another way than the approach that is taken in my articles. The 2nd edition comes with some minor corrections in form of updated references to the literature. A sign and notation misprints have also been corrected in Eq. (3.1), (3.2) and page 45. In addition, Fig. (5.12) has been updated with the latest research after my PhD.
\\

\noindent
{\bf Recommendation by the opponents}\\
The PhD thesis deals with a very essential, yet extremely difficult problem of quantum physics: determining the quantum state of few interacting particles in a confining potential. The candidate focuses his efforts in the derivation of an analytical solution to the few-particle problem. The theoretical framework introduced by him is very elegant and the results are rigorously derived, novel, and well discussed.
The candidate has excellent theoretical skills and shows great maturity. In addition, he shows a very good knowledge of the main experimental techniques in the field of ultracold atom physics, where the predictions of his theoretical work can be tested. The results of the thesis are highly interesting and are likely to be quite important in future developments of this research line, both for a better theoretical comprehension of the physics of interacting quantum systems, and for the design and interpretation of new cold-atoms experiments aimed at the verification of those fundamental effects.

Based on the above assessment we conclude that the PhD thesis of Amin Salami Dehkharghani clearly fulfils the requirements for the award of the PhD degree.

\newpage				% Create empty back of side
\thispagestyle{empty}
\mbox{}
 %print and assertation

%-----%DEDICATION
\cleardoublepage

\thispagestyle{empty}
\vspace*{\fill}
\begin{center}
\it Dedicated to Jalil, Mahin, Leili, Foreman, and Diana.\\

%, Khosro, \\ Leili, Behnam, Aidan, Ava, \\ Foreman, Sarah, \\ and Diana.\\

Kheyli doosetun daram!
\vspace{3cm}
\end{center}
\vspace*{\fill}

\newpage				% Create empty back of side
\thispagestyle{empty}
\mbox{}

%-----% ABSTRACT
\cleardoublepage

\thispagestyle{plain}			% Supress header 
\section*{English Summary}
Ever since the realization of the Bose-Einstein Condensate (BEC) in 1995, remarkable studies of the cold atomic gases have been developed both experimentally and theoretically. Especially, the low dimensional quantum systems have been of particular interest due to their simplicity and exotic physics in contrast to higher dimensions. Many state-of-the-art experiments have been conducted ever since, such that one can now setup a very fine one-dimensional geometry and have full control over the particles. The precision and control has become so sophisticated that one can simply adjust the interaction between the particles by just turning a button. In this way, the experimentalists are able to study few-body dynamics or build a Fermi sea one atom at a time and therefore investigate the transition between few- to many-body systems. Recently, it was possible to verify some of the old and exact analytical results such as the Tonks-Girardeau gas and super-Tonks-Girardeau Bose gases in one-dimensional quantum systems. However, many other quantum systems in different regimes are still uncovered and the knowledge about these systems can help us to understand the quantum properties of the particles in nature and in the near future maybe design our own quantum materials one particle at a time.

In this thesis, I will start by the well-known solutions to the one and two-particle systems trapped in a quantum harmonic oscillator and then continue to the three, four and many-body quantum systems. This is done by developing new analytical models and numerical methods both for the few- and many-body systems. One-dimensional systems are very interesting in a sense that particles aligned on a line can only change seats by going through each other. This property can be exploited in the strongly interacting regime, where particles are forced to sit in a specific configuration, which can be easily manipulated. The knowledge of how and where the particles are can be exploited in future quantum applications. In short, the thesis is about establishing a solid knowledge about everything that one needs to know about the one-dimensional few- and many-component interacting quantum systems trapped in harmonic oscillator potentials.

\newpage				% Create empty back of side
\thispagestyle{empty}
\mbox{}
\newpage				% Create empty back of side

\thispagestyle{plain}			% Supress header 
\section*{Dansk Resumé}
Siden virkeliggørelsen af Bose-Einstein Kondensat (BEC) i 1995, er der blevet lavet mange eksperimentelle og teoretiske studier inden for kolde gasser. Specielt, har lav dimensionale kvantesystemer tiltrukket megen opmærksom til sig på grund af deres enkelthed og eksotiske egenskaber i kontrast med højere dimensioner. I nyere tid er der blevet udført mange avancerede eksperimenter, hvor man kan forberede en dimensionale gasser og samtidig have fuld kontrol over partiklerne. I dag er præcisionen og kontrollen over eksperimentet blevet så sofistikeret, at man bare kan justere interaktionen mellem partiklerne ved at dreje på en knap. På denne måde, er eksperimental fysikere nu i stand til at studere få-legeme dynamik eller bygge en Fermi sø ét atom ad gangen. Derfor er det muligt at undersøge overgangen mellem få- og mange-legeme fysik ren eksperimentelt. Fornyligt har det været muligt at bekræfte nogle af de gamle og eksakte analytiske resultater såsom Tonks-Girardeau gas og super-Tonks-Girardeau Bose gas i et dimensional kvantesystemer. Men mange andre kvantesystemer for andre forskellige parametre og systemer er stadigvæk uopklarede og ikke undersøgt nogensinde. Netop viden om disse systemer kan hjælpe os til at forstå kvante-egenskaber om partikler i naturen og i fremtiden hjælpe os til at designe vores egen kvante materialer én partikel ad gangen.

I denne afhandling, vil jeg starte med velkendte resultater om en og to partikel systemer fanget i en kvante harmonisk fælde og bagefter bevæge mig hen til tre, fire og mange-legeme kvantesystemer. Det vil jeg gøre ved at udvikle analytiske modeller og numeriske metoder både for få- og mange-legeme systemer. Én dimensionale systemer er yderst interessante fordi partikler på en række kun kan bytte plads med hinanden ved at gå igennem hinanden. Denne egenskab kan blive udnyttet i den stærkt vekselvirkende grænse, hvor partikler er tvunget til at sidde i en specifik konfiguration, som man let kan manipulere. Viden om hvordan og hvor disse partikler er kan blive udnyttet i fremtidige kvante applikationer. Kort sagt, afhandlingen handler om at etablere en solid viden om alt det man behøver at vide om én dimensional få- og mange-legeme vekselvirkende kvantesystemer fanget i harmoniske fælder.

\newpage				% Create empty back of side
\thispagestyle{empty}
\mbox{}

%-----% ACKNOWLEDGEMENTS
\newpage

\thispagestyle{plain} % Supress header
\section*{Acknowledgements}
Recently, I was asked: ``{\it On a scale from one to ten, how lucky have you been in life so far?}''\\

I answered: ``{\it Ten... Definitely ten!}''\\

\noindent
And here is the tricky part: I am actually one of those people who don’t believe in luck, but in hard work. I also believe that sometimes it takes even harder work to achieve your goals. So why did I answer ten? Let me elaborate on this: Looking back at my life and the last couple of years at Aarhus University, I have been very fortunate to work with some of the most inspiring and skilled people that I have ever known. Even though I have really tried to be a hardworking person and design my own future, I cannot stop thinking of how it would have been without the presence of certain people in my life who have encouraged, supported, motivated and believed in me. Therefore I am very grateful and lucky to have these people in my life and it would be my pleasure to acknowledge them here.\\

\noindent
First, I would like to start by sending a big thank-you to all my teachers from the elementary and high school (Møllevangskolen, Vestergaardsskolen and HTX Viby), who have all always challenged me, helped me (even in their overtime) and all have done a great job in inspiring me to go after my dreams.

Second, I must also send a thank-you note to all the professors that I have met during my Bachelor's and Master's years both from the Mathematics and Physics department. I must say that I have had a lot of entertaining and inspiring lectures. Special thanks go to Ulrik Uggerhøj and Axel Svane for giving me guidance and admirable recommendation to follow a PhD career. Even though Axel Svane is not with us anymore, he will always have my recognition for being one of the {\it best} professors at all times.\\

\noindent
Third, I should definitely mention the people that I have been working with, which have resulted in so many great discussions, papers and friendships. In order of first appearance on \href{https://arxiv.org/find/all/1/all:+dehkharghani/0/1/0/all/0/1}{arXiv}: Artem G. Volosniev, Jonathan Lindgren, Jimmy Rotureau, Christian Forss{\'e}n, Niels-Jakob. S. Loft, Nirav P. Mehta, Miguel A. Garc\'{\i}a-March, Molte E. S. Andersen, Rafael E. Barfknecht, Filipe F. Bellotti, Daniel P\k{e}cak, Tomasz {Sowi\'{n}ski}, Enrique Rico, Antonio Negretti, Nathan L. Harshman, Maxim Olshanii and Steven G. Jackson.

Special thanks go to Jonathan Lindgren, Jimmy Rotureau and Christian Forss{\'e}n for their hospitality and support during my time in G{\"o}teborg and afterwards. Personal thanks go to Niels-Jakob S. Loft and Molte E. S. Andersen for their hard work and teamwork on our joint projects. I must not forget to thank Miguel A. Garc\'{\i}a-March for his collaboration and great ideas on a paper we did together. Personal obrigados go to Rafael E. Barfknecht and Filipe F. Bellotti for their friendship and excellent teamwork - Filipe, we should definitely implement our genius and original idea on how to generate electricity using gravitational forces. Another special thanks go to Daniel P\k{e}cak and Tomasz {Sowi\'{n}ski} for their invitations and hospitality in Warsaw - Daniel, thank you for showing me around in Warsaw and I won't forget the funny Mexican restaurant. I must also thank Enrique Rico and Antonio Negretti for their patience and the {\it many} hours of Skype meetings and discussions that we had. You guys have been like second supervisors to me. Antonio, thank you for the postdoc offer at the very early stage and your hospitality in Hamburg. I really appreciate your help and I will never forget it. I also had the chance to work very closely with Nathan L. Harshman and Maxim Olshanii. There have been some very fruitful and interesting topics they each individually have brought to the table. Nathan, for the record beer doesn't count when you compete with me on who can eat a whole Bøfsandwich and my team won over your team in the escape-factory-game. Thanks for the great times, but remember, hygge is not something you do, it is something you are. Finally, thank you, Artem G. Volosniev for your great ideas and inspiration. It has resulted in many wonderful and awesome projects, and a great friendship. And thanks for your quick responses even when you were on vacation.\\

\noindent
Aside from collaborating with very competent people, I have also been a part of an admirable group with great and sometimes crazy lunch discussions. Therefore I would like to thank all the people at the Few-Body and Subatomic Experimental Group at Aarhus University. Special thanks go to Dmitri V. Fedorov, Aksel S. Jensen and Nikolaj T. Zinner for their full support from day one, whether there has been a programming issue or discussion of new ideas. Also a big thank-you goes to Karsten Riisager, especially for his guidance during my PhD-writing, and Hans Fynbo and all the Postdocs, PhDs, Master's and Bachelor's students who I have had the chance to meet during my studies and who have contributed to a great group dynamic.

I have also been a member of a great board at the Aarhus University Motion Center with very nice people who have never said ``no'' to my ideas and given me unique opportunities to develop myself both personally and professionally. Thank you for the fun meetings.

Another special thank-you goes to Melek, our great cleaning lady, who has visited my office once a week and asked me how I was feeling each week, and put my Turkish language skills on a test. Ben teşekkür ederim.\\

\noindent
It is also relevant to thank three of our Bachelor's students, who I have had the chance to work with and supervise. Mathias Thomsen, Matias Wallenius and Thorbjørn Lindgren are all individually quick learners, skilled, independent and very great to work with. Their next-generation brainy minds surely did supervising easy from my part - also a little bit annoying every time they found a mistake in my work.

\noindent
Furthermore, I would like to thank my supervisor, Nikolaj T. Zinner for his guidance, advice and support all the way. I have always followed your advice in different situations and it has only resulted in excellent outcomes. And thank you for the road trip we had together to Hamburg. It truly shows how much you really care about your student's projects and their career. However, Nikolaj, you must admit that it was because of me, we made it all the way to the finals in the 2016 football tournament and even won the ``Styrkeprøven''. My time as your PhD student has been really great and awesome! As you know I don't use the word ``awesome'' very often, only when something is really awesome! You have always endorsed my strengths and challenged me even more with different kind of projects and big responsibilities. Thank you for being like Mr. Miyagi to me!\\

\noindent
I would also like to spend a few lines to thank Aarhus University's career advisor, Vibeke Broe, who has been a big help in my choice of career after PhD and prepared me well while seeking a career in the real world.

Further thanks go to Nathan L. Harshman and Nikolaj T. Zinner for proofreading my thesis. If the opponents find any mistakes in my thesis, I'll let them know how to find you guys. No worries.\\

\noindent
Lastly, I must mention my parents, family and close friends for their support throughout the years. To my mom, thank you for all the sacrifices you have made. You have always been there for me with guidance, advice and delicious Persian food. To my dad, thank you for being such a great role model both personally and professionally, which I will never forget. To my brother and sister, thank you for teaching me about science from as early as I remember and encouraging my interest towards it. To Khosro, thank you for your full support, help and interest in my work. To my uncles and aunts, thank you for your encouragement and brace. To my close friends, thank you for being dependable, sincere and honest. To my fiancée, Diana, thank you for always being there for me and supporting me. Thank you for the great and very enjoyable trips we have had together so far and I am looking forward to spending the rest of our lives together.

I have learned a lot from you guys and your support means the world to me.\\

\noindent
{\it Thank you.}

\vspace{1.5cm}
\hfill Amin Salami Dehkharghani

\hfill Aarhus, July 2017

\newpage % Create empty back of side
\thispagestyle{empty}
\mbox{}

%-----% List of publications
\newpage
\thispagestyle{plain}			% Supress header
\section*{List of Publications}
\vspace{0.2cm}

\begin{enumerate}[leftmargin=-0.0cm,label=(\Roman*)]
\item \textbf{A.~S. Dehkharghani}, A.~G. Volosniev, E.~J. Lindgren, J. Rotureau, C. Forss{\'e}n, D.~V. Fedorov, A.~S. Jensen and N.~T. Zinner, \textit{``Quantum Magnetism in Strongly Interacting One-Dimensional Spinor Bose Systems''}, Sci. Rep. {\bf 5}, 10675 (2015); doi: 10.1038/srep10675; URL \url{http://www.nature.com/articles/srep10675}  \\[-0.3cm]

\item N.~J.~S. Loft, \textbf{A.~S. Dehkharghani}, N.~P. Mehta, A.~G. Volosniev, and N.~T. Zinner, \textit{``A Variational Approach to Repulsively Interacting Three-Fermion Systems in a One-Dimensional Harmonic Trap''}, Eur. Phys. J. D {\bf 69}, 65 (2015); doi: 10.1140/epjd/e2015-50845-9; URL \url{http://link.springer.com/article/10.1140\%2Fepjd\%2Fe2015-50845-9}  \\[-0.3cm]

\item \textbf{A.~S. Dehkharghani}, A.~G. Volosniev and N.~T. Zinner, \textit{``Quantum Impurity in a One-Dimensional Trapped Bose Gas''}, Phys. Rev. A {\bf 92}, 031601(R) (2015); doi: 10.1103/PhysRevA.92.031601 URL \url{http://journals.aps.org/pra/abstract/10.1103/PhysRevA.92.031601}  \\[-0.3cm]

\item \textbf{A.~S. Dehkharghani}, A.~G. Volosniev and N.~T. Zinner, \textit{``Impenetrable Mass-Imbalanced Particles in One-Dimensional Harmonic Traps''}, J. Phys. B: At. Mol. Opt. Phys. {\bf 49}, 085301 (2016); doi: 10.1088/0953-4075/49/8/085301; URL \url{http://iopscience.iop.org/article/10.1088/0953-4075/49/8/085301/meta}  \\[-0.3cm]

\item M.~A. Garc\'{\i}a-March, \textbf{A.~S. Dehkharghani} and N.~T. Zinner, \textit{`` Entanglement of an Impurity in a Few-Body One-Dimensional Ideal Bose System''}, J. Phys. B: At. Mol. Opt. Phys. {\bf 49}, 075303 (2016); doi: 10.1088/0953-4075/49/7/075303; URL \url{http://iopscience.iop.org/article/10.1088/0953-4075/49/7/075303/meta}  \\[-0.3cm]

\item M.~E.~S. Andersen, \textbf{A.~S. Dehkharghani}, A.~G. Volosniev, E.~J. Lindgren and N.~T. Zinner, \textit{`` An Interpolatory Ansatz Captures the Physics of One-Dimensional Confined Fermi Systems''}, Sci. Rep. {\bf 6}, 28362 (2016); doi: 10.1038/srep28362; URL \url{http://www.nature.com/articles/srep28362}  \\[-0.3cm]

\item R.~E.~Barfknecht, \textbf{A.~S. Dehkharghani}, A.Foerster and N.~T. Zinner, \textit{``Correlation Properties of a Three-Body Bosonic Mixture in a Harmonic Trap''}, J. Phys. B: At. Mol. Opt. Phys. {\bf 49}, 135301 (2016); doi: 10.1088/0953-4075/49/13/135301; URL \url{http://iopscience.iop.org/article/10.1088/0953-4075/49/13/135301/meta}  \\[-0.3cm]

\item \textbf{A.~S. Dehkharghani}, N.~J.~S. Loft and N.~T. Zinner, \textit{``Kvantemagneter - Magneter når de er allermindst''}, Aktuel Videnskab {\bf 5}, (2016); URL \url{http://aktuelnaturvidenskab.dk/find-artikel/nyeste-numre/5-2016/}   \\[-0.3cm]

\item F.~F.~Bellotti, \textbf{A.~S. Dehkharghani} and N.~T. Zinner, \textit{``Comparing Numerical and Analytical Approaches to Strongly Interacting Two-Component Mixtures in One Dimensional Traps''}, Eur. Phys. J. D {\bf 71}, 37 (2017); doi: 10.1140/epjd/e2017-70650-8; URL \url{http://link.springer.com/article/10.1140\%2Fepjd\%2Fe2017-70650-8}   \\[-0.3cm]

\item \textbf{A.~S. Dehkharghani}, F.~F.~Bellotti and N.~T. Zinner, \textit{``Analytical and Numerical Studies of Bose-Fermi Mixtures in a One-Dimensional Harmonic Trap''}, J. Phys. B: At. Mol. Opt. Phys. {\bf 50}, 144002 (2017); doi: 10.1088/1361-6455/aa7797; URL \url{http://iopscience.iop.org/article/10.1088/1361-6455/aa7797/meta}  \\[-0.3cm]

\item \textbf{A.~S. Dehkharghani}, \textit{``Spiller Gud terninger med universet?''}, Videnskab.dk, {\bf 28}, 3 (2017);  URL \url{http://videnskab.dk/naturvidenskab/spiller-gud-terninger-med-universet}   \\[-0.3cm]

\item D.~P\k{e}cak, \textbf{A.~S. Dehkharghani}, N.~T. Zinner and T.~Sowi{\'n}ski, \textit{``Four Fermions in a One-Dimensional Harmonic Trap: Accuracy of a Variational-Ansatz Approach''}, Phys. Rev. A {\bf 95}, 053632 (2017); doi: 10.1103/PhysRevA.95.053632 URL \url{https://journals.aps.org/pra/abstract/10.1103/PhysRevA.95.053632}  \\[-0.3cm]

\item \textbf{A.~S. Dehkharghani}, E.~Rico, N.~T. Zinner and A.~Negretti, \textit{``Quantum Simulation of Abelian Lattice Gauge Theories via State-Dependent Hopping''}, Phys. Rev. A {\bf 96}, 043611 (2017); URL \url{https://journals.aps.org/pra/abstract/10.1103/PhysRevA.96.043611}   \\[-0.3cm]

\item N.~L~Harshman, M.~Olshanii, \textbf{A.~S. Dehkharghani}, A.~G.~Volosniev, S.~G. Jackson and N.~T. Zinner, \textit{``Integrable families of hard-core particles with unequal masses in a one-dimensional harmonic trap''}, Phys. Rev. X {\bf 7}, 041001 (2017); URL \url{https://journals.aps.org/prx/abstract/10.1103/PhysRevX.7.041001}   \\[-0.3cm]

\item \textbf{A.~S. Dehkharghani}, A.~G. Volosniev and N.~T. Zinner, \textit{``Interaction-driven Coalescence of Two Impurities in a One-dimensional Bose Gas''}, arXiv:1712.01538; URL \url{https://arxiv.org/abs/1712.01538} \\[-0.3cm]

\end{enumerate}

\newpage				% Create empty back of side
\thispagestyle{empty}

{\hypersetup{linkcolor=black}
%-----% TABLE OF CONTENTS
\newpage
\tableofcontents
}

%-----% START OF MAIN DOCUMENT
\cleardoublepage
\setcounter{page}{1}
\pagenumbering{arabic}			% Arabic numbering starting from 1 (one)

%-----% INTRODUCTION
% !TeX root = ../Main_publish.tex

\chapter{Introduction}\label{ch:introduction}
\epigraph{\it “Nobody ever figures out what life is all about, and it doesn't matter. Explore the world. Nearly everything is really interesting if you go into it deeply enough.”}{\rm ---Richard Feynman}

\begin{figure}[h]
\centering
\includegraphics[width=0.3\linewidth, trim=2cm 2cm 2cm 2cm]{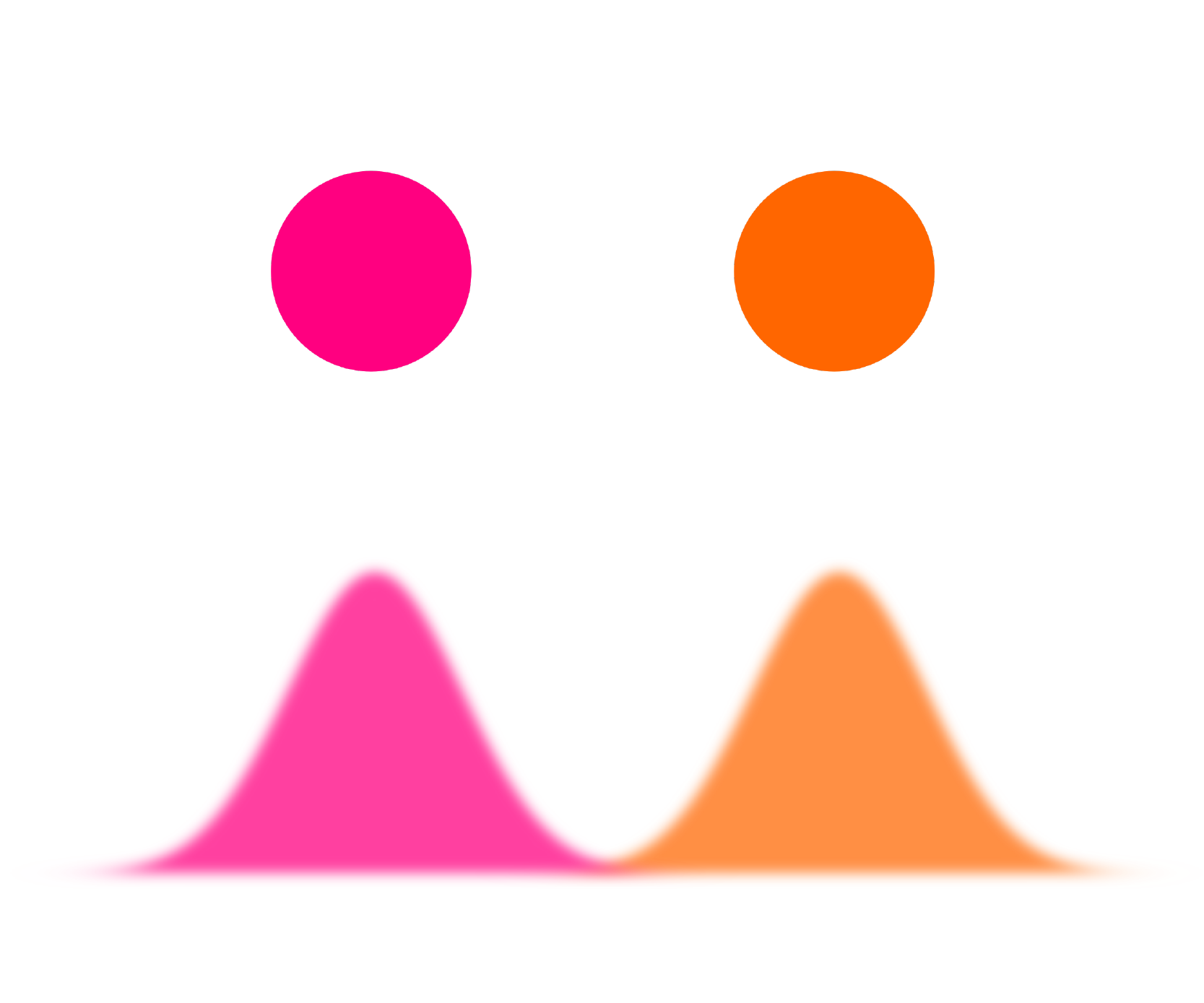}
\caption{Particles with their corresponding probability-waves}
\label{particle_wave}
\end{figure}

Allow me to start by giving a short introduction to quantum mechanics for the non-experts: One of the most fascinating facts in physics is that everything in the Universe behaves like both a particle and a wave at the same time! This might sound very strange and bizarre for most people due to the very distinguishable properties of both waves and particles, but nevertheless it is all very well described by a successful theory called quantum mechanics in modern physics. In short, quantum mechanics describes the laws relating to the very small at the atomic scales and differs a lot from classical physics and everyday life. Fig.~\ref{particle_wave} illustrates such a co-existing particle-wave property, which is characterized by quantum mechanics. 

In our everyday life when we watch a football match, we can observe and follow the position of the ball at all times. If we wanted, we could even predict the trajectory of the ball for later times. On the other hand, waves, like sound waves or water waves, can diffract and go through each other without notice. I am sure you could come up with many more examples that illustrate how different the two properties are from one another. So why would we combine such two distinct properties in one theory like quantum mechanics?

It all started with Albert Einstein \index{Albert Einstein}, when he suggested that light consists of a stream of particles, called photons \cite{einsteinAJP1905}. He was convinced that light, which had a lot of wave properties, could also behave like particles and many phenomena like the photoelectric effect could easily be explained by this analogy, and he got the Noble price for it in 1921. But then a few years later, in 1924 a French physicist by the name, Louis de Broglie \index{Louis de Broglie}, proposed in his PhD thesis that if light could have particle properties then why could it not be the other way around; particles having wave properties?

This question is the core-nature of quantum mechanics and has been developed in such a way that the wave-property of the particles is only expressed in the form of information about the particles. In other words, everything we can gain about the particle; for example its position and velocity, is described by a {\it probability-wave}\index{Probability-Wave}. The probability-wave can be described by a glass of water, where the water inside the glass represents the probability-wave for {\it one} particle. If we want to find the particle, we have to make an observation in order to see it. The observation could be in form of looking at the glass or by touching inside the glass in order to feel the particle. Since the whole probability-wave (that is water) is inside the glass we would find the particle inside the glass every time we make an observation. Let's make things a little more interesting; if we, on the other hand, pour some of the water into a second glass, then the particle is two places at once! But as mentioned earlier, there is only one particle and we can find this particle by making an observation of both glasses simultaneously.

Just like a dice, where it is highly improbable to throw a six every time, similarly with our quantum particles, we cannot be sure of finding the particle in the original glass every time. By making a series of observations, we would find the particle sometimes in the original glass, and sometimes in the second glass. The probability of finding the particle would correspond to exactly of how much water (that is probability-wave) there is in each glass. Now you can easily imagine that if we spill the water everywhere, the particle is basically everywhere. The particle collapses into a certain position when we make an observation.

Albert Einstein was not happy with this indeterministic and probability idea and strongly protested that He [God] does not play dice with the Universe and some {\it hidden} variable might be missing in quantum mechanics. But today, we know this is not true and quantum mechanics has proven to be one of the most successful theories in modern physics. The reason of why we do not encounter this phenomenon in everyday life is because quantum mechanics governs at the very small level and quickly becomes negligible as soon as one considers objects bigger than nano-scales. This is also why we do not see a ball going through the football players in a match, although it would be cool to watch.

\section{Quantum Nature \index{Quantum Nature}}\label{ch:intro:sec:quantumnature}
Ever since the development of quantum mechanics one has tried to control these indeterministic particles in order to say something about their position, understand them better and start exploiting their nature. However, the probability nature makes the particles behave like ghosts and very hard to grasp.

\begin{figure}[h]
\centering
\includegraphics[width=\linewidth, trim=2cm 2cm 2cm 2cm]{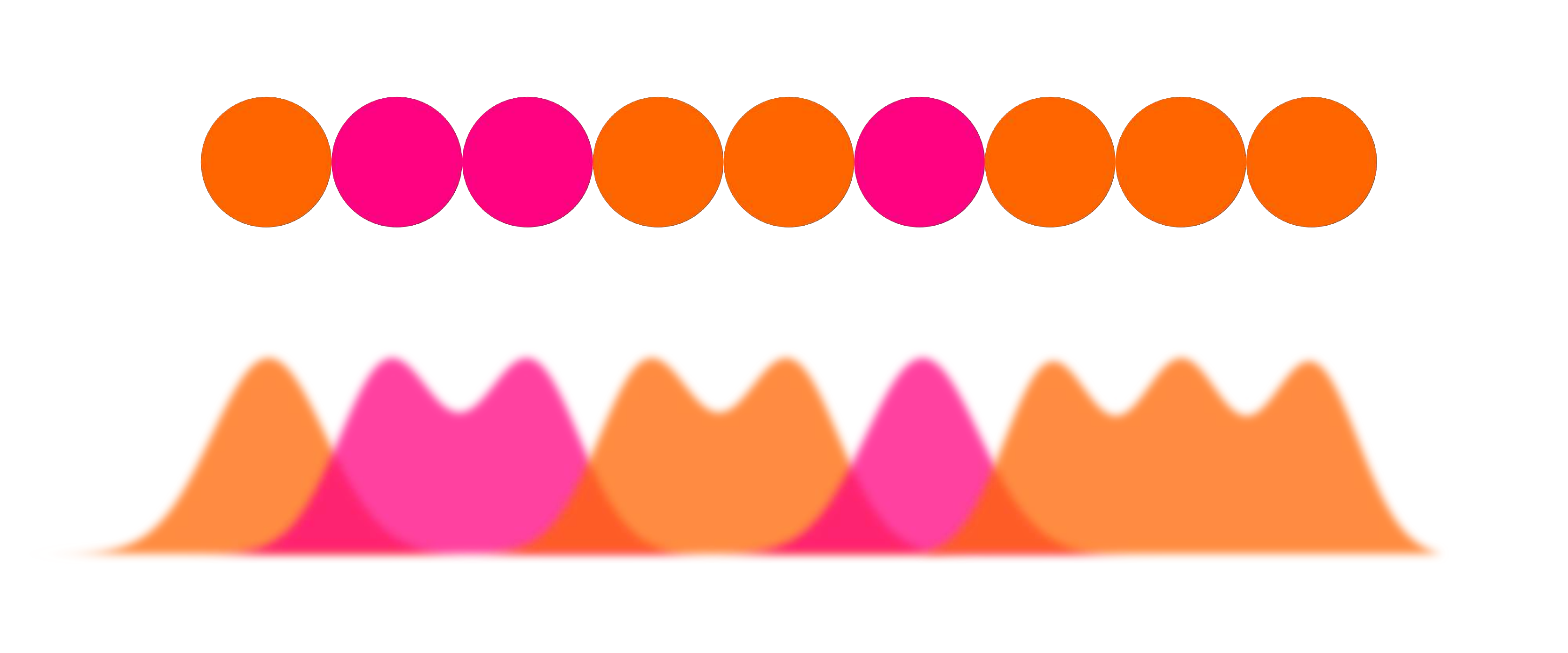}
\caption{A two-component one-dimensional system illustrating particles aligned next to each other and can only exchange position with the neighboring particles by going through each other.}
\label{particles_in_line}
\end{figure}

One of the main purposes of my PhD project behind this thesis has been to investigate some special kind of quantum systems and develop new numerical and mathematical models in order to understand them better. Closely followed by some of the state-of-the-art experiments performed around the world, the purpose has been to understand and exploit the quantum effects of the particles. It has also been the goal to not only understand but also predict the quantum behavior in other unexplored regimes and setups. In order to do this, my method has been built on a very simple geometry; By confining the particles to only move in {\it one dimension} (1D)\index{One Dimension}, just like Fig.~\ref{particles_in_line}, particles can only exchange position with their neighbors by allowing their wave function to interact with and go through each other. This simplifies the complexity in a problem in such a way that one can easily define and study a specific order of the particles and even derive some analytical solutions to a given 1D system. It also paves the way for a variety of applications that I will discuss later.

\section{One-Dimensional (1D) Systems} \label{ch:intro:sec:onedimensionalsystems}
Allow me now to shift gear and get more technical. By now, all physicists know that one-dimensional physics is not something new at all. In fact, it began as early as 1931 when Bethe wrote down his famous and exact solution to the Heisenberg model of ferromagnetism \cite{Bethe1931}. The Bethe ansatz, as it is called today, is widely used to study one-dimensional systems of both bosons and fermions. The same techniques have been used in different many-body models ever since \cite{PhysRev.130.1605, doi:10.1063/1.1704291, mcguire1964, PhysRevLett.19.1312, PhysRevLett.20.1445}. However, the Bethe ansatz \index{Bethe Ansatz} is not applicable for typical microscopic applications, where particles are externally confined to a given region of space. This is especially the case for atomic gases cooled to extremely low temperatures and confined to low dimensions.

Ever since the realization of the Bose-Einstein Condensate\index{Bose-Einstein Condensate} \cite{PhysRevLett.75.3969, PhysRevLett.75.1687}, remarkable developments in the study of cold atomic gases\index{Cold Atomic Gases} confined to low dimensions have been made. Low dimensions are of particular interest due to their simplicity in contrast to higher dimensions. But one has to understand that the realization of one-dimensional setups with cold atoms has only been just recently accomplished and hence a big interest from the theoreticians has also arisen along with the experiments. These state-of-the-art experiments are so accurately controlled, that one can setup a very fine geometry and adjustable interaction between the particles. The tuning has become so precise that one can now study few-body dynamics or build a Fermi sea one atom at a time \cite{SerwaneS2011, zurnPRL2012, zurnPRL2013, WenzS2013, MurmannPRL2015a, MurmannPRL2015b} and therefore investigate the transition between few- to many-body systems \cite{ZinnerEWoC2016, HofmannPRA2016}.

With the realization of one-dimensional (1D) cold systems \cite{blochRMP2008,lewensteinAiP2007,esslingerARoCMP2010,moritzPRL2003,
stoferlePRL2003, KinoshitaS2004, KinoshitaN2006, ParedesN2004, HallerS2009, haller2010pinning, PaganoNP2014}, one is also now able to verify new and old theories for such setups. The Tonks-Girardeau gas\index{Tonks-Girardeau Gas} \cite{TonksPR1936, GirardeauJoMP1960} and super-Tonks-Girardeau Bose gases\index{Tonks-Girardeau Gas!Super-TG Bose Gas} \cite{HallerS2009} are only some of a few well studied and tested theories that have been successfully demonstrated \cite{KinoshitaS2004, ParedesN2004, HallerS2009, GuanRMP2013, VingSR2016}. Apart from confining the particles in 1D setups, one is also able to mix different kinds of particles and species with different masses \cite{MurmannPRL2015a,Spethmann2012}. Same type of particles can additionally be tuned to different internal hyperfine- or spin-states making them different from one another. This introduces two-component systems as well as many-components that can also be investigated. Sophisticated setups like in \cite{PaganoNP2014} have illustrated the ability to do experiments with a tunable number of spin components. Pure two-component spin 1/2 fermionic systems have on the other hand shown to be spin-charge separated \cite{PhysRevLett.90.020401, DeuretzbacherPRA2014}, which is usually due to the Pauli principle\index{Pauli Principle}. Recently, it was also shown that fermions could also go in a transition from a non-magnetic to a magnetic phase \cite{GharashiPRL2013}. On the other hand, two-component bosonic systems can show even richer phenomenon because of different inter- and intra-species interactions. One important question is of course how these pure and mixed systems align themselves when they are balanced or imbalanced and heavier or lighter from one another while being trapped in a harmonic oscillator.

\section{Experimental Techniques\index{Experimental Techniques}} \label{ch:intro:sec:experimentaltechniques}
{\it This section contains some updated parts from my qualifying exam report.}

Quantum mechanical properties of atoms become significant and measurable when atoms are cooled down to very low temperatures. As mentioned briefly in the first section, in 1924 Louis de Broglie\index{Louis de Broglie} postulated that matter has a build-in wave behavior, characterized by the de Broglie wavelength\index{Louis de Broglie!Wavelength} $\lambda=h/p$, where $h$ is Planck's constant and $p$ is the momentum. Classically, the average kinetic energy per particle for an ideal gas is given as $E_{kin,ave}=\frac{3}{2}k_bT$, where $m$ and $k_b$ are the mass of the particle and the Boltzmann's constant, respectively. After a quick and simple calculation using the wavelength and the temperature, one can show that the wavelength of the particles is given as $\lambda={h}/{\sqrt{3k_bmT}}$. This clearly indicates the importance of temperatures in order to enhance quantum mechanical nature in gases and finally the formation of the Bose-Einstein Condensate (BEC)\index{Bose-Einstein Condensate} or ultracold Fermi gases.

Typically, in an experimental setup one has to go down to a few $\mu$K in order to for example form a BEC, but low temperature gases have not always been easy to prepare. This has to be done in such a way that avoids gas to condensate into solid. Therefore typical experimental cooling techniques are done in several steps. First step is usually laser cooling or sometimes called Doppler cooling\index{Doppler Cooling}. The idea is to tune the laser slightly below the absorption-resonance of stationary atoms (red detuning), such that the atoms that move towards the light will absorb more photons, due to the Doppler effect. While, the other atoms that move along the light or are stationary are unaffected by this tuning. By absorbing photons the atoms become excited and slowed down in the direction opposite to the light. However, these are unstable and after a while the atoms emit a photon spontaneously and due to the conservation of momentum they will be kicked with the same amount of momentum, \textit{but} in a random direction. Considered over many absorption-emission cycles, the average speed of the atoms is reduced within a matter of ms \cite{RevModPhys.70.721}. In addition, a magnetic trap is created by adding a spatially varying magnetic quadrupole field to the red detuned optical field. This causes the atoms to be kicked towards the center of the trap and hence forming a cloud. The setup is called Magneto-Optical Trap (MOT)\index{Magneto-Optical Trap}. In order to further cool down the atoms, the next step is to use evaporating cooling\index{Evaporating Cooling}. By using a magnetic field gradient one creates a confining potential such that more energetic atoms will escape the trap, hence removing energy from the system and reducing the temperature. This step removes a lot of particles from the condensate, but there are still many particles left at the desired temperature.

A typical three-dimensional (3D) BEC consists of a few 10.000 atoms. To confine the atoms in 1D one uses a 2D optical lattice. This freezes out the transversal motion, and in return forms an array of vertically oriented elongated tubes. Within each tube there are typically 8-25 atoms \cite{hallerPRL2010}. The one-dimensional interaction strength, $g_{\text{1D}}$, is related and controlled by the magnetic tuning of the three-dimensional scattering length, $a_{\text{3D}}$. The tuning is done by Feshbach Resonances (FRs)\index{Feshbach Resonances}. A FR occurs when the energy of two scattering atoms comes into resonance with a bound molecular state. However, it is shown that in 1D and 2D setups a new type of scattering resonance, the so-called confinement-induced resonance (CIR)\index{Confinement-Induced Resonance}, occurs \cite{OlshaniiPRL1998}. CIR arises when the $a_{\text{3D}}$ approaches the transversal confinement, i.e. the harmonic oscillator length $a_\perp=\sqrt{\hbar/(m\omega_\perp)}$, where $m$ is the mass of the particle, $\omega_\perp$ is the transversal trapping frequency and $\hbar$ is the reduced Planck constant or Dirac constant. More specifically, the 1D coupling parameter is defined as:
\begin{equation}
g_{\text{1D}}=-\frac{2\hbar^2}{m a_{\text{1D}}}=\frac{2\hbar^2a_\text{3D}}{m a_{\perp}^2}\frac{1}{1-Ca_{\text{3D}}/a_{\perp}},
\label{g1D}
\end{equation}
where $a_{\text{1D}}$ is the 1D scattering length defined by the above equation and $C=-\zeta(1/2)/\sqrt{2}\approx1.0326$ is a constant \cite{HallerS2009,hallerPRL2010,PhysRevLett.91.163201}. As shown in Eq.~(\ref{g1D}) the CIR plays a crucial role in controlling of interactions in low-dimensional system, since it is possible to tune $g_{\text{1D}}$ from being strongly repulsive to strongly attractive by just varying $a_{\perp}$. The modification of scattering length in 1D has for example been measured for fermions \cite{PhysRevLett.91.163201}. One of the recent experiments motivating this theoretical work is the experiment done by Selim Joachim's group in Heidelberg \cite{WenzS2013}. Although the preparation and trapping methods used by the group are slightly different than the typical setup as described here, the final interested one-dimensional system is the same. Here, the group has been able to not only access and prepare the system experimentally, but also fully control the size of the setup at the single-particle level. Particularly, they are able to prepare an experiment deterministically with a few-particle system of ultracold $^6$Li atoms.

\section{Exotic Physics}\label{ch:intro:sec:exoticphysics}
Even though one-dimensional geometry is simpler than the higher dimensions, there are a lot of interesting and exotic\index{Exotic Physics} phenomena that are very interesting to investigate and pay attention to. One major difference is that the 1D nature prohibits any exchange of particles without the particles going through each other and therefore necessarily must interact with each other. 1D geometry also allows one to build a chain of atoms and align the atoms in a specific order. One can then try to change the order or move one atom from one place to another by just changing the environment \cite{zurnPRL2012,ZuernPRL2013}. Tuning the interactions between different components through the Feshbach Resonances \cite{WenzS2013, chinRMP2010, OlshaniiPRL1998} is another option that one can play with in order to stress the system to be adjusted in different orderings.

This particular research can help future studies to understand the properties of technologically relevant materials that show magnetic and superconductive effects. In the future one might even be able to start building and design materials from scratch one atom at a time. It can also be used to understand a variety of technological applications in the low-dimensional structures such as quantum wires and carbon nanotubes\index{Carbon Nanotubes} \cite{Altomare2013}, and finally some potential applications in quantum information \cite{Endres1024}.

\section{One and Two Particles in a 1D Harmonic Trap}\label{ch:intro:sec:oneandtwo}
\begin{figure}[h]
\centering
\includegraphics[width=0.9\linewidth, trim=1.5cm 0.5cm 0.9cm 0cm]{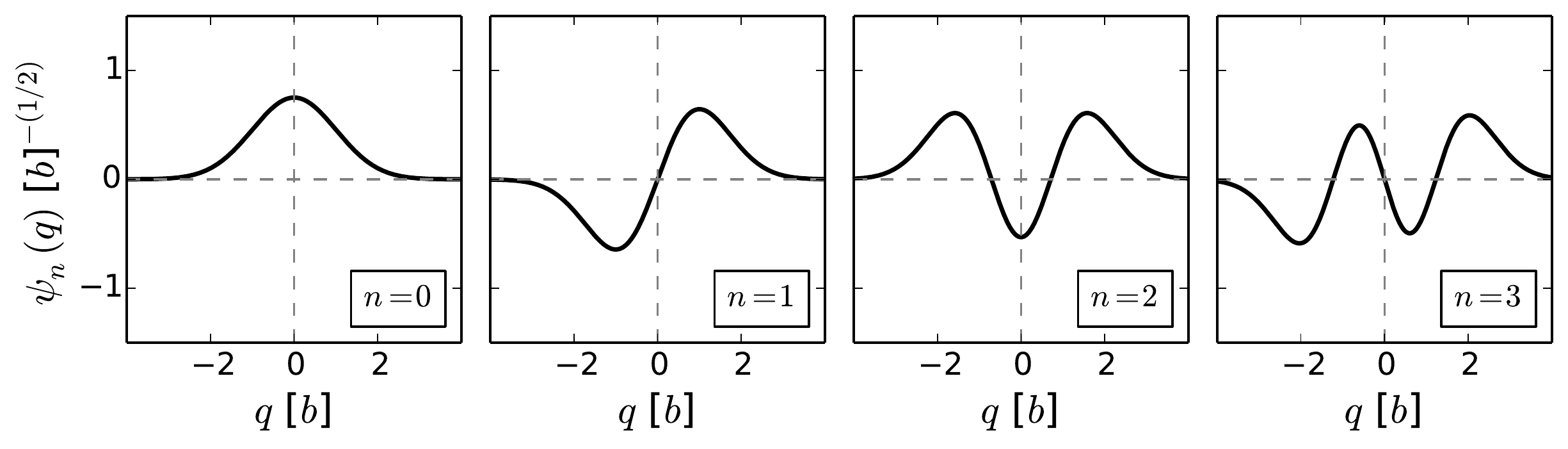}
\caption{The four lowest eigenstates for a single particle trapped in a one-dimensional harmonic oscillator.}
\label{ho_wave}
\end{figure}

\noindent
Apart from the Bethe ansatz and classical one-dimensional topics, the physicists are also familiar with one-dimensional quantum physics through many books such as \cite{griffiths2016introduction,plischke2006equilibrium,baxter2016exactly, lieb2013mathematical}. The one-dimensional harmonic oscillator\index{One Dimensional Harmonic Oscillator} is only one of the many examples that one can mention in this context. The one-dimensional harmonic oscillator Hamiltonian, $\mathcal{H}$, is written as,
\begin{equation}
\mathcal{H}\psi=\left(-\frac{\hbar^2}{2m}\frac{d^2}{dq^2}+\frac{1}{2}m\omega^2q^2\right)\psi=E\psi,
\label{HO1}
\end{equation} 
where $\hbar$ is the Dirac constant and $m$ is the mass of a particle trapped in a 1D harmonic trap with trap frequency $\omega$ and coordinate $q$. The eigenfunction with the corresponding eigenenergy are denoted as $\psi$ and $E$, respectively. The analytical solution to this kind of problem can be handled both algebraically and analytically as shown in most books for instance in \cite{griffiths2016introduction}. The solutions are given as,
\begin{equation}
\psi_n(q) = \frac{1}{\sqrt{2^n\,n!}} \cdot \left(\frac{m\omega}{\pi \hbar}\right)^{1/4} \cdot e^{
- \frac{m\omega q^2}{2 \hbar}} \cdot H_n\left(\sqrt{\frac{m\omega}{\hbar}} q \right), \qquad n = 0,1,2,\ldots.
\label{HO1_eigenfunction}
\end{equation} 
with corresponding energies:
\begin{equation}
E_n = \hbar \omega \left(n + {1\over 2}\right),
\label{HO1_eigenenergy}
\end{equation} 
where $H_n(q)$ is the Hermite polynomial\index{Hermite Plynomials} and $n$ is the quantum number. The four lowest states are plotted in Fig.~\ref{ho_wave}. When it comes to notations, it is very convenient to work in the {\bf harmonic oscillator units}\index{Harmonic Oscillator Units}, where energies are in units of $[\hbar\omega]$ and lengths are in units of $[b]\equiv[\sqrt{\frac{\hbar}{m\omega}}]$. These units will be used in most of the thesis.\\

\begin{figure}[t]
\centering
\includegraphics[width=\linewidth, trim=0cm 0.5cm 0cm 0cm]{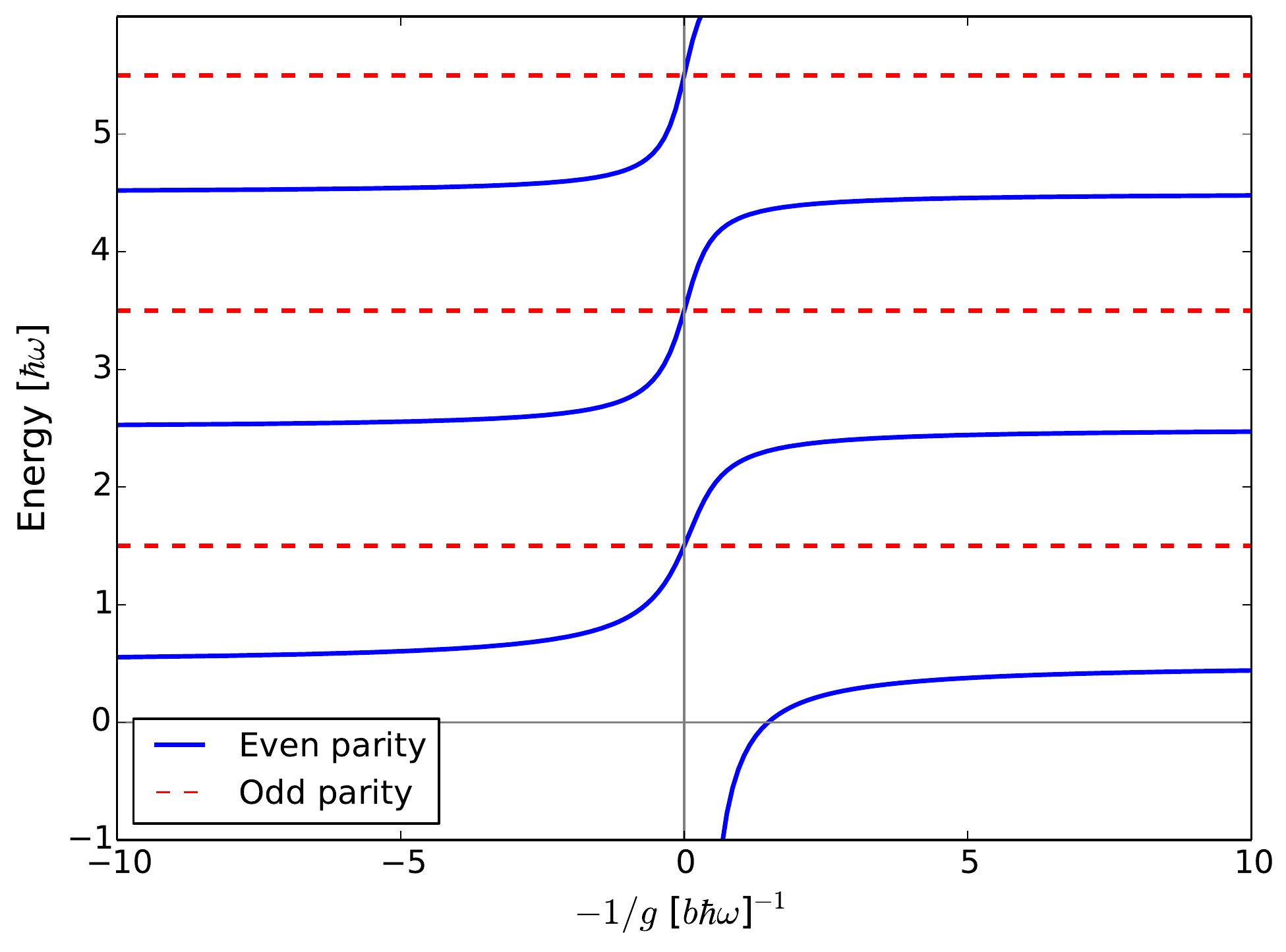}
\caption{The energy spectrum in the two particle system in a one-dimensional harmonic oscillator. The odd states are totally unaffected, while the even states vary in energy as $g$ changes.}
\label{busch}
\end{figure}

\begin{figure}[t]
\centering
\includegraphics[width=0.9\linewidth, trim=1.5cm 0.5cm 0.9cm 0cm]{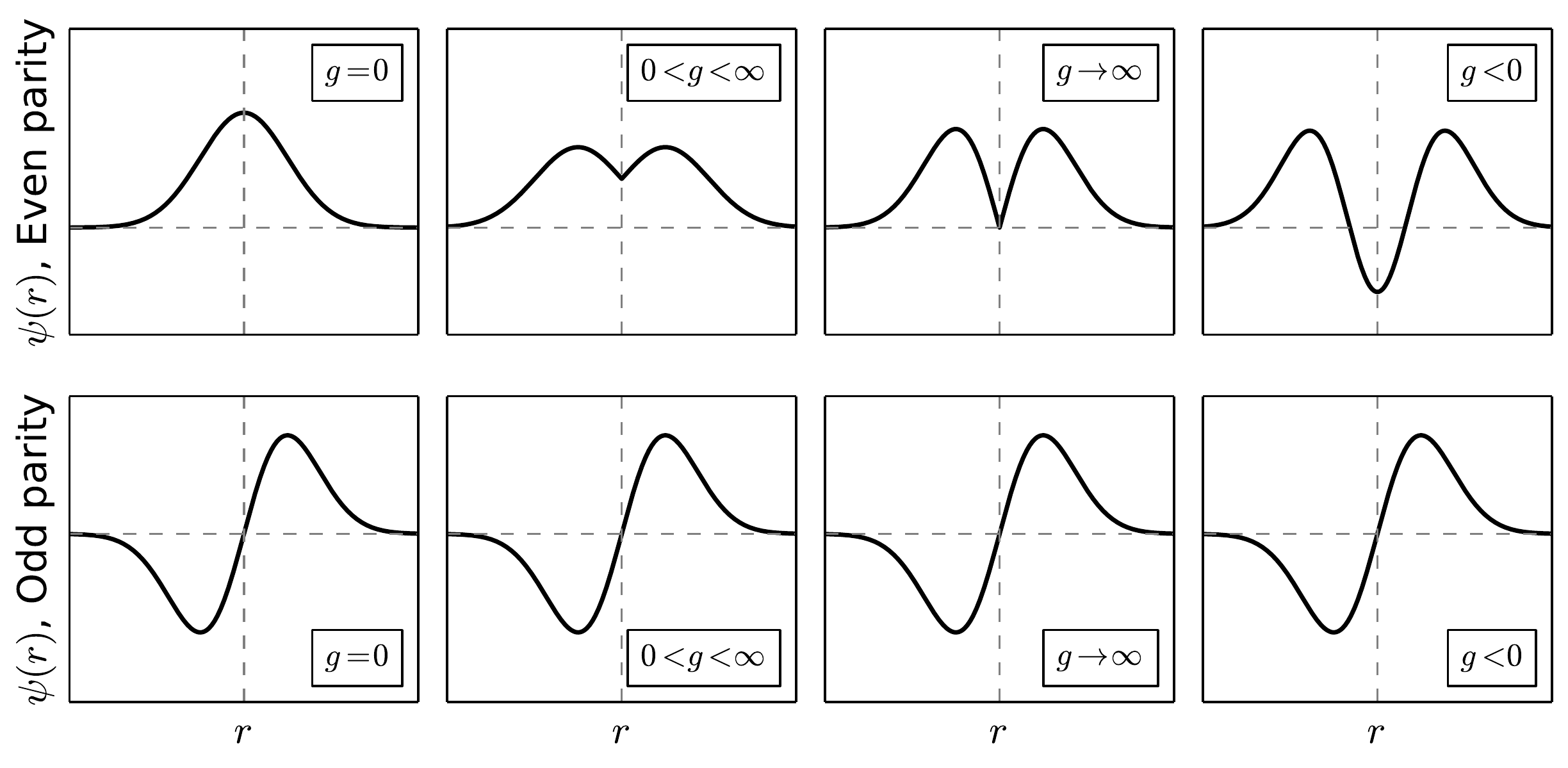}
\caption{The two lowest eigenstates in the two particle system for the repulsive interactions in a one-dimensional harmonic oscillator. Lower panels: Odd states, as the energy also indicates, are unaffected due to the relative wave function naturally being zero at $r=0$. Upper panels starting from left to right: the even state starts from a nice Gaussian wave function to starting to get a cusp to finally become fermionized. For attractive interactions, $g<0$, it looks more and more like the second excited harmonic state, $n=2$, as in Fig~\ref{ho_wave}.}
\label{busch_wave}
\end{figure}

The 1D harmonic oscillator trap\index{One-Dimensional Harmonic Oscillator} becomes quickly more interesting when one adds another distinguishable particle to the Hamiltonian and introduces an interaction in form of a short-range interaction between the two particles. This type of systems are usually called $N+M$ systems, indicating $N$ particle of one type and $M$ particles of another type. Obviously, in this case I am talking about the $1+1$\index{1+1 Systems} system, which is the most trivial example for interacting particles in a one-dimensional harmonic oscillator with contact interactions. The Hamiltonian for this system is given as (in units of harmonic oscillator),
\begin{equation}
\left(-\frac{d^2}{dq_1^2}-\frac{d^2}{dq_2^2}+\frac{1}{2}q_1^2+\frac{1}{2}q_2^2+g'\delta(q_1-q_2)\right)\psi=E\psi,
\label{HO2}
\end{equation} 
where $\delta$ is the Dirac $\delta$-function and $g'$ is the interaction strength in units of $[b\hbar\omega]$. This problem can easily be separated into a center-of-mass coordinate, $R=(q_1+q_2)/\sqrt{2}$, and a relative coordinate, $r=(q_1-q_2)/\sqrt{2}$ as,
\begin{equation}
\left(-\frac{d^2}{dR^2}+\frac{1}{2}R^2\right)\psi+\left(-\frac{d^2}{dr^2}+\frac{1}{2}r^2+g \delta(r)\right)\psi=E\psi,
\label{HO2_relative}
\end{equation} 
where $g\equiv g'/\sqrt{2}$. The solution to the center-of-mass part is easily obtained from our knowledge about the one particle in a harmonic oscillator, Eq.~(\ref{HO1}). The relative part of the Hamiltonian, on the other hand, can be obtained by expanding the relative wave function into a complete set of harmonic wave functions, $\psi_n(x)$, as in Eq.~(\ref{HO1_eigenfunction}). After some calculations one can show that in one dimension the relation between the energy, $E$, and the interaction strength, $g$, is given as the transcendental equation\index{Transcendental Equation} below \cite{busch1998}:
\begin{equation}
\frac{\Gamma(-E/2+1/4)}{\Gamma(-E/2+3/4)}=-\frac{2}{g},
\label{HO2_energy}
\end{equation} 
which is illustrated in Fig.~\ref{busch}. Notice that the odd parity states are not affected since the interaction is zero range. This argument is best illustrated as in Fig.~\ref{busch_wave}, where I have plotted the two lowest states from the repulsive ($g>1$) side for different values of $g$. As shown in the figure, the odd relative wave function is totally unaffected by the changing value of $g$, since it is naturally zero for $r=0$ and therefore the two particles cannot feel each other and interact. On the other hand, the even state undergoes a transformation as shown in the upper panels in Fig.~\ref{busch_wave}. For $g\rightarrow\infty$ the state becomes a {\it fermionized}\index{Fermionized} state, which is basically the absolute value of the odd state in the lower panel. As one starts to add more and more particles to the system, the dynamics become more complex and therefore more interesting. When $N,M>1$, apart from the combinatorics and how the particles start to align themselves in a 1D setup, important features like the quantum statistics has to be taken into account as interesting physics start to arise.

\section{Outline}\label{ch:intro:sec:outline}
\noindent
Despite big investigations and investments in 1D systems, there are still many questions and topics that have not been explored yet. One of the questions is, can we develop any analytical and numerical models to describe few-body system in full details and use this knowledge to say something about the many-body dynamics? What are the correlations of ground state and excited states in 1D systems with balanced or imbalanced number of fermions, bosons or a mixture? What are the favorite orderings of these systems in the strongly interacting regimes and can we manipulate this in such a way to obtain another desired ordering? Are there any universal properties in the dynamics of few- and many-body systems? Can we understand integrable and ergodic solutions for the few-body systems and last but not least, can we apply our knowledge about 1D system to simulate lattice quantum gauge?\\

\noindent
During this dissertation, I will answer the above questions based on my published and in-process work \cite{DehkharghaniSR2015, LoftTEPJD2015, DehkharghaniPRA2015, DehkharghaniJPB2016, Garcis-MarchJoPBAMaOP2016, AndersenSR2016, BarfknechtJPB2016, BellottiPJD2017, 2017arXiv170301836D, 2017arXiv170308720P, 2017arXiv170400664D, 2017arXiv170401433H} that have been conducted during my PhD program starting from August 2013 to July 2017.\\

\begin{figure}[t]
\centering
\includegraphics[width=\linewidth, trim=0cm 0cm 0cm 0cm]{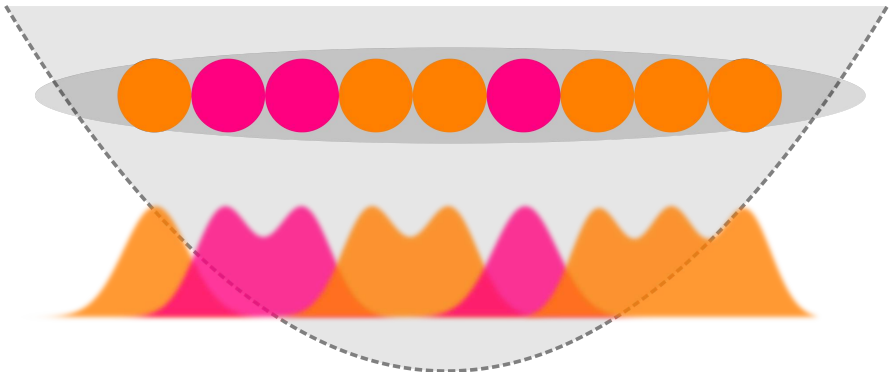}
\caption{Sketch of a two-component $3+6$ system trapped in a one-dimensional geometry, where particles can only exchange position with the neighboring particles by going through each other.}
\label{particles_in_line_trapped}
\end{figure}

\noindent
If I had to summarize the whole thesis by only one equation, then I can surely say that the main equation of interest is the following equation,
\begin{equation}
\mathcal{H}=\sum_i^{N}\big(-\frac{\hbar^2}{2m}\frac{\partial^2}{\partial q_i^2} + V(q_i)\big)+\sum_{i<j}^N g_{ij}\delta(q_i-q_j),
\label{equation_of_interest}
\end{equation}
where $q_i$ is the coordinate of the $i$'th particle, $V$ is the potential that confines the particles in a 1D geometry (for most part of the dissertation it is a 1D harmonic oscillator potential, but in Chapter 6, I will also present a generalized tilted double harmonic oscillators). The particles are assumed to interact with each other as contact interaction, which is only valid in dilute gases\index{Dilute Gases}. The $g_{ij}$ is the interaction strength between the $i$'th and $j$'th particle. The $\delta$ function is the usual Dirac function\index{Dirac Function}. Particles confined in a harmonic trap, while aligned in a one-dimensional setup, can be illustrated as in Fig.~\ref{particles_in_line_trapped}. It is worth noting that if there were {\it no interaction} between the particles, one would obtain the single particle solutions as in Eq.~(\ref{HO1_eigenfunction}).

The project has been both analytically and numerically, where new methods have been developed in order to solve the above equation for different kind of parameters and interaction regimes. In addition, there will also be some on-going work and never published results in the dissertation, which I will comment on and present as future research. The results, whether published before or brand new will be clearly stated. \\

\noindent
The dissertation is built as follows:\\

In {\bf Chapter 2} I will start by adding one more particle to the $1+1$ system and therefore start to investigate the $1+2$ fermionic or bosonic systems. Here, I will present the analytical results for this kind of systems in the strongly interacting regime. I will discuss the mass-imbalanced case and how to solve this system analytically in the same regime. I will also mention another method, which is used to solve this kind of systems. Finally, I will introduce a variational ansatz, which was developed by one of my co-authors, to say something about the intermediate interacting regime for the $1+2$ systems. All the results will be compared with experimental and numerical exact results that are available today. I have developed a few numerical methods of my own in order to solve these kind of systems and they will be explained in Chapter 7.\\

{\bf Chapter 3} will continue in the same manner as Chapter 2, where I will add another particle and form the $2+2$ or the $1+3$ system. This system can be {\it semi}-analytically solved by using similar ideas used for the $1+2$ systems for strongly interacting particles. In addition, I will illustrate the mass-imbalanced cases. The semi-analytical method here can be generalized to solve the $N+M$ systems. Different fermionic and bosonic systems as well as a mixture of fermions and bosons will also be discussed. In addition, the semi-analytical results will be compared with the developed numerical methods in the same mass case, and I will dig into the integrability and chaotic behavior of the energy spectrum for the four particles $1+1+1+1$ system with different kind of species and masses.\\

{\bf Chapter 4} is the chapter, where I will mostly discuss the $N+M$ systems and the relatively fast numerical method, which I have contributed to develop and solve the bosonic systems. Moreover, the physics in few- and many-body systems will be explored here. Particularly, I will discuss the formation of ferro- and antiferromagnetic states. In addition, the intrinsic properties and how one can use them in quantum technologies will be mentioned in this chapter.\\

{\bf Chapter 5} is dedicated to a very interesting subject, namely quantum impurities. Here, I will present the $1+N$ bosonic systems and how I have managed to capture the physics in these kind of systems by using a very effective method. It turns out that the method gets even better as $N$ increases, but I will mostly present the results for the $1+8$ systems. Furthermore, I will study the case with arbitrary inter-species and small intra-species interaction strengths, where I use the Gross-Pitaevskii equation to simulate the intra-species interactions. Finally, I will show the on-going results for the $2+N$ system and how this can be split into a three-body problem.\\

In {\bf Chapter 6} I will present a recently developed setup that can be used to simulate lattice quantum gauge. Here, I will start by presenting the analytical results for the generalized tilted double harmonic oscillator. Afterwards, I will describe the dynamics where a boson in the tilted double well interacts with an ion or a fermion in the middle of the potential and how this interaction allows the boson to jump from one side to another side. \\

{\bf Chapter 7} is about the different numerical methods that I have developed and used. Here I will first present the Effective Exact Diagonalizing Method (EEDM), which is an effective diagonalizing method for 1D $N+M$ systems with same mass and trapping frequency. Second, the Correlated Gaussian Method (CGM), which is based on the Gaussian functions and variationally converged to the true results. And finally, the Density Matrix Renormalization Group (DMRG), which is one of the most used methods in discrete quantum physics.\\

Finally, in {\bf Chapter 8} I will summarize some of the important results that were presented in each previous chapter. However, this chapter will be very short and not fully covered as the main derivations and points can be found in each chapter. Furthermore, I will give a personal remark on what one can expect from the one-dimensional quantum gases in the future.

\newpage % Create empty back of side
\thispagestyle{empty}

%-----% 1+2 SYSTEMS
% !TeX root = ../Main_publish.tex

\chapter{Three Particles in a 1D Harmonic Trap}\label{ch:threeparticles}
\epigraph{\it “One of the great things about science is that it is an entire exercise in finding what is true. ... When you have an established scientific emergent truth it is true whether or not you believe in it...”}{\rm ---Neil deGrasse Tyson}

After briefly discussing the $1+1$ system in the introductory part the question is what happens when you add one more particle to the setup? {\it This} is the topic of this chapter, the so-called $1+2$ systems, which are the most trivial two-component few-body systems. Although, the three strongly interacting and harmonically trapped particles have been briefly examined before in \cite{harshmanPRA2012}, the different mass-imbalanced and intermediate regimes have remained untouched. Here, I will go in more details with the $1+2$ systems \index{1+2 Systems} and investigate the different properties of the setup. For the $2$ identical particles in one component, which are identical but different from the third particle in the other component, one has to take the quantum statistics into account. If one has fermions, then one would consider spinless (spin-polarized) fermions with spin ($\uparrow$) and ($\downarrow$). This notation is used throughout the thesis to denote the fermionic particles, which obey the Pauli principle, while for bosons, the ($A$)- and ($B$)-type notation is used, which are symmetric under any exchange of their position. In experiments these represent different hyperfine-states of an atom.

In the following sections, I will start by considering an equal-mass case where both types of particles have the same mass, $m$. Under the assumption that the interaction strength between the particles from different components is infinite, I will show that one can construct the fermionic or bosonic wave function in this limit. Afterwards, I will generalize the method to the mass-imbalanced case and present the corresponding results. After constructing the wave function in the strong regime, I will present a variational method that constructs a very good approximate solution in the intermediate regime.

\section{Strongly Interacting Three-Particle Systems}\label{ch:threeparticles:sec:strongthree}
{\it This section contains some updated parts from my qualifying exam report.}

The total Hamiltonian, $\mathcal{H}$, of the three-particle system trapped in a 1D harmonic trap can be written as (in harmonic oscillator units\index{Harmonic Oscillator Units} $b=\sqrt{\frac{\hbar}{m\omega}}$):
\begin{align}
\mathcal{H}=\sum_{i=1}^{N=3}\big(-\frac{1}{2}\frac{\partial^2}{\partial q_i^2} + \frac{1}{2} q_i^2\big)+\sum_{i=1,i<j}^{N=3} g_{ij}\delta(q_i-q_j),
\label{ch1:eq:hamiltonianfor1+2}
\end{align}
where $q_i$ is the scaled and unit-less coordinate of the particle $i$. The $g_{ij}$ is the interaction strength (in units of $b\hbar\omega$) between the two particles situated at $q_i$ and $q_j$ for $i,j\in \{1,2,3\}$. $\omega$ is the angular frequency of the one-dimensional harmonic oscillator trap and finally, $m$ is the mass of each particle, which in this section is assumed to be the same. I will refer to the last sum in Eq.~(\ref{ch1:eq:hamiltonianfor1+2}) as the interaction-term and the first two terms in the first sum as the (non-interacting) harmonic oscillator term.

It is possible to separate the center-of-mass motion\index{Center-of-mass Motion}, $z$, from the relative motions\index{Relative Motion}, $x$ and $y$, of the particles by transforming the coordinates $\mathbf{q}=[q_1,q_2,q_3]^T$ into a standard normalized Jacobi coordinates\index{Jacobi Coordinates}, $\mathbf{r}=[x,y,z]^T$. This is done through the transformation $\mathbf{q} \rightarrow \mathbf{r}=\mathbf{J}\mathbf{q}$, where $\mathbf{J}\in SO(3)$ is given by:
\begin{align}
\begin{bmatrix} x \\ y \\ z\end{bmatrix}=\mathbf{r}=\mathbf{J}\mathbf{q}=\begin{bmatrix} \frac{1}{\sqrt{2}} & -\frac{1}{\sqrt{2}} & 0 \\ \frac{1}{\sqrt{6}} & \frac{1}{\sqrt{6}} & -\frac{\sqrt{2}}{\sqrt{3}} \\ \frac{1}{\sqrt{3}} & \frac{1}{\sqrt{3}} & \frac{1}{\sqrt{3}}\end{bmatrix}\begin{bmatrix} q_1 \\ q_2 \\ q_3\end{bmatrix}.
\label{ch1:eq:jacobicoordinates}
\end{align}
Since $\mathbf{J}\in SO(3)$ the transformation is just a rotation matrix with the property that $\mathbf{q}^2=\mathbf{r}^2$.\index{Jacobi Coordinates!Matrix} While the non-interacting part of the Hamiltonian is transformed into $-\frac{1}{2}(\frac{\partial^2}{\partial x^2}+\frac{\partial^2}{\partial y^2}+\frac{\partial^2}{\partial z^2}) + \frac{1}{2} (x^2+y^2+z^2)$, the interacting part of the Hamiltonian can now be written in terms of the new variables as follows:
\begin{align*}
\sum_{i<j}^N g_{ij}\delta(q_i-q_j)&=\frac{1}{\sqrt{2}}\Big[g_{12}\delta\big(x\big)+g_{23}\delta\big(-\frac{1}{2}x+\frac{\sqrt{3}}{2}y\big)+g_{31}\delta\big(-\frac{1}{2}x-\frac{\sqrt{3}}{2}y\big)\Big].
\end{align*}
Notice that the delta-functions now only depend on the $x$ and $y$ coordinates, and therefore the $z$-coordinate is easily separated and solved by our knowledge of the single particle solutions in the Harmonic oscillator, Eq.~(\ref{HO1}). Denoting the energy eigenbasis for the Jacobi coordinate system with $\left|\nu,\mu,\eta\right\rangle$, the basis gets separated as $\left|\nu,\mu\right\rangle \otimes \left|\eta\right\rangle$ with the separated solutions given by Eq.~(\ref{HO1_eigenfunction}). The corresponding energies are given as $E=(\eta+1/2)$ (in units of $\hbar\omega$) with $\eta\in\{0,1,2\dots\}$. Since the ground state is the primary goal to find in this thesis, $\eta$ is set to zero. What remains are the relative coordinates.

\begin{figure}[t]
\centering
\includegraphics[width=\textwidth]{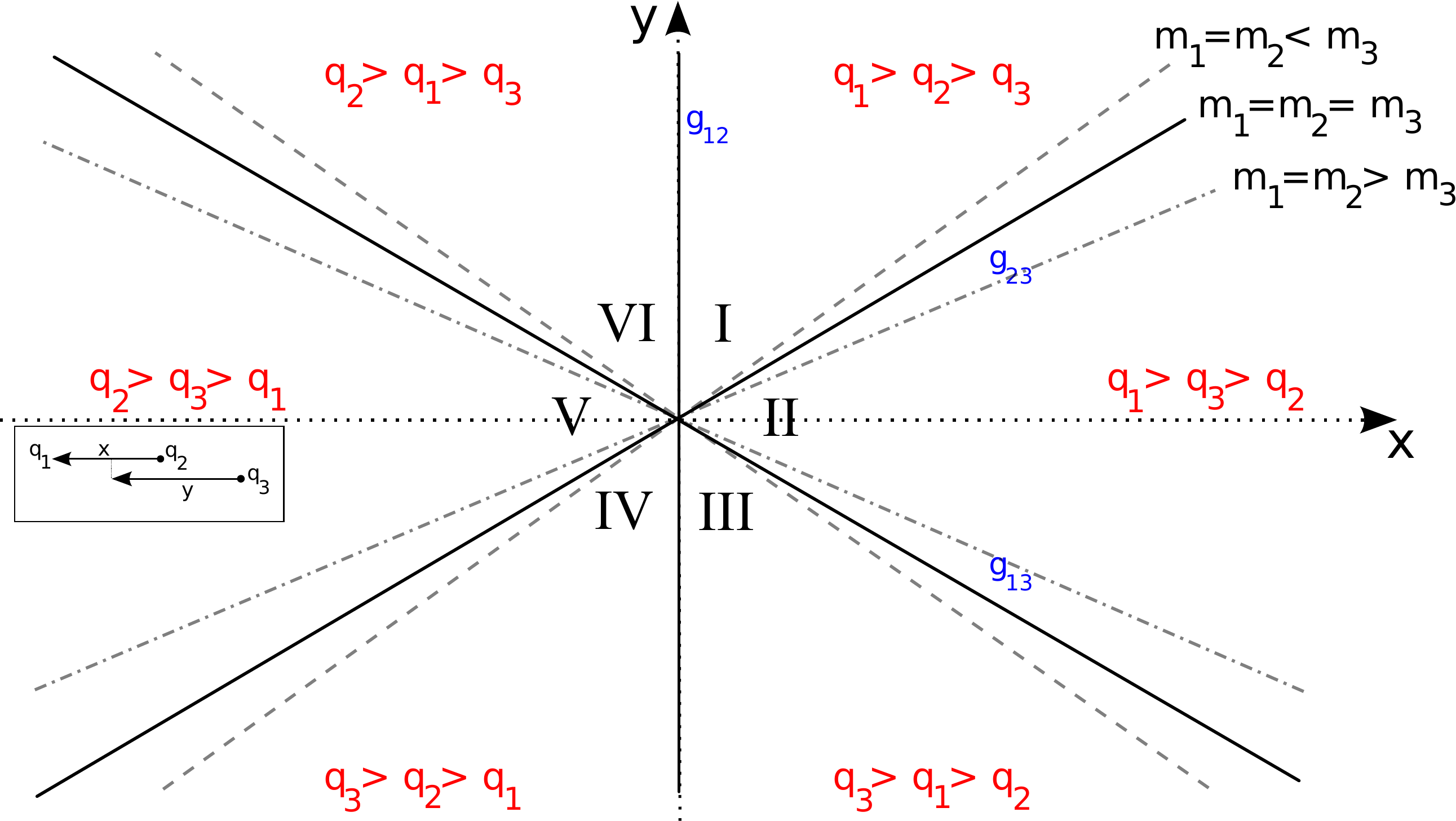}
\caption{Interaction space for the $2+1$ particle system. There are $3!=6$ ways to order the particles and hence 6 regions. The inset box shows how $x$ and $y$ are defined. Solid lines represent the argument of the delta function. The dashed and dot-dashed lines represent the boundary conditions for unequal mass cases.}
\label{3particleposition}
\end{figure}

Even though the delta-functions are not totally separable, they come with some boundary conditions in the $x,y$-space. For example when the particles all have the same mass, then the particles meet at $x=0$, $y=\frac{1}{\sqrt{3}}x$ or $y=-\frac{1}{\sqrt{3}}x$. The boundary lines are illustrated in Fig.~\ref{3particleposition}. In the same figure one can see the position of the particles situated relative to each other. For the relative motion, which is a plane perpendicular to the $z$-coordinate, one defines additionally the hyperspherical coordinates\index{Hyperspherical Coordinates}, $\rho=\sqrt{x^2+y^2}$ and $tan(\phi)={y}/{x}$. This changes the interacting part into $\sum_{i<j}^N g_{ij}\delta(q_i-q_j)=\frac{g}{b\rho\sqrt{2}}\sum_{j=1}^{6}\delta[\phi-{(2j-1)\pi}/{6}]$ assuming $g_{ij}=g$ for $\forall i,j$. The solutions for the relative motion are then given by \cite{harshmanPRA2012}:
\begin{equation}
\psi_{\nu,\mu}(\rho,\phi) = \sqrt{\frac{2~\nu!}{(\nu+\mu)!}}~L_\nu^\mu(\rho^2)~e^{-\rho^2/2} \ \rho^\mu \ f_\mu(\phi),
\end{equation}
with $ \nu,\mu= 0,1,2,\ldots$ and the functions, $L_\nu^\mu(\rho^2)$, are the Laguerre polynomials\index{Laguerre polynomials}. The function $f_\mu(\phi)$ is some function dependent on the quantum number $\mu$ and coordinate $\phi$, which has to be determined later. The corresponding energy spectrum is given as $E=2\nu +\mu+1$ (in units of $\hbar\omega$).

Noting that the Hamiltonian is invariant under the exchange of $q_1 \leftrightarrow q_2$ or $x\leftrightarrow-x$, means that $\psi(x,y)\rightarrow\psi(-x,y)=\pm \psi(x,y)$ (+) for bosons \index{Bosons} (symmetric) and (-) for fermions \index{Fermions}(antisymmetric due to Pauli principle\index{Pauli Principle}). This also means that $f_\mu(\pm\pi/2)$ must be some constant for bosons and absolutely zero for fermions. Similarly, one can use the parity \index{Parity} operator, which takes $q_i \rightarrow -q_i$, to deduce that $\psi(x,y)\rightarrow\psi(-x,-y)=\pm\psi(x,y)$, where ($+$) is for even parity and ($-$) for odd parity in the $x,y$-plane. In this way, one is able to classify the different solutions that will appear. Using the just mentioned properties along with the continuity and delta-function conditions for the wave functions at the boundary lines one can then get a set of solutions for $f_\mu(\phi)$. More specifically, the delta-function boundary requires that the difference in derivatives of the wave function, for a given $\rho$ (say $\rho_0$), from both side of the contact point (say $\phi=\pi/6$) must be equal to the value of the wave function at that point:
\begin{equation}
\lim_{\varepsilon\rightarrow 0} \big(\left.\partial \psi \over \partial \phi \right|_{+}-\left.\partial \psi \over \partial \phi\right|_{-}\big) = G\psi_{\nu,\mu}(\rho_0,\frac{\pi}{6}),
\label{griffith_criteria}
\end{equation}
where all the constants are redefined as one; $G\equiv{\sqrt{2}g\rho_0}$. However, this means that the interaction strength, $G$, is $\rho$-dependent and hence the model can only be valid for $g=0$ and $g\rightarrow\infty$. Therefore one would need another method to solve the intermediate region, where $0<g<\infty$. I will come back to this later in this chapter.

\subsection{Fermions and Bosons}\label{ch:threeparticles:subsec:fermionsandbosons}
{\it This section contains some updated parts from my qualifying exam report.}

\begin{figure}
\centering
\includegraphics[width=\textwidth]{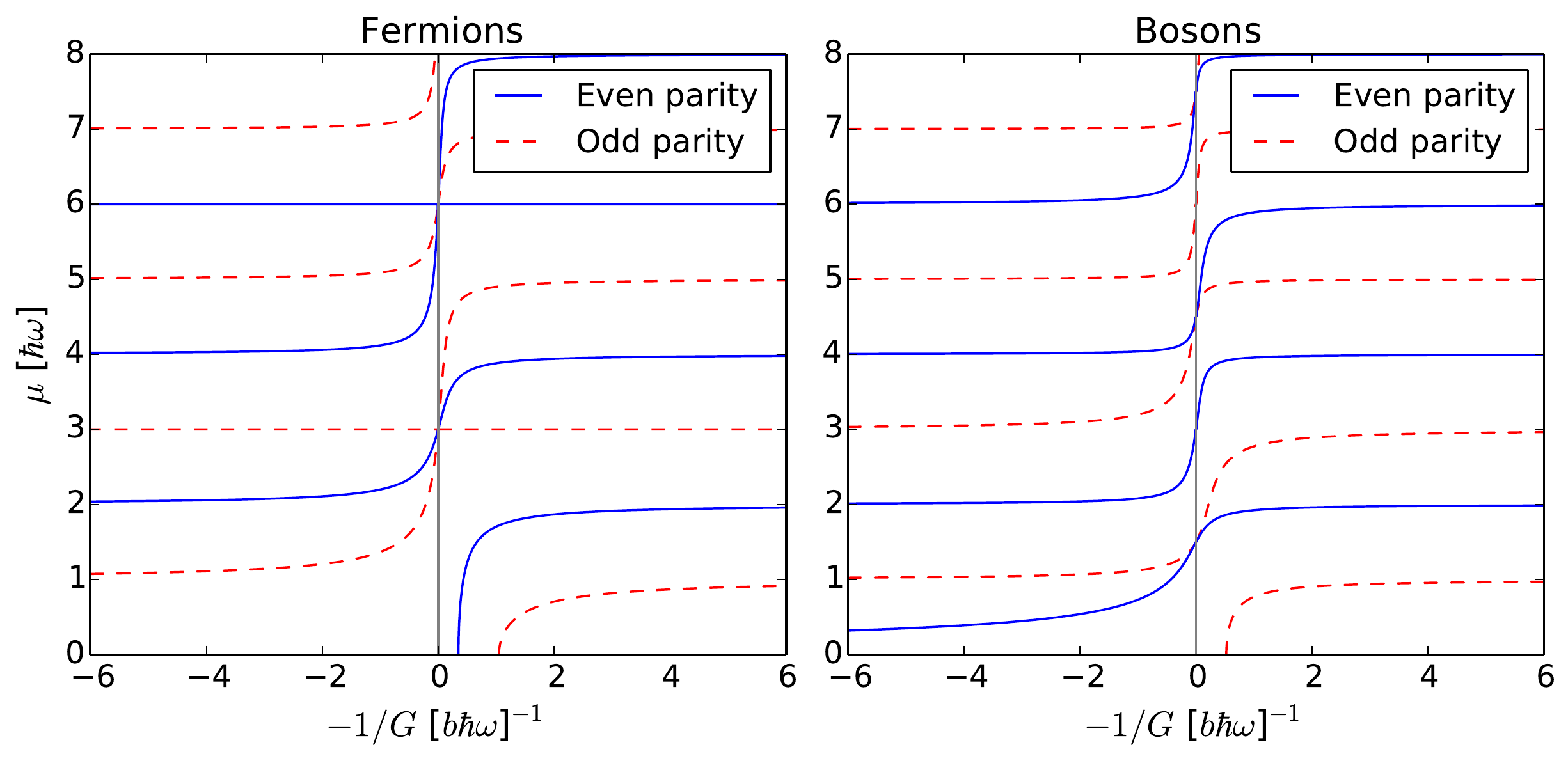}
\caption{Energy-spectrum for the $2+1$ system, where inter-species interaction strength is denoted G, while the intra-species interaction is set to zero. Left panel: Energy-spectrum for fermions where the value of $\mu$ is exact for $G=0$ and $G\rightarrow\pm\infty$. Right panel: shows the same plot for bosons. Notice, here the values at $G\rightarrow\pm\infty$ can be non-integer. Interactions are repulsive if $G>0$ and attractive otherwise.}
\label{energyspectrum1+2}
\end{figure}

In order to find the corresponding $\mu$ quantum number one needs to make a general ansatz of the $\phi$-coordinate in form of $f_\mu(\phi)=A\cdot cos(\mu\phi)+B\cdot sin(\mu\phi)$ in every six regions of Fig.~\ref{3particleposition}, with different constants $A$ and $B$. The different constants can then be found by applying the conditions mentioned before (parity, symmetry, continuity and delta-function). In the following, I will present the results for the 2+1 systems, where the interaction between the intra-species particles are set to zero, while the interaction between the inter-species particles is strong. This produces four equations (fermionic or bosonic with even or odd parity) from which $\mu$ can be obtained:
\begin{align}
0=&\mu\cdot cos(\mu\pi/2) + G \cdot sin(\mu \pi/3)cos(\mu \pi/6) && \text{for odd fermions,}
\label{eq:muequationssamemass1}\\
0=&\mu\cdot sin(\mu\pi/2) +G \cdot sin(\mu \pi/3)sin(\mu \pi/6) && \text{for even fermions,}
\label{eq:muequationssamemass2}\\
0=&\mu\cdot cos(\mu\pi/2)+G \cdot cos(\mu\pi/3)sin(\mu\pi/6) && \text{for odd bosons,}\\
0=&\mu\cdot sin(\mu\pi/2)-G \cdot cos(\mu\pi/3)cos(\mu\pi/6) && \text{for even bosons.}
\label{1+2_equations}
\end{align}
It is clear from Fig.~\ref{energyspectrum1+2} that whenever $G=0$ we have the solutions $\mu=2,4,6,\dots$ for even parity and $\mu=1,3,5,\dots$ for odd parity solutions. This gives the trivial solutions for $f_\mu(\phi)=1/\sqrt{\pi}~cos(\mu\phi)$ for odd parity and $f_\mu(\phi)=1/\sqrt{\pi}~sin(\mu\phi)$ for even parity, which can be reduced to the non-interacting single-particle harmonic oscillator solutions Eq.~(\ref{HO1_eigenfunction}). However, when $G\rightarrow\infty$ the calculated $\mu$'s become different than before. For fermions one still gets integer values but different from the values in $G=0$, while for bosons the values are sometimes found to be non-integers, as illustrated in Fig.~\ref{energyspectrum1+2}. In the next section, these values are used to construct the wave function.

\subsection{Wave Functions}\label{ch:threeparticles:subsec:wavefunctions}
{\it This section contains some updated parts from my qualifying exam report.}

In case of fermions one concludes from Fig.~\ref{energyspectrum1+2} left panel that, $\mu=3, 6, 9, \dots$ at $G\rightarrow\infty$ with an even-odd oscillating parity. In addition, the first three states become degenerated at this point, thus the wave functions must be orthogonal to each other here. Notice that the parity \index{Parity}is a conserved property and therefore remains the same throughout the spectrum. The angular part of the wave function can be written as: 
\begin{equation*}
f_\mu(\phi)=N\cdot
\begin{cases}
a_1 & \text{for $\phi\in \mathbf{III}$}\\
a_2 & \text{for $\phi\in \mathbf{II}$}\\
a_3 & \text{for $\phi\in \mathbf{I}$}
\end{cases},
\end{equation*}
where $N$ is some normalization factor and $a$'s are some independent functions dependent on $\mu$ and $\phi$. Because of symmetry it is sufficient to only consider region $\mathbf{I}$, $\mathbf{II}$ and $\mathbf{III}$ of Fig.~\ref{3particleposition}. One can now construct the wave functions for the three lowest states on the repulsive side, $G>0$. One solution that is known, which is not dependent on the value of $G$ is the non-interacting case (the lowest horizontal red line in Fig.~\ref{energyspectrum1+2} left panel), where the wave function naturally vanishes at the contact points with the delta-boundaries. This is the $a_1=a_2=a_3=cos(3\phi)$ case with $\mu=3$ and very similar to the two-particle case discussed earlier in Fig.~\ref{busch}. Using the fact that the degenerated wave functions have to be orthogonal to each other and have the correct parity, one can then obtain the coefficients for the ground and 1st excited state with ($a_1=a_3=cos(3\phi)$ and $a_2=-2cos(3\phi)$) and ($a_1=-a_3=cos(3\phi)$ and $a_2=0$), respectively. The normalization factor can then be calculated afterwards.\\

\begin{figure}
\centering
\includegraphics[width=\textwidth]{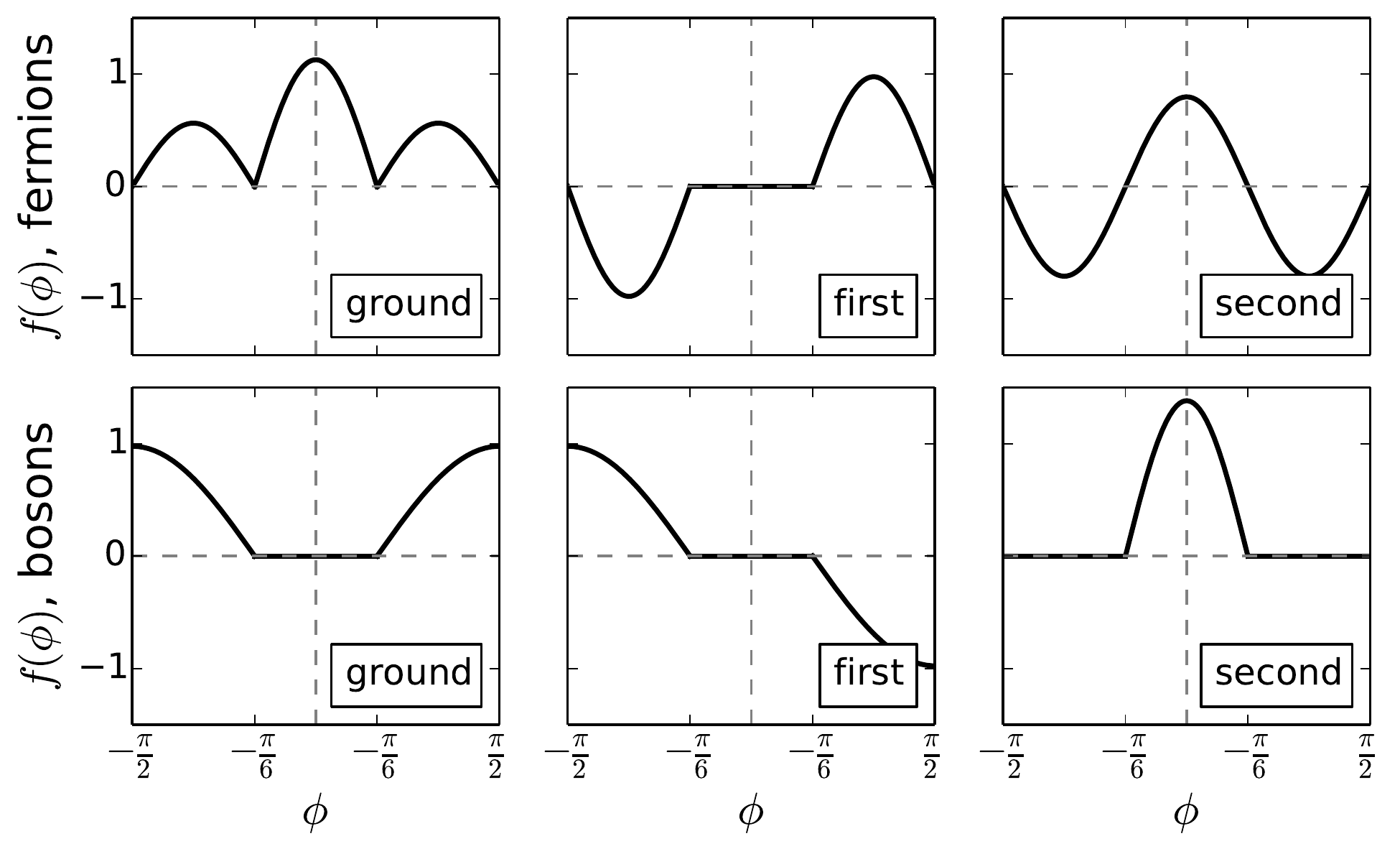}
\caption{Upper panel: The three lowest solutions, which correspond to the fermionic states in the 1+2 system. From left, I have plotted the lowest line on the repulsive side from Fig.~\ref{energyspectrum1+2} left panel. Then in middle the first excited and to the right the second excited states. Lower panel: corresponds to the bosonic case.}
\label{plot_three_wavefunctions}
\end{figure}

\noindent
For the case of bosons, one obtains the following non-integer $\mu$-values ($\mu=1.5, 3, 4.5, 6,\dots$). In addition, the bosonic states are only doubly degenerated for half-integer and non-degenerated for integer values of $\mu$. Hence, one can only use parity and orthogonality condition to built the bosonic wave functions. The ground state in this case is given as: $a_1=-sin(\frac{3}{2}(\phi -\frac{\pi}{6}))$, $a_3=sin(\frac{3}{2}(\phi +\frac{\pi}{6}))$ and $a_2=0$). The 1st excited state is the odd-parity version of the ground state, that is $a_1=-sin(\frac{3}{2}(\phi -\frac{\pi}{6}))$ and $a_3=-sin(\frac{3}{2}(\phi +\frac{\pi}{6}))$. Finally, the 2nd non-degenerate excited state is given as $a_1=a_3=0$ and $a_2=cos(3\phi)$. The remaining higher excited states can easily be obtained from here because they have the same structure with only different $\mu$-values. The three lowest solutions are summed and plotted in Fig.~\ref{plot_three_wavefunctions}.\\

\begin{figure}
\centering
\includegraphics[width=\textwidth]{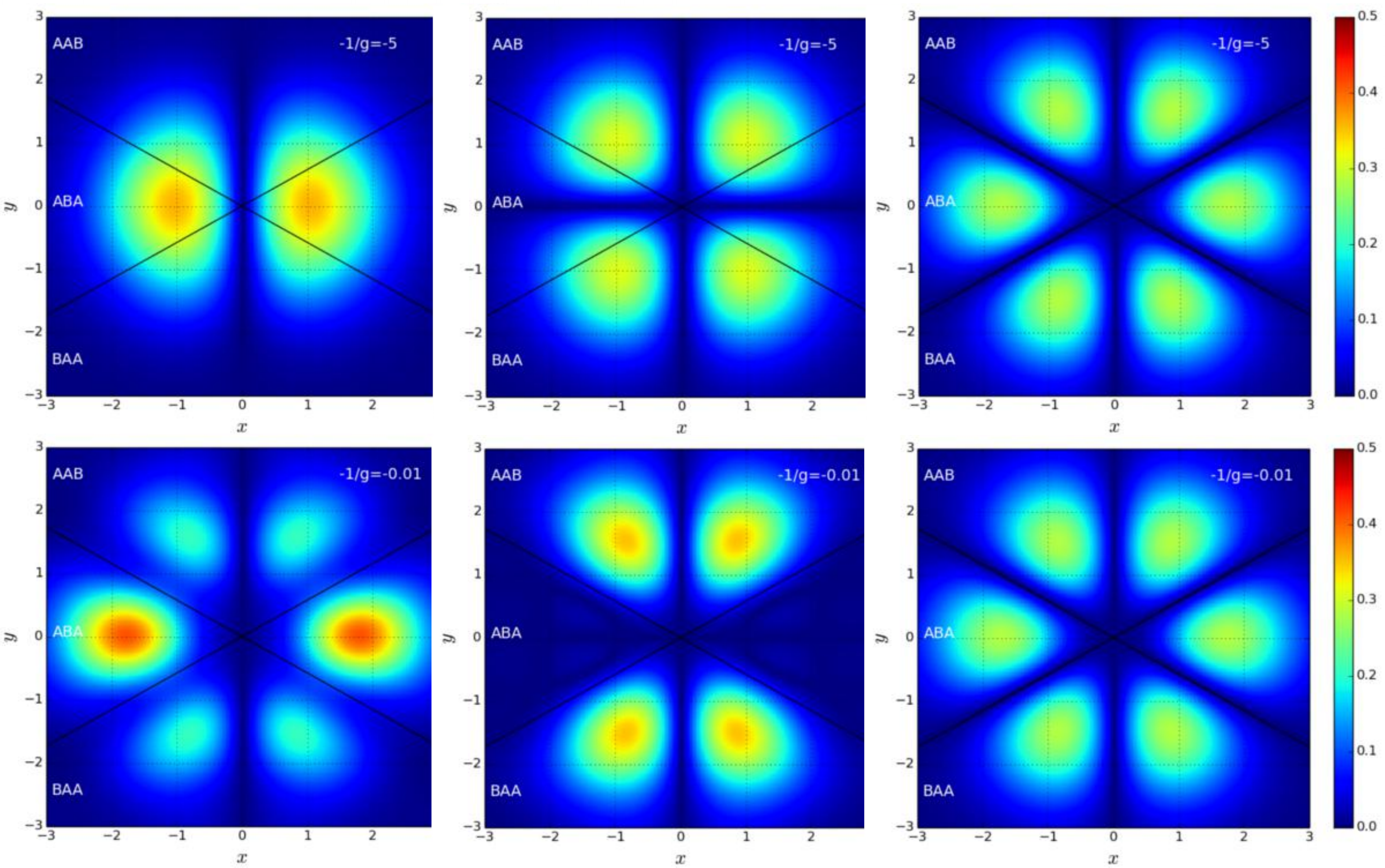}
\caption{Numerical EEDM results: Upper panels show the non-interacting fermionic density for the three lowest states. Notice that it is fermionic because at $x=q_1-q_2=0$ the wave function is zero in contrast to bosons in upper panels of Fig.~\ref{plot_threebosons_xyplane}. Lower panels show the corresponding strongly interacting density in the $x,y$-plane. Each region is also labeled with the corresponding configuration and the delta boundaries are drawn with black lines as sketched also in Fig.~\ref{3particleposition}}.
\label{plot_threefermions_xyplane}
\includegraphics[width=\textwidth]{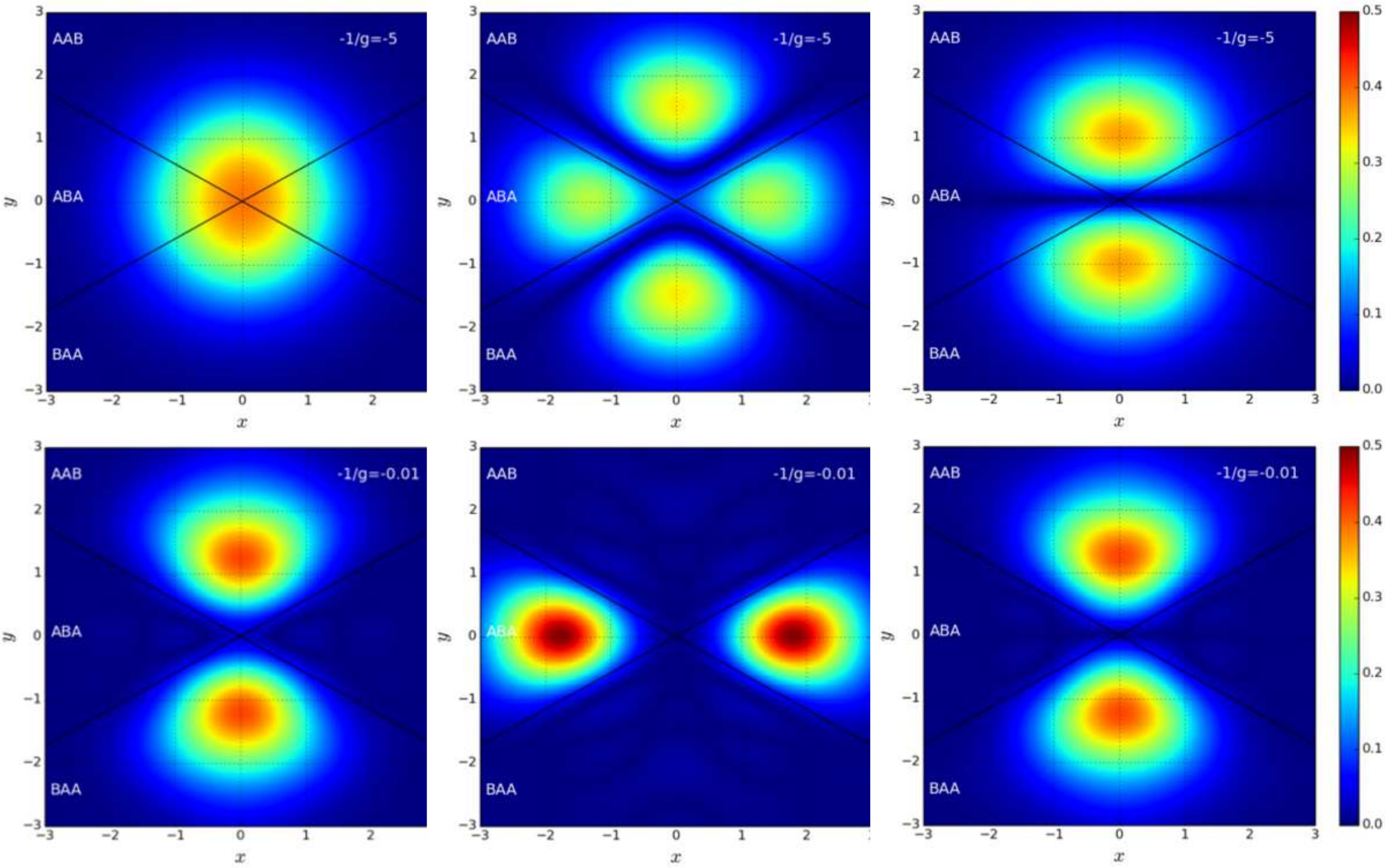}
\caption{Upper and lower panel: same as in Fig~\ref{plot_threefermions_xyplane} but for bosonic symmetries.}
\label{plot_threebosons_xyplane}
\end{figure}

Fig.~\ref{plot_threefermions_xyplane} and Fig.~\ref{plot_threebosons_xyplane} show how the three-particle systems look like in the $x,y$-plane, which are calculated with the Effective Exact Diagonalizing Method (EEDM). I will explain more about this numerical method in Chapter 7. For now, one can see the very nice similarity between these results and the ones obtained analytically in Fig.~\ref{plot_three_wavefunctions}. For both results, one striking fact is that in the limit of strong interaction, the bosons and fermions behave very differently. While the ground state of $1+2$ fermionic state, Fig.~\ref{plot_threefermions_xyplane} lower left panel, becomes a linear combination of $ABA$ and $AAB$ (or $BAA$) ordering, it is clear that the bosonic ground state, Fig.~\ref{plot_threebosons_xyplane} lower left panel, is purely a $AAB$ ordering, indicating a perfect ferromagnetic\index{Ferromagnetic} structure \cite{LoftTEPJD2015, ZinnerEEL2014}. The excited states behave in a different manner but certainly different from each other. Magnetic correlations are therefore easily accessible for these kind of systems and this makes it easy to engineer ferro- and antiferromagnetic\index{Antiferromagnetic} states \cite{SerwaneS2011, zurnPRL2012}.

The non-trivial distribution of the ordering violates with the fact that one can construct the wave functions with a Bose-Fermi mapping\index{Bose-Fermi Mapping} of Girardeau \cite{GirardeauJoMP1960} and one must therefore treat this with care, since the coefficients $a_1$, $a_2$ and $a_3$ can be non-trivial. In other words, the bosons and fermions can be very different even in the hard-core limit. Let me elaborate on this: one naive ansatz for the wave function of the $2+1$ fermionic system with $g\rightarrow\infty$ is to use the Tonks-Girardeau\index{Tonks-Girardeau Gas! State} state to construct an ansatz like,
\begin{align}
\psi_{(1+2f)}(q_1, q_2, q_3) \propto (q_1-q_2) \cdot | q_2-q_3| \cdot | q_1-q_3 | \cdot \exp(-q_1^2-q_2^2-q_3^2),
\end{align}
where $q_1$ and $q_2$ are the coordinates of the identical fermions and the $q_3$ is the coordinate of the impurity. But as shown previously, this state is far from the correct state that is adiabatically connected to other states in the strongly but finite interaction strength, which is also discussed here \cite{LevinsenSA2015}. In case of two identical bosons instead of fermions, the wave function can be constructed by replacing $(x_1-x_2)$ with $|x_1-x_2|$. However, even though this might turn out to produce the exact results in some cases (all particles regardless species strongly interacting with each other), one should proceed with care in other cases such as identical particles within each components not interacting.

\subsection{Mass-Imbalanced}\label{ch:threeparticles:subsec:strongthree}
{\it This section contains some updated parts from my qualifying exam report.}

\begin{figure}
\centering
\includegraphics[width=\textwidth]{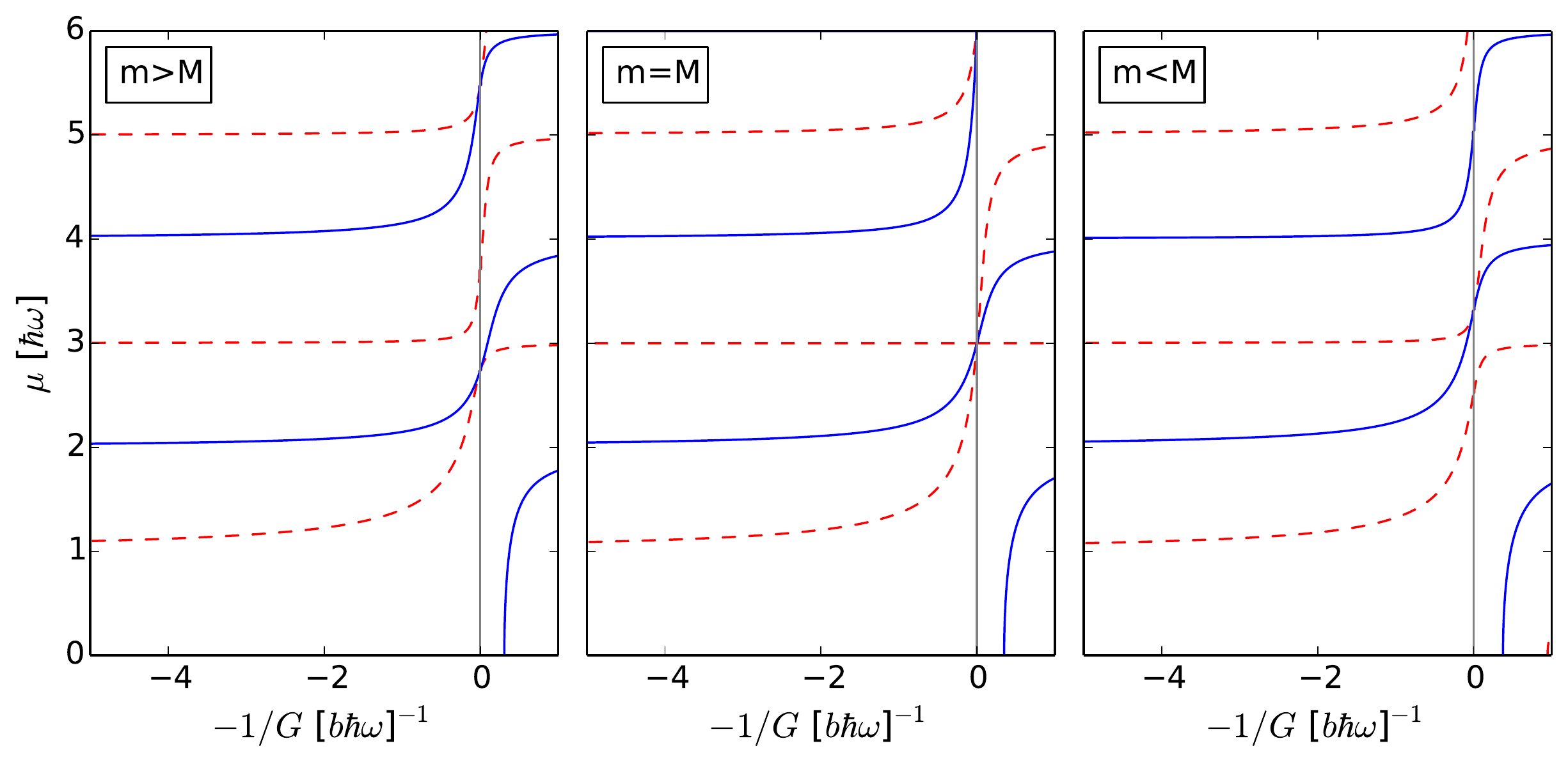}
\caption{Left panel: The third particle is lighter than the other two. The degeneracy is broken. Middle panel: the equal mass case with triple degeneracy. Right panel: the third particle is heavier than the others. Again the degeneracy is broken.}
\label{plot_three_wavefunctions_diffmass}
\end{figure}

In case of different masses, where the Bose-Fermi mapping definitely fails \cite{GirardeauJoMP1960,GirardeauPRL2007,GirardeauPRA2004}, one can apply some of the same techniques as before to solve the system. Because the different masses mix the terms in the Hamiltonian, one has to choose another length unit defined as $b=\sqrt{\frac{\hbar}{\mu\omega}}$, where
$\mu=\sqrt{m_1m_2m_3/(m_1+m_2+m_3)}$. The corresponding rotation matrix, $\mathbf{J}$, that rationalizes the coordinates (although not unique) is defined as \cite{mcguire1964}:
\begin{equation}
\mathbf{J} = \frac{1}{\sqrt{\mu}}
\begin{bmatrix}
\mu_{12} & -\mu_{12} & 0 \\[0.5em]
\frac{\mu m_1}{\mu_{12}M_{12}} & \frac{\mu m_2}{\mu_{12}M_{12}} & -\frac{\mu}{\mu_{12}} \\[0.5em]
\frac{m_1}{\sqrt{M_{123}}} & \frac{m_2}{\sqrt{M_{123}}} & \frac{m_3}{\sqrt{M_{123}}}
\end{bmatrix},
\label{ch2:eq:jacobicoordinates_diffmass}
\end{equation}
where $M_{12}=m_1+m_2$, $M_{123}=m_1+m_2+m_3$ and
$\mu_{12}=\sqrt{m_1m_2/M_{12}}$. The procedure is the same as before; the Hamiltonian separates into the center of mass and relative motions, $\mathcal{H}=H_{CM}+H_{rel}$, with:
\begin{equation}
H_{rel}=\frac{1}{2}(p_x^2+p_y^2)+\frac{1}{2}(x^2+y^2)+V,
\label{ch2:eq:jacobihamiltonian_diffmass}
\end{equation}
where
\begin{align*}
V=g\left[\frac{\mu_{12}}{\sqrt{\mu}}\delta(x)+\frac{\sqrt{\mu}}{\mu_{12}}\delta(\frac{\mu}{m_1}x+y)+\frac{\sqrt{\mu}}{\mu_{12}}\delta(-\frac{\mu}{m_2}x+y)\right].
\end{align*} 
Then hyperspherical coordinates\index{Hyperspherical Coordinates}, $\rho=\sqrt{x^2+y^2}$ and $tan(\phi)={y}/{x}$, are introduced. In case of $1+2$ systems, the masses are denoted as $m_1=m_2=m$ and $m_3=M$. The interacting part of the Hamiltonian is then given as:
\begin{align*}
V=\frac{g}{\rho}\sqrt{\frac{2\gamma}{\gamma^2+1}} \sum_\pm \Big( \delta\big(\phi\pm\theta_0\big)+\delta\big(\phi\pm\theta_0-\pi\big)\Big) \; ,
\end{align*}
where $\gamma \equiv \frac{\mu_{123}}{m} = \sqrt{\frac{M}{2m+M}}$ and
$\theta_0=\arctan \gamma$ is the angle between the $x$-axis and the
$q_2=q_3$ delta-line. It is worth noting that the interaction strength, $G$, is now redefined as $G\equiv {2\rho g} \sqrt{\frac{2\gamma}{\gamma^2+1}}$. One can then apply the conditions exactly as the previous section and obtain a new set of equations to obtain $\mu$. For example for fermions (compare this with Eq. (\ref{eq:muequationssamemass1}) and (\ref{eq:muequationssamemass2})) one obtains:
\begin{align*}
0=&\mu\cdot cos(\mu\pi/2) + G \cdot sin\big(\mu (\pi/2-\theta_0)\big)~cos(\mu \theta_0) && \text{for odd fermions,}\\
0=&\mu\cdot sin(\mu\pi/2) + G \cdot sin\big(\mu (\pi/2-\theta_0)\big)~sin(\mu \theta_0) && \text{for even fermions.}
\end{align*}\\
\noindent
Notice that when $m=M$ then $\gamma = \frac{1}{\sqrt{3}}$ yielding
$\theta_0=\tfrac{\pi}{6}$. This is the exact same result as
calculated in subsection \ref{ch:threeparticles:subsec:fermionsandbosons}. It turns out that only when $M=m$, it is possible to construct a ground state whose angular part can exist in
all regions (region I, II, III, IV, V and VI - see Fig.~\ref{3particleposition}) though with different weight. However, when $M<m$ the wave function vanishes in region II and V and the lowest lying states are doubly degenerated in the strong interacting regime, see Fig.~\ref{plot_three_wavefunctions_diffmass} left panel. Whenever $M>m$ the wave function vanishes in regions I, III, IV and VI and the ground state is non-degenerated, see Fig.~\ref{plot_three_wavefunctions_diffmass} right panel. This is due to the change in spatial areas, as illustrated in Fig.~\ref{3particleposition}, which breaks the symmetry instantaneously when $M\neq m$. In other words, when $M>m$ the ($\uparrow\downarrow\uparrow$) ordering is favored in the ground state with $a_1=a_3=0$ and $a_2=cos(\mu\phi)$. Notice that the regions for $a_1$, $a_2$ and $a_3$ are no longer equally distributed. For the opposite case, $M<m$, the ordering ($\uparrow\uparrow\downarrow$) and ($\downarrow\uparrow\uparrow$) are preferred with $a_1=sin(\mu(\phi-\theta_0))$, $a_3=sin(\mu(\phi+\theta_0))$ and $a_2=0$ for the ground state.

\section{Intermediate interaction strengths}\label{ch:threeparticles:sec:intermediate}
So far in this chapter, I have solved the $1+2$ problem exactly in the strongly interacting limit. The question is therefore, can one say something about the intermediate region of interaction strengths? It is important to note again, that the previous method is not exact in the intermediate region\index{Intermediate Regime}, because here $g$ depends on the hyperradius, $\rho$, which has to be taken into account, if one wants to solve it for this regime. In our group, we have attacked this problem in several attempts. In the following, I will briefly discuss two of them.

\subsection{Pair-Correlated Wave Function}\label{ch:threeparticles:subsec:paircorrelated}
Since interactions between the particles are short contact interactions (typically much shorter than the interparticle spacings), one can write the relative part of the wave function, $\psi_r$, as \cite{BarfknechtJPB2016};
\begin{equation}
\psi_r=C_r\prod_{i<j}^3 D(\beta_{ij} r_{ij}|\mu_{ij}),
\end{equation}
where $C_r$ is a normalization factor and $D$ is the parabolic cylinder function,\index{Parabolic Cylinder Function} dependent on the relative distance between the two particles $r_{ij}=|q_j-q_i|$. Furthermore, $\mu_{ij}$ is a constant, which is obtained by using the condition for two particles interacting with each other. Just like Eq.~(\ref{griffith_criteria}), one obtains $2\beta_{ij} D'(0|\mu_{ij})=g_{ij}D(0|\mu_{ij})$, which can be reduced to the following relation:
\begin{equation}
\frac{g_{ij}}{\beta_{ij}}=-\frac{2^{3/2}\Gamma(\frac{1-\mu_{ij}}{2})}{\Gamma(\frac{-\mu_{ij}}{2})},
\end{equation}
where $\Gamma$'s are the Gamma functions and $\mu_{ij}$ takes some value between 0 and 1 as $g_{ij}$ goes from 0 to $\infty$. As $g$ is zero or infinitely strong, the method captures the two limiting solutions successfully by definition, because it is build on the solution for a pair of bosons in a trap. While for the intermediate case, one needs to build the wave function variationally by varying the constant $\beta_{ij}$. This is therefore not exact, however, the method captures the qualitative behavior of this kind of system. The work was performed primarily by R.~E.~Barfknecht and discussed in \cite{BarfknechtJPB2016}.

\subsection{Interpolatory Ansatz\index{Interpolarory Ansatz}}\label{ch:threeparticles:subsec:interpolatory}
\begin{figure}
\centering
\includegraphics[width=\textwidth]{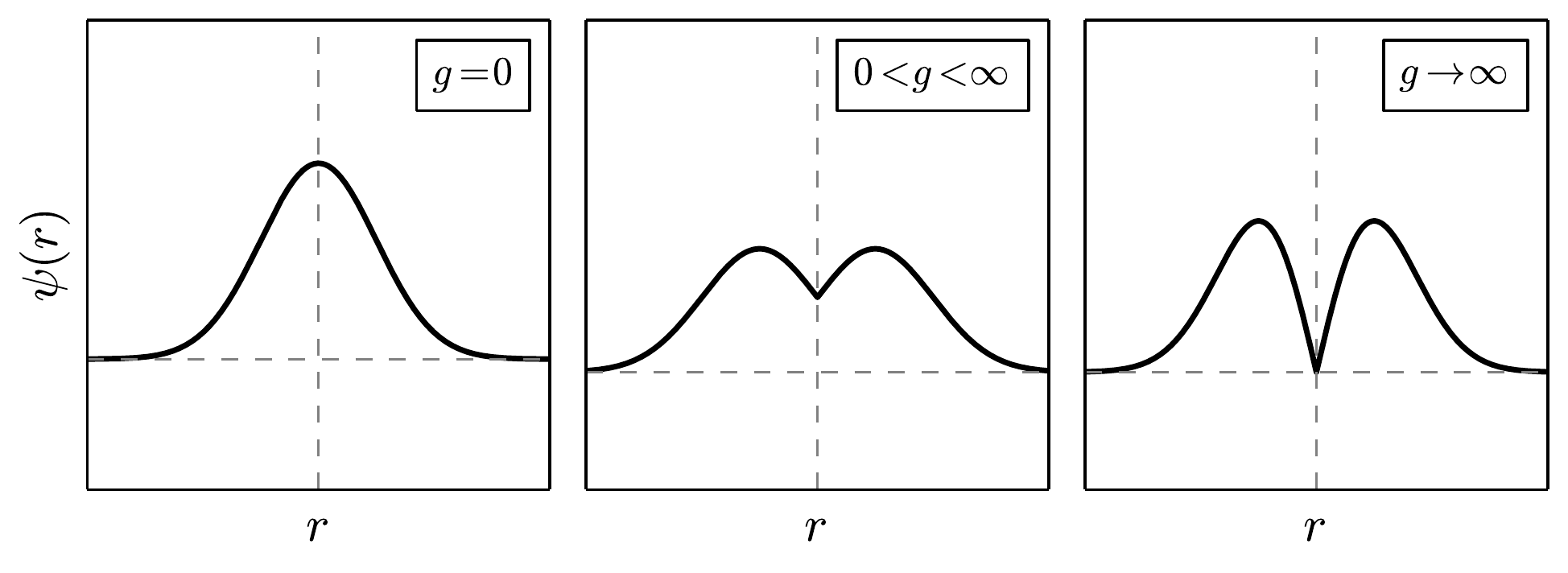}
\caption{Left panel: non-interacting regime. Middle panel: intermediate regime. Right panel: strongly interacting regime. Knowing the exact solution of the non-interacting and strongly interacting regimes can be used to say something about the intermediate region.}
\label{plot_1+2interpolatory}
\end{figure}

Another method that we tried, builds on a very simple idea; since we know the solutions for the non-interacting and strongly interacting regimes, can we capture the intermediate regime by making a clever linear combination of the two limiting solutions? The reader might remember some sketches of the intermediate case back in the Introductory chapter in Fig.~\ref{busch_wave}. Here, the idea is to find a smart way to combine the state in Fig.~\ref{plot_1+2interpolatory} left panel with right panel to get a good description of the middle panel. Notice that even though the figure is sketched for the two-body problem, the idea is the same in the many-body problem where one would have several zero-nodes like the one in the right panel of Fig.~\ref{plot_1+2interpolatory} or zero-lines and -planes. By looking at the figures, it is apparent that the intermediate regime might be a little bit tricky to capture in full details. Furthermore, numerical calculations \cite{DehkharghaniSR2015, PecakNJoP2016} tell us that the wave function is not easy to write and one needs several single particle eigenstates to construct one solution. However, the question can be reduced to how well the ansatz works by comparing it to numerical exact solutions.

Our ansatz, $|\psi\rangle$, is constructed in the following way;
\begin{equation}
|{\psi}\rangle = \alpha_0 |{\psi_0}\rangle + \alpha_\infty |{\psi_\infty}\rangle,
\label{variational_ansatz}
\end{equation}
with $\alpha_0$ and $\alpha_\infty$ being real parameters, and $|{\psi_0}\rangle$ and $|{\psi_\infty}\rangle$ are the well-constructed non- and strongly interacting exact solutions with the corresponding energies $E_0$ and $E_\infty$, respectively. As the reader might remember, the Hamiltonian is given as $\mathcal{H}=H_0+V$, where $H_0=\sum_i \frac{1}{2}\frac{\partial^2}{\partial q_i^2}+\frac{1}{2}q_i^2$ and $V=g\sum_{i<j}\delta(q_i-q_j)$. In the previous subsections, I showed how one can obtain the fully analytical exact solution in the strongly interacting regime, $|{\psi_\infty}\rangle$, for the $1+2$ system with any mass ratio. In the next chapter I will show the exact solutions for the $2+2$ systems, which makes the following method also applicable there, but I will come back to this later.

The variational energy\index{Variational Energy} of the trial state becomes \cite{AndersenSR2016},
\begin{align*}
E &= \frac{\langle{\psi|\mathcal{H}|\psi}\rangle}{\langle{\psi|\psi}\rangle} \notag \\
&= E_0 + \frac{\langle{\psi_0|V|\psi_0}\rangle\alpha_0^2 + \Delta E\alpha_\infty^2}{\alpha_0^2 + \alpha_\infty^2 + 2\langle{\psi_0|\psi_\infty}\rangle\alpha_0\alpha_\infty}, \end{align*}
where $\Delta E \equiv E_\infty - E_0$ and the wave functions,$|{\psi_0}\rangle$ and $|{\psi_\infty}\rangle$, are normalized. Notice that $|{\psi_\infty}\rangle$ is unaffected by $V$ since it is zero at these contact points by definition. By identifying the stationary points of the ansatz, one can derive a relation between the coefficients, $a_0$ and $a_\infty$,
\begin{equation}
\left(\frac{\alpha_0}{\alpha_\infty}\right)_\mathrm{opt}^{(\pm)} = \frac{\Delta E -\langle{\psi_0|V|\psi_0}\rangle
\mp\sqrt{\left(\Delta E -\langle{\psi_0|V|\psi_0}\rangle\right)^2+4\langle{\psi_0|V|\psi_0}\rangle\Delta E\langle{\psi_0|\psi_\infty}\rangle^2}}{2\langle{\psi_0|V|\psi_0}\rangle\langle{\psi_0|\psi_\infty}\rangle}.
\label{variational_constant_relation}
\end{equation}
In addition, one can apply the normalization criteria to obtain the full value of $a_0$ and $a_\infty$ for a given $g$. Furthermore, one can show that the variational energy can be reduced to,
\begin{equation}
\begin{split}
&E^{(\pm)} = E_0 +\\
&\frac{\langle{\psi_0|V|\psi_0}\rangle + \Delta E
\pm \sqrt{(\langle{\psi_0|V|\psi_0}\rangle + \Delta E)^2 - 4\langle{\psi_0|V|\psi_0}\rangle\Delta E\left(1-\langle{\psi_0|\psi_\infty}\rangle^2\right)}}{2\left(1-\langle{\psi_0|\psi_\infty}\rangle^2\right)},
\end{split}
\label{variational_energy}
\end{equation}
where $(\pm)$ solutions are the maximum and minimum energy solutions, respectively. When $g>0$ the minimum solution is chosen, while when $g<0$ the maximum solution is chosen as the correct energy due to the sign of $g$. As simple as it looks, it is also important to note that the method is quite general. It is independent of the external potential, masses of the particles and the size of the system. As long as you know the solutions to the non- and strongly interacting regimes the method is easily applicable.

\begin{figure}
\centering
\includegraphics[width=\textwidth]{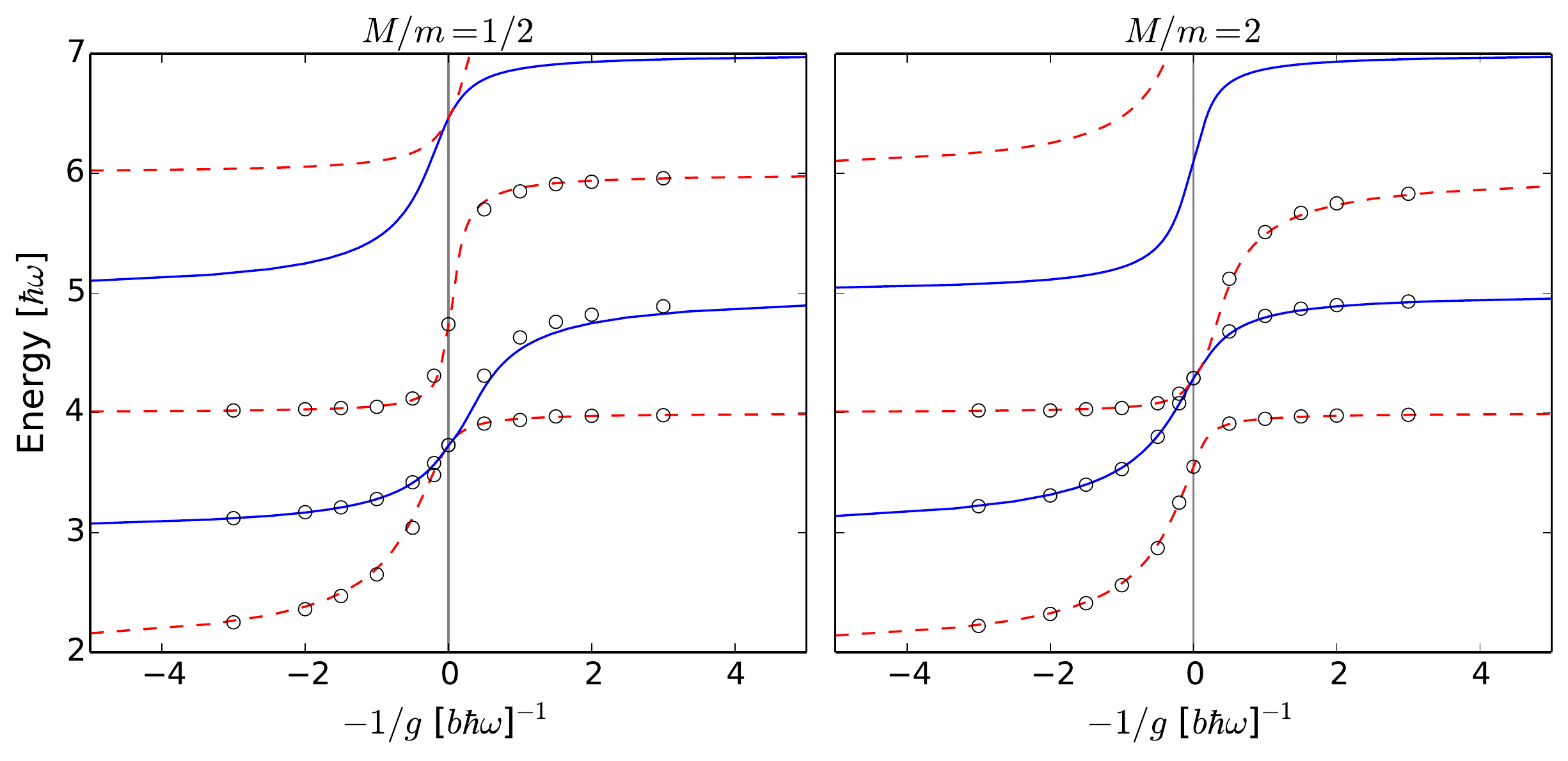}
\caption{Left panel: Energy spectrum calculated with Correlated Gaussian Method (CGM) for the mass imbalanced, $M/m=1/2$, 1+2 fermionic system (red and blue lines) compared with the variational method (black circles). Right panel: Same as left panel, but for the other case $M/m=2$. Figure is adapted from \cite{AndersenSR2016}.}
\label{three-particles-energy-spectrum-unequal-masses}
\end{figure}

We tested the method by comparing it to the numerical results from Effective Exact Diagonalizing Method (EEDM) \index{Effective Exact Diagonalizing Method}and Correlated Gaussian Method \index{Correlated Gaussian Method} (CGM). For more details about the numerical methods go to Chapter 7. The investigation was done for the $1+N$ cases with $N$ up to $6$ particles and the energy results were very close as shown in \cite{AndersenSR2016}. However, we found out that we could get even closer by {\it modifying} \index{Interpolarory Ansatz!Modified Ansatz}the ansatz. 

The modification was necessary, because further investigations showed that the initial ansatz did not generally reproduce the correct first-order energy as $g\rightarrow\infty$. Therefore one could modify the ansatz by forcing it to have the correct energy slope as a function of $-1/g$ as derived in \cite{VolosnievNC2014}. Here, it is shown that the slope of the energy, $K_\mathrm{opt}^\infty$, up to first order is given as,
\begin{equation}
K_\mathrm{opt}^\infty = \frac{\partial E_\mathrm{opt}}{\partial (-1/g)}\bigg|_{g\rightarrow\infty} = \frac{(\Delta E)^2}{K^0} \langle{\psi_0|\psi_\infty}\rangle^2,
\end{equation}
where $K^0 = \langle{\psi_0|V|\psi_0}\rangle/g$. Now, since the original ansatz kept producing the wrong $K_\mathrm{opt}^\infty$, the modified ansatz was constructed by looking at it the other way around. Since $K_\mathrm{opt}^\infty$ is known from \cite{VolosnievNC2014} one can obtain the value of $\langle{\psi_0|\psi_\infty}\rangle^2$ from the above equation and use this in the Eq.~(\ref{variational_energy}) to obtain a much better estimation of the ground state energy. However, it comes with the cost that one no longer knows how the wave function is given (only the overlap with the non-interacting solution is given).\\

\noindent
In order to illustrate the precision of the modified ansatz, in Fig.~\ref{three-particles-energy-spectrum-unequal-masses} I have plotted the energy spectrum calculated numerically with CGM for the 1+2 fermionic system with different masses compared with the modified ansatz energies. The figure is redesigned and adopted from \cite{AndersenSR2016}. As one can see the comparison between the modified variational method and the numerical results is very good, having in mind that the method only makes use of a simple linear combination of only two states. However, the use of modified ansatz hides the information about the wave function at $g\rightarrow\infty$, which was the starting point of this chapter. In conclusion, it is therefore important to note that the initial ansatz can be used to approximate the wave functions, while the modified ansatz can be used to estimate energies.

\begin{figure}
\centering
\includegraphics[width=\textwidth]{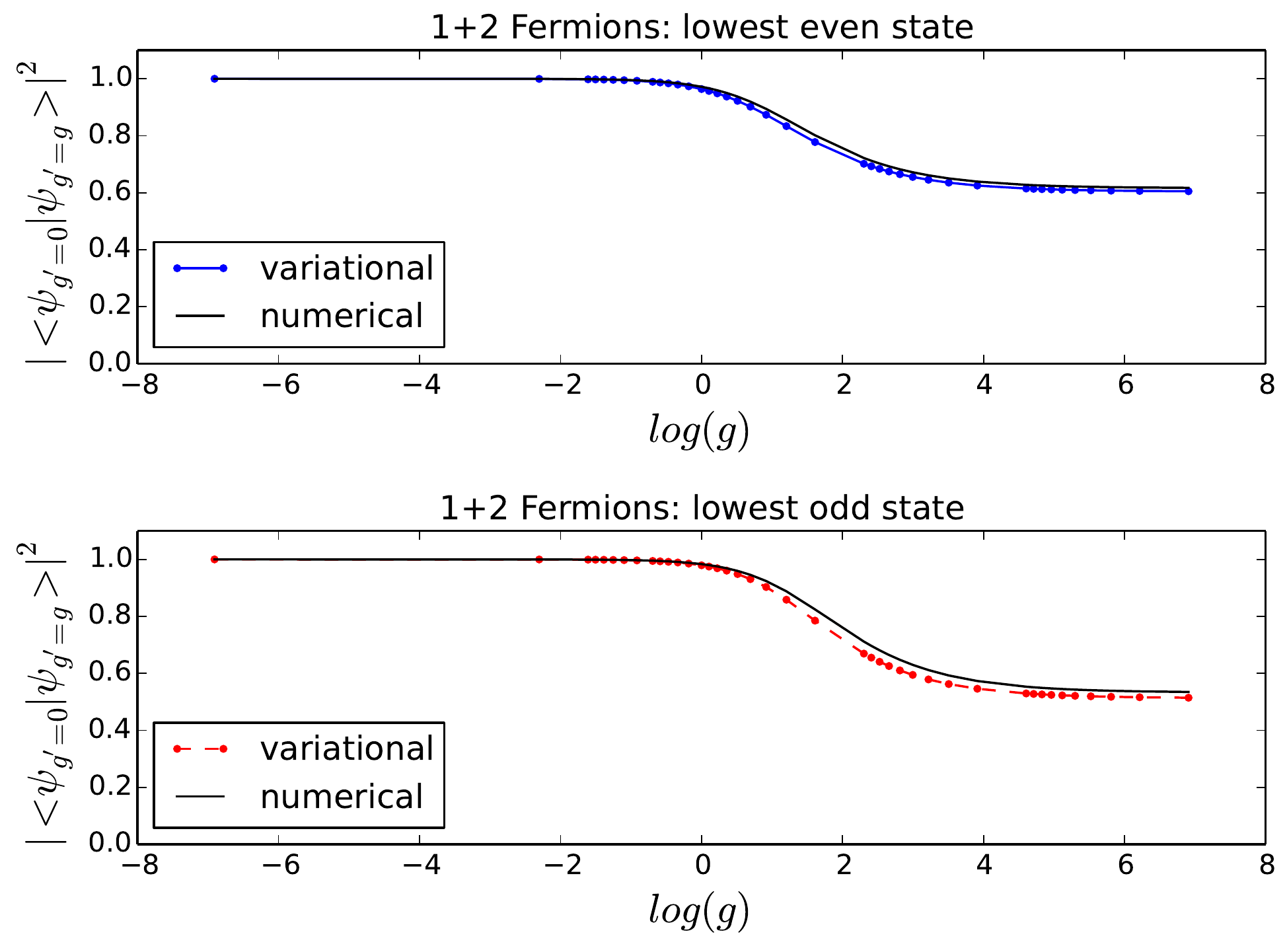}
\caption{Upper panel: The Anderson overlap for the lowest even state in the $1+2$ fermionic system. Lower panel: Same as upper panel but for the lowest odd state. Figure is adapted from \cite{AndersenSR2016}.}
\label{anderson_1+2}
\end{figure}

Further, one could also investigate the Anderson overlap\index{Anderson Overlap}, which is an overlap of the wave function between non-interacting and strongly interacting solutions, $\langle\psi_0|\psi_\infty\rangle$. This quantity relates to the Anderson orthogonality catastrophy\index{Anderson Overlap!Orthogonality Catastrophy} \cite{andersonPRL1967}, which is proven to go to zero for $N\rightarrow\infty$, particularly for $g\rightarrow\infty$. Fig.~\ref{anderson_1+2} shows the comparison between the numerically calculated results (EEDM) and the variational method with the initial ansatz for the same mass case. There is a clear decrease in the graph for the overlap of both states, however, as the 1+2 system is not a big system with $N\gg1$, the overlap is therefore not close to zero at all. However, the comparison between the two methods is good enough to say that the variational method is indeed able to reproduce the numerically exact calculated results with only small deviation. Again, it is important to notice that the method is only a simple linear combination of two states. The overlap for higher number of particles with the variational method is done in \cite{AndersenSR2016} for $N$ up to $6$ particles. Here it is shown that there is a tendency for the overlap to approach zero as the interaction strength and number of particles increase.\\

\noindent
Even though the presented results turn out to capture the energy quite well, one could argue that the wave function might be not very well reconstructed in the intermediate case. In addition, one could also be concerned that when the number of particles in both components is more than just a single particle the correlation in the system could become more complex such that one no longer can capture the physics with just a simple linear combination. All these questions are legitimate and I will answer them in the next chapter when I investigate the four-particle systems.

\newpage % Create empty back of side
\thispagestyle{empty}

%-----% 2+2 SYSTEMS
% !TeX root = ../Main_publish.tex

\chapter{Four Particles in a 1D Harmonic Trap}\label{ch:fourparticles}
\epigraph{\it “Divide each difficulty into as many parts as is feasible and necessary to resolve it.”}{\rm ---René Decartes}

In this chapter I will dig into the next simplest case, namely the four-body system. In the previous chapter I illustrated how the three-body system could be analytically treated in the strongly interacting regime. Some of the same methods and ideas can also be used for the four-body case, or even many-body case. However, in the many-body case the analytics become more complex and harder to illustrate, even though in theory, it is possible to solve them.

In the following subsections, I will start by analyzing the four-body system\index{Four-Body Systems} and illustrate how one can derive the solution in the strong regime for a two-component $2+2$ system. As the reader might have already guessed, quantum statistics become more important as the number of particles increases in each component. Having obtained the solution in this regime and knowing the trivial solutions in the non-interacting regime, I will apply the variational method here and see how well the method does in this case. Later, I will analyze the $1+3$ and the so-called four-component $1+1+1+1$ system with different masses. For the latter case, I will investigate the behavior of the energy spectrum as a function of mass ratios.

\section{Strongly Interacting Mass-Imbalanced Four-Particle System.} \label{ch:fourparticles:sec:strongfour}
{\it This section contains some updated parts from my qualifying exam report.}

\begin{figure}[t]
\centering
\includegraphics[width=\columnwidth]{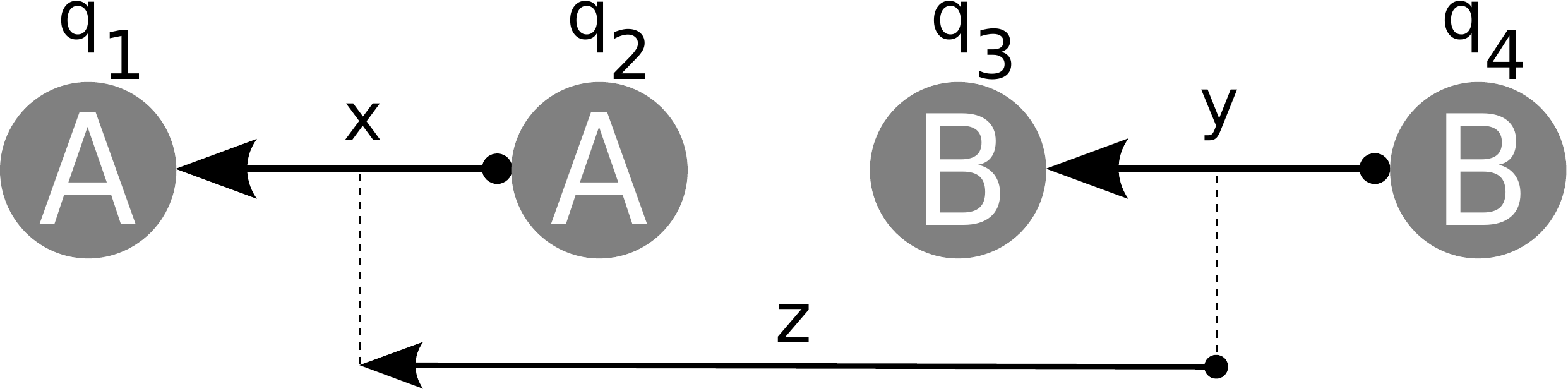}
\caption{Relative coordinates, $q_i$, for a four-body system and the Jacobi coordinates, $x$,$y$ and $z$. Note, that the center-of-mass coordinate, $R$, is not shown.}
\label{fourparticles_position}
\end{figure}

In the following paragraph I consider a two-component system with strong inter-species and zero intra-species interactions. In other words, only particles from different components interact strongly with each other, while particles in the same component are non-interacting. In what follows, the calculations can be easily generalized, so the same species interactions can also be taken into account. Bosons are denoted by $(A)$ and $(B)$, and similarly $(\uparrow)$ and $(\downarrow)$ for fermions. Different species are obtained by exciting the atoms into different hyperfine-states. The particles with arbitrary masses are confined in the same trap with the same frequency, $\omega$. Notice, that this choice allows one to separate the center-of-mass from the system. Although same trapping frequency might not be a realistic choice in experiments for particles with different masses, it is a starting point to see how the system behaves under these circumstances. In addition, I choose to position the particles in such a way that the first $N_A$ coordinates describe the $A$ particles. For example, for $2+2$ system $q_1, q_2$ are coordinates of the two $A$ particles and $q_3, q_4$ are for $B$ particles, see Fig.~\ref{fourparticles_position}. 

Since the particles can have any masses that are not necessary the same, another unit of length is introduced here. In this case $b=\sqrt{\hbar/(\mu\omega)}$, where for concreteness I take $\mu=\sqrt{m_1m_2m_3m_4/(m_1+m_2+m_3+m_4)}$. However, energies are still measured in units of $\hbar\omega$. Accordingly, the Hamiltonian can be written as:
\begin{align}
\mathcal{H}&=\frac{1}{2} \sum_{i=1}^{N=4}\left(-\frac{\mu}{m_i}\frac{\partial^2}{\partial q_i^2}+\frac{m_i}{\mu} q_i^2\right)+\sum_{i=1,i<j}^{N=4}g_{ij}\delta(q_i-q_j) \; .
\end{align}
The corresponding wave function, $\Psi(q_1,q_2,q_3,q_4)$, solves the eigen-equation, $\mathcal{H} \Psi = E \Psi$. Following the same ideas of the previous chapter I introduce a transformation of the coordinates $\mathbf{q}=(q_1,q_2,q_3,q_4)^T$ through $\mathbf{r}=(x,y,z,R)^T=\mathbf{J}\mathbf{q}$ where $\mathbf{J}$ is given as:
\begin{align*}
\mathbf{J}=
\begin{bmatrix}
\mu_{12} & -\mu_{12} & 0 & 0 \\
0 & 0 & \mu_{34} & -\mu_{34} \\
\frac{\mu~m_1}{\mu_{12}\mu_{34}M_{12}} & \frac{\mu~m_2}{\mu_{12}\mu_{34}M_{12}} & \frac{-\mu~m_3}{\mu_{12}\mu_{34}M_{34}} & \frac{-\mu~m_4}{\mu_{12}\mu_{34}M_{34}} \\
\frac{m_1}{\sqrt{M_{1234}}} & \frac{m_2}{\sqrt{M_{1234}}} & \frac{m_3}{\sqrt{M_{1234}}} & \frac{m_4}{\sqrt{M_{1234}}}
\end{bmatrix}\cdot \frac{1}{\sqrt{\mu}}.
\end{align*}
Here, $M_{ij}=m_i+m_j$ is a mass of two atoms, the total mass is denoted with $M_{1234}=M_{12}+M_{34}$, and $\mu_{ij}=\sqrt{{m_im_j}/{M_{ij}}}$. Notice, that the four-dimensional volume element changes upon the transformation in the following way: $ \mathrm{d}q_1\mathrm{d}q_2\mathrm{d}q_3\mathrm{d}q_4\rightarrow\frac{\mu}{\sqrt{M_{1234}}}\mathrm{d}x \mathrm{d}y \mathrm{d}z \mathrm{d}R$. The factor in front of the volume element is the determinant of the Jacobi matrix\index{Jacobi Coordinates!Matrix}.

With this transformation, the Hamiltonian becomes separable where in one part there is the center-of-mass\index{Center-of-mass Motion} that contains coordinate $R$, and in the other part there is the intrinsic motion \index{Relative Motion}part with $x,y,z$. Denoting the eigenbasis for the Jacobi coordinate \index{Jacobi Coordinates}system with $\left|\nu,\tau,\eta,\chi\right\rangle$, the separation becomes as follows: $\left|\nu,\tau,\eta\right\rangle \otimes \left|\chi\right\rangle$. In terms of the wave function it is given as, $\Psi=\psi_{\nu,\tau,\eta}(x,y,z)_{relative}\cdot \psi_\chi(R)_{CM}$. Notice that this was not possible if $\omega$ was not the same for all particles.

Below the Hamiltonian is written explicitly, and here one can clearly see how the coordinate $R$, is easily separable, since the delta-boundaries are only dependent on $x,y,z$:
\begin{align}
\begin{split}
\mathcal{H}=&\frac{\hbar \omega}{2}(\mathbf{r}^2+\nabla^2)+\\
\frac{1}{b}\frac{\sqrt{\mu}}{\mu_{12}\mu_{34}}&\Big\{\frac{\mu_{12}^2\mu_{34}}{\mu}g_{12}\delta\big(x\big)+\frac{\mu_{12}\mu_{34}^2}{\mu}g_{34}\delta\big(y\big)+\\
&g_{13}\delta\Big(\sqrt{\frac{m_2(m_3+m_4)}{m_1M}}x-\sqrt{\frac{m_4(m_1+m_2)}{m_3M}}y+z\Big)+\\
&g_{14}\delta\Big(\sqrt{\frac{m_2(m_3+m_4)}{m_1M}}x+\sqrt{\frac{m_3(m_1+m_2)}{m_4M}}y+z\Big)+\\
&g_{23}\delta\Big(-\sqrt{\frac{m_1(m_3+m_4)}{m_2M}}x-\sqrt{\frac{m_4(m_1+m_2)}{m_3M}}y+z\Big)+\\
&g_{24}\delta\Big(-\sqrt{\frac{m_1(m_3+m_4)}{m_2M}}x+\sqrt{\frac{m_3(m_1+m_2)}{m_4M}}y+z\Big)\Big\}.
\end{split}
\label{fourparticle_hamiltonian}
\end{align}

\begin{figure}[t]
\centering
\includegraphics[width=\columnwidth]{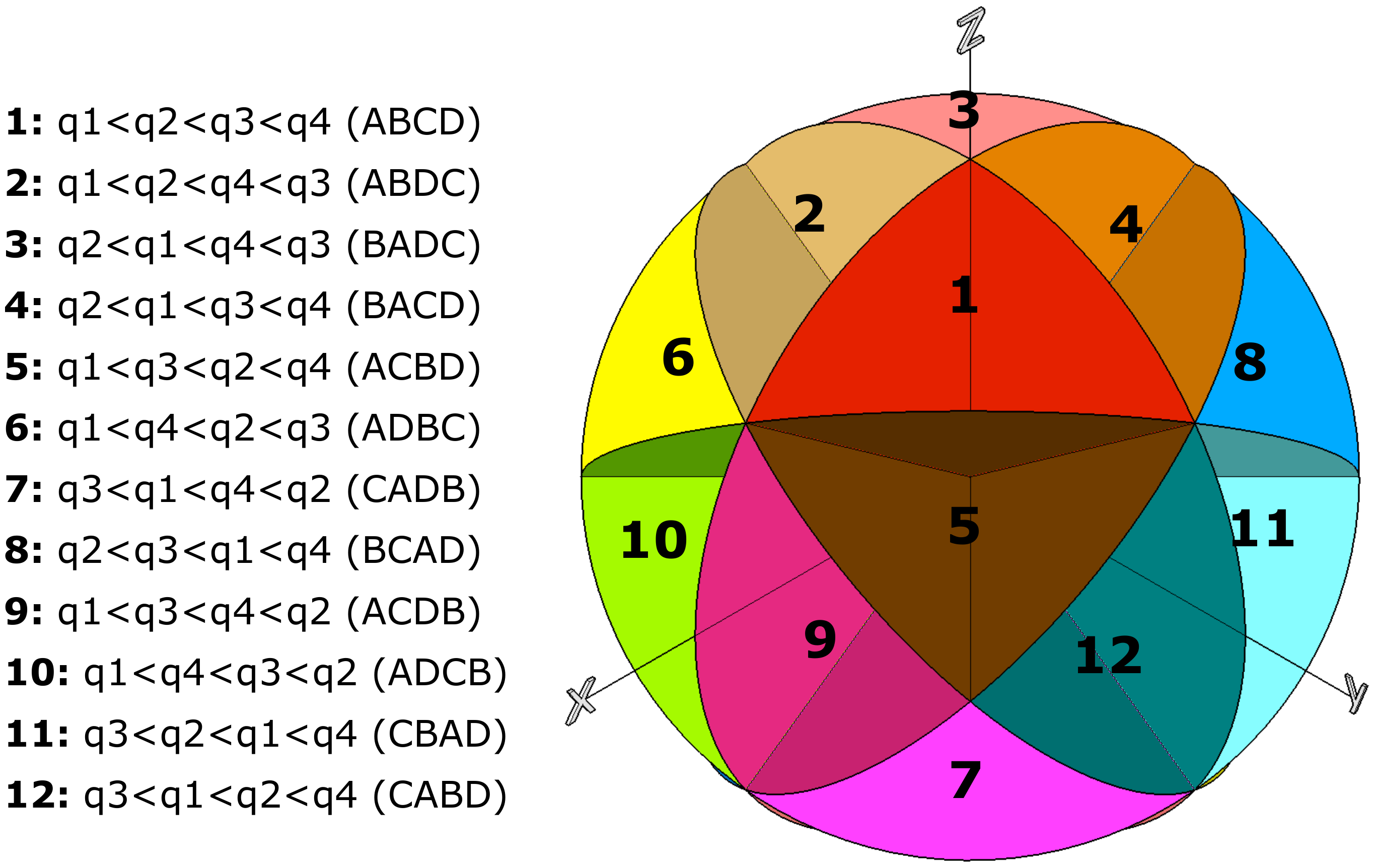}
\caption{Jacobi space for the general four particle systems with $A$, $B$, $C$ and $D$ type of particles. The delta-boundary planes divide the space. The solid planes represent the interacting planes where the wave function must be zero in the strong regime. The region is divided into 24 regions, but due to parity there are only 12 distinguishable orderings, which are listed to the left of the figure.}
\label{jacobi_sphere_1+1+1+1}
\end{figure}

\noindent
The wave function for the center-of-mass part with coordinate $R$ and quantum number $\chi$, are the well-known one-dimensional normalized harmonic oscillator states, $\psi_\chi(R)_{CM}$, mentioned back in Chapter 1, Eq.~(\ref{HO1_eigenfunction}). What remains to be solved is the wave function, $\psi_{\nu,\tau,\eta}(x,y,z)_{relative}$, for the relative motions supplemented by the corresponding transformed normalization condition, $\int\left|\psi_{relative}\right|^2 \frac{\mu}{\sqrt{M_{1234}}}\mathrm{d}x \mathrm{d}y \mathrm{d}z=1$. This wave function must satisfy the boundary conditions, which are given as boundary-{\it planes} in the Jacobi space, in the $\delta$-functions. This is very similar to the boundary-{\it lines} that were presented in Chapter 2 for three-particle systems. However, the space created by the $\delta$-functions in the general four-particle case is a three-dimensional space, which can be illustrated by a sphere as in Fig.~\ref{jacobi_sphere_1+1+1+1}.
Finding a solution to the Hamiltonian in this space requires that one takes one region at a time and analyzes that region independently. In the following I will go into the details of how one can find a specific solution to the $2+2$ systems.\\

\begin{figure}[t]
\centering
\includegraphics[width=0.6\columnwidth]{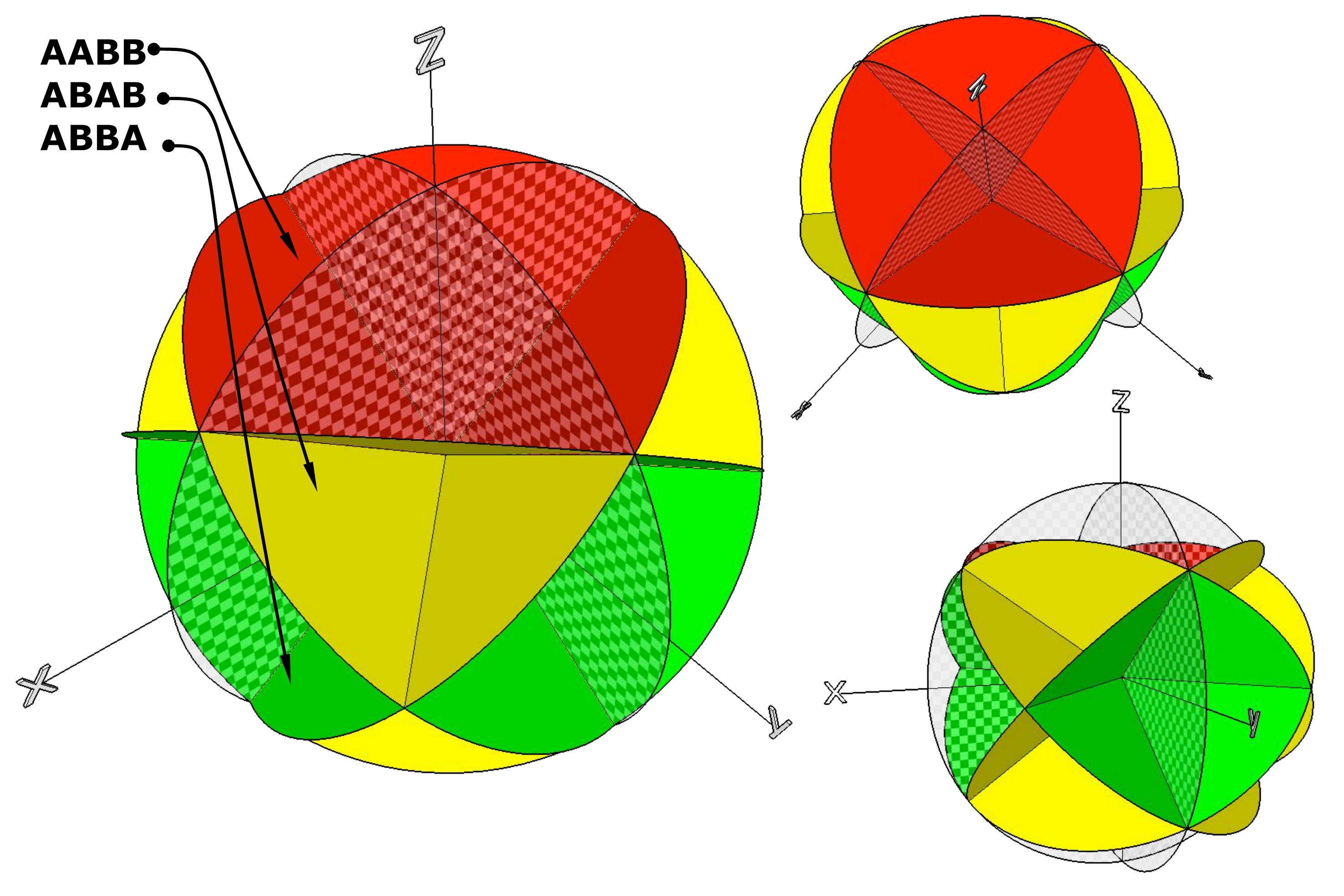}
\caption{Jacobi space for the 2+2 system seen from different perspectives. The space is divided by the delta-boundary conditions. The solid planes are the interacting planes where the wave function must be zero in the strongly interacting regime. The chess-transparent planes at $x=0$ and $y=0$ are the planes where the two particles meet. The wave function of the fermionic particles must be zero on these planes due to Pauli principle. Red region corresponds to a $AABB$ or $BBAA$ ordering of the particles, while yellow is for $ABAB$ or $BABA$ and green is for $ABBA$ or $BAAB$.}
\label{jacobi_sphere_2+2}
\includegraphics[width=0.9\columnwidth, trim= 0cm 10cm 0cm 11cm]{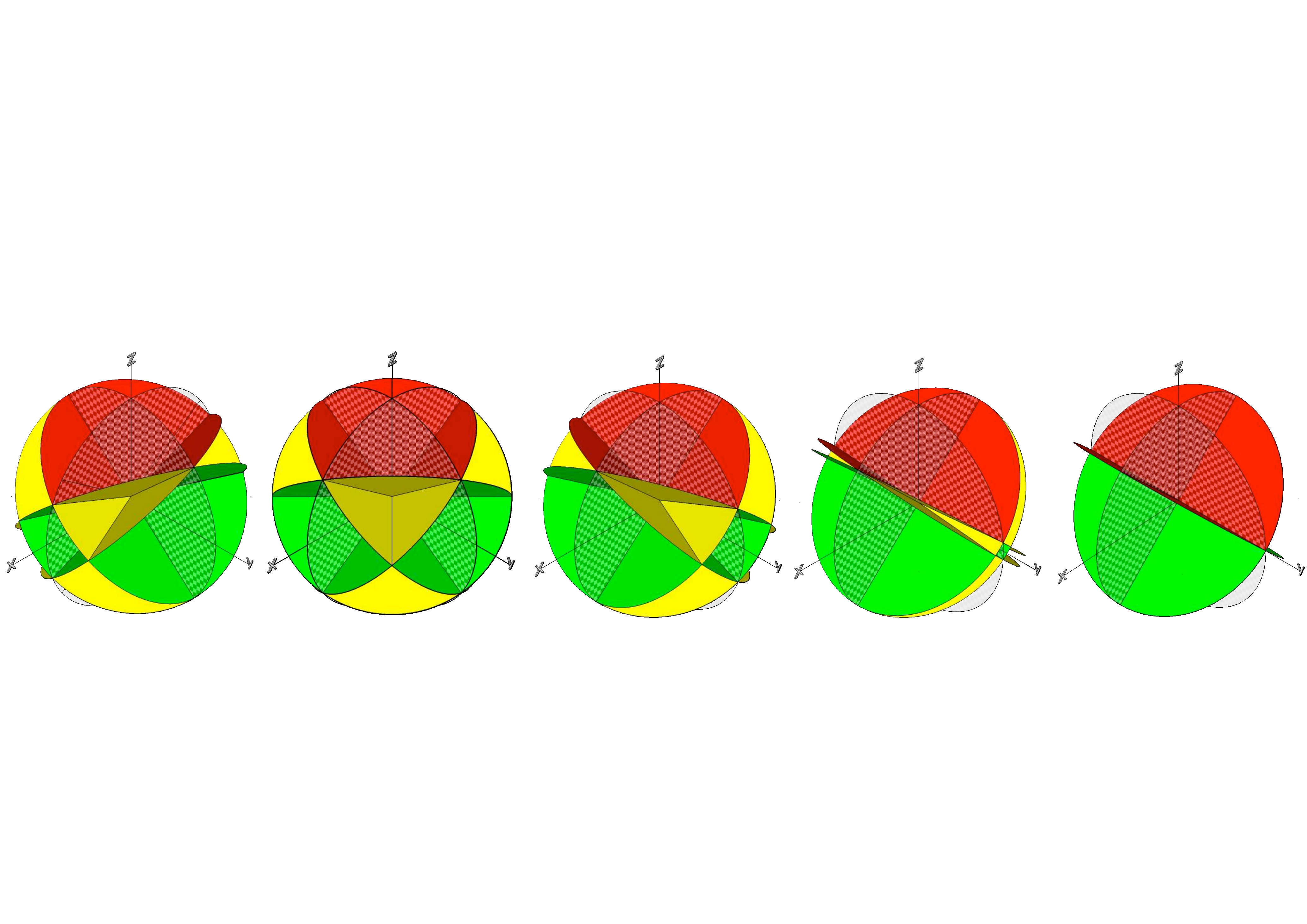}
\caption{Jacobi space for the 2+2 system for different mass ratios: $\beta/\alpha=1/5, 1, 5, 100, \infty$ plotted from left to right. As the $B$ particles get heavier the space of $AABB$ (red) and $ABBA$ (green) get equally big, while $ABAB$ (yellow) and $BAAB$ (green) vanish, since the particles would require a lot of energy to position themselves in those configurations.}
\label{jacobi_sphere_2+2_2}
\end{figure}

\subsection{2+2 Two-Component Systems}
{\it This section contains some updated parts from my qualifying exam report.}

In the $2+2$ system\index{2+2 Systems} the space is divided as in Fig.~\ref{jacobi_sphere_2+2}. Fig.~\ref{jacobi_sphere_2+2_2} shows how the space and the planes move as one changes the mass in one of the components. The intra-species interactions are set to zero and the inter-species interactions are hold equally strong, that is $g_{12}=g_{34}=0$ and $g_{13}=g_{14}=g_{23}=g_{24}=g$. Accordingly, the masses are defined as $m_1=m_2=\alpha$ and $m_3=m_4=\beta$. Since the quantum statistics are going to play an important role in the found solutions, I will introduce a notation such as 2b+2f -- indicating that there are 2 bosons each with mass $\alpha$ and 2 fermions each with mass $\beta$. In the same manner 2f+2b indicates the same but reversed masses. At all times the last mentioned particles are the ones that I will vary in mass, that is $\beta$. To proceed further, I will define spherical coordinates\index{Hyperspherical Coordinates}, $\rho$, $\phi$ and $\theta$ as: $x=\rho~\sin\theta \cos\phi$, $y=\rho~\sin\theta \sin\phi$, $z=\rho~\cos\theta$, where $\rho\in[0,\infty[$, $\phi\in[0,2\pi[$ and $\theta\in[0,\pi]$. This transformation reduces the Hamiltonian into the following form (remember, center-of-mass is separated),
\begin{align}
\mathcal{H}_r=\frac{1}{2}(\rho^2-\Delta_{sph})+G\sum_{\pm}\delta[\cos\theta\pm \sin(\phi \pm \xi)\sin\theta],
\end{align}
where $G\equiv \frac{g}{\rho} \frac{2^{3/4}}{(\alpha+\beta)^{1/4}}$ is an `overall strength' and $\xi\equiv \mathrm{atan}(\sqrt{{\beta}/{\alpha}})$ is the interaction `plane angle'. Note $G$ is again $\rho$ dependent and therefore the method is only valid when $G\rightarrow\infty$ as here the wave function must be zero on the planes. Note that if the masses are the same, i.e. $\alpha=\beta$, then $\xi=\pi/4$. The volume element transforms accordingly as: $\mathrm{d}x\mathrm{d}y\mathrm{d}z\rightarrow\rho^2\sin(\theta)\mathrm{d}\rho \mathrm{d}\phi \mathrm{d}\theta$, and the Laplacian is,
\begin{align*}
\Delta_{sph}=&\left( -\frac{1}{\rho^2} \frac{\partial}{\partial \rho}\left(\rho^2 \frac{\partial}{\partial \rho} \right) - \frac{1}{\rho^2} \Delta_{ang} \right)\\
=&{1 \over \rho^2}{\partial \over \partial \rho}\!\left(\rho^2 {\partial \over \partial \rho}\right)
\!+\!{1 \over \rho^2\!\sin\theta}{\partial \over \partial \theta}\!\left(\sin\theta {\partial \over \partial \theta}\right)
\!+\!{1 \over \rho^2\!\sin^2\theta}{\partial^2 \over \partial \phi^2}.
\end{align*}
The solution to the above differential equation with the harmonic trap is the well-known 3D harmonic oscillator states:
\begin{align*}
\psi_{\nu,\tau,\eta}(\rho,\theta,\phi)_{relative}=R_{\nu,\tau}(\rho)Y_{\tau\eta}(\theta,\phi)=N_{\nu \tau}\rho^le^{-\rho^2/2}L_{(\nu-\tau)/2}^{\tau+1/2}(\rho^2)Y_{\tau\eta}(\theta,\phi),
\end{align*}
in units of harmonic length $b=\sqrt{\frac{\hbar}{\mu\omega}}$, where $L_{(\nu-\tau)/2}^{\tau+1/2}(\rho^2)$ are the usual Laguerre polynomials\index{Laguerre polynomials} and $Y_{\tau\eta}(\theta,\phi)$ are the harmonic spherical solutions to the eigenvalue problem in case of $G=0$. However, since the space is restricted by the delta-interactions, at $G\rightarrow\infty$ another set of solutions can be found in the new angular dependent functions. For this reason I will replace $Y_{\tau\eta}(\theta,\phi)$ by $f_{\tau\eta}(\theta,\phi)$ as the new angular function, which must satisfy the eigenvalue problem,
\begin{equation}
\Delta_{ang}f=\tau(\tau+1)f,
\end{equation}
with the $\delta$-boundary planes. The corresponding eigen-energies are given as $E_{hyperradius}+E_{hyperangular}+3/2=(2\nu+\tau+3/2)$, with $\nu=0,1,\dots$ and $\tau$ to be determined later. $\eta$ is just another label to distinguish any possible degeneracy in the angular part. In the non-interacting case $\eta$ corresponds to the magnetic quantum number, $m_l$ and $\tau$ is the orbital quantum number, usually called $l$. The total energy of the four-particle system is given as:
\begin{align*}
E_{tot}&=E_{CM}+E_{hyperradius}+E_{hyperangular}+3/2\\
&=(\chi+1/2)+(2\nu+\tau+3/2),
\end{align*}
with $\chi,\nu=0,1,2,\dots$. In the following sections I will go through the details of finding $\tau$ in the red region.

\noindent
In order to find $f$ in the red region a set of transformations are needed. Due to the delta-boundary planes no general and full analytical solutions exist in the mass-imbalanced case at the moment, so what I will do is to project the red region into a two-dimensional area, which then can be solved by a set of complete basis. For this I perform the following two-step transformations: step i) $a=\cos\phi~\tan\theta$, $
b=\sin\phi~\tan\theta$, and step ii) $\lambda=a~\sin\xi - b~\cos\xi$, $\gamma=a~\sin\xi + b~\cos\xi$. In terms of the original coordinates, $q_i$, the last variables are given as: $\lambda=\sqrt{1+\frac{\beta}{\alpha}}\cdot\frac{ \sqrt{{\alpha}/{\beta}}({q_1-q_2})\sin\xi-({q_3-q_4})\cos\xi}{q_1+q_2-q_3-q_4}$ and $\gamma=\sqrt{1+\frac{\beta}{\alpha}}\cdot\frac{ \sqrt{{\alpha}/{\beta}}({q_1-q_2})\sin\xi+({q_3-q_4})\cos\xi}{q_1+q_2-q_3-q_4}$.
In the final form the equation for the angular part reads:
\begin{equation}
\begin{aligned}
\Delta_{ang}f=&\Big(1+&\frac{\lambda^2+\gamma^2+2\lambda\gamma \cos(2\xi)}{\sin(2\xi)^2}\Big) \cdot \Big\{(1+\lambda^2){\partial^2 f \over \partial \lambda^2}+(1+\gamma^2){\partial^2 f \over \partial \gamma^2}\\ &&+(2\lambda\gamma-2\cos(2\xi)){\partial^2 f \over \partial \lambda\partial \gamma}+2\lambda{\partial f \over \partial \lambda}+2\gamma{\partial f \over \partial \gamma}\Big\}.
\end{aligned}
\label{2+2_red}
\end{equation}
Notice that the radial part is chosen to satisfy the following normalization condition: $\int\left|R(\rho)\right|^2 \frac{\mu}{\sqrt{M_{1234}}}\mathrm{d}\rho=1$, while the angular part is chosen to satisfy $\int|f|^2 \sin(\theta)\mathrm{d}\phi \mathrm{d}\theta=1$. To implement the latter condition in terms of $a$ and $b$ one should transform the volume element accordingly: $\sin(\theta)\mathrm{d}\phi \mathrm{d}\theta\rightarrow \frac{1}{(1+a^2+b^2)^{3/2}}\mathrm{d}a \mathrm{d}b$. But this volume has to be transformed once more in terms of $\lambda,\gamma$: $\rightarrow {\sin(2\xi)^2}/{[\sin(2\xi)^2+\lambda^2+\gamma^2+2\lambda\gamma \cos(2\xi)]^{3/2}} \mathrm{d}\gamma \mathrm{d}\lambda$. These transformations lead to a very simple boundary conditions, namely that the wave function must vanish at $\lambda=\pm1$ or $\gamma=\pm1$. This is a square, which can be expanded in a complete basis in form of a Fourier series: 
\begin{equation} f_\tau(\lambda,\gamma)= \sum_{n,m} C_{n,m}~\sin\left[\frac{\pi n}{2}(\lambda-1)\right]~\sin\left[\frac{\pi m}{2}(\gamma-1)\right]\; ,
\end{equation} 
defined on a square $\lambda,\gamma\in\{-1,1\}$. Notice that the transformations only work on the top red region of the coordinate space and hence the obtained results are valid only for the $AABB$ or $BBAA$ combinations. This choice was made, when I chose to introduce spherical coordinates, where $z=\rho cos(\theta)$. For the red region $cos(\theta)>0$ for all $\theta$. When I introduced the $(a,b)$-coordinates I divide by $cos(\theta)$ which is allowed because $cos(\theta)>0$. If one wants to investigate for example the green area, then one must make a rotation of the coordinate system in such a way that when spherical coordinates are introduced the $cos(\theta)>0$ for all $\theta$.

\begin{figure}[t]
\centering
\includegraphics[width=\columnwidth]{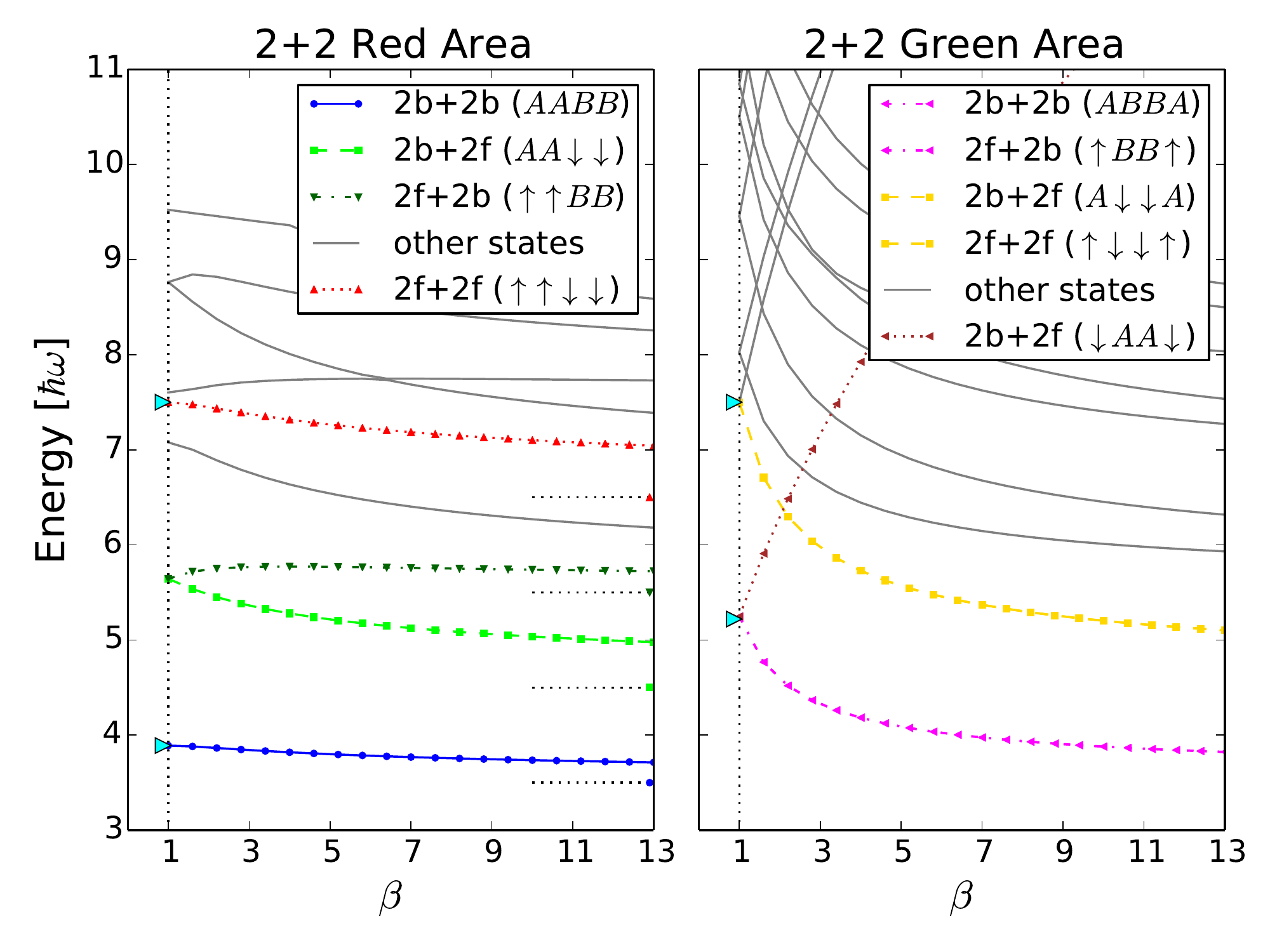}
\caption{Energy plot without the center-of-mass energy, that is the angular quantum number, $\tau$ plus $3/2$, found by the sine-function expansion. Left panel: shows numerically found values for 2+2 system in the red region as a function of $\beta$, which is the mass of the $B$ particles. The mass of $A$ particles is set to be one, $\alpha=1$. Right panel: shows the same graph but for the green region. The asymptotic value for $\beta\rightarrow\infty$ is shown with dashed lines for some states. The cyan triangles at $\beta=1$ are the numerically exact calculated energies obtained with EEDM (See Chapter 7).}
\label{energy_as_mass_2+2}
\end{figure}

\subsection{Red Region}
Focusing back on the red region, one can find $C_{n,m}$ coefficients by inserting the expansion into Eq.~(\ref{2+2_red}) and then use the fact that the basis is orthonormal, i.e. $\int_{-1}^{1} \sin[{\pi n}/{2}(\lambda-1)]~\sin[{\pi m}/{2}(\gamma-1)]=\delta_{nm}$. This produces a simple matrix eigenvalue problem whose eigenvectors and eigenvalues are $C_{n,m}$ and $\tau(\tau+1)$, respectively. In my calculations, I let $n$ and $m$ to run up to 30, and with this basis cut, the energy for the five lowest states were converged up to 3rd decimal. This approximation is the reason I call it semi-analytical approach to the four-body problem. After the set of $C_{n,m}$'s are established, all information about the system is obtained. The total wave function with quantum numbers $\chi=0$ and $\nu=0$ is therefore given as:
\begin{align}
\Psi(q_1,q_2,q_3,q_4)=N\rho^\tau e^{-(\rho^2+R^2)/2}f_{\tau}(\lambda,\gamma).
\end{align}

\begin{figure}[t]
\centering
\includegraphics[width=\columnwidth]{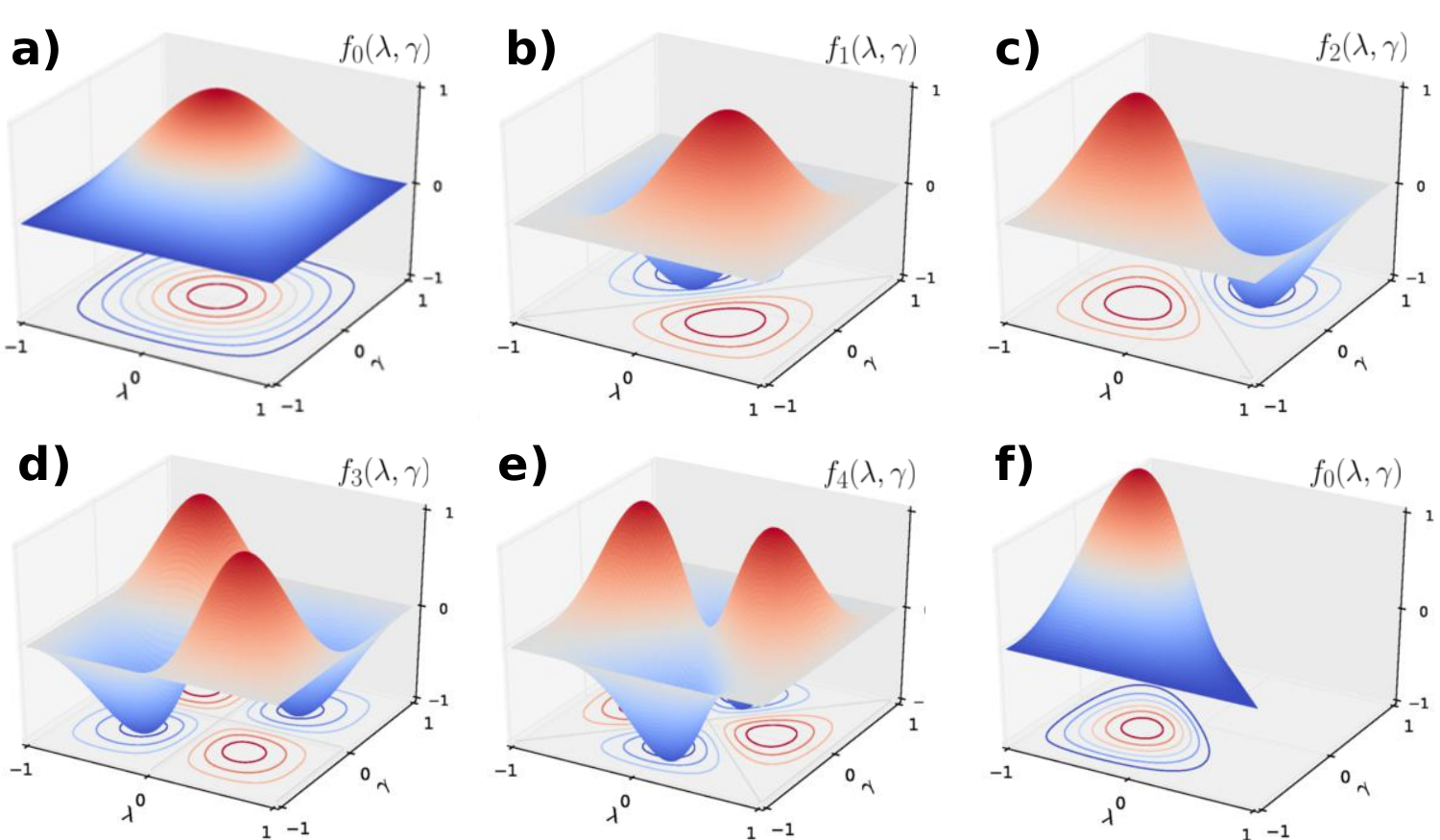}
\caption{a)-e) shows the different calculated eigenfunctions for the quadratic region in the $\lambda,\gamma$-space. Different symmetries are found and recognized with the different two-component four particles. a) shows the bosonic $2b+2b$ ($AABB$) configuration. b) is the $2b+2f$ ($AA\downarrow\downarrow$) and c) is the $2f+2b$ ($\uparrow\uparrow BB$) configuration. d) does not satisfy any symmetry that is relevant for our system. e) is the fully antisymmetric case, namely the $2f+2f$ ($\uparrow\uparrow\downarrow\downarrow$). f) shows the lowest state in a triangle\index{Spherical Triangle} instead of a square, which can also be solved in the same manner. This triangular region is used to solve the $1+1+1+1$ system, which is discussed later in this chapter.}
\label{fourparticles_angular2D_eigenstates}
\end{figure}

Numerically found values of $\tau+3/2$ are plotted in Fig.~\ref{energy_as_mass_2+2}. By looking at the symmetries of angular wave functions one can determine which system can be described with such solutions. Among all these eigenfunctions, one can recognize bosonic, fermionic or mixture symmetry between the particles. This gives the four possible combinations: 2b+2b, 2b+2f, 2f+2b and 2f+2f, which are labeled in Fig.~\ref{energy_as_mass_2+2}. Notice that the 2f+2b and 2b+2f are two independent systems if $\beta\geq1$, otherwise, due to symmetry, they are the same system if one allows $\beta\in ]0,\infty[$. Allow me to remind the reader that $\beta$ is the notation for the mass of the 3rd and 4th particle, while $\alpha\equiv1$ is the mass of the 1st and 2nd particle. However, in the following I will only vary $\beta$ from $1\rightarrow\infty$. Notice that just like three-body system the degeneracy present at $\beta=\alpha$ in Fig.~\ref{energy_as_mass_2+2} breaks immediately as soon as $\beta>1$. This degeneracy however appears later for some other ratios in the higher excited states. The limiting case of $\beta\rightarrow\infty$ is also shown in the figure. For example in the $2b+2b$ case the limiting value on Fig.~\ref{energy_as_mass_2+2} is $3.5\hbar\omega$. The total energy with the center of mass energy is then $4\hbar\omega$. This energy corresponds exactly to the fact that the two heavy bosons sit in the middle of the condensate, that is Gaussian with energy $2\cdot 1/2\hbar\omega$, and the other two light bosons, which interact strongly with the heavy ones, sit antisymmetric like the 1st excited state in a harmonic oscillator potential, with energy $2\cdot 3/2\hbar\omega$, and hence a total energy of $4\hbar\omega$.

The two-dimensional eigenvectors in the $(\lambda,\gamma)$-space are shown in Fig.~\ref{fourparticles_angular2D_eigenstates}. Here one can see how the different symmetries appear, which can be associated to the different bosonic, fermionic and mixture solutions. For instance, whenever the wave function is zero at $\lambda=\gamma$ and/or $\lambda=-\gamma$, a fermionic antisymmetric wave function can be related to these solutions. \\

\begin{figure}[t]
\centering
\includegraphics[width=1.05\columnwidth,trim=2cm 0cm 0cm 0cm]{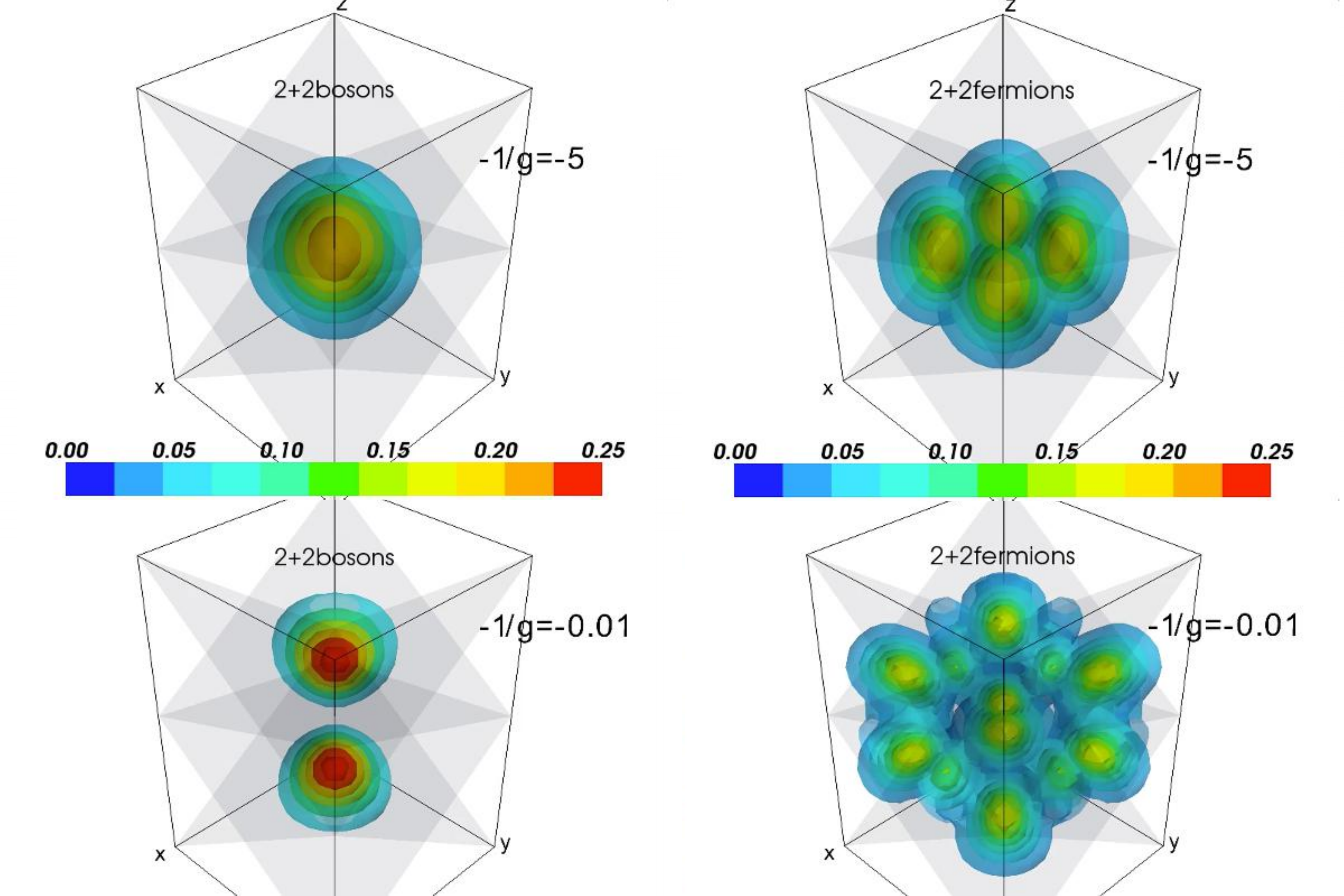}
\caption{The bosonic and fermionic $2+2$ ground state in the Jacobi space obtained numerically with the EEDM. Left panels: the non- and strongly interacting $2b+2b$ particles. The bosons start to occupy only the red region from Fig.~\ref{jacobi_sphere_2+2}. Right panels: the same as left but this time for $2f+2f$ fermions.}
\label{numerical_contour_2+2}
\end{figure}

\noindent
In Fig.~\ref{numerical_contour_2+2} I have plotted the wave function in the Jacobi space. The results here are obtained by the EEDM and not by the analytical method. The EEDM is explained in more details in Chapter 7. The point with the figure is to show how the bosonic and fermionic ground states occupy the different regions as the interaction strength grows. It is notable that the bosonic state only occupies the red region, while the fermionic state distributes itself everywhere when it is strongly interacting. This can be justified by arguing that the red region is the biggest region for this case, while all the regions are equally big in the fermionic case for equal masses. One can also compare the similarities in the red bosonic region in Fig.~\ref{numerical_contour_2+2} with Fig.~\ref{fourparticles_angular2D_eigenstates} a).\\

\subsection{Results - 2+2 Systems}
Having established the correct symmetries and therefore found the correct wave functions and energies in all regions, one can talk about the densities and how the particles align themselves in the ground state as the mass of the particles changes. The analysis for excited states with this method is also possible, however, in the following I am only focusing on the ground states.\\

\noindent
$\mathbf{2b+2b}$ \index{2+2 Systems!2b+2b Systems} denotes the symmetric case of two-component bosons. When the masses are all equal the configuration $AABB$ (or $BBAA$) is highly favored, which is due to the big volume (the red region in Fig.~\ref{jacobi_sphere_2+2}) that allows the wave function to expand itself as much as possible and therefore reducing the energy of the system. Furthermore, the density and pair-correlation plots in Fig.~\ref{density_2b+2f} support this fact as $A$ particles are in one side and $B$ particles in the other side. As the mass of the $B$ particles grows (denoted by $\beta$), the green area of configuration $ABBA$ in the Jacobi space also grows. The evolution of the coordinate space is illustrated in Fig.~\ref{jacobi_sphere_2+2_2} for $\beta=1/5, 1, 5, 100, \infty$. This makes the region also favorable in that limit, which also makes sense because of the heavy particles moving into the middle. Therefore, in the limit of $\beta\rightarrow\infty$ the system will be doubly degenerated with $ABBA$ and $BBAA$ (or $AABB$) being the ground state configurations.\\

\begin{figure}[t]
\center
\includegraphics[width=\columnwidth]{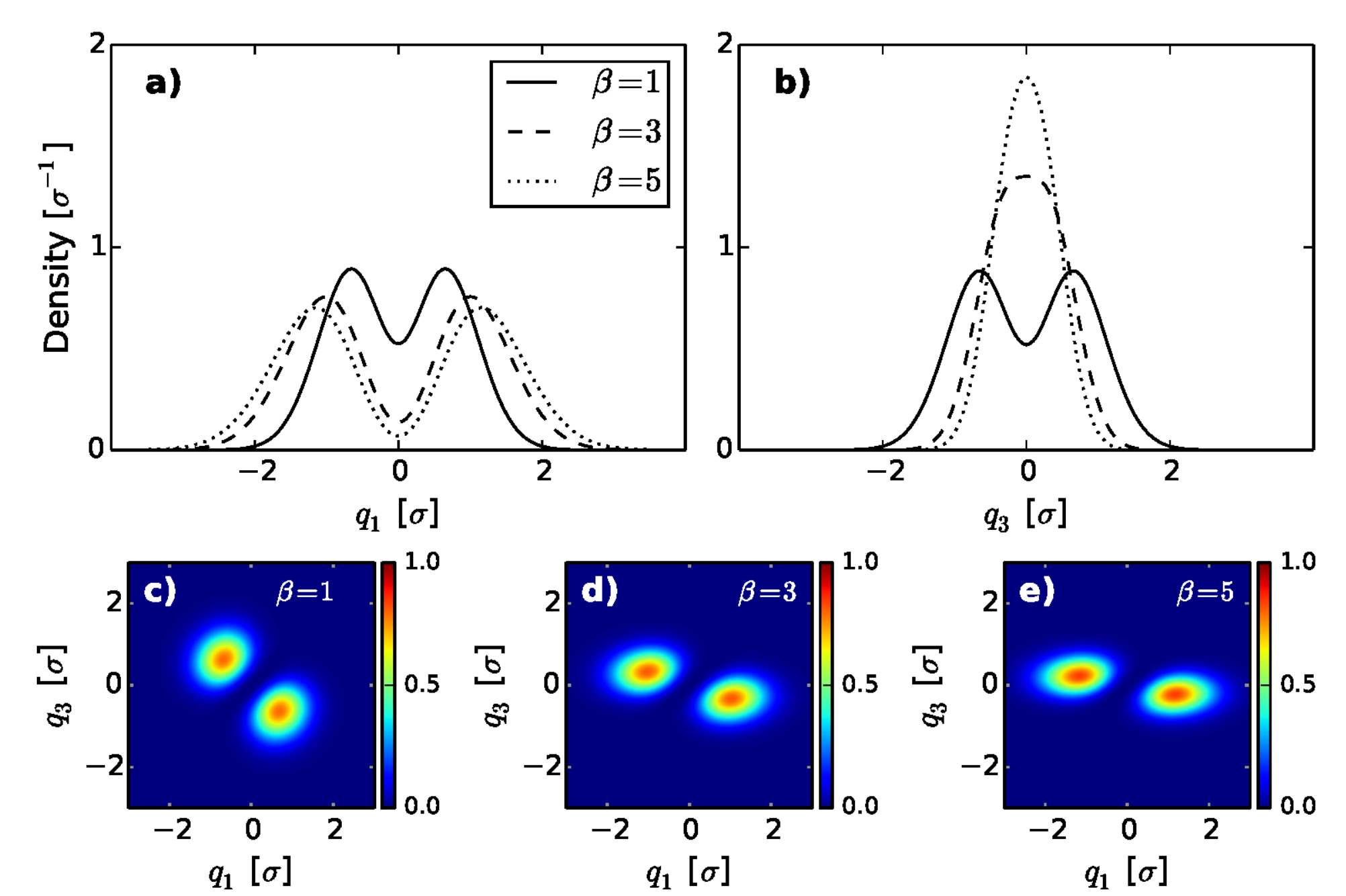}
\caption{Density and paircorrelation for $\mathbf{2b+2b}$ ($BAAB$) system. {\bf a)} and {\bf b)} show the density for respectively $A$ and $B$ particles with different masses $\beta=1,3,5$. {\bf c)}, {\bf d)} and {\bf e)} show the corresponding paircorrelation of the system. Figure is adapted from \cite{DehkharghaniJoPBAMaOP2016}.}
\label{density_2b+2b}
\end{figure}

\noindent
Results for $\mathbf{2f+2b}$ \index{2+2 Systems!2f+2b Systems}systems show that it is favorable for the system to be in a $\uparrow BB\uparrow$ configuration for $\beta>1$ as the heavy particles want to move to the middle, while the $\uparrow\uparrow BB$ is suppressed and not favored due to the Pauli principle that increases the energy for the system because of the two fermions sitting right next to each other. In contrast to $2b+2b$ system, the $2f+2b$ system is therefore only favored as $\uparrow BB\uparrow$ configuration for $\beta>1$. The density and paircorrelation plots for this configuration can be found in \cite{DehkharghaniJoPBAMaOP2016}.

\begin{figure}[t]
\center
\includegraphics[width=\columnwidth]{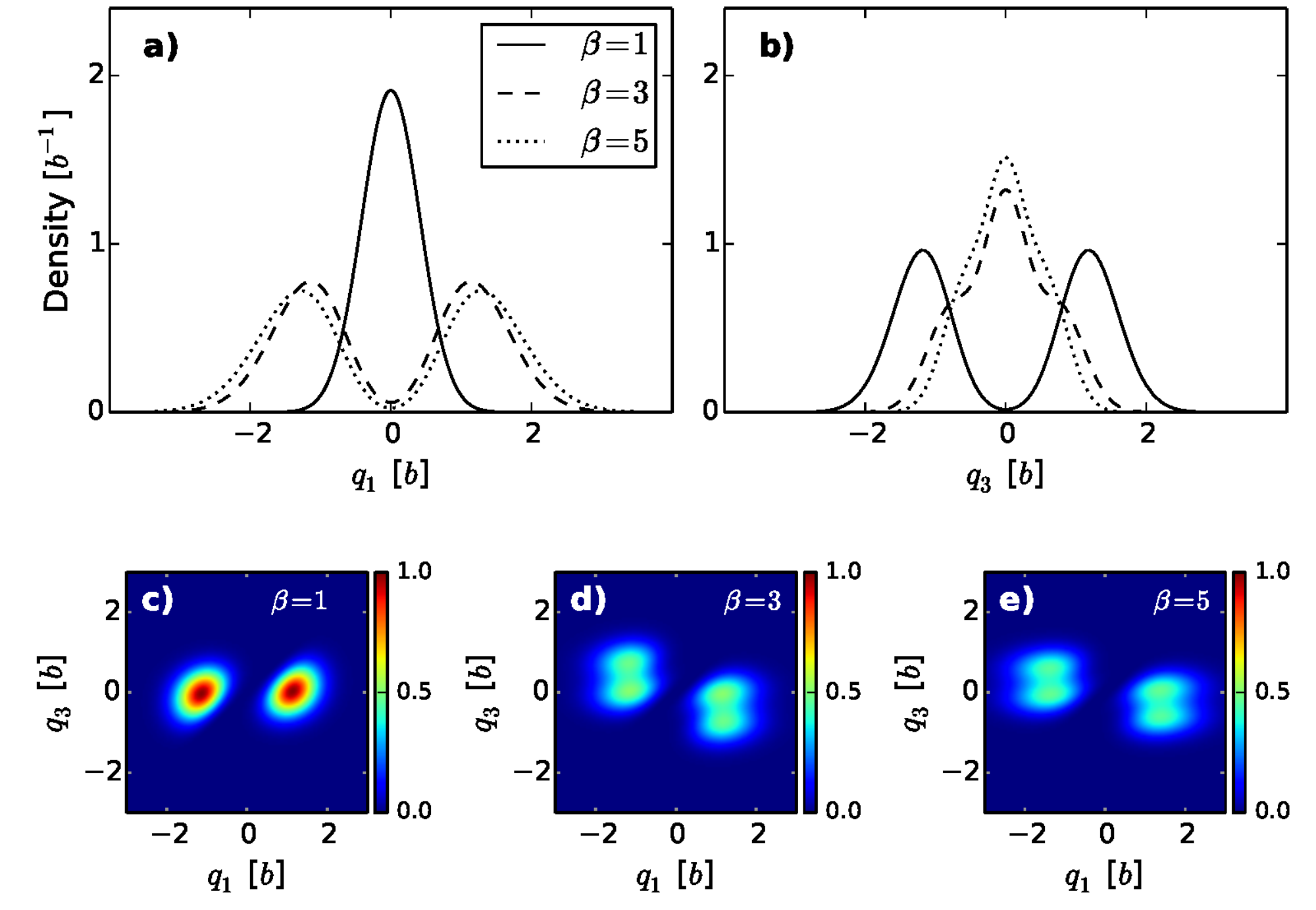}
\caption{Density and paircorrelation for $\mathbf{2b+2f}$ ($\downarrow AA\downarrow$ to $\downarrow \downarrow AA$) system. {\bf a)} and {\bf b)} show the density for respectively $A$ and $\downarrow$ particles with different masses $\beta=1,3,5$. {\bf c)}, {\bf d)} and {\bf e)} show the corresponding paircorrelation of the system. The transition happen at $\beta\approx 1.3$. Figure is adapted from \cite{DehkharghaniJoPBAMaOP2016}.}
\label{density_2b+2f}
\end{figure}

On the other hand, I have plotted the results for the $\mathbf{2b+2f}$ in Fig.~\ref{density_2b+2f} adopted from \cite{DehkharghaniJoPBAMaOP2016}, because of the interesting fact that the system goes from a $\downarrow AA \downarrow$ ordering to $\downarrow\downarrow AA $ configuration as the mass increases.   Further investigation shows that the critical mass where the transition from $\downarrow AA \downarrow$ ordering to $\downarrow\downarrow AA $ happens at a mass ratio $\beta\approx1.3$ and this is due to two facts: one, the Pauli exclusion principle and second, the external trap. After this mass ratio the Pauli principle is suppressed by the mass of the fermionic particles, as it costs more energy to move the particles to the sides than force them to stay close to each other.\\

\begin{figure}[t]
\center
\includegraphics[width=\columnwidth]{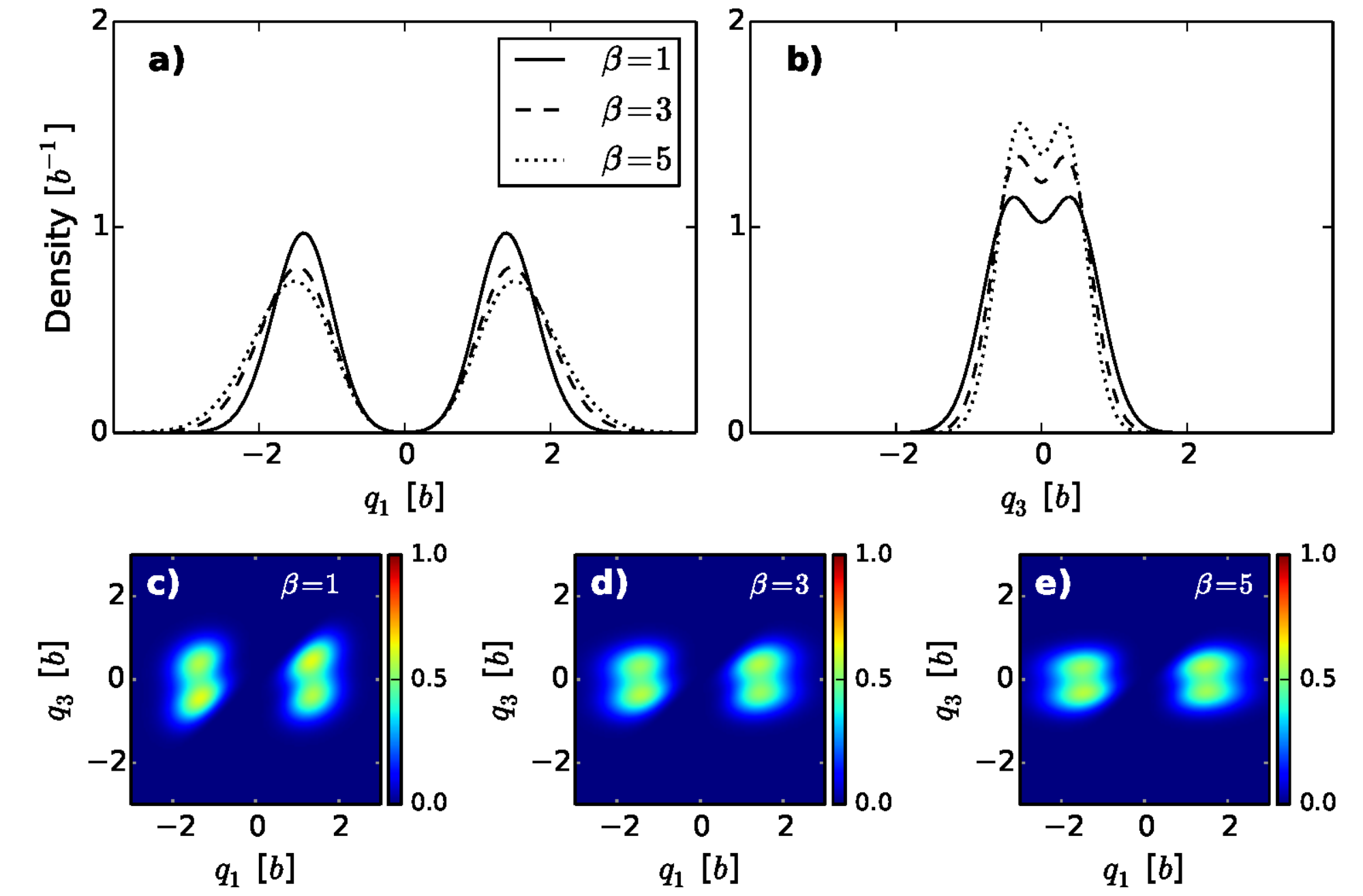}
\caption{Density and paircorrelation for $\mathbf{2f+2f}$ ($\uparrow\downarrow\downarrow\uparrow$) system. {\bf a)} and {\bf b)} show the density for respectively $\uparrow$ and $\downarrow$ particles with different masses $\beta=1+\epsilon,3,5$. {\bf c)}, {\bf d)} and {\bf e)} show the corresponding paircorrelation of the system. Figure is adapted from \cite{DehkharghaniJoPBAMaOP2016}.}
\label{density_2f+2f}
\end{figure}

\noindent
For the $\mathbf{2f+2f}$ \index{2+2 Systems!2f+2f Systems}systems the configuration $\uparrow\downarrow\downarrow\uparrow$ is the most favored one for $\beta>1$ as shown in Fig.~\ref{density_2f+2f}. This is not surprising because the volume in the Jacobi space is the biggest here for this configuration. I should highlight that the volume spaces for $\beta$ exactly one are equally big in the Jacobi sphere and therefore one would find the fermions in all regions and configurations with some weights that can be calculated as in \cite{VolosnievNC2014, LoftCPC2016}. But as soon as you make $\beta>1$ only one configuration is favorable and that is shown here in Fig.~\ref{density_2f+2f}.\\

\noindent
It should be noted that in the (semi)-analytical results I use the $4$-body wave function, $\psi(q_1,q_2,q_3,q_4)$, to obtain the density of the system, for example to obtain the density for particle $q_1$ one calculates the following: $n_A(q_1)=\int{|\psi|^2\mathrm{d}q_{2}\mathrm{d}q_{3}\mathrm{d}q_{4}}$. The pair-correlation function for a $A$-type-$B$-type pair is defined as follows: $n_{AB}(q_1,q_3)=\int{|\psi|^2 dq_{2} dq_{4}}$. The latter reveals the probability of finding a $B$-type particle at $q_3$ when the position, $q_1$, of a $A$-type particle is given. Similarly, even though not shown here, one can find the momentum distribution by defining Fourier transform\index{Fourier Transform} of the wave function as $\psi(p_1,...,p_{4})=({1}/{\sqrt{2\pi}})^{4}\int{\psi(q_1,...,q_{4})e^{ip_1q_1}... e^{ip_{4}q_{4}}}dq_1... dq_{4}$ where $p_i$ is the momentum of the particle. The momentum distribution is interesting from the experimental point of view, which can be found in \cite{DehkharghaniJoPBAMaOP2016}.\\

\noindent
What I have not commented on is the yellow region for the $2+2$ system in Fig.~\ref{jacobi_sphere_2+2}. This was intentionally avoided due to the geometry of the volume. As it is clear, if one did a projection onto a 2D region, the region would appear as a triangle. Later in this chapter, I will discuss how one can solve these kind of triangular regions, when I analyze the $1+1+1+1$ system.

\section{Interpolatory Ansatz\index{Interpolarory Ansatz} for four particles}\label{ch:fourparticles:sec:interpolatory}
The obtained knowledge about the wave function in the strong interacting limit for the $2+2$ systems can be used to say something about the intermediate regime. Specifically, it could be interesting to see how the $2+2$ systems behave in the intermediate region by using the interpolatory method as I discussed back in Chapter 2. However, after the success in the $2+1$ system it is now very interesting to find out whether or not the interpolatory method can be used in such a correlated system like the $2+2$, which has more than one particle in each component.\\

\begin{figure}[h!]
\center
\includegraphics[width=0.85\columnwidth, trim=0cm 0.4cm 0cm 0cm]{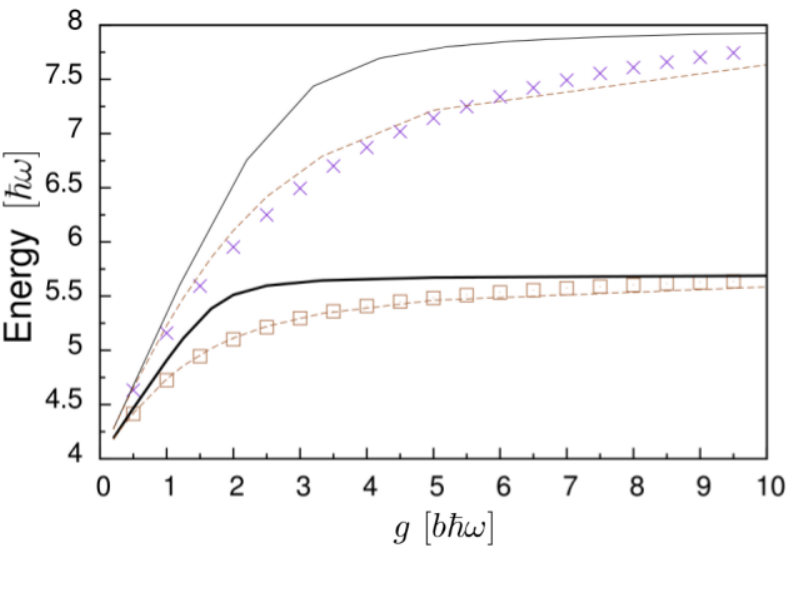}
\caption{Ground-state energy for the $2f+2f$ system as a function of $g$. Solid thin and solid thick lines show the energies for $\beta=1$ and $\beta=10$ calculated by the initial variational ansatz. Crosses ($\beta=1$) and squares ($\beta=10$) show numerically exact diagonalization of the Hamiltonian. Dashed lines are the corresponding energies predicted by the modified ansatz. Figure is adapted from \cite{2017arXiv170308720P}}
\label{interpolatory_2+2_energies}
\includegraphics[width=0.85\columnwidth,trim=0cm 0.4cm 0cm -0.2cm]{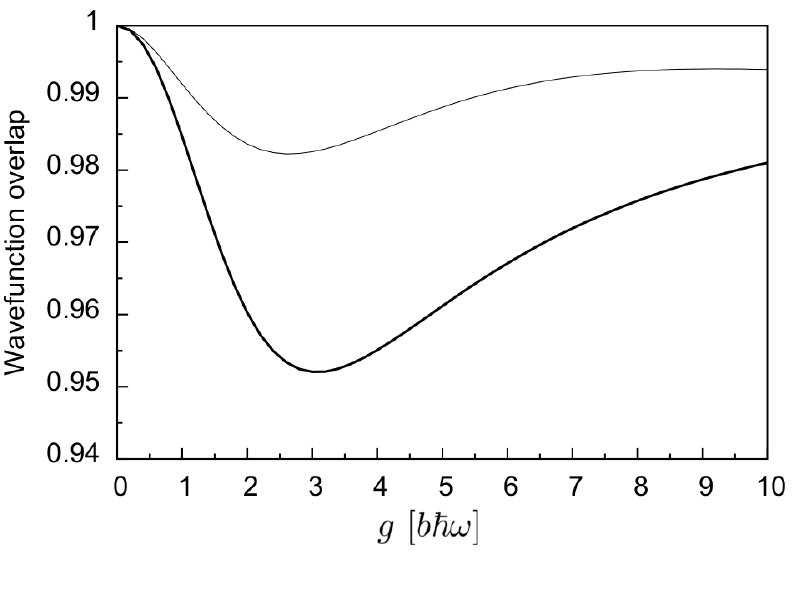}
\caption{The overlap between the interpolatory ansatz and numerically exact calculated ground state.
Thin line is for $\beta=1$ and Thick for $\beta=10$. The limiting cases are by construction 1, while for heavy masses the fidelity is damped a little bit. Figure is adapted from \cite{2017arXiv170308720P}}
\label{interpolatory_2+2_wavefunction}
\end{figure}

\noindent
This work was conducted in collaboration with T.~Sowi{\'n}ski and D.~P\k{e}cak in Warsaw. Fig.~\ref{interpolatory_2+2_energies} shows the energy spectrum for the $2f+2f$ fermionic system as a function of interaction strength, $g$. The energies found via the initial interpolatory ansatz and modified ansatz are compared with numerical exact calculations in \cite{2017arXiv170308720P,PecakNJoP2016}. We actually did compare the energies for pure bosons, $2b+2b$ as well as for a mixture $2b+2f$, and found excellent agreement in both cases. However, the pure fermionic case was the worse case among these system, which we decided to investigate further. As the figure shows, the modified ansatz is much better at capturing the energy as expected. It is also surprising to see how well the interpolatory reproduces the whole range of the ground states.\\

\noindent
It is naturally easy to say that variational methods are by construction built in such a way that they capture the energy, but how about the wave function? Can the variational method also reproduce the wave function of the system? To answer this question, we took the overlap between the interpolatory ansatz and the numerically calculated wave function. This was done for both equal masses and a mass ratio of $\beta=10$. The result for this calculation is shown in Fig.~\ref{interpolatory_2+2_wavefunction}, and here it is again clear that the interpolatory ansatz, despite the small deviations, actually does very good. The deviation is worse for $\beta=10$. This can be explained by the construction of the variational wave function. Since the variational method only depends on the limiting cases ($g=0$ and $g\rightarrow \infty$), where the heavy particles are in the middle of the trap, the light particles are spatially much more spread out from the middle of the harmonic trap. Hence, it requires more single-particle orbitals to reconstruct the true wave function in this particular case.

\begin{figure}[t]
\center
\includegraphics[width=0.85\columnwidth]{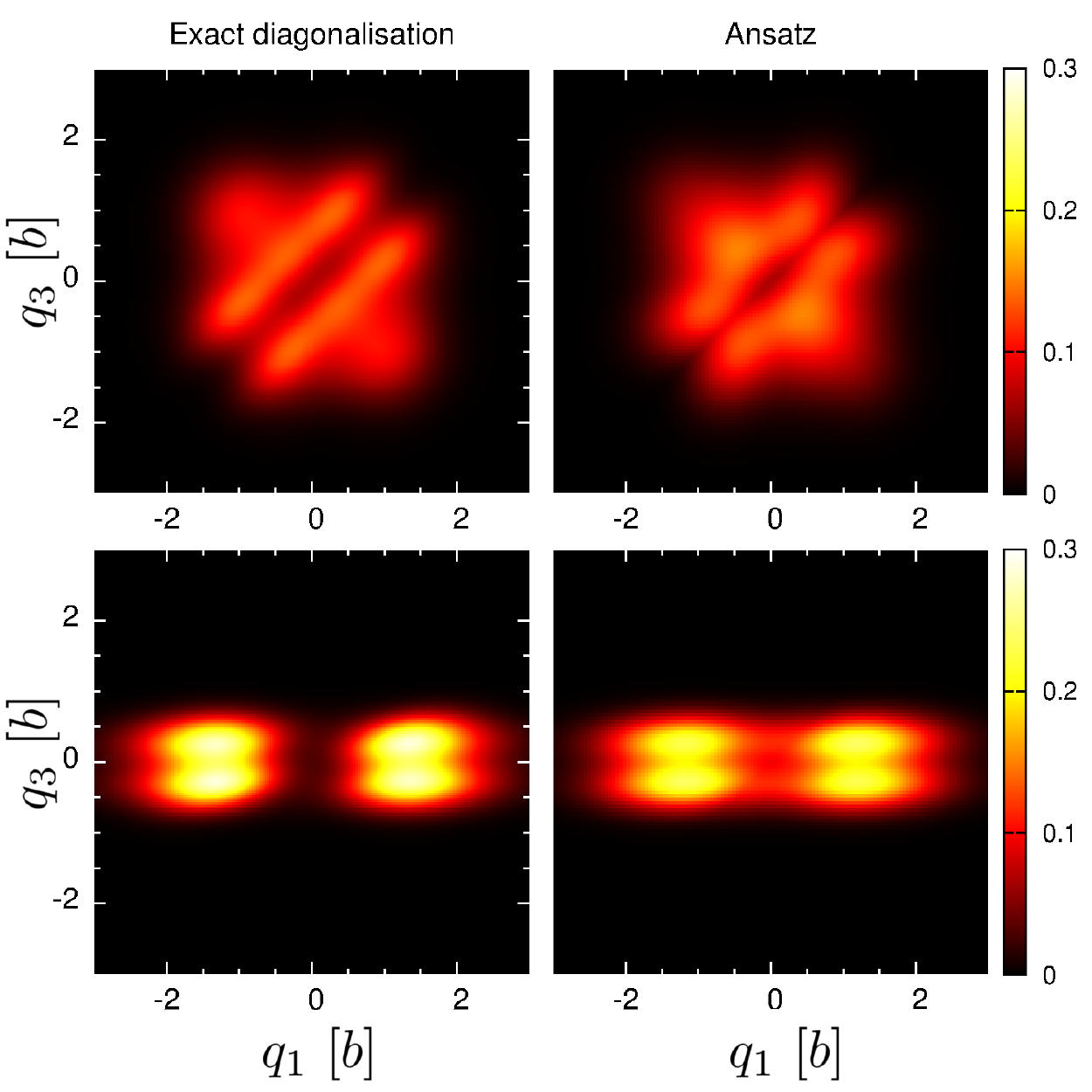}
\caption{Pair-correlation plot for the $2f+2f$ system. Upper panels are for $\beta=1$ and lower panels for $\beta=10$. The overall pattern is captured and reconstructed by the variational method, as seen on the right panels. Lower panels also reveal a $\uparrow\downarrow\downarrow\uparrow$ ordering, which was already discussed and commented in Fig.~\ref{density_2f+2f}. In that figure the ordering was already apparent at a mass ratio of $\beta=5$. Figure is adapted from \cite{2017arXiv170308720P}.}
\label{interpolatory_2+2_paircorrelation}
\end{figure}

Finally, Fig.~\ref{interpolatory_2+2_paircorrelation} shows the pair-correlation plot for the interpolatory ansatz and the numerically exact calculated results. Again, it is clear that the ansatz has captured the overall pattern in the system. Upper panels show the pair correlation for $\beta=1$ calculated by the exact diagonalization method (left) and the variational ansatz (right). At this mass ratio the solution is a linear combination of all the possible configuration there is, as the Jacobi space is divided into 24 equally big regions. Lower panels in Fig.~\ref{interpolatory_2+2_paircorrelation} reveal a $\uparrow\downarrow\downarrow\uparrow$ ordering, which was discussed and predicted back in Fig.~\ref{density_2f+2f} for $\beta>1$.

\section{$\mathbf{1+1+1+1}$ Systems}\label{ch:fourparticles:sec:1111systems}
The $2+2$ systems are only one combination of the four-particle systems that I have analyzed earlier in this chapter. Another set of systems that are of particular interest are the $3+1$ and the $1+1+1+1$ systems. Next, I will show the details of calculating the energy of a $1+1+1+1$ system\index{1+1+1+1 Systems}, as the $3+1$ can be obtained in the same manner. As a side note, I can also reveal that the $3+1$ has been investigated in details by Matias Wallenius and Mathias Thomsen in their unpublished Bachelor's thesis and reports.

One of the biggest differences in the $1+1+1+1$ (and $3+1$) system from the $2+2$ system is that the $1+1+1+1$ system can only be mapped into a triangle, where as in the $2+2$ system the projection was always into a square (except the yellow region in Fig.~\ref{jacobi_sphere_2+2}), making it easy to expand in the Fourier expansion. It is worth noting, that the triangular region can also be solved as a square by only considering solutions that satisfy the triangular boundary condition where it vanishes on a diagonal of the square. However, in the following I will go in details on how exactly one can solve the triangular region. Let me elaborate on this: After arriving to the simplified Hamiltonian:
\begin{align*}
\mathcal{H}=&\frac{\hbar \omega}{2}(\mathbf{r}^2+\nabla^2)+\\
&\frac{1}{b}\frac{\sqrt{\mu}}{\mu_{12}\mu_{34}}\Big\{\frac{\mu_{12}^2\mu_{34}}{\mu}g_{12}\delta\big(x\big)+\frac{\mu_{12}\mu_{34}^2}{\mu}g_{34}\delta\big(y\big)+\\
&g_{13}\delta\Big(\sqrt{\frac{m_2(m_3+m_4)}{m_1M}}x-\sqrt{\frac{m_4(m_1+m_2)}{m_3M}}y+z\Big)+\\
&g_{14}\delta\Big(\sqrt{\frac{m_2(m_3+m_4)}{m_1M}}x+\sqrt{\frac{m_3(m_1+m_2)}{m_4M}}y+z\Big)+\\
&g_{23}\delta\Big(-\sqrt{\frac{m_1(m_3+m_4)}{m_2M}}x-\sqrt{\frac{m_4(m_1+m_2)}{m_3M}}y+z\Big)+\\
&g_{24}\delta\Big(-\sqrt{\frac{m_1(m_3+m_4)}{m_2M}}x+\sqrt{\frac{m_3(m_1+m_2)}{m_4M}}y+z\Big)\Big\},
\end{align*}
similarly to the $2+2$ system, one can separate the center-of-mass and end up with three coordinates. However, there are now 12 distinct regions as illustrated in Fig.~\ref{jacobi_sphere_1+1+1+1} instead of three different regions in the $2+2$ system (red, green and yellow in Fig.~\ref{jacobi_sphere_2+2}). Therefore one must solve each 12 regions individually. In the following I will focus on region 9, which goes along the x-axis in Fig.~\ref{jacobi_sphere_1+1+1+1}. The other regions can be solved in almost the same manner, however one should be careful when defining the hyper-spherical coordinates, as some singularities might occur if they are not chosen carefully. I will point this step out as I go through region 9.

\noindent
{\bf Region 9} with the $ACDB$ ordering is limited by the \{$g_{34}, g_{13},g_{24}$\}~$\delta$-planes. So the planes of interest, where the wave function has to be zero, are:
\begin{align*}
y&=0\\
\sqrt{\frac{m_2(m_3+m_4)}{m_1M}}x-\sqrt{\frac{m_4(m_1+m_2)}{m_3M}}y+z&=0\\
-\sqrt{\frac{m_1(m_3+m_4)}{m_2M}}x+\sqrt{\frac{m_3(m_1+m_2)}{m_4M}}y+z&=0.
\end{align*}
I choose to isolate $x$ in the above equations and introduce spherical coordinates with $x=\rho~cos(\theta)$, $y=\rho~sin(\theta) cos(\phi)$ and $z=\rho~sin(\theta) sin(\phi)$. This part is the crucial and important part, because in this way $cos(\theta)$ is positive for all $\theta$ in that region. If I was considering another region, at this part one should introduce the spherical coordinates in such a way that allows $cos(\theta)>0$ for all $\theta$.

\noindent
So isolating $x$ in the first step gives:
\begin{align*}
y&=0\\
x-a y+b z&=0\\
x-c y-d z&=0,
\end{align*}
with $a=\sqrt{\frac{(m_1+m_2) m_1 m_4}{(m_3+m_4) m_3 m_2}}$, $b=\sqrt{\frac{(m_1+m_2+m_3+m_4) m_1}{(m_3+m_4) m2}}$, $c=\sqrt{\frac{(m_1+m_2) m_2 m_3}{(m_3+m_4) m_1 m_4}}$ and $d=\sqrt{\frac{(m_1+m_2+m_3+m_4) m_2}{(m_3+m_4) m1}}$ and 
then with the spherical coordinates the equations become:
\begin{align*}
sin(\theta)cos(\phi)&=0\\
cos(\theta)-a~sin(\theta)cos(\phi)+b~sin(\theta)sin(\phi)&=0\\
cos(\theta)-c~sin(\theta)cos(\phi)-d~sin(\theta)sin(\phi)&=0,
\end{align*}
where I have neglected $\rho$ and taken it outside the delta-functions. By dividing with $cos(\theta)$ which is $>0$ on the region one obtains:
\begin{align*}
tan(\theta)cos(\phi)&=0\\
1-a~tan(\theta)cos(\phi)+b~tan(\theta)sin(\phi)&=0\\
1-c~tan(\theta)cos(\phi)-d~tan(\theta)sin(\phi)&=0.
\end{align*}
Accordingly, $\nabla^2$ has been transformed to the usual Laplacian in spherical coordinates, and one can separate $\rho$ from the delta-planes and only focus on the $\theta$ and $\phi$ angles. By introducing new coordinates: $\tilde{y}=y/x=tan(\theta)cos(\phi)$ and $\tilde{z}=z/x=tan(\theta)sin(\phi)$ one can show that:
\begin{align}
\tilde{y}&=0 \label{transformtilde1}\\
1-a~\tilde{y}+b~\tilde{z}&=0 \label{transformtilde2}\\
1-c~\tilde{y}-d~\tilde{z}&=0, \label{transformtilde3}
\end{align}
and correspondingly:
\begin{align}
\begin{split}
\Delta_{ang} f =&(1+\tilde{y}^2+\tilde{z}^2)\cdot\\ 
&\left\{ (1+\tilde{y}^2)\frac{\partial^2 f}{\partial \tilde{y}^2}+(1+\tilde{z}^2)\frac{\partial^2 f}{\partial \tilde{z}^2}+(2\tilde{y}\tilde{z})\frac{\partial^2 f}{\partial \tilde{y}\partial\tilde{z}}+(2\tilde{y})\frac{\partial f}{\partial \tilde{y}}+(2\tilde{z})\frac{\partial f}{\partial \tilde{z}} \right\}.
\end{split}
\end{align}
\noindent
Until this part, I have projected the spherical triangle into a 2D triangle, but this triangle is very irregular. Since I am interested in a isosceles right triangle with corners in $(-1,-1)$, $(-1,1)$ and $(1,-1)$, I need to transform the above set of equations once more. Therefore, I introduce some new coordinates:
\begin{align*}
\lambda&=\frac{2d}{(b+d)}(-a\tilde{y}+b\tilde{z})-\frac{(b-d)}{(b+d)}\\
\gamma&=\frac{2b}{(b+d)}(-c\tilde{y}-d\tilde{z})-\frac{(d-b)}{(b+d)},
\end{align*}
or inversely: $\tilde{y}(\lambda,\gamma)=-\frac{(b+d)}{2(ad+bc)}(\gamma+\lambda)$ and $\tilde{z}(\lambda,\gamma)=-\frac{a(b+d)}{2b(ad+bc)}(\gamma+\lambda)+\frac{(b-d)+(b+d)\lambda}{2bd}$. Notice that with this choice of $\lambda$ and $\gamma$ the delta-planes transform into:
\begin{align*}
\text{Eq.~(\ref{transformtilde1})} &\rightarrow \gamma=-\lambda\\
\text{Eq.~(\ref{transformtilde2})} &\rightarrow 1+\lambda=0\\
\text{Eq.~(\ref{transformtilde3})} &\rightarrow 1+\gamma=0,
\end{align*}
which is a triangle \index{Spherical Triangle}and correspondingly the differential parts transforms as:
\begin{align}
\begin{split}
&\Delta_{ang} f =\\&\textstyle\left( \left[1+\frac{(b-d)^2}{4b^2d^2}\right] +\lambda^2\left[\frac{(b+d)^2(c^2+d^2)}{4d^2(ad+bc)^2)}\right]+ \gamma^2\left[\frac{(b+d)^2(b^2+a^2)}{4b^2(ad+bc)^2)}\right]+\right.\\
&\textstyle\left.\lambda\left[\frac{c(b^2-d^2)}{2bd^2(ad+bc)}\right]+ \gamma\left[\frac{a(d^2-b^2)}{2db^2(ad+bc)}\right] +\lambda\gamma \left[\frac{(b+d)^2(bd-ac)}{2db(ad+bc)^2}\right] \right) \cdot\\
&\textstyle\left\{ \left[\left(\frac{2ad}{b+d}\right)^2+\left(\frac{2bd}{b+d}\right)^2+\left(\lambda+\frac{b-d}{b+d}\right)^2\right]\frac{\partial^2 f}{\partial {\lambda}^2}\right.\\ 
&\textstyle\left.
+\left[\left(\frac{2cb}{b+d}\right)^2+\left(\frac{2bd}{b+d}\right)^2+\left(\gamma+\frac{d-b}{b+d}\right)^2\right]\frac{\partial^2 f}{\partial {\gamma}^2}\right.\\ 
&\textstyle\left.
+2\left[\left(\frac{-2ad}{b+d}\right)\left(\frac{-2cb}{b+d}\right)+\left(\frac{2bd}{b+d}\right)\left(\frac{-2bd}{b+d}\right)+\left(\lambda+\frac{b-d}{b+d}\right)\left(\gamma+\frac{d-b}{b+d}\right)\right] \frac{\partial^2 f}{\partial {\gamma}\partial {\lambda}}\right.\\ 
&\textstyle\left.
+2\left[\left(\lambda+\frac{b-d}{b+d}\right)\right] \frac{\partial f}{\partial {\lambda}}\right.\\ 
&\textstyle\left.
+2\left[\left(\gamma+\frac{d-b}{b+d}\right)\right] \frac{\partial f}{\partial {\gamma}}\right\}.
\end{split}
\label{laplace1+1+1+1}
\end{align}
Despite being long, the terms have some kind of symmetric constants built inside, which reveals the actual symmetry that there is if the masses were the same. On the other hand, if the masses were different one can see how some terms are more weighted than the others.

\noindent
The above differential equation can be solved by making the following ansatz:
\begin{equation}
f(\lambda,\gamma)=\sum_{n,m}C_{n,m}\cdot h_{n,m}(\lambda,\gamma),
\end{equation}
where $n<m$ with $\{n,m\}\in\{1,2,...,N_{max}\}$ and:
\begin{align*}
\begin{split}
h_{n,m}(\lambda,\gamma)=\frac{1}{4}~\big\{
&e^{i\pi/2[(+n)(-\lambda-1)+(+m)(\gamma-1)]}
-e^{i\pi/2[(+n)(-\lambda-1)+(-m)(\gamma-1)]}\\
&~e^{i\pi/2[(-n)(-\lambda-1)+(-m)(\gamma-1)]}
-e^{i\pi/2[(-n)(-\lambda-1)+(+m)(\gamma-1)]}\\
-&~e^{i\pi/2[(+m)(-\lambda-1)+(+n)(\gamma-1)]}
+e^{i\pi/2[(+m)(-\lambda-1)+(-n)(\gamma-1)]}\\
-&~e^{i\pi/2[(-m)(-\lambda-1)+(-n)(\gamma-1)]}
+e^{i\pi/2[(-m)(-\lambda-1)+(+n)(\gamma-1)]}\big\}.
\end{split}
\end{align*}
Note that $h(-1,\gamma)=h(\lambda,-1)=h(\lambda,-\gamma)=0$ for $\forall n,m$, which is the triangle that was needed in the beginning with corners in $(-1,-1)$, $(-1,1)$ and $(1,-1)$ from $\gamma=-\lambda$; $1+\gamma=0$; $1+\lambda=0$. In addition, $h_{n,m}(\lambda,\gamma)$ is an orthonormal basis in the triangular-region and satisfy the following integral: $\int_{-1}^1\int_{-1}^{-\lambda}d\gamma d\lambda h_{n,m}(\lambda,\gamma) h_{n',m'}(\lambda,\gamma)=\delta_{n,n'}\delta_{m,m'}$.

\begin{figure}[hbt!]
\center
\includegraphics[width=0.93\columnwidth]{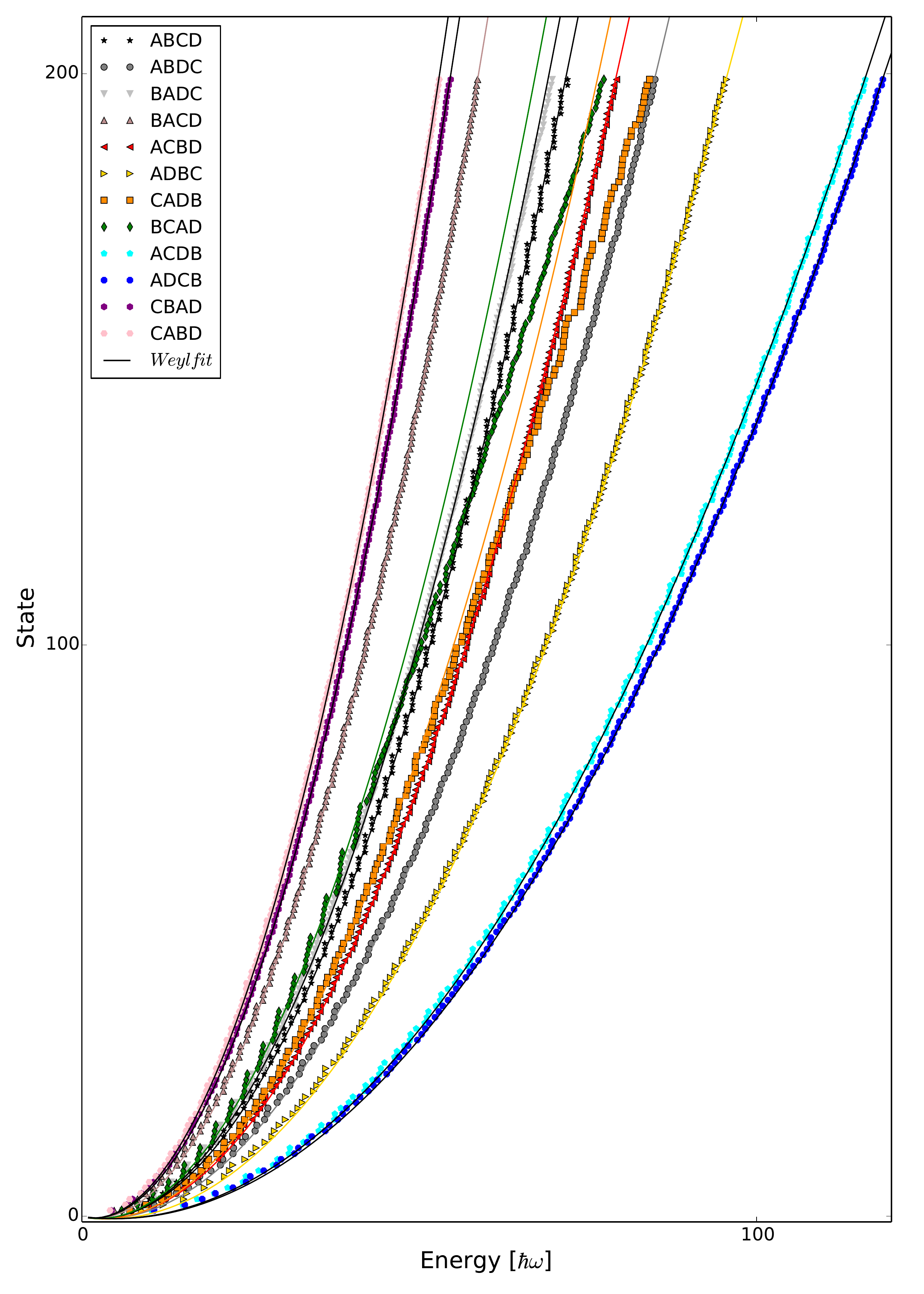}
\caption{Energy spectrum plot for the first 200 states. Notice the center-of-mass and the hyperradius energies are taken out as they are not relevant for this plot. The solid line shows the Weyl's law as described in the main text. The dots are the energy eigenvalues for each 12 regions. Note the degeneracy in $ABCD$ region (black stars). Note also that the green and orange data do not follow the Weyl's prediction for large energies. This is because of the slow convergence in these regions.}
\label{chaos_a3_energy_spectrum}
\end{figure}

The matrix elements of Eq.~(\ref{laplace1+1+1+1}) can be calculated by this ansatz and then diagonalized. Fig.~\ref{fourparticles_angular2D_eigenstates} f) shows the ground state solution in the triangular region, for equal mass case. The energy spectrum for each region is plotted in Fig.~\ref{chaos_a3_energy_spectrum} for a special kind of mass ratio. I will come back to this mass ratio and energy plot in the next section. However, the calculations are made with a $N_{max}=80$ basis. With this value the first 300 energies are quite converged up to the second decimal, where the calculations take approx. 10 days. In addition, some regions are slower to converge than the others. The slow convergence in some regions, for instance region 5 ($ACBD$) in Fig.~\ref{jacobi_sphere_1+1+1+1}, is due to the fact that this region does not align with any of the Jacobi axes and for some ratio it might be below the $x,y$-plane. This means that one needs an additional but simple transformation to rotate the region in such a way that it aligns with one of the axes and has a value of $cos(\theta)>0$ for all $\theta$. This additional transformation, introduces more variables that complicates the Laplacian even more that in turn converges more slowly than the other regions.

\subsection{Quantum Billiards\index{Quantum Billiards}} \label{ch:fourparticles:subsec:quantum}

In the last section, I analyzed how one can solve the energy spectrum of the four-particle system in a one-dimensional harmonic trap with hard-core interactions for any mass. This was done semi-analytically, which introduced some numerical approximations that are crucial in the precision of the energies. However, it turns out that for a certain unequal mass ratios between the particles the model has an exact analytical solution. Recently, similar observations were made for hard-core interactions in free space \cite{OlshaniiNJP2015,scoquartSPP2016}. In what follows, I will present the results for harmonically trapped four-body systems with different masses. Similar to the Tonks-Girardeau gas \cite{GirardeauJoMP1960} and Lieb Liniger model \cite{PhysRev.130.1605}, the present model for any mass can provide insight about the dynamics of quantum systems, which should be controllable and testable in modern experiments \cite{GuanRMP2013,OlshaniiPRL1998, RevModPhys.83.1405}. The four-body mass imbalanced system is additionally interesting due to the ability to play with any mass, which only very few models can allow to my knowledge.\\

\noindent
It is shown that for a specific order of four particles on a line there exist a set of symmetries in three dimensions similar to the Platonic solids \index{Platonic Solids} that have three families of solvable masses \cite{10.2307/4145238}. The three families are denoted by $A_3$, $C_3$ and $H_3$, which correspond to a Tetrahedron, Cube and Icosahedron symmetries, respectively. The symmetries can be classified by a reflection group generally called Coxeter group\index{Coxeter Group}. For each Coxeter group there is a set of mass ratios of the particles in a specific order, dependent on one-parameter only, for which one can solve and find exact solutions to. One special kind of mass ratio in $H_3$ has been investigated in details in \cite{2017arXiv170401433H} and the general analytical solutions are also given there. In this thesis I will present another set of solutions from the $A_3$ symmetry and investigate the system by my semi-analytical method derived in the previous chapter. These results have not been published anywhere before.

The set of mass ratios of $A_3$ that have produce an analytical solution to the problem can be shown to obey the following one-parameter relations \cite{2017arXiv170401433H}: $m_1=\frac{u+1}{3u-1}m_2$, $m_3=u~m_2$, $m_4=\frac{u(u+1)}{3-u}m_2$ with $1/3\leq u\leq 3$ with the ordering $ABCD$. I call these masses ``magic numbers'', even though there is no magic about them and they are obtained through the group algebra. I have chosen to investigate the magic mass ratio that is produced with $u=0.5$, that is $\{m_1;m_2;m_3;m_4\}=\{3.0;1.0;0.5;0.3\}$. As already revealed, the energy spectrum of this ratio is plotted in Fig.~\ref{chaos_a3_energy_spectrum} for each region that represent a specific order. Notice that the energy spectrum is not the total energy since the center-of-mass and hyperradius energies are taken out as they are not relevant for this section, but can easily be considered if necessary. These are also not important when it comes to the behavior of the energy spacing. In addition, notice that the energy spectrum for the $ABCD$ region (the black stars) is degenerate and this degeneracy increases as one looks at the higher end of the spectrum. \\

\noindent
The method derived in the previous chapter can easily be compared with quantum billiard problems, which are studied extensively during the last decade \cite{RICHENS1981495, PhysRevE.89.042918, PhysRevE.87.062902, PhysRevLett.83.4729, PhysRevE.55.6384}. The dynamics of these problems depend very much of the shape of the triangle and the domain. Similarly, I have projected the spherical triangles \index{Spherical Triangle}into a flat isosceles right triangle. The projection transforms the spherical Laplacian into another set of differential operator that is solved within the triangle's hard-walls as discussed earlier. Diagonalizing the final Laplacian-like problem produces Fig.~\ref{chaos_a3_energy_spectrum}, which shows the similar kind of dynamics as the quantum billiards. Here the mass ratio is the key factor because it is incorporated in the differential operator. As one can see in the figure there seems to be a systematic set of solutions for the $ABCD$ region (black stars) with degeneracy that is growing as the energy is increased, while for the other regions with incorrect configuration there seem to be a randomly distributed energy spectrum. Of course, these conclusions should be analyzed more quantitatively, which I will come into right below.

The solid black line plotted in Fig.~\ref{chaos_a3_energy_spectrum} shows the Weyl's law\index{Weyl's Law} \cite{Ivrii2016}. The Weyl's Law for a spherical triangle, $\Omega$, is given as:
\begin{equation}
N(\tilde{E})=\frac{area(\Omega)}{4\pi}\tilde{E}-\frac{circumference(\Omega)}{4\pi}\sqrt{\tilde{E}},
\label{Weylslaw}
\end{equation}
where $N(\tilde{E})$ is the total number of energy eigenvalues below a given scaled energy, $\tilde{E}=(2mR^2/\hbar^2)E$ with $R$ being an arbitrary radius of the sphere. $area(\Omega)$ is the area, and $circumference(\Omega)$ is the circumference of the spherical triangle. Apart from the numerical precision in some regions, which I mentioned earlier was due to the more complicated Laplacian and therefore needs a higher $N_{max}$ and more calculation time, all the regions obey this law very well.

\begin{figure}[t]
\center
\includegraphics[width=0.9\columnwidth]{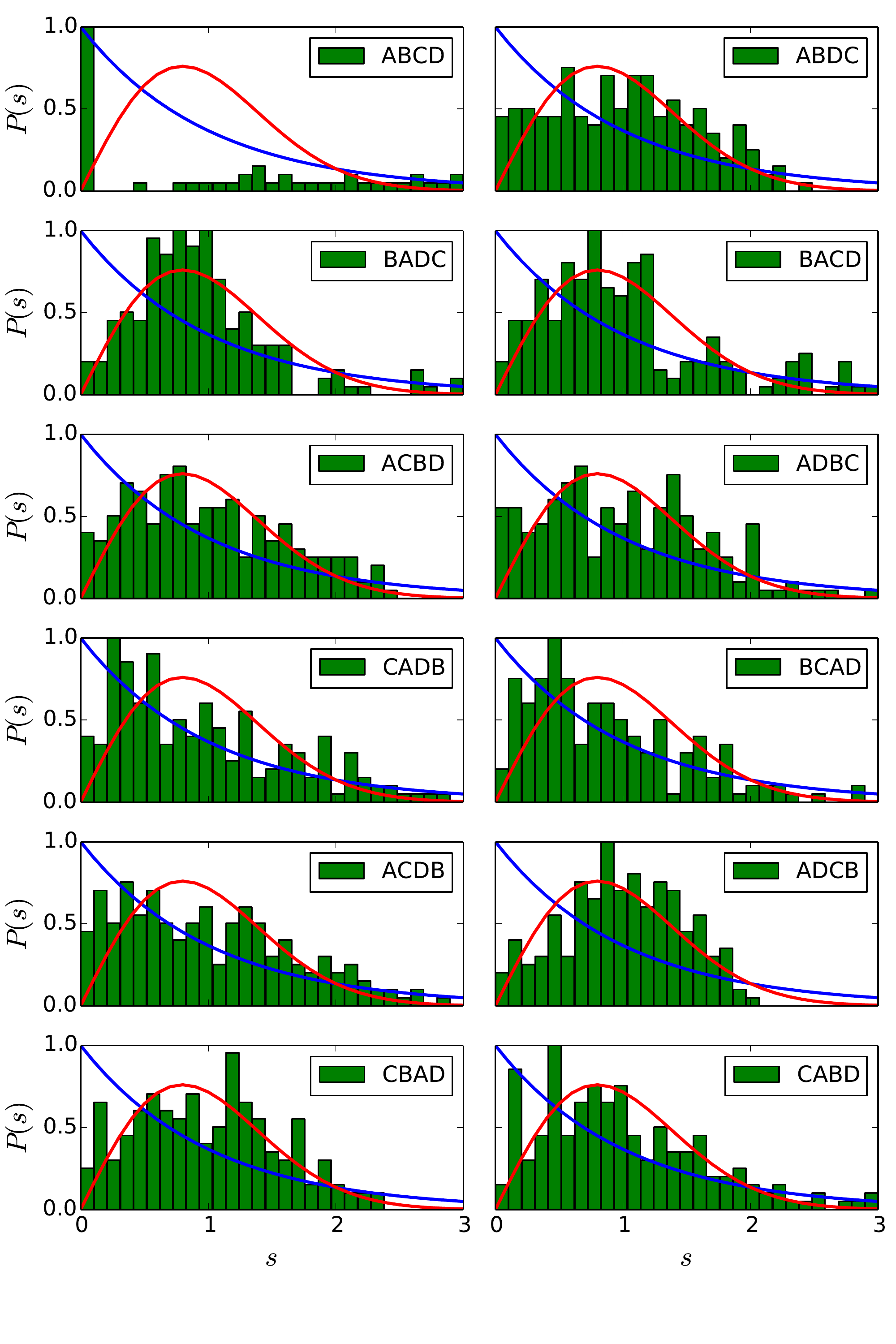}
\caption{Unpublished results: Unfolded spectrum statistics for $A_3$ with mass ratios $\{3.0;1.0;0.5;0.3\}$. There are 12 distinguishable regions with different configurations. The variable $s$ is the measure of the unfolded energies levels. The blue line shows the Poisson distribution, which is expected for analytically solvable models, while the red line shows the Wigner-Dyson distribution, which is expected for ergodic systems. The unfolding has been done over the first 200 states.}
\label{chaos_a3_energy_unfolded}
\end{figure}

In order to be more detailed and quantitative in terms of whether there is an ergodic\index{Ergodicity} behavior in the energy spectrum or not, one can do a level spacing statistics by unfolding the spectrum. The unfolding is standard and can be found in \cite{stockmann2006quantum, giannoni1991chaos}, but it basically is a measure of how far the energy spacing between two neighboring energy levels are from the mean value. This is measured by a variable called $s$, which is the scaled level spacing parameter. More technically, $s$ is found by first fitting a polynomial line to the index vs. energy plot and then calculate the energy spacing between the energies found from this fit. Finally, a histogram can be made of the energy spacings as shown in Fig.~\ref{chaos_a3_energy_unfolded}. Since there are 12 distinguishable regions, the unfolding has been applied on each region. As it is apparent, the first region with the $ABCD$ ordering is quite different from the others. This region is in fact an analytical region, whose analytical solutions can be found by the group algebra derived in \cite{2017arXiv170401433H}. These kind of systems are expected to follow a Poisson distribution \index{Poisson Distribution} illustrated by the blue line. However, due to the extra degeneracy that there is in the system, the distribution is even more peaked for $s=0$. On the other hand, it is seen that all the other regions more or less closely follow the Wigner-Dyson distribution \index{Wigner-Dyson Distribution}(red line). Numerically, these sectors show a clear ergodic behavior in a quantum sense. Therefore they are believed to be not integrable and not analytically solvable, since no symmetries can be associated with these orderings and therefore energies come in a random order \cite{stockmann2006quantum}. 

One could argue that the magic mass ratios are not experimentally feasible and the question then becomes how the spectrum would look like if the masses are just slightly different from the magic numbers. Further investigation has shown that the unfolding graphs still show some Poisson distribution, even though the peak close to zero becomes a little bit more spread. This is mostly because the system is close to the analytical one. Therefore some small random noise start to appear. In conclusion, the transition between the ergodic and systematic solutions is believed to be continuous.

\newpage % Create empty back of side
%\thispagestyle{empty}

%-----% N+M SYSTEMS
% !TeX root = ../Main_publish.tex

\chapter{Many Particles: $\mathbf{N_A+N_B}$ Systems in a 1D Harmonic Trap\index{N+M Systems}}\label{ch:nm}
\epigraph{\it “ipsa scientia potestas est.”}{\rm ---Sir Francis Bacon}

As the number of particles grows it becomes harder to analyze the two-component systems analytically due to the increasing degrees of freedom. The calculations become more complex and the Jacobi space becomes harder to illustrate and grasp. It is therefore sometimes helpful to setup a numerical code to investigate systems with few to many particles. Actually, I have already revealed some of the few-body numerically calculated results in the previous chapters in Fig.~\ref{plot_threefermions_xyplane}, Fig.~\ref{plot_threebosons_xyplane} and Fig.~\ref{numerical_contour_2+2}. However, in this chapter I will continue to the many-body limits. Another advantage of numerical calculations is that it can be used to confirm and test some of the few-body properties in the many-body limits. And of course, some new properties could also appear, that is not necessary present in the few-body limits. This is the topic of this chapter.

I have contributed with numerical calculations that rely on exact diagonalization of the many-body Schr{\"o}dinger equation. This is done by projecting the full Hamiltonian onto a finite basis of single-particle harmonic oscillator states, $|n\rangle$ or Eq.~(\ref{HO1_eigenfunction}). Each many-body basis state is then written as a tensor product of a symmetrized harmonic oscillator state of the $N_A$ type-A bosons and the $N_B$ type-B bosons, $|n_1\dots n_{N_A}\rangle \otimes |n_1\dots n_{N_B}\rangle$. These are used as a basis in order to write a matrix representation to be diagonalized afterwards. The interaction between the particles has been modeled by an effective two-body short-range interaction, which speeds up the convergence in the calculations. This method was first used in \cite{LindgrenNJoP2014} for two-component {\it fermionic} many-body states, which I then have contributed and expanded to also be applicable for {\it bosonic} many-body two-component systems. In Chapter 7, you will find more details about the numerics, however, in the following I will discuss some of the interesting results obtained in the few- and many-body limit using this code.

\section{Results}\label{ch:nm:sec:results}
The method is based on the two-component system with $N_A$ particles in one component and $N_B$ particles in another, the so-called $N_A+N_B$ system. In addition, all the particles are assumed to have the same mass, $m$, and trapping frequency, $\omega$. The interactions between particles from {\it different} components, the so-called interspecies interactions are denoted by $g$, while interactions between particles from {\it same} component, the so-called intraspecies interactions are neglected. The general Hamiltonian is therefore written as (in units of harmonic oscillator, $b=\sqrt{\frac{\hbar}{m\omega}}$):
\begin{equation}
\mathcal{H}=\sum_{i=1}^{N_A}\frac{1}{2}\left(p_{A,i}^2+q_{A,i}^2\right)+\sum_{i=1}^{N_B}\frac{1}{2}\left(p_{B,i}^2+q_{B,i}^2\right)+\sum_{i_A=0}^{N_A}\sum_{i_B=0}^{N_B}V_{i_A,i_B},
\label{manyhamiltonian}
\end{equation}
where $V_{i_A,i_B}=g\delta(x_{i_A}-x_{i_B})$ is the interaction terms between the particles from different components. $p_{k,i}$ and $q_{k,i}$ are the usual momentum and coordinate operators, where each operate in their own subspace $k\in\{A,B\}$, respectively.
After diagonalizing the above equation, one can investigate the wave function for any $N_A+N_B$ systems.

\subsection{Few Body\index{Few Body}}\label{ch:nm:subsec:fewbody}
\begin{figure}[t!]
\center
\includegraphics[width=1\columnwidth]{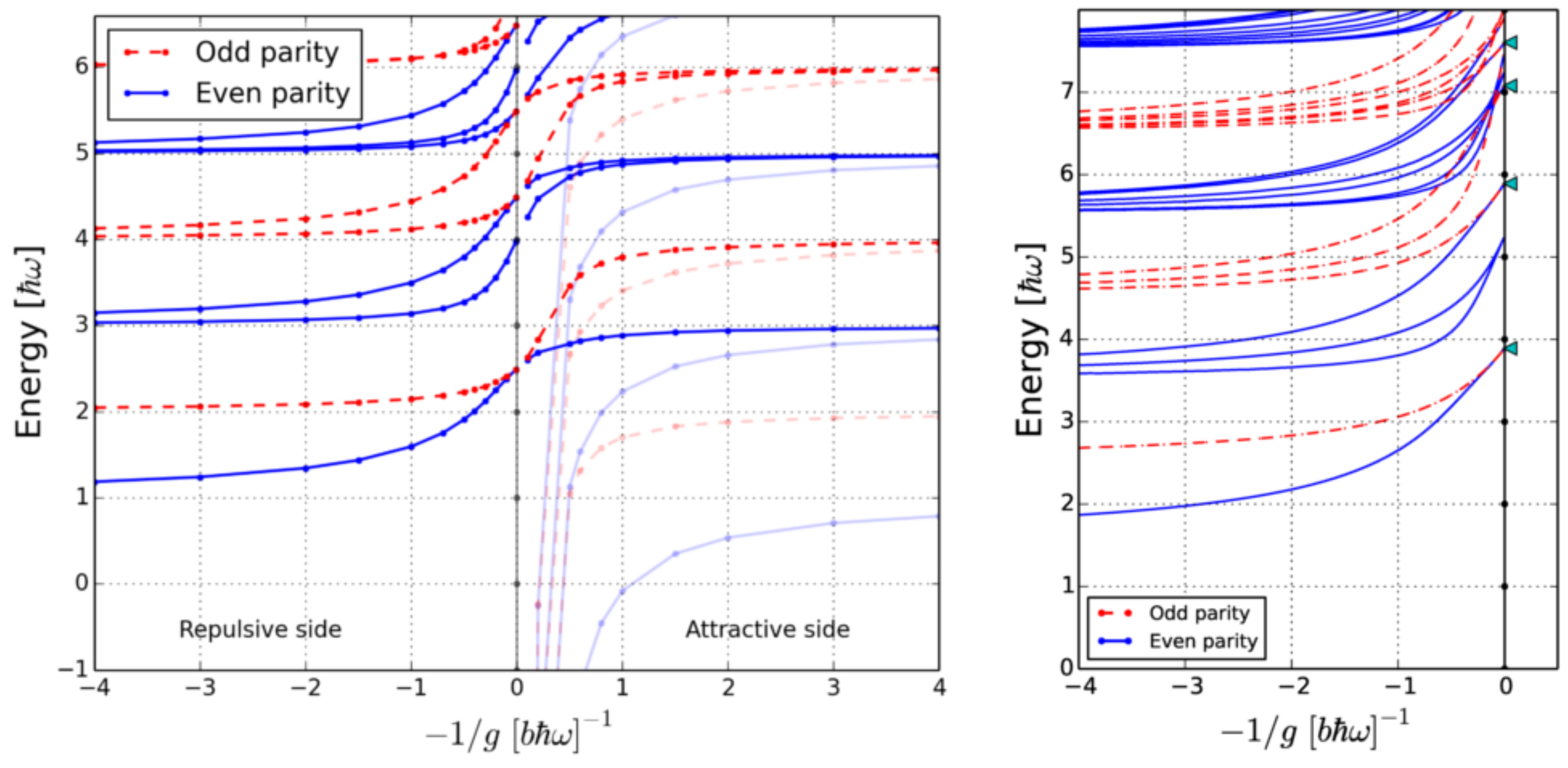}
\caption{ Left panel: Energy spectrum for the $1+2$ bosonic system obtained by numerical exact calculations. At strong interaction some of the states become doubly degenerated and have half-integer energies. Right panel: Energy spectrum for the $2+2$ bosonic system in the repulsive side. The triangles are analytically known energies in the strong interaction limit. Figures are adapted from \cite{DehkharghaniSR2015}}
\label{numeric1+2}
\includegraphics[scale=0.5,clip=true]{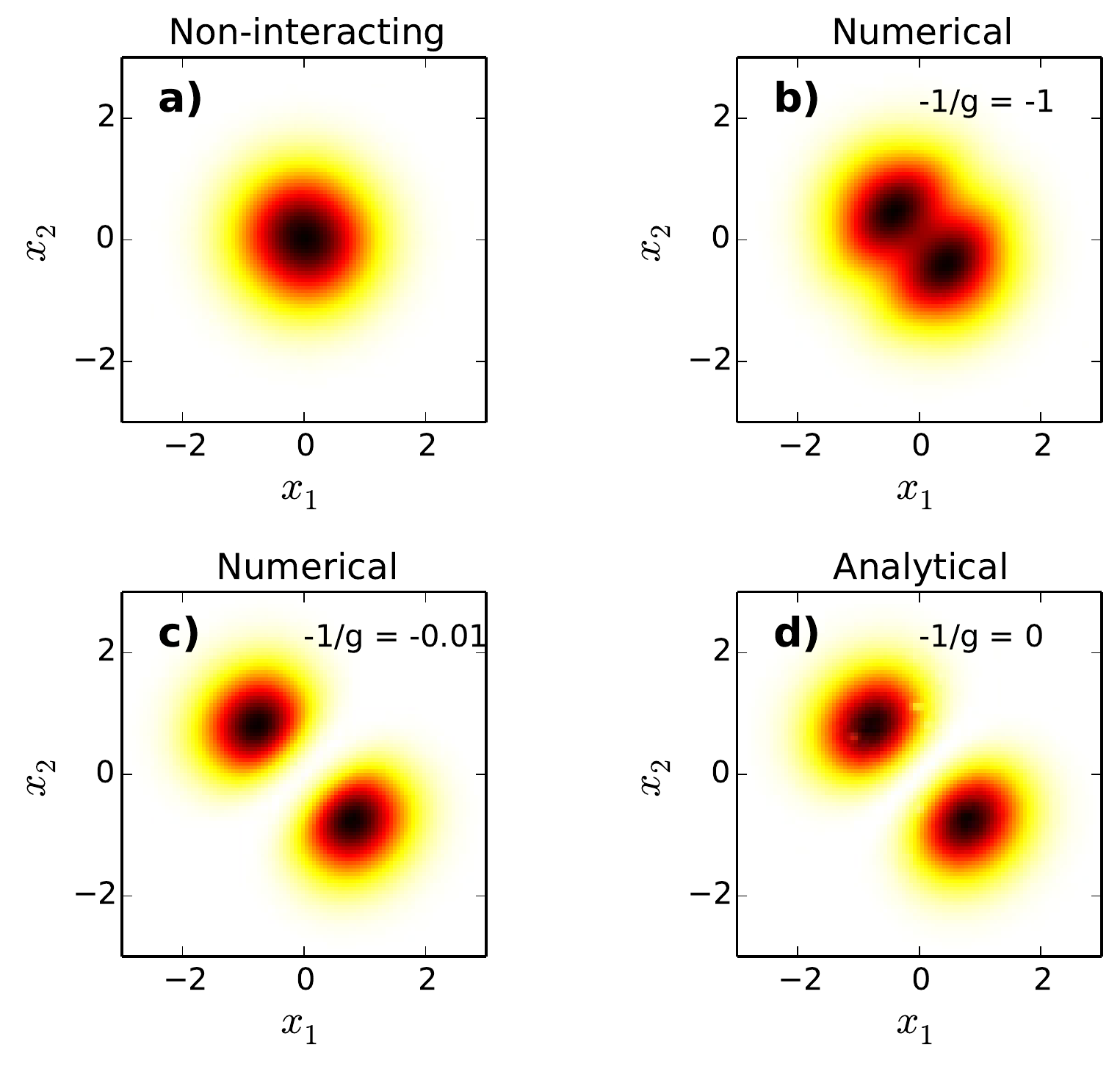}
\caption{Pair-correlation plot for the $2+2$ bosonic ground state with $x_1\in N_A$ and $x_2\in N_B$ in the a) non-interacting, b) intermediate and c) strong interacting regime. The separation into ferromagnetic ordering is seen in the strong regime, which agrees very well with analytical result in d). Figure has length-units of [$b$] and is adapted from \cite{DehkharghaniSR2015}.}
\label{numerical_2+2pair}
\end{figure}

The energy spectrum for the $1+2$\index{1+2 Systems} bosonic system is shown in Fig.~\ref{numeric1+2} left panel. The numerical results also show the bounded attractive states, which are dimmed a little bit. The center of mass contributions, on the other hand, have been removed. The solid lines, however, are very familiar and were also plotted back in Chapter 2 in Fig.~\ref{energyspectrum1+2} right panel. But remember, even though Fig.~\ref{energyspectrum1+2} shows the same pattern, the figure is only valid for $g=0$ and $g\rightarrow\infty$ due to the radial dependency of interaction strength. For the intermediate case, one would need the interpolatory method to reproduce the results in Fig.~\ref{numeric1+2}. One of the interesting results here is that the odd and even state become degenerate at a half-integer energy, while fermions are always whole integer at $g\rightarrow\infty$.

Looking at the next simplest few body case, Fig.~\ref{numeric1+2} right panel shows the energy spectrum for the $2+2$ \index{2+2 Systems}bosonic systems. Even though the two-fold degenerate ground state is present in the strong interaction limit, the limiting energy is non-integer. This is due to the bosonic properties  (symmetric under the exchange of any two particle labels) that distribute the particles in such a way that lowers the energy to a non-integer value. The ground state of this system also shows a fully spatially ferromagnetic\index{Ferromagnetic} quantum state, which is shown in the pair-correlation plot, Fig.~\ref{numerical_2+2pair}. This plot shows how the particles in one component ($x_1\in N_A$) align themselves depending on where particles from the other component are ($x_2\in N_B$). Analytical results from Chapter 3 are also shown here in the strong regime.

\subsection{Many-Body \index{Many Body}Balanced Systems\index{Balanced Systems}}\label{ch:nm:subsec:balancedsystems}
\begin{figure}
\centering
\includegraphics[width=0.76\textwidth]{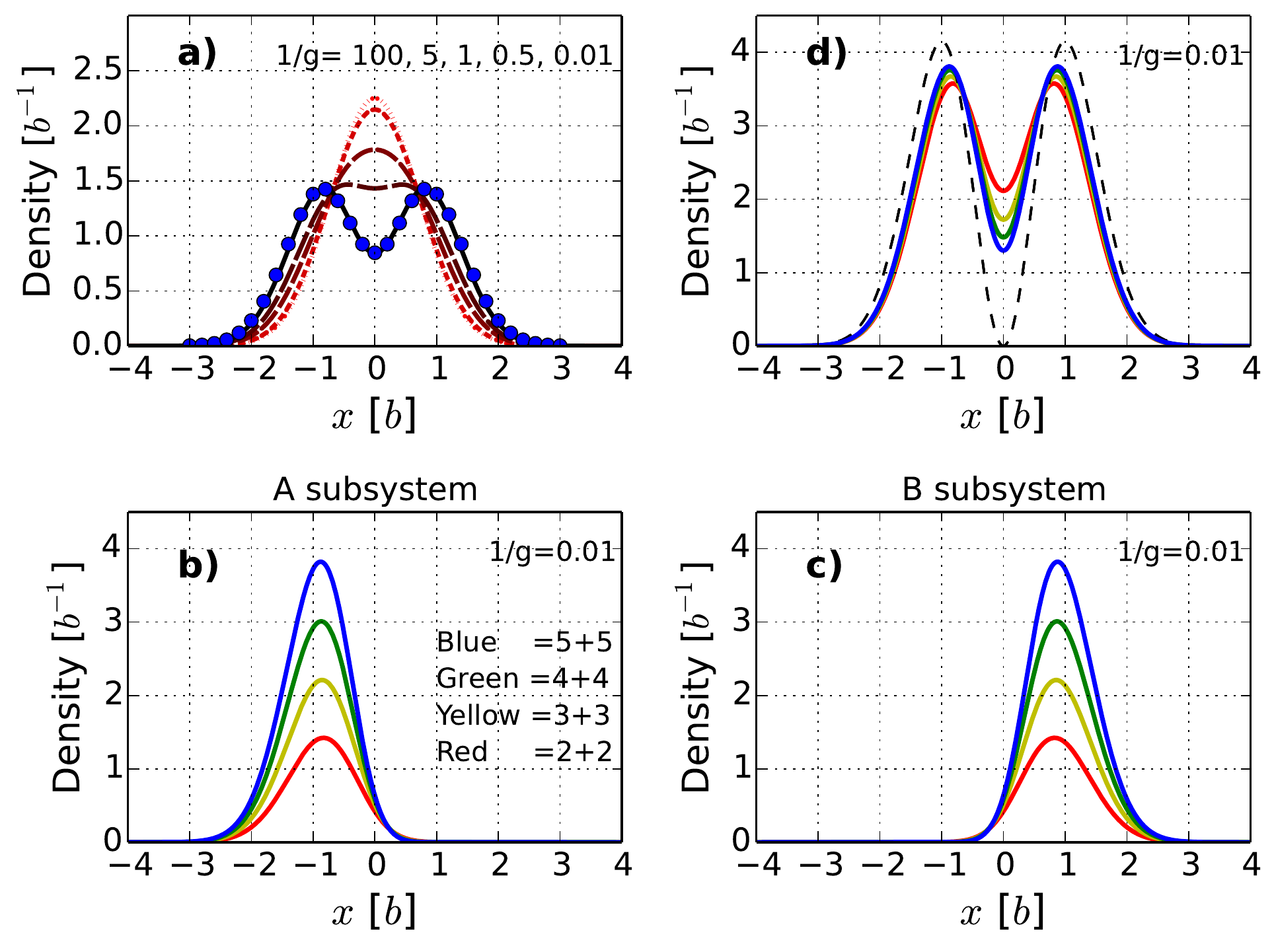}
\caption{ Ground state densities of balanced systems. a) shows the density plot for the balanced systems as a function of interaction strength, $g$. As the interaction becomes strong, the particles tend to separate in a $A\dots A\pm B\dots B$ configuration. This is compared with the analytical results shown with dots. To test the separation, a linear combination of these two states, $A\dots A\pm B\dots B$, is plotted in b) and c). Finally, d) shows the rescaled total density separation of the particles into two separate condensates. The limiting case for many particles with $N_A=N_B\gg 1$, is shown with dashed line. Figure is adapted from \cite{DehkharghaniSR2015}.}
\label{numerical_density_balanced}
\end{figure}

After observing the perfect ferromagnetic state, the question becomes whether or not this behavior is also present for the many-body balanced systems. To answer this question I investigated the $N_A=N_B\leq 5$ bosonic systems and the density plots can be found in Fig.~\ref{numerical_density_balanced} a). As it is shown, the ferromagnetic ground state becomes clear in the strong regime and matches very well with the analytical results for the $2+2$ system. Furthermore, energy spectrums for all the balanced systems show that the two lowest states become degenerated in the infinitely interacting limit. Therefore it is tempting to postulate that these two degenerated states are in form of $A\dots A\pm B\dots B$ at strong interactions. Of course, this can be easily verified by taking the even-odd superposition of those states and the result is shown in Fig.~\ref{numerical_density_balanced} b) and c). It is clear now that the $A\dots A\pm B\dots B$ combination is indeed the case in this limit. One could even argue that in the limit of many particles one would get two separate condensates, which is shown in Fig.~\ref{numerical_density_balanced} d). The dashed line shows the many body limits where particles tend to separate into two distinct condensates. As a side note, it should be noted that the term ``condensate'' used in a 1D context is really a quasi-condensate with all bosons in the ground state as there are no long-range orders here.

\subsection{Many-Body Imbalanced Systems\index{Imbalanced Systems}}\label{ch:nm:subsec:imbalancedsystems}
\begin{figure}
\centering
\includegraphics[width=0.8\textwidth]{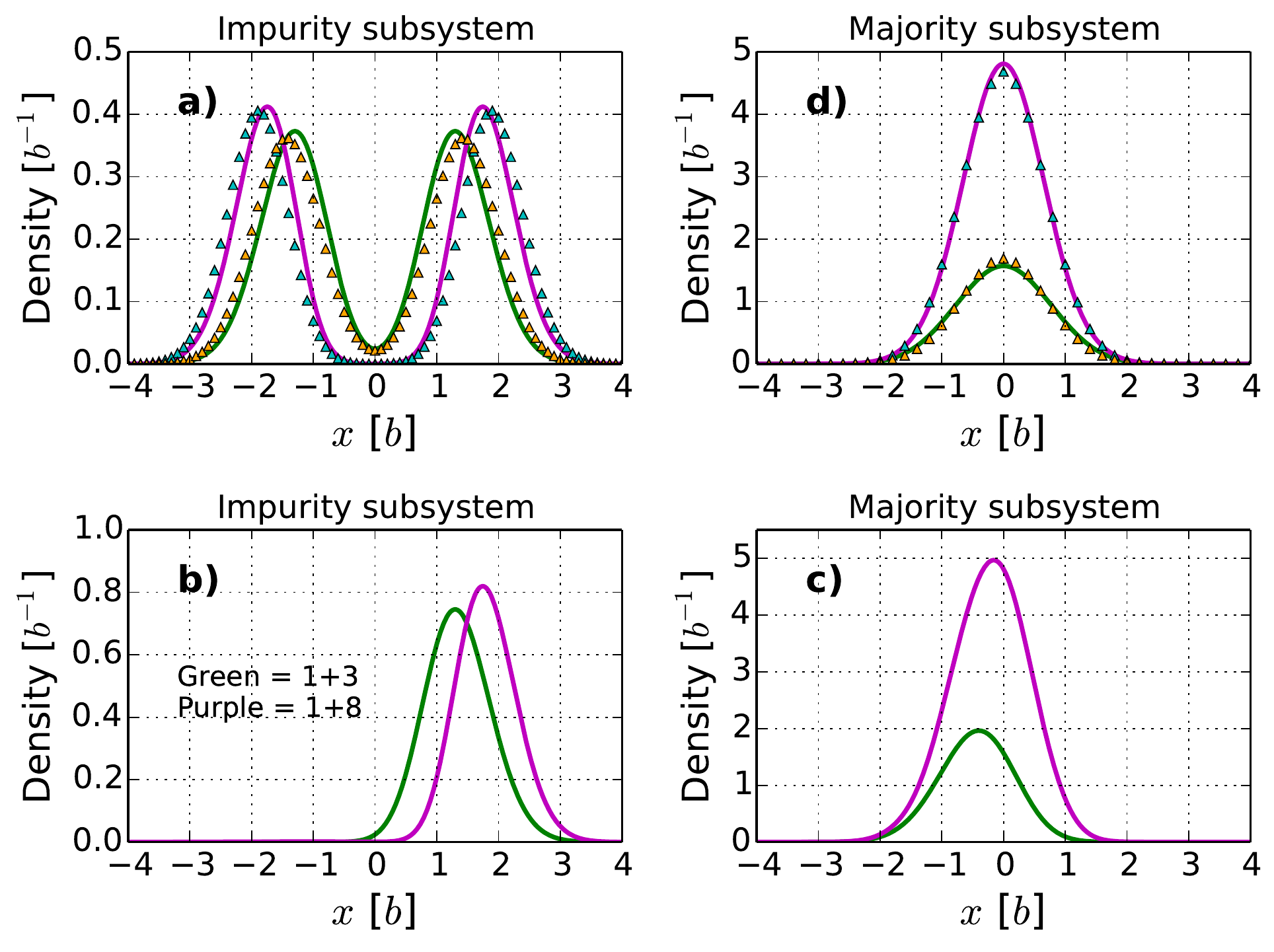}
\caption{Ground state densities of imbalanced systems. {a)} shows the density distribution of the impurity for a $N_B=3$ or $N_B=8$ system with $N_A=1$. The analytical results for $1/g=0$ are shown with triangles, which agree very well. Panel {b)} and {c)} are as in {Fig.~\ref{numerical_density_balanced}~b)} and {c)} but for $N_A=1$ with $N_B=3$ or $N_B=8$ systems. Panel {d)} shows the density of the majority particles($N_B$). These are almost unaffected by the present of the impurity. All numerical results have been obtained with $1/g=0.01$. Figure is adapted from \cite{DehkharghaniSR2015}.}
\label{numerical_density_imbalanced}
\end{figure}

Another kind of systems that are interesting to investigate are the so-called Polaron systems or simply called quantum impurities. Here $N_A=1$ and is usually called the impurity, while $N_B$ is varied and is usually called the number of majority particles. Fig.~\ref{numerical_density_imbalanced} a) and d) show the density plots of $1+3$ and $1+8$ bosonic systems. The triangles in the density plots are the obtained results from another semi-analytical method, which I will explain in more details in the next chapter. From the density plots one can see that the majority particles are mostly found in the middle of the trap, while the impurity is pushed to the side. These kind of systems also show a two-fold degeneracy in the strong regime where one suspects a $A\pm B\dots B$ configuration. Again this turns out to be the case when one plots the linear combination of the two configurations as in Fig.~\ref{numerical_density_imbalanced} b) and c). This means also that when the position of the impurity is known, then the majority particles are to the opposite side. This is in a huge contrast to identical fermions where it has been shown that the impurity is actually mostly found in the middle of the condensate \cite{LindgrenNJoP2014}.

\subsection{Many-Body Properties}\label{ch:nm:subsec:properties}
\begin{figure}[t]
\centering
\includegraphics[width=1\textwidth, trim=1.5cm 0cm 1.5cm 0cm]{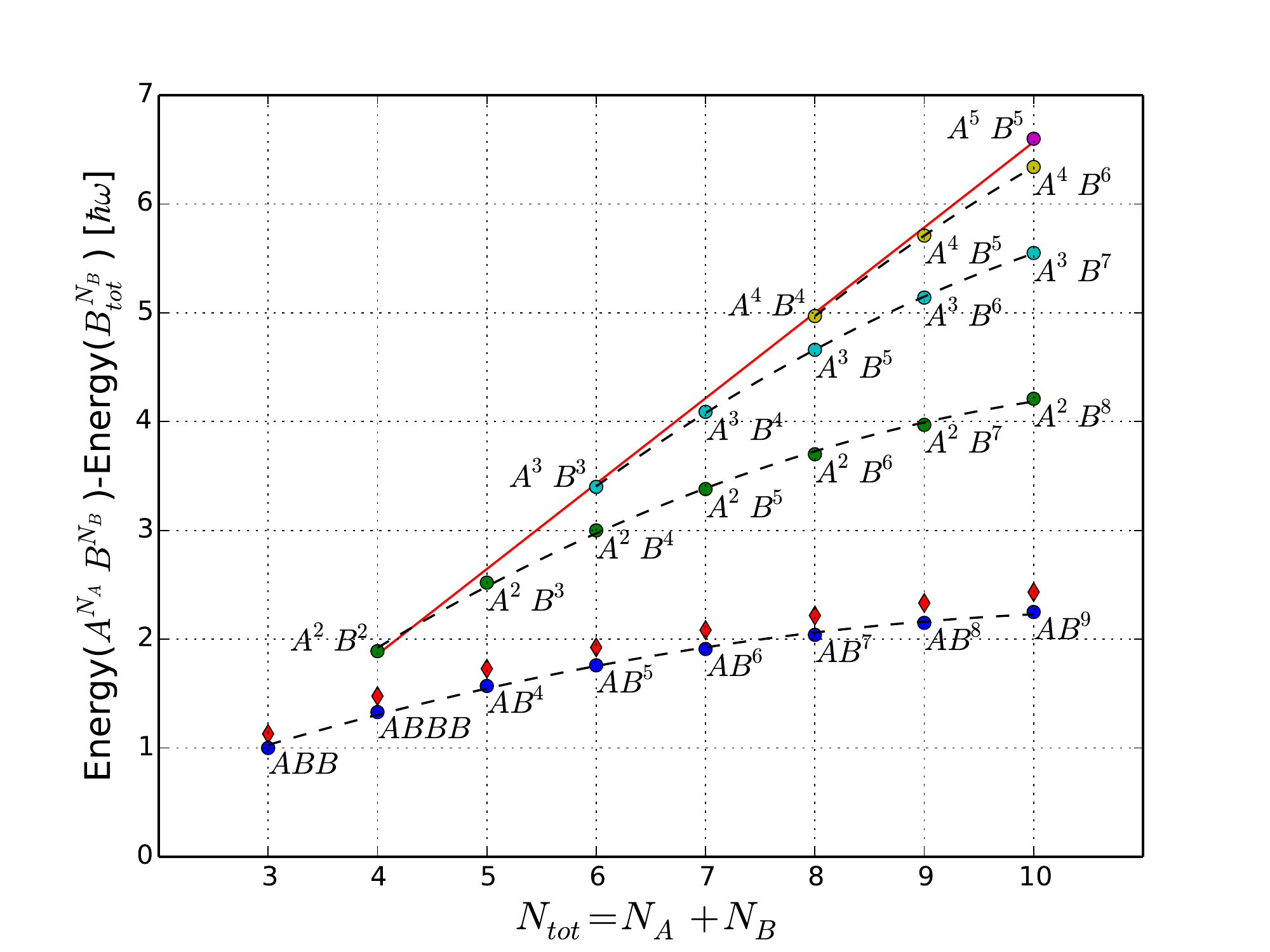}\caption{Ground state energy plot for different combination of number of particles, $A^{N_A} B^{N_B}$, up to $N=10$ particles. Notice that the energy has been subtracted by the zero-point energy, which is given by a single-component system of the same size, $B^{N}$. The dashed lines shows a quadratic interpolation for fixed number of $B$ particles, which clearly saturates. The solid red line is a linear interpolation of the energy for the balanced systems. The red diamonds in the figure for $1+N_B$ systems are results calculated by a semi-analytical method, which will be explained in more details in the next chapter. Figure is adapted from \cite{DehkharghaniSR2015}.}
\label{enetot1}
\end{figure}

Ten-particle systems have been investigated as well and a map of the energy spectrum for all the systems has been plotted in Fig.~\ref{enetot1}. Notice that the energy spectrum for each system $A^{N_A}B^{N_B}$ shows only the ground state energy subtracted by the ground state energy of $B$-species of the same size, $B^{N_A+N_B}$. In other words, the figure shows the energy increase in the system if one turns $N_A$ of the $B$-particles, $B^{N_A+N_B}$, into $A$-particles. While the balanced systems increase linearly in energy as the number of particles increases, shown with the red line in the figure, the results for other kind of systems show a clear saturation for larger systems (dashed lines). The saturation becomes faster for more imbalanced systems indicating that the impurity becomes negligible and pushed more to the side. The red diamonds in the figure for $1+N_B$ systems are results calculated by a semi-analytical method, which will be explained in more details in the next chapter.

\begin{figure}[t]
\centering
\includegraphics[width=\textwidth, trim=0cm 0cm 0cm -1cm]{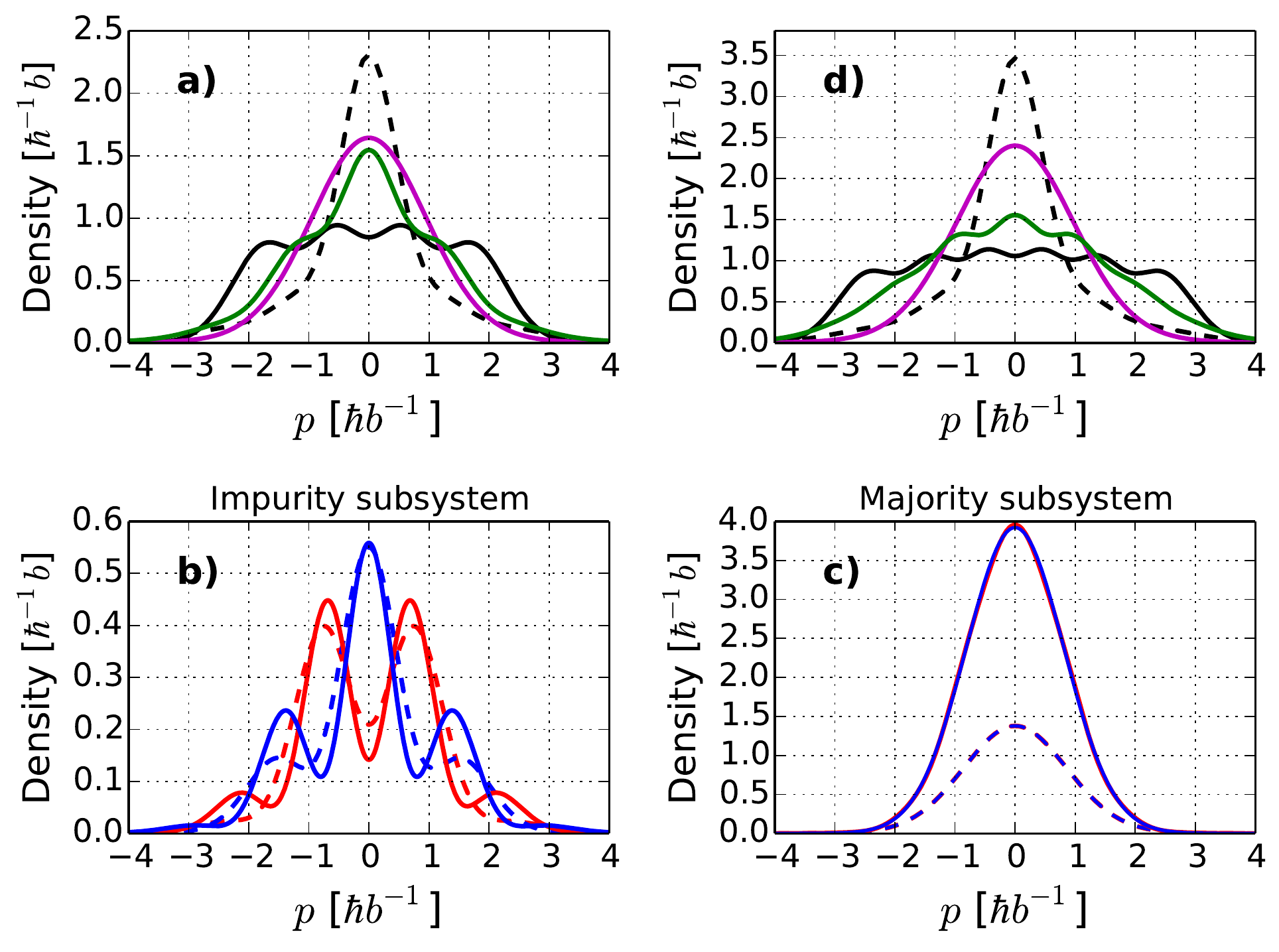}
\caption{Momentum distributions \index{Momentum Distribution}plot in the strong two-folded degenerate regime, where in a) and d) the $2+2$ and $3+3$ systems are shown, respectively. The even parity ground and 1st excited states are shown in purple and green, respectively. In addition, the fermionic (solid black line) and Tonks-Girardeau hard-core bosons (dashed black line) are plotted for comparison. On the other hand, b) and c) show the same distribution for the even and odd parity states, but this time for $1+3$ and $1+8$ bosonic imbalanced systems. c) reveals a clear notable pattern that can easily be confirm both for the odd and even parity state, while the majority particles are Gaussian-shaped. Figure is adapted from \cite{DehkharghaniSR2015}.}
\label{enetot2}
\end{figure}

In order to detect the different states with different ferro- and antiferromagnetic\index{Antiferromagnetic} configurations, a momentum distribution plot has also been created in order to show the difference in such systems. Fig.~\ref{enetot2} shows such a plot, where in a) and d) the momentum distribution for the $2+2$ and $3+3$ systems has been plotted, respectively. Both plots show the lowest even-parity ground state in purple, while the first excited antiferromagnetic state is shown in green. In addition, the fermionic (solid black line) and Tonks-Girardeau \index{Tonks-Girardeau Gas}hard-core bosonic state (dashed black line) are also plotted in order to show how different they are from the particular case. Therefore, it should be easy to distinguish these states from one another in experiments. The perfect antiferromagnetic state in the bosonic $2+2$ systems was already seen in Fig.~\ref{numeric1+2} right panel at a value of energy $7.5$. This energy fits well with the fermionic $2+2$ systems where the ground state and the perfect anti-ferromagnetic state all have the energy $7.5$ with each particle occupying the four lowest states in the harmonic oscillator. However, the momentum distribution reveals the difference between symmetric and antisymmetric particles.

In fig.~\ref{enetot2} b) the same momentum plot is shown for both $1+3$ (dashed lines) and $1+8$ (solid lines) bosonic system. The even parity (blue) and odd parity (red) show a clear and very different oscillatory distribution. The oscillatory behavior becomes sharper as the number of the particles increases. The majority particles, on the other hand, are quite untouched and distributed as a Gaussian.

\newpage % Create empty back of side
%\thispagestyle{empty}

%-----% 1+N AND 2+N POLARONS
% !TeX root = ../Main_publish.tex

\chapter{Quantum Impurities: $\mathbf{1+N}$ and $\mathbf{2+N}$ Systems} \label{ch:polarons}
\epigraph{\it “We cannot solve our problems with the same thinking we used when we created them.”}{\rm ---Albert Einstein}
In this chapter I will go through the semi-analytical method that treats the $1+N$ bosonic systems. These systems have recently attracted a lot of attention due to the experimental realization in cold atoms \cite{Spethmann2012, schirotzekPRL2009, PalzerPRL2009} and the interesting physics behind the polarons \index{Polarons}\cite{zollnerPRA2011, kohstall2012, koschorreck2012, massignanRoPiP2014}. The single particle different from the rest of the condensate is usually called the impurity, and the rest are called the majority particles or the condensate. The method builds on a simple approximation that is very easy to implement. The ideas behind the method works for any trapping potentials with any particle masses or trapping frequencies. In addition, it can handle any number of majority particles and one could even introduce a small intra-species interaction in form of the Gross-Pitaevskii equation\index{Gross-Pitaevskii Equation} among the bosons. Since the method is just an approximation and gives upper limit energy, I will show how well it does when compared with numerical exact results. Afterwards, I will extend the ideas of the above method and apply them to double quantum impurities, the so-called $2+N$ systems.

\section{Single Quantum Impurity\index{Quantum Impurity}\index{Imbalanced Systems}: $\mathbf{1+N}$\index{1+N Systems}}\label{ch:polarons:sec:singlequantumimpurity}
Imagine having a single particle, call it type-$A$, interacting with a bunch of other particles of another type, $BB\dots B$, which are all trapped in a 1D harmonic oscillator potential. Even though the details of the external trapping potential are not crucial for the method, I will consider a harmonic oscillator as an example in order to compare the obtained results with the numerical exact calculations. In addition, consider that the impurity has mass $m_A$ and the majority particles have mass $m_B$. The interaction is short range between the particles and the majority particles can be considered as an ideal gas or slightly interacting and modeled by the 1D Gross-Pitaevskii equation. The trapping frequencies are assumed to be different here and therefore they are called $\omega_A$ and $\omega_B$. The overall Hamiltonian of the system can be written as:
\begin{align}
H&=H_{A}(x) + \sum_{i=1}^{N_B} H_{B}(y_i)+\sum_{i=1}^{N_B}g\delta(x-y_i) + \sum_{i<k}^{N_B}g_{BB}\delta(y_i-y_k),\\
H_{A}(x)&=\frac{p_{x}^2+ m^2_{AB}\omega_{AB}^2 x^2}{2m_{AB}},\\
H_{B}(y)&=\frac{p_{y}^2 + y^2}{2},
\label{ch5:eq:hamiltonian}
\end{align}
where $x$ and $p_x$ are the coordinate and momentum of the impurity, respectively, and $y$ and $p_y$ are the corresponding parameters for the majority particles. The interaction strength between the different species is called $g$, while the self-interaction is called $g_{BB}$. The mass ratio and frequency ratio are denoted $m_{AB}=m_A/m_B$ and $\omega_{AB}=\omega_A/\omega_B$, respectively. Notice that the units are chosen slightly different in this chapter, because the two species can have different trapping frequencies. Since the system is mostly made of the majority particles, I choose to measure the length in units of $b=\sqrt{\hbar/m_B\omega_B}$ and energy in units of $\hbar\omega_B$.

\begin{figure}[t]
\centering
\includegraphics[width=\columnwidth]{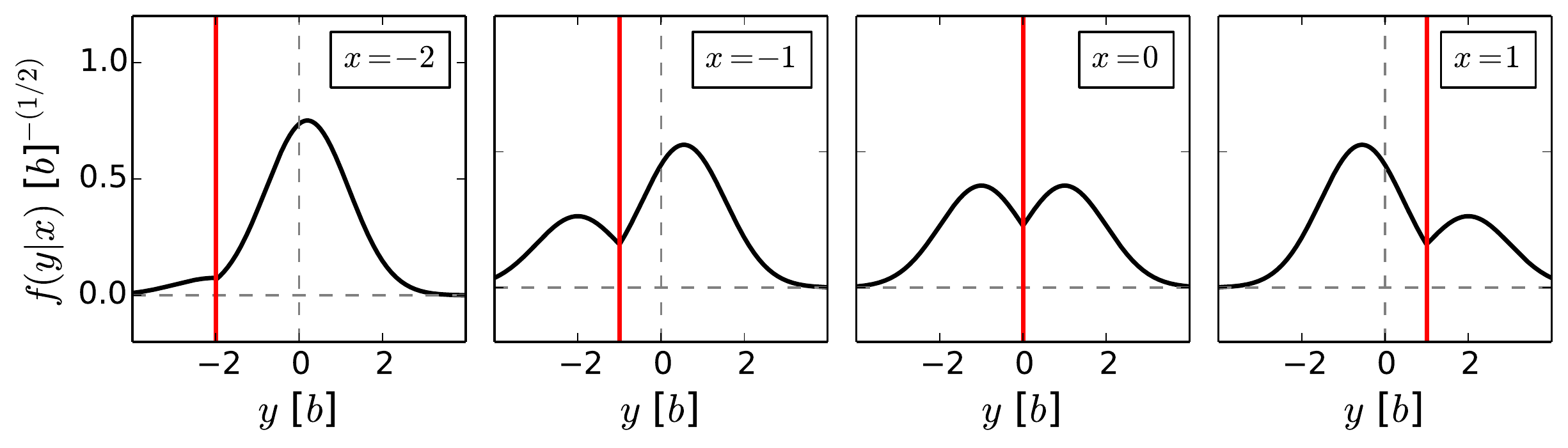}
\caption{Sketch of how the ground state of the single boson (black solid line) trapped in a harmonic oscillator trap (not sketched here) reacts to the present of the impurity (red vertical line) throughout its changing position. The wave function and the corresponding energy of the majority particles that solves the Hamiltonian, $H_B(y_i)+g\delta(x-y_i)$, are then stored for a given impurity position $x$. This is used afterwards as an effective potential that contributes to the solution of the impurity.}
\label{polaronsketch1+N}
\end{figure}

The idea behind the method is as following: for a given position of the impurity, $x$, in the harmonic trap, find the wave function and the corresponding energy of the majority particles. By solving this one obtains an effective potential for the impurity, which can be solved afterwards. In other words, the wave function of the majority particles is dependent on where the impurity is, and the impurity is solved by the potential that is created by the majority particles. Fig.~\ref{polaronsketch1+N} shows how the wave function of the majority particles is affected by the present of a given position of the impurity. The method moves the impurity one step and at each step it finds the corresponding wave function and energy for the majority particles. Similar considerations are known from the Born-Oppenheimer approximation, \index{Born-Oppenheimer Approximation} where here the impurity is considered as the ``slow'' parameter, however the physics is different. While this assumption must be justifiable for heavy impurities, one could also check how the method works for equal masses. I will check this in the following. Using an adiabatic decomposition\index{Adiabatic Decomposition}, the total wave function for the whole system is written as:
\begin{equation}
\Psi(x,y_1,\dots,y_{N_B})=\sum_{j=1} \phi_j(x)\Phi_j(y_1,\dots,y_{N_B}|x),
\label{ch5:eq:adiabatic}
\end{equation}
where $\Phi_j$ is the $j$th eigenstate to the following eigenvalue-problem:
\begin{equation}
\left(\sum_{i=1}^{N_B} H_B(y_i)+g\delta(x-y_i)\right)\Phi_j=E_j(x)\Phi_j,
\label{ch5:eq:majorityparticles_eigenvalue}
\end{equation}
with a given fixed position, $x$, of the impurity. The bosons are assumed to be non-correlated in the beginning, and therefore one can write the total wave function as a symmetric product of single wave functions:
\begin{align}
\Phi_j(y_1,\ldots,y_{N_B}|x)=\hat S\prod_{i=1}^{N_B}f_{k^j_i}(y_i|x),
\label{ch5:eq:product}
\end{align}
with corresponding energies: $E_j(x)=\sum_{i=1}^{N_B} \epsilon_{k^j_i}(x)$. Here $\hat S$ is the symmetrization operator and $f_{k^j_i}(y_i|x)$ is the $k_i^j$th single boson solution to the single particle Hamiltonian: $H_B(y_i)+g\delta(x-y_i)$. The solution for a harmonically trapped single boson wave function in the presence of an impurity is sketched in Fig.~\ref{polaronsketch1+N} for the ground state. In the following I will omit the ${k^j_i}$ index for convenience.

Depending on the interaction strength, the single boson wave function must satisfy the following delta-function boundary condition (which was also discussed before back in Eq.~(\ref{griffith_criteria})):
\begin{align}
f(x|x)=\frac{1}{2g}\left\{\frac{\partial f(y|x)}{\partial y}\Bigg|_{y=x+\varepsilon}-\frac{\partial f(y|x)}{\partial y}\Bigg|_{y=x-\varepsilon} \right\},
\label{ch5:eq:condition}
\end{align}
which indicates a discontinuity in the derivative of the wave function from right, $y=x+\varepsilon$ and left, $y=x-\varepsilon$, to the position of the impurity, $x$. Note that the same condition requires the $f(x|x)=0$ for $g\rightarrow\infty$. One can find the wave function, $f(y|x)$, either numerically by discretizing the space or analytically, which have been studied before in \cite{peresEJP2011} for a quantum harmonic oscillator plus a delta-function potential at the origin of the trap. These solutions contain a linear combination of Tricomi and Kummer confluent hypergeometric functions\index{Confluent Hypergeometric Function} multiplied by a Gaussian function. However, for the quantum impurity case, one needs to find the appropriate solutions for an impurity positioned anywhere on the x-axis. This requires solving the problem with appropriate conditions such as: square-integrability, discontinuity of the derivative at $x=y$ that is in Eq. (\ref{ch5:eq:condition}), and continuity everywhere else. A quick but thorough calculation shows that for $x>0$ the wave function is given as:
\begin{align}
\begin{split}
&f(y|x)=\\
&A(x) e^{-y^2/2}U\left(a(x),\frac{1}{2},y^2\right)\cdot
\begin{cases}
1 & y\le 0\\
-1 + 2 \frac{U\left(a(x),\frac{1}{2},0\right)M\left(a(x),\frac{1}{2},y^2\right)}{M\left(a(x),\frac{1}{2},0\right)U\left(a(x),\frac{1}{2},y^2\right)} & 0<y<x\\
-1 + 2 \frac{U\left(a(x),\frac{1}{2},0\right)M\left(a(x),\frac{1}{2},x^2\right)}{M\left(a(x),\frac{1}{2},0\right)U\left(a(x),\frac{1}{2},x^2\right)} & y>x,
\end{cases}
\end{split}
\label{ch5:eq:analytical_wavefunction}
\end{align}
where $a(x)=1/4-\epsilon(x)/2$ and $A(x)$ is a normalization factor. The Tricomi and Kummer confluent hypergeometric functions are denoted by $U$ and $M$, respectively. The discontinuity condition on the derivative leads to:
\begin{align}
\begin{split}
\frac{-2x a(x)}{g} \cdot \frac{U\left(a(x),\frac{1}{2},0\right)}{M\left(a(x),\frac{1}{2},0\right)} \cdot&\\ \Bigg( \frac{M\left(a(x),\frac{1}{2},x^2\right)}{U\left(a(x),\frac{1}{2},x^2\right)} \Bigg. U\left(a(x)+1,\frac{3}{2},x^2\right) + 2 M\left(a(x)+1,\frac{3}{2},x^2\right) \Bigg)+&\\
U\left(a(x),\frac{1}{2},x^2\right)-2\frac{U\left(a(x),\frac{1}{2},0\right)}{M\left(a(x),\frac{1}{2},0\right)}M\left(a(x),\frac{1}{2},x^2\right) =&0,
\end{split}
\label{ch5:eq:conditiona}
\end{align}
which for $x=0$ simply reduces to:
\begin{align}
\frac{2}{g}+\frac{\Gamma\left(-\epsilon(0)/2+1/4\right)}{\Gamma\left(-\epsilon(0)/2+3/4\right)}=0,
\label{ch5:eq:reduced_x0}
\end{align}
by using the facts that in the limit of ${x\rightarrow0^+}$: $U(a(x),\frac{1}{2},x^2)=\frac{\sqrt{\pi}}{\Gamma(1/2+a(0))}$, $U'(a(x),\frac{1}{2},x^2)=\frac{-2\sqrt{\pi}}{\Gamma(a(0))}$, $M(a(x),\frac{1}{2},x^2)=1$, and $M'(a(x),\frac{1}{2},x^2)=0$.

Note that Eq.~(\ref{ch5:eq:reduced_x0}) is exactly the same equation I presented in Eq.~(\ref{HO2_energy}) back in Chapter 1 for the two-particles in a harmonic oscillator and in \cite{busch1998}. This is not surprising because at the end, the impurity-boson interaction is simply a two-body interaction with contact interaction. Finally, this means that for a given $x$ and $g$ one is able to find the value of $a(x)$ (hence $\epsilon(x)$) and write down the wave function for the bosons, $f(y|x)$. Condition (\ref{ch5:eq:conditiona}) along with Eq.~(\ref{ch5:eq:analytical_wavefunction}) can now be used to setup an effective coupled equation for the impurity:
\begin{align}
\left[H_A(x)+E_i(x)\right]\phi_i=\frac{1}{m_{AB}}\sum_{j=1} \left(Q_{ij}(x)\phi_j+P_{ij}(x)\frac{\partial \phi_j}{\partial x}\right),
\end{align}
where $P_{ij}(x)=\langle \Phi_i|\frac{\partial}{\partial x}|\Phi_j\rangle_{y}$ and $Q_{ij}(x)=\frac{1}{2}\langle\Phi_i|\frac{\partial^2}{\partial x^{2}}|\Phi_j\rangle_{y}$ and the subindex $y$ indicates that the integral is over all bosons $y_1,\ldots,y_{N_B}$. It has been shown that the ground state overlaps of these quantities gives $P_{ii}=0$ and $Q_{ii}<0$ \cite{nielsenPR2001}. For $Q_{ij}$ and $P_{ij}$, which are not necessarily zero, one can use physical intuition to argue that they can be neglected for $N\gg1$. Physical intuition and the results obtained in Chapter 4 for ideal Bose gases reveal that in the ground state the impurity is mostly found at the edge of the condensate in order to minimize the energy if $g_{BB}\ll g$ and $N_B\gg1$. This fact is mostly due to the interaction strength, which in the strong regime dominates all the other terms in the Hamiltonian. Therefore, it is reasonable to say that the quantity $\frac{\partial f}{\partial x}$ is negligible as the impurity is mostly found at the edge. Let me remind the reader that $f$ was introduced back in Eq.~(\ref{ch5:eq:product}) for being the single boson solution to the single particle quantum harmonic oscillator. Hence, the $P_{ij}$'s are not the important and leading terms in the above equation. 

In the same manner, one can expect that the contribution from the excited states may be minimal and the presence of the impurity does a poor job in coupling the ground state to the excited states of the bosons.  In fact, as the number of bosons increases, the method works even better, because the contribution from $\frac{\partial f}{\partial x}$ and $\frac{\partial^2 f}{\partial x^2}$ for the excited states becomes negligible compared to the ground state. Another property of bosons is that excited states have at least some gap of $\hbar\omega$ due to their trap, so even from an energy point of view the contributions are small. Therefore, the leading term in the above equation is $Q_{ii}$, which can be simply written as: $Q_{ii}=-\frac{1}{2}N_B\left\langle\left(\frac{\partial f_{gs}(y|x)}{\partial x}\right)^2\right\rangle_y$ assuming the bosons being in the ground state and writing $\Phi$ as a $N_B$ product of $f_{gs}(y|x)$'s.

\begin{figure}[t]
\centering
\includegraphics[width=0.9\columnwidth]{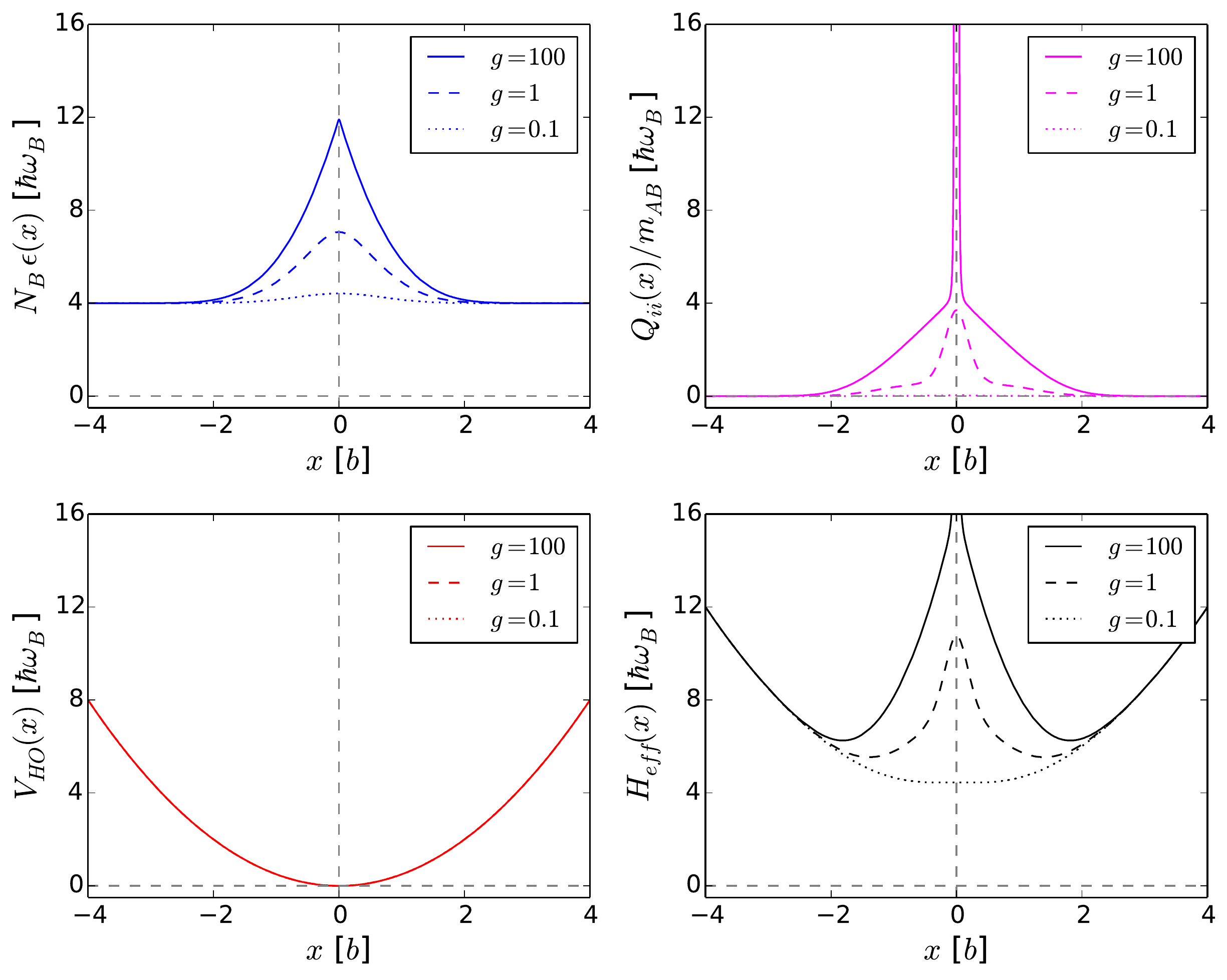}
\caption{Upper left panel shows the total bosonic energy in a harmonic oscillator for the $1+8$ system. This energy is a function of the position of the impurity, $x$, fixed in addition to the trap and is interacting with the bosons. Notice that it starts from $4\hbar\omega_B$ since there are 8 bosons in the ground state. Upper right panel shows the $Q_{ii}/m_{AB}$ as a function of $x$, which is a measure of the change in the wave function of the bosons as the impurity changes. Lower left panel shows the trap for the impurity, which is a harmonic potential for any interaction strength. Lower right panel shows the total effective Hamiltonian for the impurity (without the kinetic term).}
\label{ch5:effectivespectrum}
\end{figure}

\noindent
Finally, this reduces the effective Hamiltonian into this:
\begin{equation}
\begin{split}
H_{eff}\phi_1=\Bigg(H_A(x)+N_B\epsilon_{gs}(x)+\frac{N_B}{2m_{AB}}\left\langle\left(\frac{\partial f_{gs}(y|x)}{\partial x}\right)^2\right\rangle_y\Bigg) \phi_1=E\phi_1,
\label{ch5:eq:numericalequation1}
\end{split}
\end{equation}
where $f_{gs}(y|x)$ is the ground state wave function of the single-boson for a given $x$ with the corresponding energy $\epsilon_{gs}(x)$ as sketched in Fig.~\ref{polaronsketch1+N}. This is the main equation for the impurity, which is a 1D eigenvalue problem. It is interesting to see what each term in the big parentheses in Eq.~(\ref{ch5:eq:numericalequation1}) looks like. Fig.~\ref{ch5:effectivespectrum} shows the contribution from each term for different interaction strengths, $g=0.1$, $1.0$ and 100. Upper left panel shows the stored bosonic energy, $\epsilon_{gs}(x)$, for a given impurity position. The plot is for the $1+8$ systems, which is also why the energy starts at $4\hbar\omega_B$. Fig.~\ref{ch5:effectivespectrum} upper right panel shows the $Q_{ii}/m_{AB}$ as a function of $x$. One interesting feature is where the graph diverges near $x=0$ for $g=100$. This effect is due to the sudden change of the wave function for the bosons as the impurity goes from $x=0+\epsilon$ to $x=0-\epsilon$. In these two cases the wave function goes from being almost to the left to almost to the right resulting in the big change in the second derivative of the $f$ as a function of the position of the impurity. Fig.~\ref{ch5:effectivespectrum} lower left panel shows the trap for the impurity, which is a harmonic potential independent of interaction strength. Fig.~\ref{ch5:effectivespectrum} lower right panel shows the total effective Hamiltonian for the impurity (without the kinetic term). It is this effective Hamiltonian that is solved for the impurity. One can tell from this potential that the impurity solution will most probably be at the edges due to the peak in the middle of the potential.\\

\noindent
Solving $H_{eff}$, with the kinetic term in Eq.~(\ref{ch5:eq:numericalequation1}), is done by discretization. The discretization builds on the Finite Difference Method\index{Finite Difference Method}, where one discretizes the domain, $x$, into a small uniform grid of space, $\Delta x$, and produces an approximation of the derivative of the function. One realization is that in this grid the Hamiltonian is diagonal except the kinetic term, which requires the $\partial^2/\partial x^2$ to be written in another form (using 2nd order central step method):
\begin{equation}
\frac{\partial^2 f}{\partial x^2}=\frac{f(x_{j+1})-2f(x_{j})+f(x_{j-1})}{\Delta x^2}.
\end{equation}
In this form the error is in order of $O(\Delta x)^2$. Written as a matrix, this becomes:
\begin{equation}
\frac{1}{\Delta x^2}
\begin{bmatrix}
-2 & 1 & 0 & \cdots & 0 & 0 \\
1 & -2 & 1 & \cdots & 0 & 0 \\
0 & 1 & -2 & \cdots & 0 & 0 \\
\vdots & \vdots & \vdots & \ddots & \vdots & \vdots \\
0 & 0 & 0 & 0 & -2 & 1 \\
0 & 0 & 0 & 0 & 1 & -2
\end{bmatrix}.
\end{equation}
Along with the rest of the terms in Eq.~(\ref{ch5:eq:numericalequation1}), which are diagonal and evaluated at each grid point, one can then obtain the energy spectrum and eigenvalue of the impurity. These results are presented in the next section.

In the case of non-ideal Bose gases, one can even use the 1D Gross-Pitaevskii equation\index{Gross-Pitaevskii Equation} (GPE) in addition to the terms in Eq.~(\ref{ch5:eq:majorityparticles_eigenvalue}). In other words, when one wants to find the solutions to the harmonic trap with an impurity fixed somewhere on the axis, then one also needs to take the self-interaction term, $N_B\cdot g_{BB}|f_{k^j_i}|^2$, into account:
\begin{align}
\mu(x) f_{k^j_i}=\left(-\frac{1}{2}\frac{\partial^2}{\partial y_i^2}+\frac{1}{2}y_i^2+N_B\cdot g_{BB}| f_{k^j_i}|^2\right) f_{k^j_i}\label{eq3},
\end{align}
where $f_{k^j_i}$ is normalized to 1, $\int f_{k^j_i}(y|x)|^2 dy=1$, and $\mu(x)$ is a chemical potential for a given $x$. The $g_{BB}$ is the one-dimensional interaction strength, which is determined through the three-dimensional boson-boson scattering length, $a_s$, as: 
\begin{equation}
g_{BB}= \frac{2 a_s}{b}\frac{\omega_\perp}{\omega_B}\frac{1}{(1-C\frac{a_s}{a_\perp})},
\end{equation}
in units of [$b\hbar\omega_B$] where $\omega_{\perp}$ is the trapping frequency along transverse directions in order to confine the particles in a 1D geometry. $a_\perp=\sqrt{\frac{\hbar}{m\omega_\perp}}$ is the size of the ground state in the transverse direction and $C$, as derived in \cite{OlshaniiPRL1998}, is a constant that is approximately $C=-\zeta(1/2)/\sqrt{2}\approx 1.0326$, as discussed in the introductory section, Eq.~(\ref{g1D}). The $(1-C\frac{a_s}{a_\perp})$ factor is approximately $0.92$ for typical values of a $^{87}$Rb gas. This gives a value of $g_{BB}=0.3\rightarrow3.0$ for $10\rightarrow100$ particles.

\subsection{Results}\label{ch:polarons:subsec:results}
After solving Eq.~(\ref{ch5:eq:numericalequation1}), it is interesting to know how well the method captures the energies of a certain system. This was tested for both $1+3$, $1+4$, $\dots$, and $1+8$ systems. It turned out that the relative precision gets better for higher number of particles. In that sense the method does not so well for few-body systems, but much better for higher number of particles. On the other hand, most experiments deal with a large number of particles, so this method would be very appropriate to use in those cases. In the following I will present the results for the $1+8$ systems.

\begin{figure}[hbt!]
\centering
\includegraphics[width=\columnwidth, trim=0cm 1cm 0cm 1.3cm]{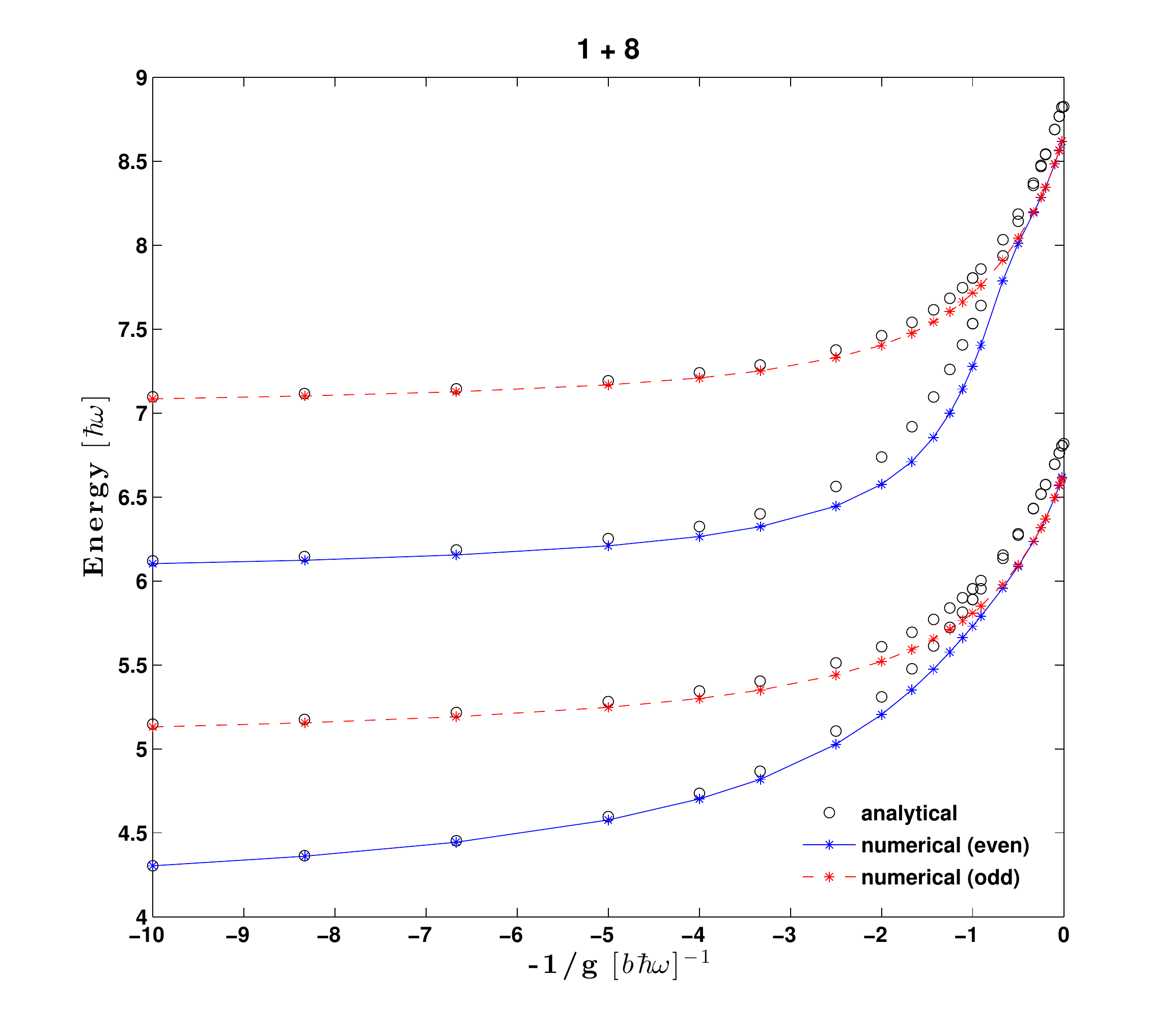}
\caption{The energy spectrum of low-lying states for a system of eight non-interacting bosons and an impurity with $m_{AB}=\omega_{AB}=1$. The semi-analytical data is compared with numerical exact calculations (EEDM). Figure is adapted from \cite{DehkharghaniPRA2015}.}
\label{polarons_energyspectrum}
\end{figure}

Fig.~\ref{polarons_energyspectrum} shows the energy spectrum for the $1+8$ system for the lowest lying energies compared with results obtained from exact diagonalization \index{Effective Exact Diagonalizing Method}(EEDM). As the figure reveals, the relative error is only about $3.5\%$ in the worst-case scenario ($g\rightarrow\infty$). As the energies seem to match very well, one would wonder how the wave functions will manage when they are compared to the numerical results.

\begin{figure}[hbt!]
\centering
\includegraphics[width=\columnwidth, trim=0cm 0cm 0cm -1.5cm]{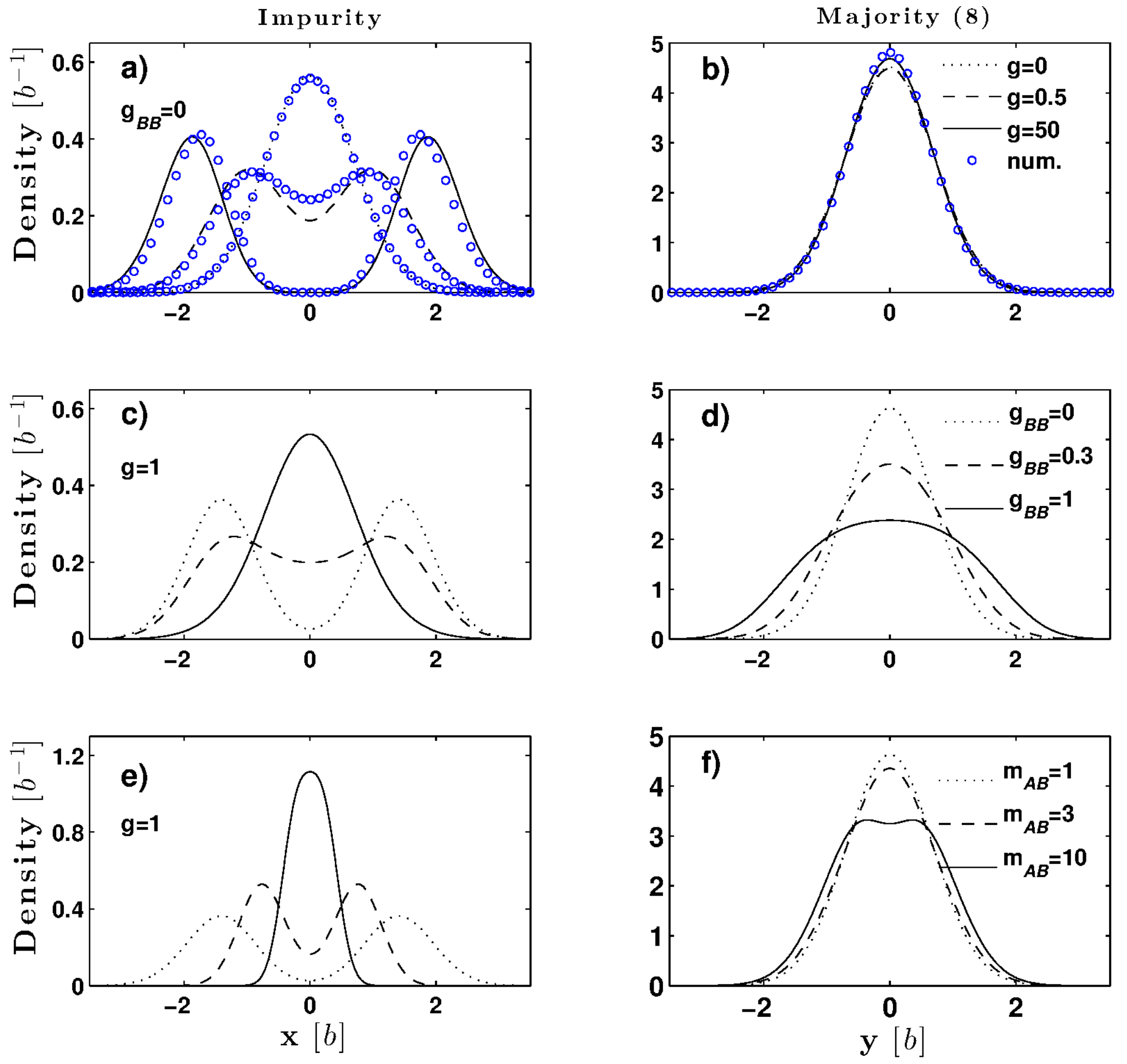}
\caption{Panels a) and b) show the density distributions for the impurity and majority, respectively. The investigated system is the $1+8$ system with varying $g$. The blue circles show the corresponding numerically exact calculated results (EEDM). Panels c) and d) show the distribution for a fixed value of $g$ but varying self-interaction strength, $g_{BB}$. Panels e) and f) show the distribution for ideal Bose gas but fixed $g=1$ and varying mass ratios, $m_{AB}=1$, $3$, and $10$. Figure is adapted from \cite{DehkharghaniPRA2015}.}
\label{polarons_densities}
\end{figure}

\begin{figure}[t]
\centering
\includegraphics[width=0.8\columnwidth]{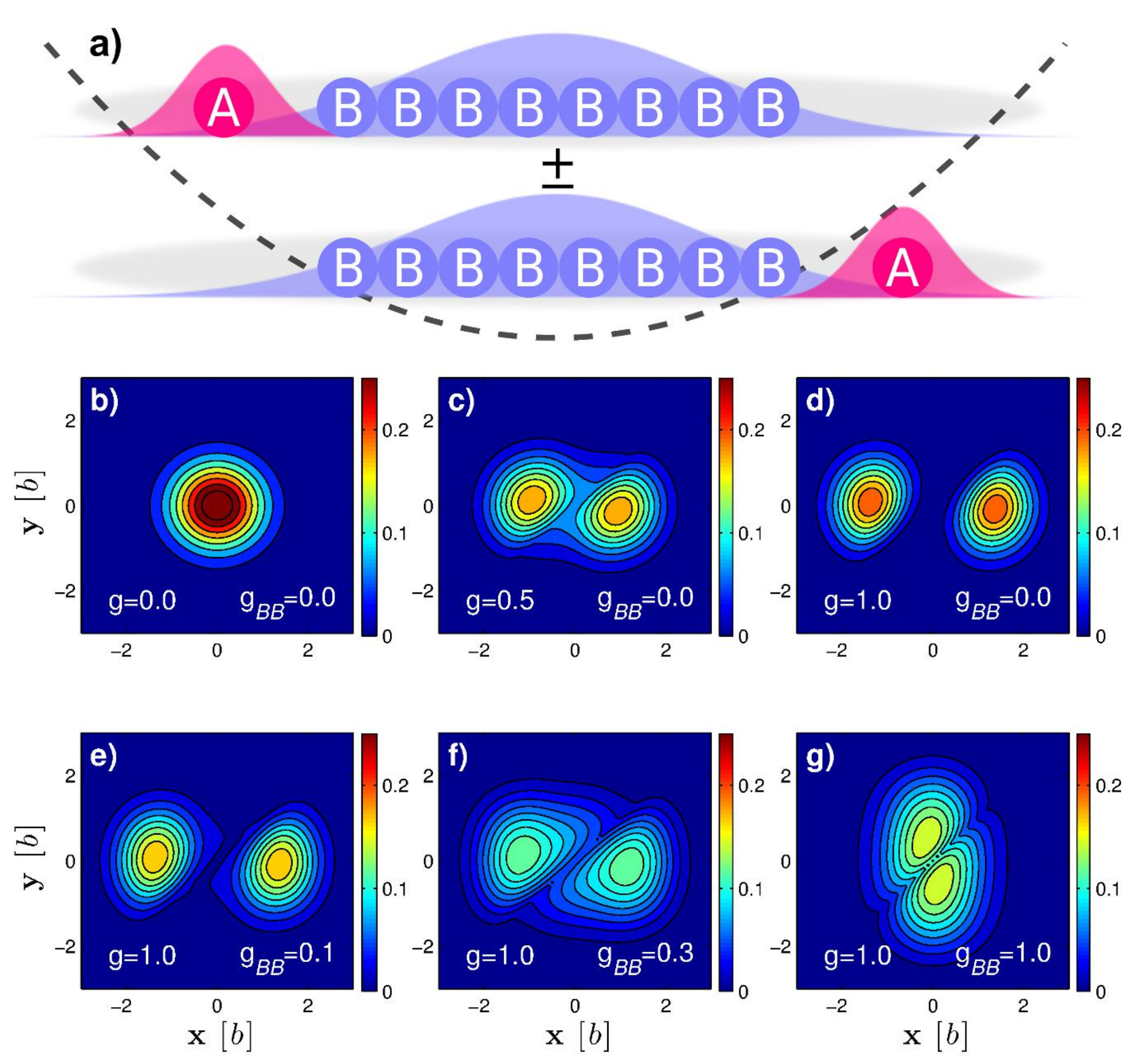}
\caption{a) shows a sketch of the $1+8$ system in a 1D harmonic potential. Due to the symmetry, the ground state becomes doubly degenerated due to the party with the impurity being in the left or right of the condensate. Panels b)-g) show the pair-correlation plot as one varies the interaction strength for fixed self-interaction, b)-d), or the other way around, e)-g). Figure is adapted from \cite{DehkharghaniPRA2015}.}
\label{polaron_paircorrelation}
\end{figure}

Fig.~\ref{polarons_densities} shows the density plots of the $1+8$ system for different kind of variables. The densities for each species is calculated as follows: $n(x)=\int{|\psi|^2\mathrm{d}y_{1}\mathrm{d}y_{2}\dots \mathrm{d}y_{N_B}}$ and $n_B(y)=N_B\int{|\psi|^2\mathrm{d}x\mathrm{d}y_{2}\mathrm{d}y_{3}\dots \mathrm{d}y_{N_B}}$. First, the upper panels, a) and b), show the density distribution for an ideal gas that is interacting with an impurity with strengths ($g=0$, $0.5$ and $50$). These distribution (dotted, dashed and solid lines, respectively) are then compared with numerical values (blue circles). Once again, it is clear that the method is quite good in capturing the physics in the system. It is apparent from the figure that as the interaction strength becomes strong, the impurity starts to move from the middle of the condensate to the edges. On the other hand, the majority particles are unaffected by this change. Looking at this from the energy perspective, this shows that the energy becomes doubly degenerated. Since the impurity is either to the right or the left of the condensate, a plus/minus linear combination of the two configurations are equally good candidates for the eigenstates. This is illustrated in Fig.~\ref{polaron_paircorrelation} a).


However, if one then observes the behavior of the condensate with non-ideal Bose gases, it becomes notable that the impurity starts to move into the middle as the Bosons interact with each other. This is shown in the middle panels, c) and d), of Fig.~\ref{polarons_densities}, where $m_{AB}=\omega_{AB}=1$. In this case the self-interaction strength is set to vary, $g_{BB}=\{0, 0.3, 1\}$ while the interaction strength with impurity is fixed to $g=1$. At $g_{BB}=1$ the impurity is in the middle, while the bosons have become more spread in space due to the self-interaction. Even though the distribution in the case of $g=g_{BB}$ shows different patterns, and yet contribute equally in the Hamiltonian, it is worth mentioning that the two are treated differently. The GPE and the developed method are not exact and they only produce qualitative results. Along with the density plots, a pair-correlation plot was also investigate and this is shown in Fig.~\ref{polaron_paircorrelation} b)-g). This is calculated by the following equation: $n_{AB}(x,y_1)=\int{|\psi|^2 dy_{2}\dots dy_{N_B}}$. The panels show how the impurity moves to the edge as $g$ varies and $g_{BB}$ is fixed to zero. Then the impurity starts to move to the middle as $g$ is fixed to one and then $g_{BB}$ is varied.

Finally, in the lower panels, Fig.~\ref{polarons_densities} e) and f), one can see how the impurity, as it gets heavy, starts to split the condensate for $g_{BB}=0, g=1$ and $\omega_{AB}=1$. Notable deformation happens at a mass ratio $m_{AB}=10$. It is again clear that if the impurity is heavy enough, then for the ground state it is more energy efficient for the impurity to be in the middle of the condensate rather than moving to the edge of the condensate. As it is clear now, it all comes down to the question of which variable is the most dominant one and at what values it is the most energy stable configuration for the system to be in. This method can qualitatively and very quickly with a decent precision produce what happens for some given variables. Therefore the method can be very helpful to do a quick calculation in experiments with a lot of bosons.\\

\begin{figure}[t]
\centering
\includegraphics[width=\columnwidth]{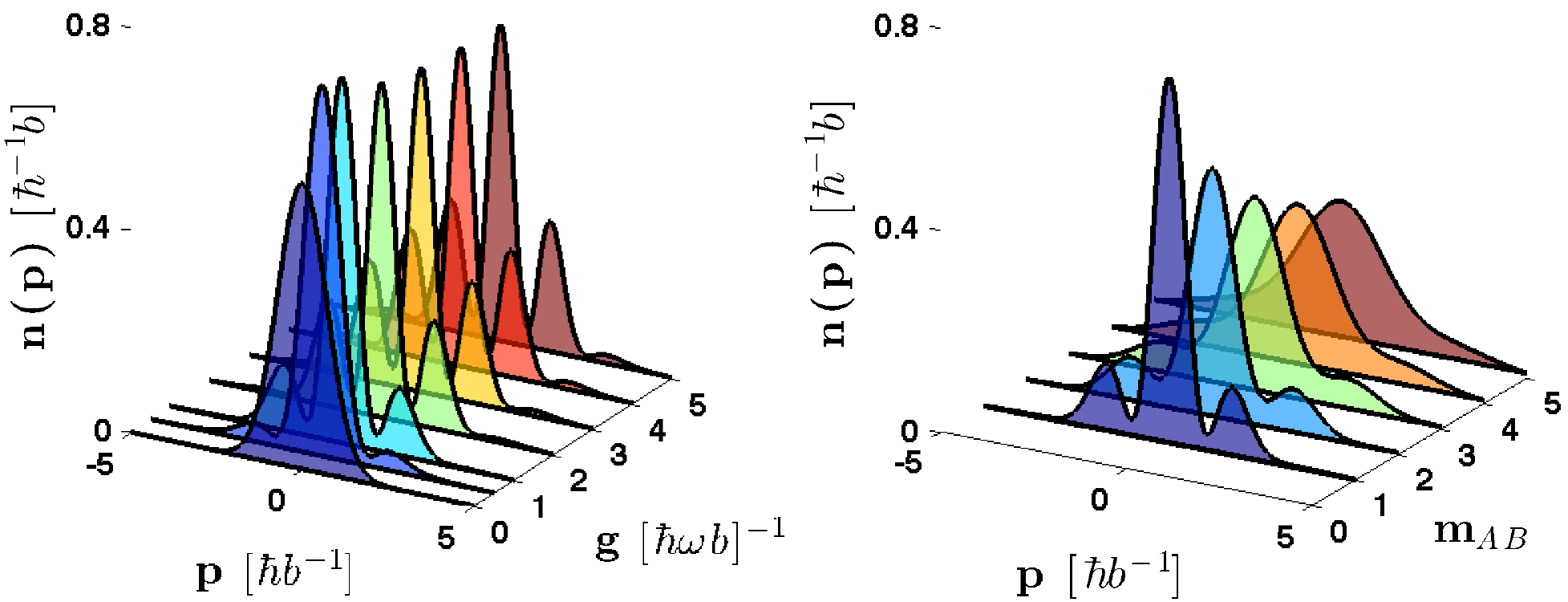}
\caption{Momentum distribution for the impurity. Left panel shows the different distributions as the mass ratio is fixed and $g$ is varied. Right panel shows the same distribution but for a fixed $g=1$ and varying mass ratio, $m_{AB}$. All the other variables are unchanged, $g_{BB}=0$ and $\omega_{AB}=1$. Figure is adapted from \cite{DehkharghaniPRA2015}.}
\label{polaron_momentum}
\end{figure}

\noindent
Experimentally it is relevant to know how the momentum distribution behaves as the variables are changed. Fig.~\ref{polaron_momentum} shows how the momentum distribution changes as a function of interaction strength, $g$, (left panel) and mass ratio, $m_{AB}$, (right panel). Both figures are for $g_{BB}=0$ and $\omega_{AB}=1$. The momentum distribution is calculated by the following equation $n(p)=\int |\psi(p, q_1,\dots,q_{N_B})|^2 dq_1,\dots,dq_{N_B}$:
\begin{align}
n(p)=\frac{1}{2\pi}\int_{all~space}dxd\tilde{x}~{\phi^*(x)\phi(\tilde{x})\left(\int dy f^*(y|x)f(y|\tilde{x}) \right)^{N_B}~e^{ip(x-\tilde{x})}}.
\end{align}
Having assumed that the wave function is a product state. In the above equation one needs to take the Fourier transform of the wave function, which is defined as following,
\begin{equation*}
\begin{split}
&\psi(p, q_1,\dots,q_{N_B})=\\
&\left(\frac{1}{\sqrt{2\pi}}\right)^{1+N_B}\int{dx_1dy_1\dots dy_{N_B}\psi(x_1,y_1,\dots,y_{N_B})e^{ipx}e^{iq_1y_1}... e^{iq_{N_B}x_{N_B}}},
\end{split}
\end{equation*}
in order to calculate the momentum distribution. As it is illustrated in the left panel, the impurity starts from a Gaussian-shape to gain some wings as the interaction strength increases. On the other hand, when the impurity starts to be the heavy particle, it goes to the middle and as it is a sharper Gaussian function in coordinate space, it becomes a wider Gaussian in the momentum space. The characteristic wings in both graphs are very distinguishable in both limits, which can be verified experimentally. The momentum distribution\index{Momentum Distribution} for majority particles is not shown here, because the distribution does not change significantly.

\subsection{Entanglement\index{Entanglement} of The Impurity}\label{ch:polarons:sec:entanglementoftheimpurity}
After establishing the approximated method for quantum impurities, one can then investigate the correlations in the system due to the interactions between the bosons and the impurity. The idea is to calculate the one-body density matrix (OBDM) \index{One-Body Density Matrix}for a boson, $\rho^{{\mathrm{B}}}$, and an impurity, $\rho^{{\mathrm{I}}}$, in the following way:
\begin{equation}
\rho^{{\mathrm{B}}}(y,y')\!=\!N_{\mathrm{B}}\!\!\int\!\! d y_2\dots d y_{N_B}d x \Psi(y,y_2,\dots, y_{N_B}, x)\Psi(y',y_2,\dots , y_{N_B}, x), 
\end{equation}
and
\begin{equation}
\rho^{{\mathrm{I}}}(x,x')\!=\!\int\!\! d y_1\dots d y_{N_{B}}\Psi(y_1,\dots,y_{N_B},x)\Psi(y_1,\dots,y_{N_B},x'),
\end{equation}
and then diagonalize the matrices to obtain natural orbitals and their occupation, $\lambda_i$. From the natural orbits occupations one can also calculate the entanglement entropy \index{Entanglement!Entropy}given by:
\begin{equation}
S(\rho^{{\mathrm{I}}})=-\mbox{Tr}[\rho^{{\mathrm{I}}}~\mbox{log}_2~\rho^{{\mathrm{I}}}]=-\sum_i \lambda_i \mbox{log}_2 \lambda_i\
\end{equation}
This work was carefully investigated in \cite{Garcis-MarchJoPBAMaOP2016} where the semi-analytical results that I have done, and numerical exact calculations provided by M.~A.~Garc\'{\i}a-March were compared and analyzed.

One of the main results here is that the largest occupation of the natural orbital for the bosons is decreasing as the interaction between the impurity and the bosons increases. On the other hand, if the number of the bosons increases, the occupation gets less disturbed by the presence of the impurity. As it comes to the entropy, it is shown that the entropy increases as a function of $g$, meaning that the interaction increases the correlation between the bosons and the impurity. Furthermore, the impact of the mass ratio has also been explored and here, it is shown that when the impurity is very heavy and localized in the middle the least possible correlation occurs between the particles. In fact, the entropy is increasing as the mass ratio is increased from 1 to 3 or 4 but then, as the heavy impurity manages to make space between the bosons and separate the bosons into two condensates, the entropy decrease and reaches a minimum. For more details please have a look at \cite{Garcis-MarchJoPBAMaOP2016}.

\section{Double Quantum Impurity: $\mathbf{2+N}$\index{2+N Systems}}\label{ch:polarons:sec:doublequantumimpurity}
\begin{figure}[t]
\centering
\includegraphics[width=\columnwidth]{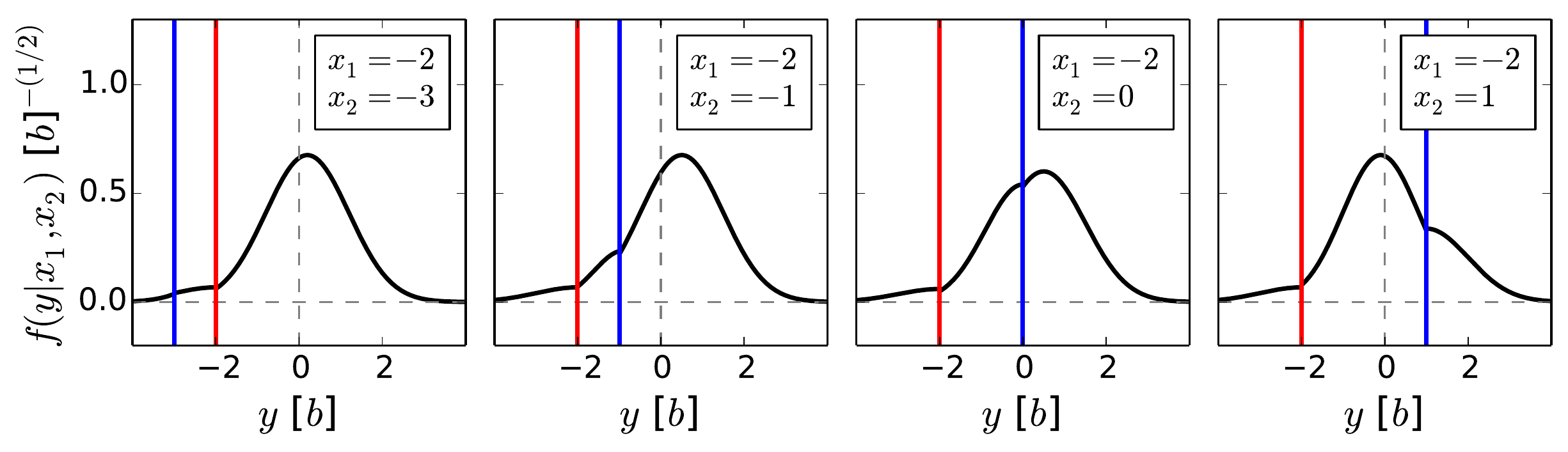}
\caption{Sweeping through the x-axis with two impurities. Red line represents one impurity and the blue line represents the other impurity. As the wave function, $f(y|x_1,x_2)$, and the corresponding energy, $\epsilon(x_1,x_2)$, are found, one can then calculate the effective potential for the impurities, which can be solved afterwards. This figure has not been published before.}
\label{polaronsketch2+N}
\end{figure}

After developing the structure and framework for the single impurity, the question becomes whether or not the same method could be applicable for two impurities. Using the same notation as the single impurity one can write the overall Hamiltonian of the system as:
\begin{align}
H&=\sum_{i=1}^2 H_{A}(x_i) + \sum_{i=1}^{N_B} H_{B}(y_i)+\sum_{j=1}^{2}\sum_{i=1}^{N_B}g\delta(x_j-y_i) + \sum_{i<k}^{N_B}g_{BB}\delta(y_i-y_k).
\label{ch5:eq:hamiltonian}
\end{align}
As it is shown in Fig.~\ref{polaronsketch2+N} instead of sweeping once (the red line), one has to sweep through with two impurities (red and blue lines). In this way, one has to find the wave function for the majority particles for every possible fixed position of the impurities. This increases the number of calculations, but yet it is possible to apply the same techniques as before in order to obtain some approximated solutions.

By applying the semi-analytical adiabatic decomposition\index{Adiabatic Decomposition} for the total wave function, where the impurities are assumed to be the slow variables, one can decompose the total wave function, $\Psi(x_1,x_2,y_1,\dots,y_{N_B})$, into a wave function for the impurities, $\phi_{II}(x_1,x_2)$, and a total wave function for the bosons for the given position of the same impurities, $\Phi_j(y_1,\dots,y_{N_B}|x_1,x_2)$. The $\Phi_j$'s are the solutions to the particular eigenvalue problem:
\begin{equation}
\left(\sum_{i=1}^{N_B} H_{B}(y_i)+\sum_{i=1}^2 \sum_{j=1}^{N_B}g\delta(x_i-y_j)\right)\Phi_j=E_j(x_1, x_2)\Phi_j,
\end{equation}
for a given impurity position $x_1$ and $x_2$, and interaction strength $g$. The bosons are assumed to be a product of the single particle solutions, $f(y_i|x_1,x_2)$ sketched in Fig.~\ref{polaronsketch2+N}, to the same eigenvalue problem with energies $E=N_B\epsilon(x_1,x_2)$. Here, I assume that only the lowest states contribute to the interactions, hence the single particle wave function is $f_{gs}(y_i|x_1,x_2)$ with corresponding energy $\epsilon_{gs}(x_1,x_2)$.\\

\begin{figure}[t!]
\centering
\includegraphics[width=\columnwidth]{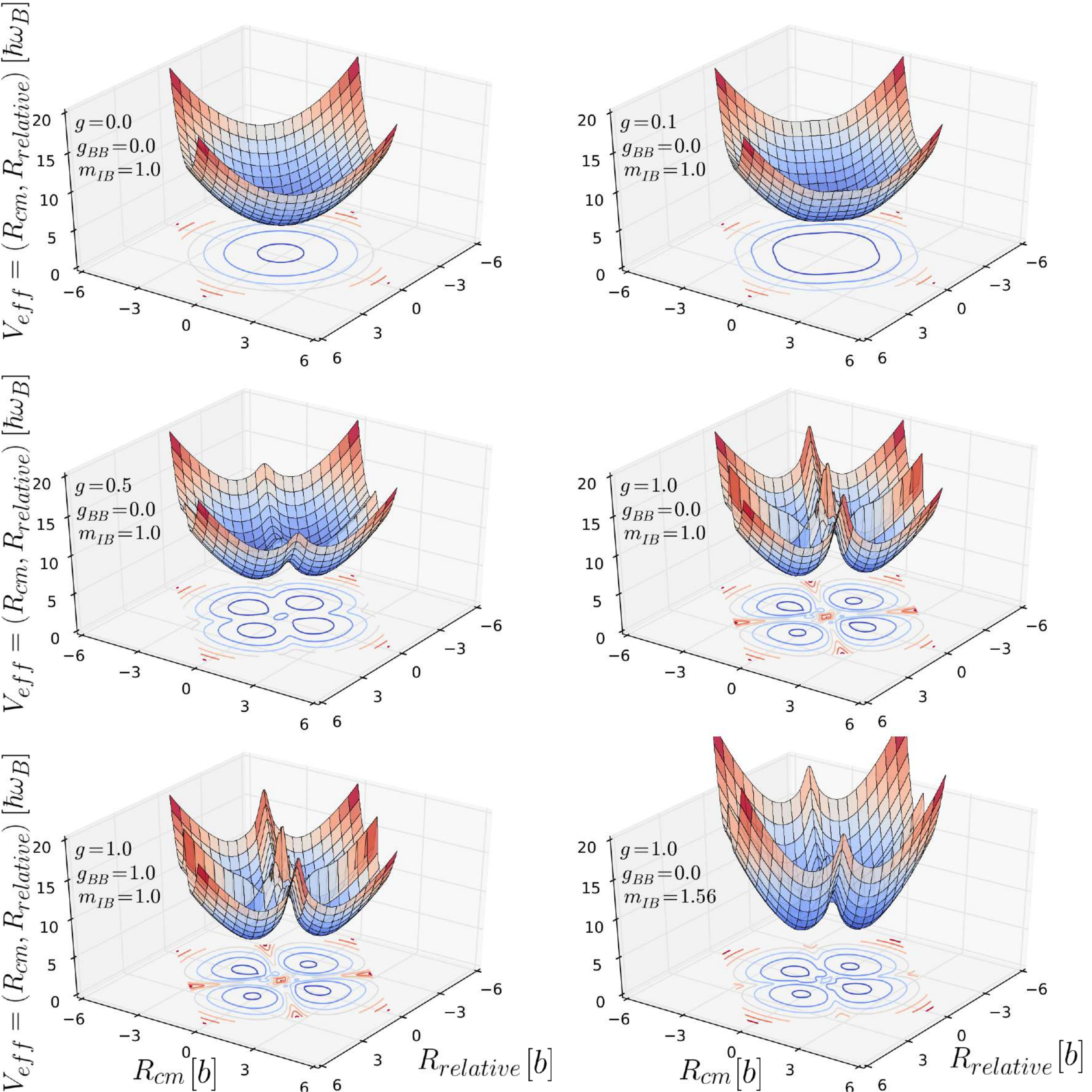}
\caption{For a given position of the impurity ($x_1,x_2$) the effective Hamiltonian without the kinetic terms, $V_{eff}\equiv\sum^2_{i=1} \left(x_i^2+\frac{N_B Q_{x_i}}{2m_{AB}}\right)+N_B\epsilon_{gs}$, can be plotted in a 3D plot as a function of the relative, $R_{relative}=(x_1-x_2)/\sqrt{2}$, and center of mass coordinates, $R_{cm}=(x_1+x_2)/\sqrt{2}$. The upper and middle panels show the effective Hamiltonian for varying interaction strength, $g$. Lower left panel shows the impact of self-interaction, which makes the wells broader allowing the impurities to separate even more, while the lower right panel shows the Cs-Rb mass imbalance ($m_{AB}=m(Cs)/m(Rb)=1.56$), which makes the wells deeper. This figure and the upcoming figures have not been published before submitting this thesis.}
\label{ch6:effectivepotential2+N}
\end{figure}

\noindent
After finding the solutions to $f_{gs}(y_i|x_1,x_2)$ and the corresponding energies $\epsilon_{gs}(x_1,x_2)$, an effective potential is obtained for the impurities $\phi_{II}(x_1,x_2)$:
\begin{equation}
\begin{split}
\Bigg(\sum^2_{i=1} \Big( H_A(x_i)+\frac{N_B Q_{ii, x_i}}{2m_{AB}}\Big)+N_B\epsilon_{gs}(x_1,x_2)\Bigg) \phi_{II}(x_1,x_2)=E\phi_{II}(x_1,x_2),
\label{eq:numericalequation}
\end{split}
\end{equation}
where $Q_{ii, x_i}=\left\langle\left(\frac{\partial f_{gs}(y|x_1,x_2)}{\partial x_i}\right)^2\right\rangle_y$ for $i\in\{1,2\}$. This equation can be solved in a 2D discretized space and the corresponding eigenenergies and wave functions are therefore easily obtained. The effective Hamiltonian without the kinetic terms in Eq.~(\ref{eq:numericalequation}) is called:
\begin{equation}
V_{eff}\equiv\sum^2_{i=1} \left(x_i^2+\frac{N_B Q_{x_i}}{2m_{AB}}\right)+N_B\epsilon_{gs},
\end{equation}
and is plotted in Fig.~\ref{ch6:effectivepotential2+N} for different parameters as a function of the relative coordinate, $R_{relative}=(x_1-x_2)/\sqrt{2}$, and the center of mass coordinate, $R_{cm}=(x_1+x_2)/\sqrt{2}$. The upper and middle panel show the effective Hamiltonian as one changes the interaction strength, $g$. Notice that in the non-interacting case, the solution is simply a 2D Gaussian wave function, as the effective potential is a smooth 2D parabola. On the other hand, it is clear how the effective potential splits in the middle as $g$ is increased. Lower left panel shows the impact of self-interaction, which makes the wells broader allowing the impurities to separate even more, while the lower right panel shows the Cs-Rb mass imbalance ($m_{AB}=m(Cs)/m(Rb)=1.56$), which makes the wells deeper. This mass ratio was recently realized in an experiment with impurity being the Cs atom and the majority particles represented by a Rb gas \cite{Spethmann2012}. From this contour plot one can have a qualitative description of where the impurity is most probably going to be found. While the harmonic potential is the dominating term for bigger distances, one could ask how the picture will be affected in the presence of no external potential. Whether or not the impurities still tend to sit together or just mix together with the bosons is a good question to ask and worth investigating, which I will discuss later in this chapter.

Although the effective Hamiltonian is a 2D potential, it is still relatively easy to solve. One quick and relatively precise method is the 2D finite discretized method. In this method the space is discretized in 2D where the eigenvalue problem is evaluated in this space. However, according to this method all the terms are diagonal except the $\partial^2/\partial x_1^2+\partial^2/\partial x_2^2$ terms in the differential equation. With a step size of $\Delta x$ one can write a 2D version of the $\partial^2/\partial x_1^2+\partial^2/\partial x_2^2$ as:
\begin{align}
\begin{split}
\frac{\partial^2 \phi}{\partial x_1^2}+\frac{\partial^2 \phi}{\partial x_2^2}=&\frac{\phi(x_{i+1,j})-2\phi(x_{i,j})+\phi(x_{i-1,j})}{\Delta x^2}\\&+\frac{\phi(x_{i,j+1})-2\phi(x_{i,j})+\phi(x_{i,j-1})}{\Delta x^2},
\end{split}
\end{align}
which has an error of order $O(\Delta x)^2$. This means that the smaller the step is, the better is the precision. It should be noted that the above discretization is the most simple central way $\{1,-2,1\}$ to discretize the space, which has a second-order accuracy. Other higher order discretization such as $\{-1/12, 4/3, -5/2, 4/3, -1/12\}$ could also be chosen for better accuracy instead of step-size, however this makes the matrix more complicated. For this purpose here, the simples second-order accuracy with a step size of $\Delta x=0.08$ is chosen. The simplest second-order accuracy in matrix form can be written as:
\begin{equation}
\frac{1}{\Delta x^2}
\begin{bmatrix}
-4 & 1 & 0 & {\color{red}1} & 0 & \cdots & 0 & 0 \\
1 & -4 & 1 & 0 & {\color{red}1} & \cdots & 0 & 0 \\
0 & 1 & -4 & 1 & 0 &\cdots & 0 & 0 \\
{\color{red}1} & 0 & 1 & -4 & 1 &\cdots & 0 & 0 \\
0 & {\color{red}1} & 0 & 1 & -4 &\cdots & 0 & 0 \\
\vdots & \vdots & \vdots &\vdots & \vdots & \ddots &\vdots&\vdots \\
0 & 0 & 0 & {\color{red}1} &0 & 1 & -4 & 1 \\
0 & 0 & 0 & 0 &{\color{red}1} & 0 & 1 & -4
\end{bmatrix},
\end{equation}
where the appearance of the last {\color{red}red} diagonals depends on the size of the discretization. In the example shown above the matrix is written for a space discretized in a $3\times3$ grid. In general these red coefficients would appear after the $n$'th discretized entry. After the diagonalization, one obtains the eigenenergy and wave function of the impurities. In my calculations I have discretized the space into $n=201$ points, which has a total calculation time of 10 minutes on a 16GB RAM computer. Higher discrete points requires a bigger RAM. As a side note, I can also reveal that the work here was repeated and tested in details by Thorbjørn Lindgren in his Bachelor's thesis.\\

\begin{figure}[t!]
\centering
\includegraphics[width=\columnwidth]{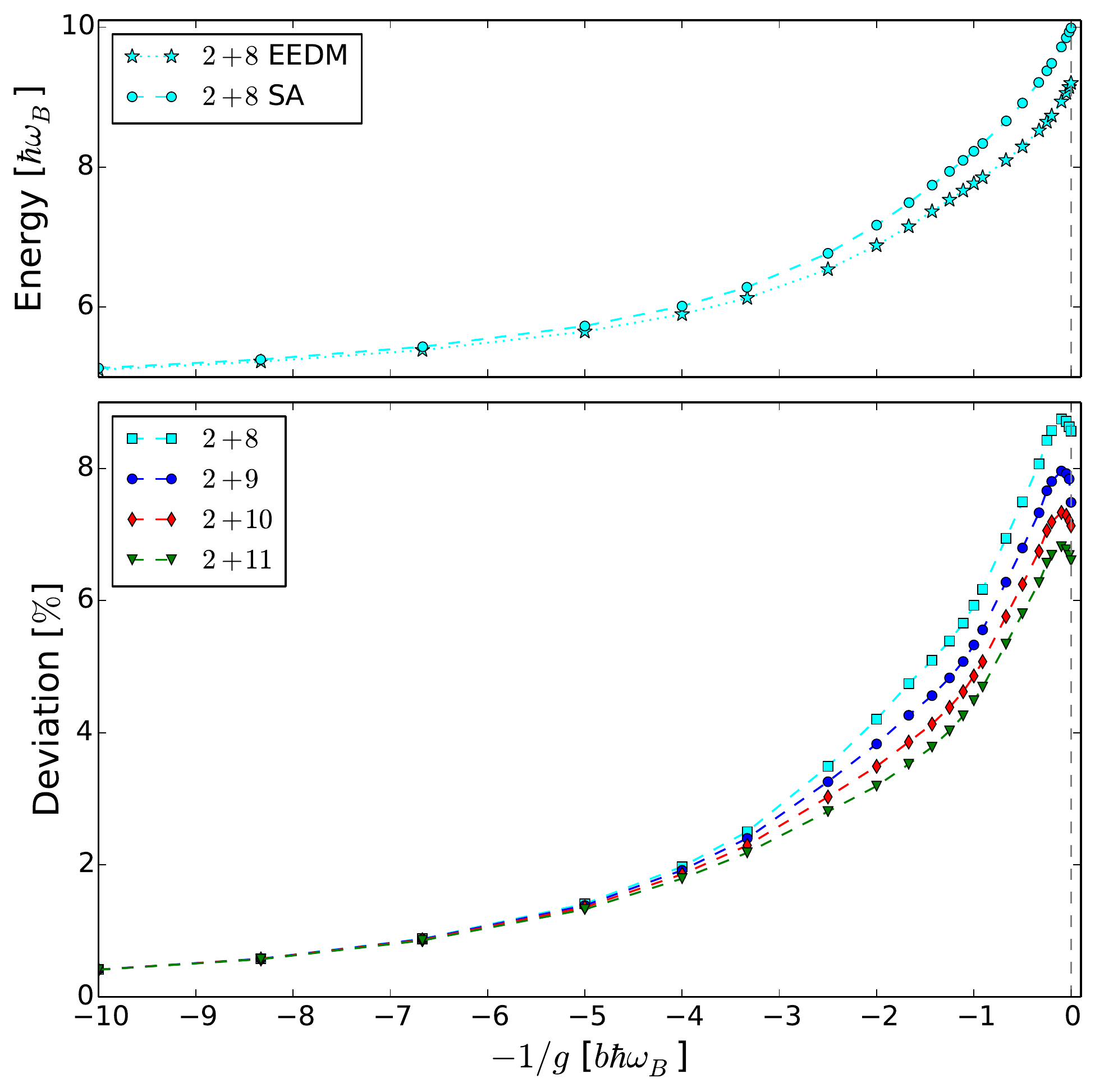}
\caption{Upper panel: Energy-spectrum of the ground state for 8 ideal bosons and two identical impurities ($2+8$) with $m_{AB}=\omega_{AB}=1$. The stars are the numerical exact calculated energies from EEDM, while cyan squares are the semi-analytical results for 8 bosons. Lower panel: The deviation from the exact results in \%. The lines are guides for the eye. Notice how the deviation decreases as the number of particles increases. This figure has not been published before.}
\label{ch6:numerics_energy_vs_semi}
\end{figure}

\noindent
So in order to see how the method does compared with exact diagonalizing method, I have calculated the total energy of the $2+8,9,10,11$ systems. Fig.~\ref{ch6:numerics_energy_vs_semi} upper panel shows the energy spectrum of the ground state in a system of two non-interacting impurities and $N_B=8$ non-interacting identical bosons. In the lower panel the deviation for each system is also shown. Despite the worst case $\approx8\%$ deviation, it is clear that the method gets better in precision as the number of bosons increases. The reason for this is found in the physical arguments made earlier for the single impurity case that the coupling between the ground state and excited states is minimal. In fact, it is shown in \cite{DehkharghaniPRA2015} that these couplings indeed are $N_B$-dependent in such a way that they become negligible as the number of bosons increases. Therefore only $Q_{ii}$ is the leading coupling term that is important to calculate in order to get a reasonable solution.

\begin{figure}[t!]
\centering
\includegraphics[width=\columnwidth]{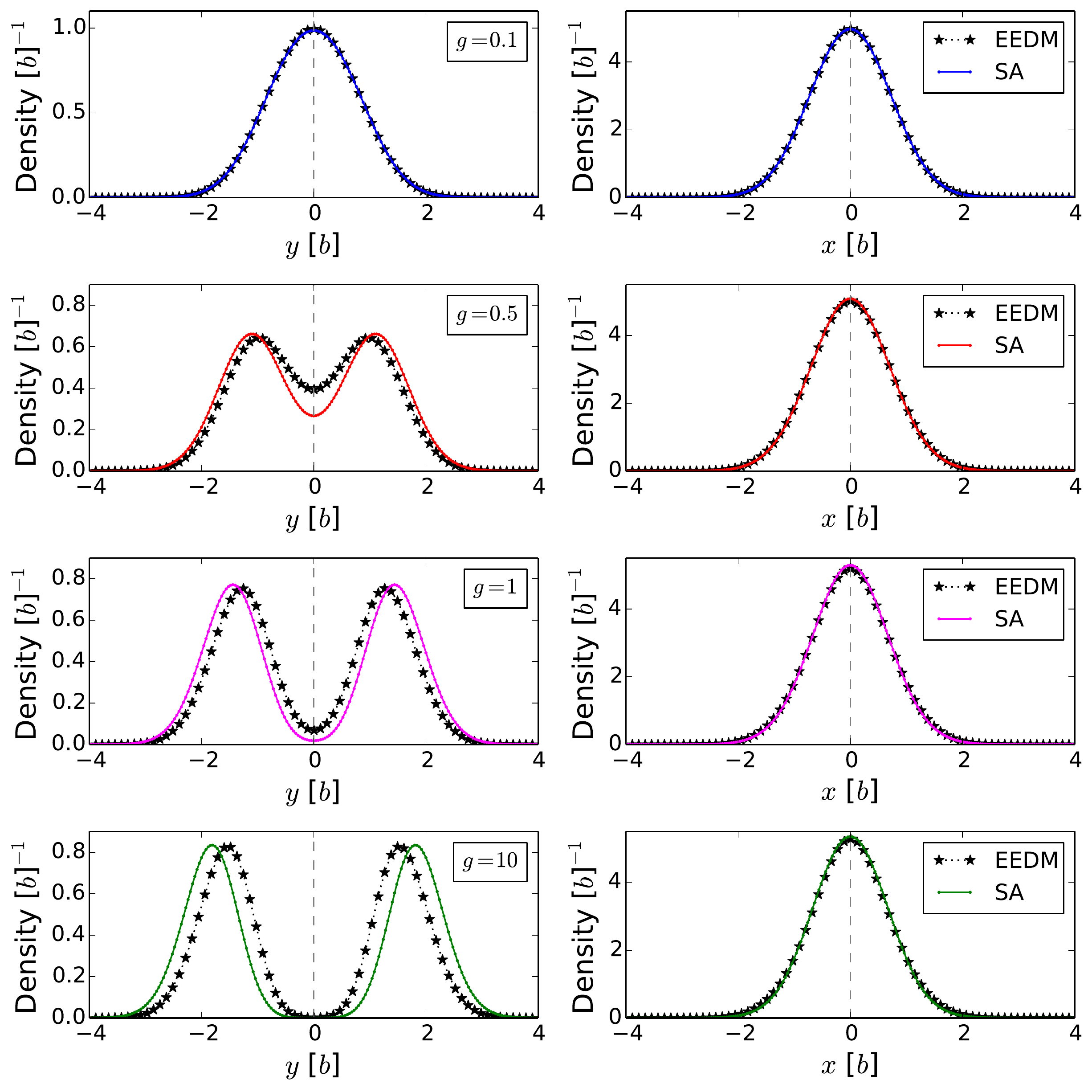}
\caption{The density plot semi-analytically calculated for both components for different interaction strengths ($g=0.1, 0.5, 1, 10$). The density distribution for the majority particles (right panels) are unaffected by the present of the impurities, while the impurities (left panels) split as the interaction increases. The black stars are calculated with EEDM in Chapter 7. This figure has not been published before.}
\label{ch6:numerics_densities_vs_semi}
\end{figure}

Having obtained the eigenvalues and the corresponding eigenfunctions, one can investigate the density. When it comes to the density plot, Fig.~\ref{ch6:numerics_densities_vs_semi} shows how well the method does in capturing the pattern of the system. The figure shows a series of different densities depending on the interaction between the impurity and the bosons only. In the figure, the left panels show the distribution for the impurities, while the right panels show the distribution for the majority particles. The majority particles are almost unaffected by the presence of the impurities, while the impurities tend to split when the interaction increases. From the density distribution of the impurities one can tell that the impurities go to the edge of the condensate.

\begin{figure}[t!]
\centering
\includegraphics[width=\columnwidth]{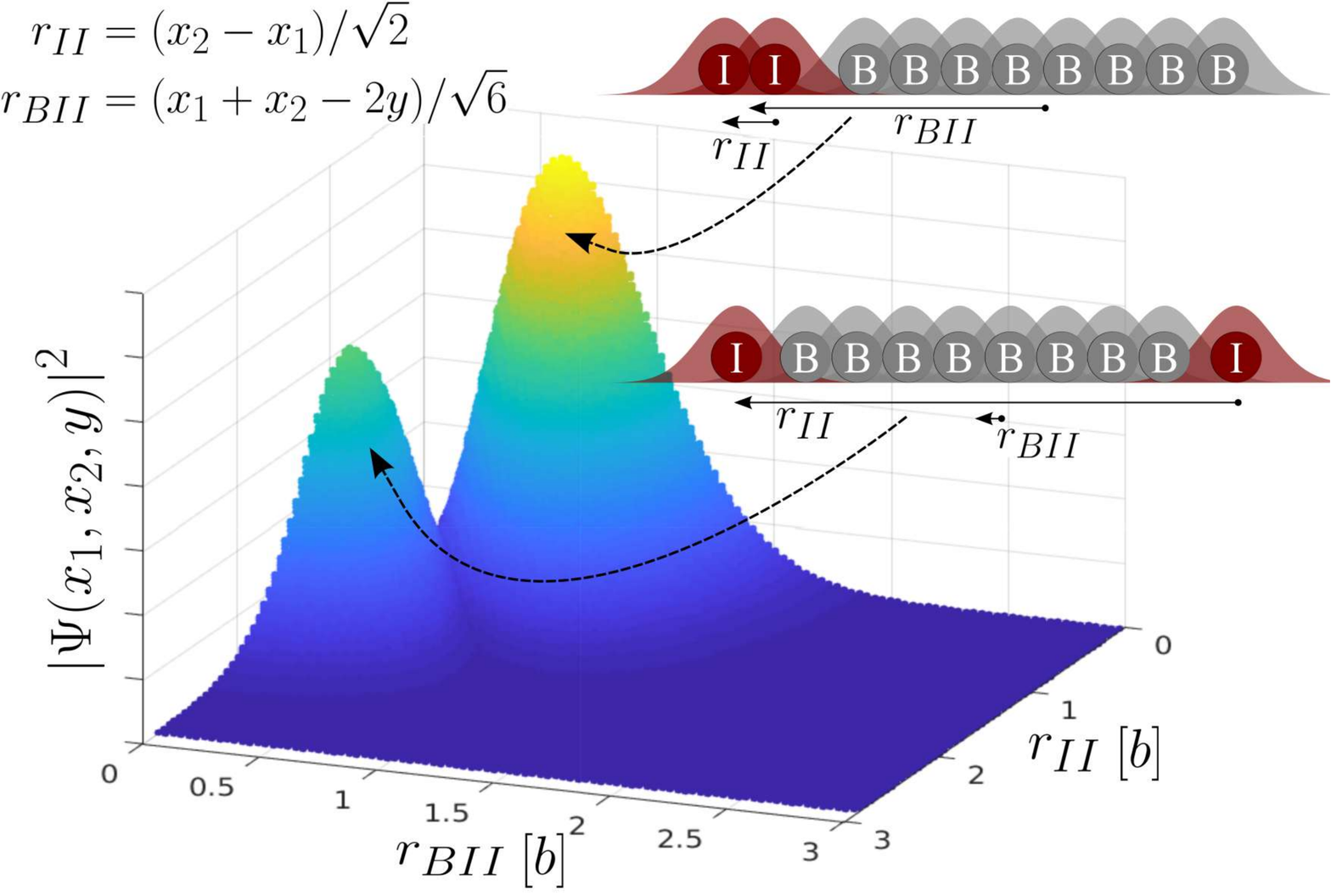}
\caption{Probability distribution of the two impurities and the bosons. $r_{II}$ is the relative distance between the two impurities, while $r_{BII}$ is the relative distance between the bosons and the center of $r_{II}$. For each $x_1, x_2$ and $y$, the corresponding $r_{II}$, $r_{BII}$ and the density is calculated and plotted as a scattering plot. This figure shows the particular probability density for $g=0.2$ and $g_{BB}=0.3$ and $m_{IB}=1$. This figure has not been published before.}
\label{ch6:relativewavefunction2+N}
\end{figure}

Another way of showing the most probable position of the impurities is the probability distribution.
Fig.~\ref{ch6:relativewavefunction2+N} shows such a graph, where the probability distribution is shown as a function of the position of the two impurities and the bosons. For each $x_1, x_2$ and $y$, the relative distance between the two impurities, $r_{II}$, and the relative distance, $r_{BII}$, between the bosons and the center of $r_{II}$ is calculated. Fig.~\ref{ch6:relativewavefunction2+N} shows a discrete plot for such a probability distribution with $g=0.2$ and $g_{BB}=0.3$ and $m_{IB}=1$. The plot is at the same time color-coded in order to show the most probable position. As is seen in the figure, for the given parameters the impurities can be found sitting either together ($r_{II}=0$ and $r_{BII}>0$) or separated ($r_{II}>0$ and $r_{BII}=0$). Further investigations have shown that for stronger interaction strengths the first configuration ($r_{II}=0$ and $r_{BII}>0$) where the impurities sit together is in fact the most probable configuration and they tend to move to the edge of the condensate.\\

\subsection{Periodic Ring Potential}
It is important to note that even though the analysis discussed so far has been with the harmonic oscillator and equal masses only, the method can easily implement any external trap, systems with different masses, and/or different trapping frequencies. The method is also general enough to include the Gross-Pitaevskii equation in description of the majority bosons. In the following I will investigate what happens when the external harmonic oscillator potential is turned off and a one-dimensional periodic ring-shape\index{One Dimensional Periodic Ring Potential} potential with length $2L$ is considered instead. The periodic boundary ring-shape potential is considered to make calculations easy and in fact, the space big enough so that the asymptotic solutions are the ones you obtain in free space and when the impurities are far from each other.

\begin{figure}[t]
\centering
\includegraphics[width=\columnwidth]{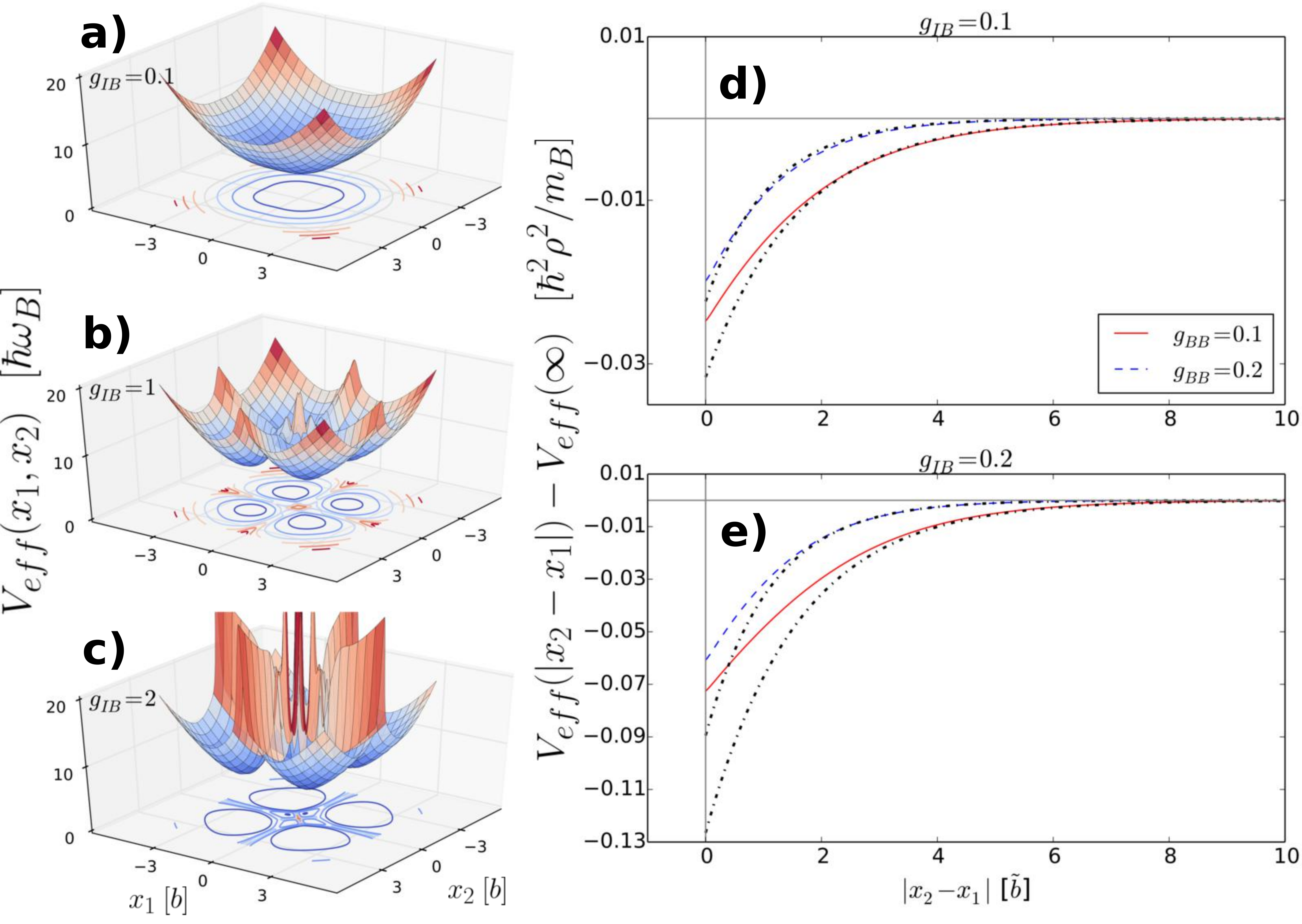}
\caption{Left panels: Effective impurity-impurity potential as a function of the impurity coordinates. Right panels: The same effective potential as a function of the relative coordinates. The plots are for converged values with $L=20$ and $N_B=40$, where $2L$ is the total length of the ring. Notice that the vertical axis is shifted such that the potential between the impurities is zero when they are as far as possible away from each other.}
\label{effectivepotential_ring}
\end{figure}

As the potential is turned off, I introduce another unit of length, $\tilde{b}=\sqrt{\hbar^2/m_B}$, which makes the energies unit less. The total Hamiltonian of the system is therefore given as,
\begin{align}
\begin{split}
\mathcal{H}&=-\sum_{i=1}^2 \frac{1}{2m_{IB}}\frac{\partial^2}{\partial x_i^2} - \sum_{i=1}^{N_B} \frac{1}{2}\frac{\partial^2}{\partial y_i^2}+\sum_{i=1}^2 \sum_{j=1}^{N_B}g\delta(x_i-y_j)+\sum_{j<k}g_{BB}\delta(y_j-y_k),
\label{eq:hamiltonian}
\end{split}
\end{align}
where $g$ and $g_{BB}$ are in units of [$\tilde{b}$]. The above equation is constrained with a periodic boundary condition from a one-dimensional ring potential with length $L$. Following the same procedure as before, one ends up with the following effective potential:
\begin{equation}
\begin{split}
V_{eff}\equiv\sum^2_{i=1} \frac{N_B Q_{x_i}}{2m_{AB}}+N_B\epsilon_{gs}.
\label{eq:numericalequation_ring}
\end{split}
\end{equation}

\noindent
This effective potential can now be plotted for different values of interaction strengths as a function of the relative distance between the impurities. The result is shown in Fig.~\ref{effectivepotential_ring}. Notice that the vertical axis is shifted such that the potential between the impurities is zero when they are as far as possible away from each other. One striking feature here is that the effective potential is negative close to $|x_2-x_1|=0$ indicating that the wave function for the impurities might have a bound solution here and therefore the impurities could stick together. Notice that this was also the case when there was an external potential. Fig.~\ref{effectivepotential_ring} shows how the interaction between the impurities and the bosons makes the effective potential deeper and bigger, while the self-interaction pushes the barrier closer to $|x_2-x_1|=0$. Intuitively this makes sense as the bosons get more dissipated and therefore leave less room for the impurities. In conclusion, the impurities find themselves in a self-trapping effective potential where they stick together. This effect shows up both in the present of an external harmonic oscillator potential and in a one-dimensional ring-shaped trap. For more detailed discussion please have a look in \cite{2017arXiv171201538D}

\newpage % Create empty back of side
\thispagestyle{empty}

%-----% GAUGE SIMULATIONS
% !TeX root = ../Main_publish.tex

\chapter{Lattice Gauge Simulation}\label{ch:latticegauge}
\epigraph{\it “An expert is a person who has found out by his own painful experience all the mistakes that one can make in a very narrow field.”}{\rm ---Niels Bohr}

Back in the introductory chapter I summarized the solutions for the single harmonic oscillator potential. In this chapter, I will start by finding the solutions to the generalized tilted double harmonic oscillator potential. It is basically a potential consisting of two single harmonic traps where one of the traps can be higher and/or more narrow than the other one.

Later, I will discuss how this model can be used as a building block of a quantum simulator\index{Quantum Simulator}, which is the main topic of this article \cite{2017arXiv170400664D}. The quantum simulator is used to study gauge theories in discretized lattices. Gauge theories appear to describe fundamental forces of Nature pretty well. They also leave the action and the classical equations of motion invariant. Therefore it is important to do some simulations of gauge theories in order to understand them even better in other connections. 

In this case I am interested in the $U(1)$ abelian gauge theory. $U(1)$ gauge theory describes the weak gauge coubling in quantum electrodynamics (QED). It has one gauge field, the electromagnetic four-potential, with the photon being the gauge boson. In order to understand QED better, one can look at the coupling by tuning the strength. For small strengths it is solved by the perturbative methods and illustrated by for instance by the Feynman diagrams, however when the strength becomes strong, for instance in strong fields such as plasma experiments or neutron stars, the perturbative models fail. Hence being able to simulate such an experiment with tunable coupling strength is desirable. In this chapter I use the lattice model to simulate such an interaction because simulations become easier to handle especially on a computer.

\section{Generalized Tilted Double Harmonic Oscillator Potential\index{Tilted Double Harmonic Trap}}\label{ch:latticegauge:sec:GTDHOP}

A generalized tilted double harmonic oscillator potential can be written as following:
\begin{equation}
V ^{\mathrm{ext}}(x)= 
\begin{cases}
\frac{1}{2}m \omega_L^2(x+d_L)^2,& \text{if } x<0\\
\frac{1}{2}m \omega_R^2(-x+d_R)^2+\Delta,& \text{otherwise,}
\end{cases}
\label{eq:tdhop}
\end{equation}
where $d_L$ and $d_R$ are the displaced centers of the harmonic oscillator potentials shifted to the left and right of the center, respectively. $\omega_L$ and $\omega_R$ are correspondingly the trapping frequencies in each trap and $x$ is the real-space coordinate for a particle with mass $m$ trapped in the potential. $\Delta$ is the shifted height of the center of the right potential with respect to the left potential. Nevertheless, the potential is required to be continuous by this relation: $\frac{1}{2}m \omega_L^2d_L^2=\frac{1}{2}m \omega_R^2d_R^2+\Delta$. There are several versions of the tilted double harmonic oscillator potentials, however, in the following I will only focus on the analytical solutions for the Eq.~(\ref{eq:tdhop}) and the applications of this model.\\

\noindent
Just like any other quantum problem, the goal is to find stationary solutions to the following eigenvalue problem:
\begin{equation}
-\frac{\hbar^2}{2m }\frac{\partial^2\psi(x)}{\partial x^2}+V ^{\mathrm{ext}}(x)\psi(x)=E\psi(x),
\label{eq:SE}
\end{equation}
and just like the previous chapters I introduce $b=\sqrt{\hbar/(m \omega_L)}$ as a measure for lengths and energies are measured in units of $\hbar\omega_L$. Notice that since there are two trapping frequencies, the frequency of the left trap is chosen to measure these units. By defining $\Delta\equiv\hbar\omega_L\delta$, in these units the continuity condition becomes $d_L^2=r^2\cdot d_R^2+2\delta$ where $r\equiv\omega_R/\omega_L$, which gives a simple relation between the shifted centers of the harmonic potentials. Notice that any potential can be produced by three independent variables: $r$, $d_R$ and $\delta$.

\begin{figure}[t]
\includegraphics[width=1.0\textwidth]{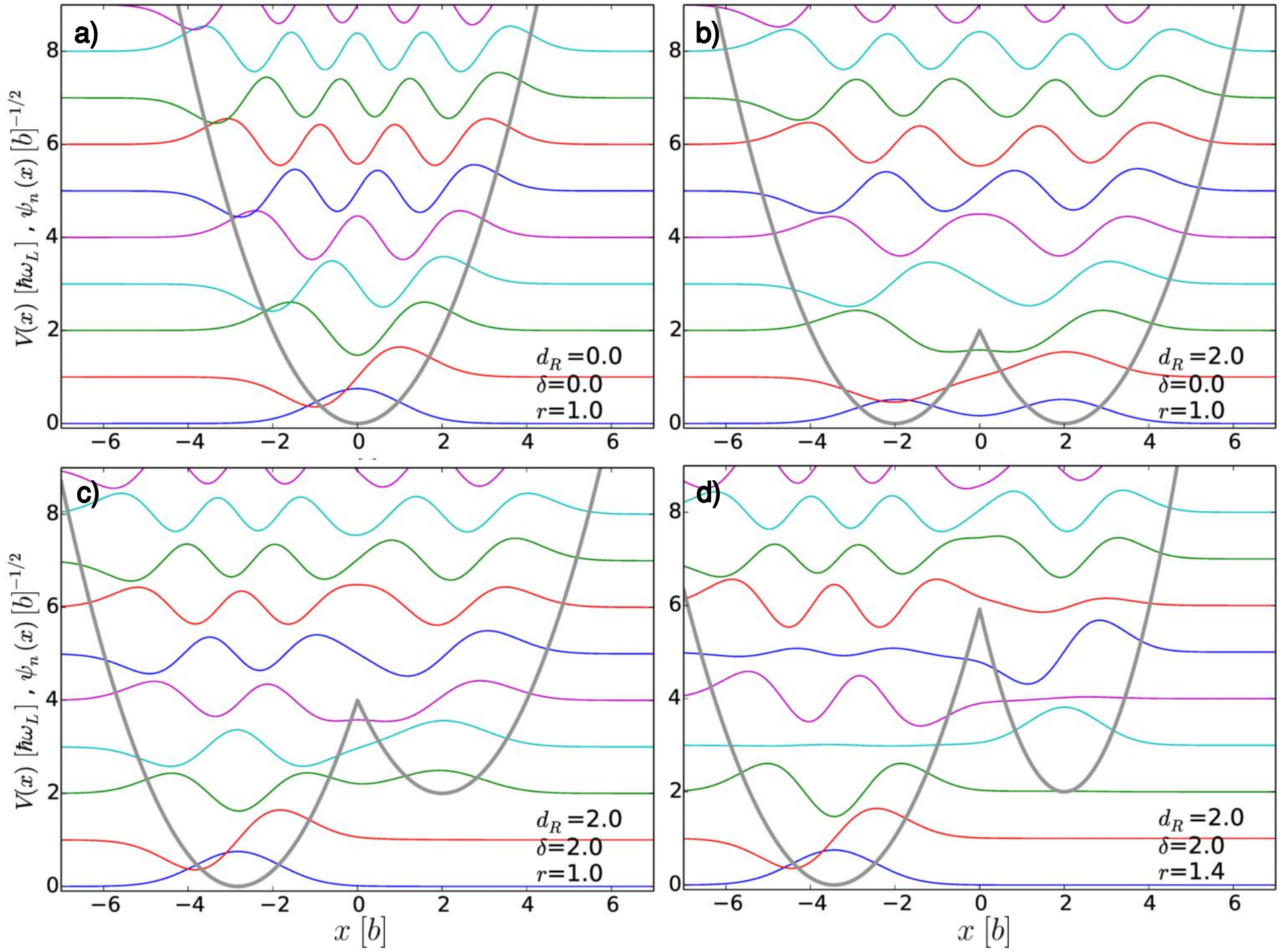}
\caption{Different scenarios in the generalized tilted double harmonic oscillator potential. a) shows the solutions to the simple single harmonic oscillator potential which can be obtained by setting $d_R=0$, $\delta=0$ and $r=1$. b) shows the same as a) but for a symmetric double harmonic potential. c) and d) are two other examples with tilted and squeezed traps. Figure is adapted from \cite{2017arXiv170400664D}.}
\label{fig:tdhop}
\end{figure}

In Fig.~\ref{fig:tdhop} a couple of different versions of the tilted double harmonic oscillator potential is illustrated. Along with the potential (gray line) the corresponding analytical solutions are also shown here (colored lines). These solutions are obtained by the transforming the Hamiltonian into something familiar. I will come back to this in a moment. For now, the interesting observation is to see how the single harmonic potential is easily captured in Fig.~\ref{fig:tdhop} a) by this generalized method. In addition, one can follow the transformation of the solutions as the trap splits into two symmetric traps, Fig.~\ref{fig:tdhop} b), and then lifted, Fig.~\ref{fig:tdhop} c) and finally squeezed, Fig.~\ref{fig:tdhop} d).

Let me now derive the solutions to the problem. The Schr\"odinger equation for $x<0$ in the scaled units is given as (the same techniques can be applied for $x>0$):
\begin{equation}
\frac{\partial^2\psi(x)}{\partial x^2}-\left[(x+d_L)^2-2E\right]\psi(x)=0,
\end{equation}
where $E$ in Eq.~(\ref{eq:SE}) has been moved to the left side of the equal sign. It is now convenient to define $E\equiv\nu_1+\frac{1}{2}=r\cdot(\nu_2+\frac{1}{2})+\delta$, where $\nu_1$ and $\nu_2$ are real numbers but not necessarily integers. These quantum numbers have to be determined later. Since the goal is to transform the differential equation to something that is familiar, I transform the $x$ coordinate by introducing $y \equiv -\sqrt{2}(x+d_L)$. The above Schr\"odinger equation is therefore given as:
\begin{equation}
\frac{\partial^2\psi(y)}{\partial y^2}-\left[\frac{y^2}{4}-\left(\nu_1+\frac{1}{2}\right)\right]\psi(y)=0.
\label{eq:transformedeq}
\end{equation}
Note that the transformation in Eq.~(\ref{eq:transformedeq}) could have also been obtained by defining $y \equiv \sqrt{2}(x+d_L)$, however, the acceptable normalized wave functions in the interval $(-\infty,0)$ are only possible through the first transformation $y \equiv -\sqrt{2}(x+d_L)$. Eq.~(\ref{eq:transformedeq}) has the exact same form as a parabolic cylinder differential equation:
\begin{equation}
\frac{\partial^2w}{\partial z^2}-\left(\frac{z^2}{4}+a\right)w=0,
\end{equation}
which on the other hand has some well-known solutions, called parabolic cylinder functions\index{Parabolic Cylinder Function}, and are usually denoted by $D_{-a-\frac{1}{2}}(z)$ \cite{abramowitz1965}. When it comes to Eq.~(\ref{eq:transformedeq}) the corresponding solutions can therefore be written as,
\begin{equation}
\psi(x)= 
\begin{cases}
c_1 D_{\nu_1}\left(-\sqrt{2}(x+d_L)\right)& \text{if } x<0\\
c_2 D_{\nu_2}\left(\sqrt{2r}(x-d_R)\right) & \text{otherwise},
\end{cases}
\end{equation}
where $c_1$ and $c_2$ are some constants. This means that for a given $r$, $d_R$ and $\delta$ one can find the full solutions to the eigenvalue problem. However, in the above solution there are four unknowns, which have to be determined by four conditions:\\

i) $\psi(x)|_{x=0}$ has to be continuous at $x=0$.\\

ii) The derivative of the wave function, $\psi(x)'|_{x=0}$, has to be continuous.\\

iii) The energy on both sides must be the same: $\nu_1+1/2=r(\nu_2+1/2)+\delta$.\\

iv) The solution must be normalized to $1=\int |\psi(x)|^2 dx$.\\

In addition, if the potential is symmetric around $x=0$, the parity has a role too: even solutions must have zero derivative at $x=0$, that is $\psi(x)'|_{x=0}=0$, while odd solutions must be zero at $x=0$, that is $\psi(x)|_{x=0}=0$. These solutions are shown in Fig.~\ref{fig:tdhop} a)-d).

\section{Odd and Even Short Range Interactions with an Impurity}\label{ch:latticegauge:sec:oddandeven}

After knowing the full analytical solutions to the generalized tilted double harmonic oscillator one can investigate how these solutions react in a present of an impurity in the middle of the potential, $x=x_{\text I}=0$. It is assumed that the impurity is stationary and has two degrees of freedom, say $|\uparrow\rangle$ and $|\downarrow\rangle$. In addition, it is assumed that the interaction between the particles in the double well is short range and modeled by the Dirac delta-function. The interaction can have an even or odd form and it is given as
\begin{align}
\label{eq:Vx}
V_{\text{p-I}}(x) = \sum_{s=\uparrow,\,\downarrow}\left[ \upsilon_{\mathrm{1D}}^{e,s}(x - x_{\text I}) + \upsilon_{\mathrm{1D}}^{o,s}(x - x_{\text I})\right],
\end{align}
where 
\begin{align}
\label{eq:pseudo-eo}
\upsilon_{\mathrm{1D}}^{e,s}(x) = g_{\mathrm{1D}}^{e,s} \hat\delta_{\pm}(x),\qquad
\upsilon_{\mathrm{1D}}^{o,s}(x) = g_{\mathrm{1D}}^{o,s} \delta^{\prime}(x)\hat\partial_{\pm},
\end{align}
and $g_{\mathrm{1D}}^{e,s} = -\hbar^2/\mu a_{\mathrm{1D}}^{e,s}$ and $g_{\mathrm{1D}}^{o,s} = -\hbar^2 a_{\mathrm{1D}}^{o,s}/\mu$ are the 1D interaction strengths dependent on the scattering lengths\index{Scattering Length} for even- and odd-waves, respectively. $\mu$ is the reduced particle-impurity mass.

The even term looks indeed very familiar as it is described before in \cite{busch1998}, but the odd is new. This term is relevant for spin-polarized fermions that can be related to $p$-wave scatterings \cite{GirardeauPRA2004}. Therefore the total interaction term is a linear combination of the even and odd terms. The operators in Eq.~(\ref{eq:pseudo-eo}) act on the wave function differently, dependent on being odd or even. The action of these operators is given as:
\begin{align}
\begin{split}
2\, \hat\delta_{\pm}(x)\psi(x) &\equiv [\psi(0^+) + \psi(0^-)] \delta(x),\\
2\, \hat\partial_{\pm}\psi(x) &\equiv [\psi^{\prime}(0^+) + \psi^{\prime}(0^-)] ,
\end{split}
\end{align}
where $\psi(0^{\pm})$ and $\psi^{\prime}(0^{\pm})$ denotes the value and the derivative of the wave function from left and right side of zero. This means that for even and odd scattering, respectively, the wave function must obey the following conditions:
\begin{align}
\begin{split}
[\psi^{\prime}(0^+) - \psi^{\prime}(0^-)] &= -\frac{1}{a_{1D}^e}[\psi(0^+) + \psi(0^-)],\\
[\psi(0^+) - \psi(0^-)] &= -a_{1D}^o [\psi^{\prime}(0^+) + \psi^{\prime}(0^-)].
\end{split}
\end{align}

\noindent
When the particle interacts with the impurity, the continuity of the derivative of the wave function is no longer a valid statement. Neither is the continuity of the wave function itself when there is odd wave interactions. Actually, one can show that the wave function must satisfy the following transcendental equation \cite{2017arXiv170400664D}:
\begin{align}
&\frac{\frac{1}{a_{\text{1D}}^e}D_{\nu_1}(z_1)+\sqrt{2}\left. \frac{\partial D_{\nu_1}}{\partial z}\right|_{z_1}}{D_{\nu_1}(z_1)+a_{\text{1D}}^o \sqrt{2}\left. \frac{\partial D_{\nu_1}}{\partial z}\right|_{z_1}}+\frac{\frac{1}{a_{\text{1D}}^e}D_{\nu_2}(z_2)+{\sqrt{2r}}\left. \frac{\partial D_{\nu_2}}{\partial z}\right|_{z_2}}{D_{\nu_2}(z_2)+a_{\text{1D}}^o \sqrt{2r}\left. \frac{\partial D_{\nu_2}}{\partial z}\right|_{z_2}}=0, 
\label{transcendental-eq}
\end{align}
where $z_1=-\sqrt{2}d_L$ and $z_2=-\sqrt{2r}d_R$. Again, for a given $r$, $d_R$ and $\delta$ one can use this condition along with the normalization condition to find the solutions to the problem with interacting impurity. Actually, one can even show, that when $r=1$, $\delta=0$, $d_L=d_R=0$, the above equation reduces to something very familiar. Particularly, in this case $\nu_1=\nu_2=\nu$, $a_{\text{1D}}^o=0$, and $-1/a_{\text{1D}}^e\equiv g^e_{\text{1D}}$, the transcendental equation for the two interacting atoms in a harmonic oscillator potential appears \cite{busch1998}, which was also seen before in Eq.~(\ref{HO2_energy}): 
\begin{align}
-\frac{2}{g^e_{\text{1D}}}=\frac{\Gamma{(-E/2+1/4)}}{\Gamma{(-E/2+3/4)}},
\end{align}\\
while for the purely odd interaction, that is $a_{\text{1D}}^o\equiv -g^o_{\text{1D}}$ and $1/a_{\text{1D}}^e=0$, one gets the following condition,
\begin{align}
-\frac{1}{2g^o_{\text{1D}}}=\frac{\Gamma{(-E/2+3/4)}}{\Gamma{(-E/2+1/4)}}.
\end{align}

\section{Building a Quantum Simulator}\label{ch:latticegauge:sec:quantummodel}
Now, after knowing the full solution to the generalized tilted double harmonic trap and how it interacts with an impurity in the middle of the trap, the question becomes whether or not this setup can be exploited in any sense. Imagine the following scenario: put a particle, boson, in one of the wells. Allow it to interact with an impurity, say a fermion, in the middle of the double well. While having a boson in one of the wells the goal is now to move the particle to the opposite well at the expense of flipping the impurity to the other state. This might sound easy to configure, but there are some criteria that must be satisfied. Let's examine the full Hamiltonian
\begin{equation}
\mathcal{\hat H}_{\text{p-I}} = \hat H_\text{p} + \hat H_\text{I} + \hat V_{\text{p-I}} + \hat H_\text{I}^{\text{int}},
\end{equation}
where

\begin{align}
\begin{split}
\hat H_{\text{I}}^{\text{int}} &= \frac{\hbar \Omega_R}{2} \hat \sigma_x;\\
\hat H_{\text{p}} & = -\frac{\hbar^2}{2m}\frac{\partial^2}{\partial x^2} + V^{\mathrm{ext}}(x)\\
\hat H_{\text{I}} & = -\frac{\hbar^2}{2m_\text{I}}\frac{\partial^2}{\partial x_\text{I}^2} + \frac{m_\text{I}\omega_\text{I}^2}{2}x_\text{I}^2\\
\hat V_{\text{p-I}} &= \!\!\sum_{s=\uparrow,\downarrow}\left[\upsilon_{\mathrm{1D}}^{e,s}(x - x_\text{I}) + \upsilon_{\mathrm{1D}}^{o,s}(x - x_\text{I})\right].
\end{split}
\end{align}

\noindent
Here $V^{\mathrm{ext}}(x)$ is the tilted double well potential for the bosons with mass $m$, while $m_\text{I}$ is the mass of the impurity, which is trapped in a single harmonic trap with frequency $\omega_\text{I}$ and coordinate $x_\text{I}$. The $H_{\text{i}}^{\text{int}}$ term is the important term that is responsible of the flipping the spin and is tuned with the frequency, $\Omega_R$.

\begin{figure}[t!]
\includegraphics[width=\linewidth]{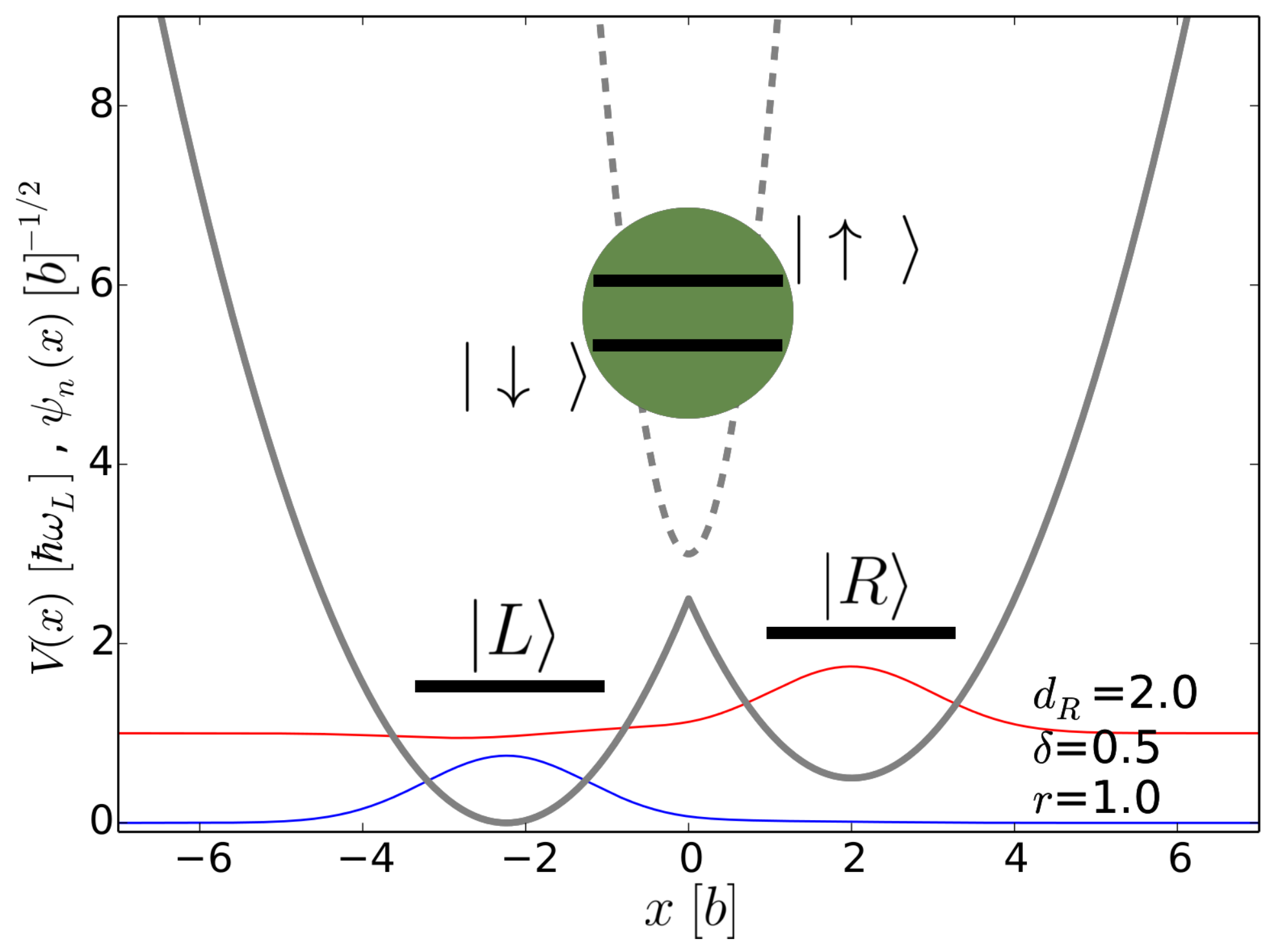}
\caption{One example of the double well with two lowest states $\{|L\rangle,|R\rangle\}$, illustrated by the blue and red line, respectively. In this example $d_L = -\sqrt{5}$, $d_R = 2.0$, and $E^0_{L}=-0.5013$ and $E^0_{R}=-0.0045$. Note that the energy levels are shifted by $2\hbar\omega_L$ for clarity purposes and energies are given i units of $\hbar\omega_L$ whereas lengths are in units of $b=\sqrt{\hbar/(m\omega_L)}$. The left (blue) Gaussian solution and the right (red) Gaussian solution represent the two lowest eigenfunctions of the tilted double harmonic trap. The impurity, shown with the green ball, is situated in the middle of the double well and has two intrinsic states, $\{|\uparrow\rangle,|\downarrow\rangle\}$. It is trapped in a single harmonic trap (gray dashed line) which has a trapping frequency $\omega_I$. Figure is adapted from \cite{2017arXiv170400664D}.}
\label{gauge_supp_fig2}
\end{figure}

The overall goal is this; can one setup a double well such that by detuning the $\Omega_R$ the particle moves from one side to the other by flipping the impurity? The situation is illustrated in Fig.~\ref{gauge_supp_fig2}. Focusing on the two lowest states,$\{|L\rangle,|R\rangle\}$, in the double well, the boson can jump between these two states by flipping the state of the impurity in the middle, $\{|\uparrow\rangle,|\downarrow\rangle\}$ (the green ball). Other than $\Omega_R$ there are no other factors that can allow the bosons to change side and the energy comes from the fermion, which changes its internal spin-state. Note that the double well is asymmetric so there is an energy difference between $|L\rangle$ and $|R\rangle$, which is why it does not hop around without the $\Omega_R$ term. In the next section I will discuss how to obtain the time evolution of an initial state.

\subsection{Dynamics of the Initial State}\label{ch:latticegauge:subsec:dynamics}
In order to describe the time evolution \index{Time Evolution}of a state, the time-dependent Schr\"{o}dinger equation is needed to be solved:
\begin{align}
i\hbar\frac{\partial}{\partial t}|\Psi(t)\rangle=\mathcal{\hat H}_{\text{p-I}}(t) |\Psi(t)\rangle.
\label{eq:timeSEgeneral}
\end{align}
This equation has the formal solution $|\Psi(t)\rangle=\hat U(t)|\Psi(0)\rangle$, where $|\Psi(0)\rangle$ is the initial state at time zero and the $\hat U(t)=e^{-i\mathcal{\hat H}_{\text{p-I}}t/\hbar}$ is the unitary time evolution operator. In practice, I solve the equation by expanding it a basis set consisting of an ansatz with a bosonic particle $\psi_n(x)$, and an impurity $\phi_m(x_{\text I})$ with spinor state $\chi_s$:
\begin{align}
\Psi(x ,x_{\text I},s;t) = \sum_{n,m,s} C_{n,m,s}(t) \psi_n(x )\phi_m(x_{\text I})\chi_s.
\label{eq:timePsitxys}
\end{align}
where $C_{n,m,s}(t)$ are the time-dependent coefficients to be determined later. These coefficients can be determined by applying the ansatz in Eq.~(\ref{eq:timePsitxys}) to the time dependent problem in Eq.~(\ref{eq:timeSEgeneral}) and projecting it into the $\langle \chi_p,\phi_m,\psi_n\vert$ state. This provides a coupled vector differential equation, $d/dt~ C_{n,m,s}(t)=M~C_{n,m,s}(t)$, where $M$ is some matrix. I have solved this differential equation numerically by using the ode45 routine in Python.

\subsection{Four-Level System}\label{ch:latticegauge:subsec:fourlevel}
Let me give an example of how the previous expansion works in the four-level case. By four-level system, I mean that I only consider the two lowest states, $n\in\{|L\rangle,|R\rangle\}$ with the impurity, $s\in\{|\uparrow\rangle,|\downarrow\rangle\}$, so all in all, there are four combinatory levels: $\{|L,\downarrow\rangle, |R,\downarrow\rangle,|L,\uparrow\rangle,|R,\uparrow\rangle\}$. Note that $|L\rangle$ is the ground state in the double well, while $|R\rangle$ is the corresponding excited state. For simplicity, I will also assume that the impurity is stationary, hence the $m$ quantum number is ignored, but the impurity still interacts with the boson through the spin terms in $\hat V_{\text{p-I}}$. So the Hamiltonian that has to be solved is simplified as the following:
\begin{align*}
\hat H &= \hat H_{\text{p}} + \hat H_{\text{I}}^{\text{int}} + \hat V_{\text{p-I}} \\
&= -\frac{\hbar^2}{2m}\frac{\partial^2}{\partial x ^2} + V^{\mathrm{ext}}(x ) + \frac{\hbar \Omega_R}{2} \hat \sigma_x+\sum_{s=\uparrow,\downarrow}\left[\upsilon_{\mathrm{1D}}^{e,s}(x-0) + \upsilon_{\mathrm{1D}}^{o,s}(x-0)\right].
\end{align*}
The ansatz is then given as:
\begin{align}
\begin{split}
\Psi(x ,s;t) &= \sum_{n,s} C_{n,s}(t) \psi_n(x )\chi_s\\
&= C_{L,\downarrow} |L,\downarrow\rangle + C_{R,\downarrow} |R,\downarrow\rangle +C_{L,\uparrow} |L,\uparrow\rangle + C_{R,\uparrow} |R,\uparrow\rangle,
\end{split}
\label{eq:exp-4states}
\end{align}
where as pointed before, the impurity is considered as being stationary. This can also be achieved by narrowing the trapping frequency of the impurity (the dashed gray line in Fig.~\ref{gauge_supp_fig2}) and therefore obtain a very peaked impurity wave function. Now, applying the ansatz and calculating the matrix elements of the above Hamiltonian gives: 
\begin{align}
M=
\begin{bmatrix}
E^0_{L}+J^{\downarrow}_{LL}& J^{\downarrow}_{LR} & \frac{\hbar\Omega_R}{2}& 0\\[0.3em]
J^{\downarrow}_{RL} & E^0_{R}+J^{\downarrow}_{RR} & 0 & \frac{\hbar\Omega_R}{2}\\[0.3em]
\frac{\hbar\Omega_R}{2} & 0 & E^0_{L}+J^{\uparrow}_{LL} & J^{\uparrow}_{LR} \\[0.3em]
0 & \frac{\hbar\Omega_R}{2} & J^{\uparrow}_{RL} & E^0_{R}+J^{\uparrow}_{RR}\\[0.3em]
\end{bmatrix},%=\mathcal{H}\quad
\nonumber
\end{align}
in the $\{|L,\downarrow\rangle, |R,\downarrow\rangle,|L,\uparrow\rangle,|R,\uparrow\rangle\}$ basis. Here, $\hat H_{\text{p}} |L\rangle= E_L^0 |L\rangle$ and $\hat H_{\text{p}} |R\rangle= E_R^0 |R\rangle$. Additionally,
\begin{align*}
\begin{split}
J^s_{NM}=&\langle N|g_{\mathrm{1D}}^{e,s}\hat{\delta}_{\pm}|M \rangle+\langle N| g_{\mathrm{1D}}^{o, s} \delta^{\prime}(x )\hat{\partial}_{\pm}|M \rangle\\
=&g_{\mathrm{1D}}^{e,s}\cdot N(0)\cdot [M(0^+)+M(0^-)]/2\\
&-g_{\mathrm{1D}}^{o,s}\cdot N^{\prime}(x )|_{x =0}\cdot [M^{\prime}(0^+)+M^{\prime}(0^-)]/2,
\end{split}
\end{align*}
with $s=\uparrow,\downarrow$ and $N,M=L,R$. Notice that the properties of the even and odd scattering lengths have been used in the above equation and the units of $g_{\mathrm{1D}}^{o,\uparrow}$ and $g_{\mathrm{1D}}^{e,\uparrow}$ are in [$\hbar\omega_Lb^3]$ and $[\hbar\omega_Lb]$, respectively. Having $M$ allows one to solve the coupled differential equation, $d/dt~ C_{n,m,s}(t)=M~C_{n,m,s}(t)$.

\section{Results}\label{ch:latticegauge:sec:results}
Starting in the initial state: $|\Psi(t = 0)\rangle=\frac{|R,{\uparrow}\rangle-|R,{\downarrow}\rangle}{\sqrt{2}}\equiv|R,-\rangle$, the goal is to end up in the target state $|L,+\rangle$ for a later time. Let me remind the reader that the $m$ quantum number is ignored in this section as it is assumed stationary while $n\in\{|L\rangle,|R\rangle\}$ and $s\in\{|\uparrow\rangle,|\downarrow\rangle\}$. I will discuss later why I have chosen the $|R,-\rangle$ state and not just purely $|R,\uparrow\rangle$ or $|R,\downarrow\rangle$ state. For now, it is clear that the only non-vanishing coefficients at time zero are: $C_{R,{\downarrow}}=-1/\sqrt{2}$ and $C_{R,{\uparrow}}=1/\sqrt{2}$. The probability of finding the particle in the target state $|L,+\rangle$, is then given as:
\begin{equation}
\mathcal{O}^+_L(t) = \vert C_{L,\uparrow}(t) + C_{L,\downarrow}(t)\vert^2/2
\end{equation}
[same for $\mathcal{O}^-_R(t)$]. The overlap is the norm-squared of the linear combination of the $\uparrow$ and $\downarrow$ states, which defines the $|L,+\rangle$ state. More technically the overlap is defined as:
\begin{equation}
\mathcal{O}_{y}^{\alpha}(t) =\int_{\mathbb{R}^y}d x \int_{\mathbb{R}}d x_\text{I}\vert\langle x,x_\text{I},\alpha_\text{I}\vert\psi_\text{p-I}(t)\rangle\vert^2,
\end{equation}
with $\alpha_\text{I}=\pm$, $y=L,R$, $\mathbb{R}^{R,L}\equiv \mathbb{R}^{\pm}$, $\psi_\text{p-I}(t)$ is the time evolved particle-impurity state. $x$ and $x_\text{I}$ are the coordinates of the bosons and impurity, respectively. Hence, $\langle x,x_\text{I},\alpha_\text{I}\vert$ is the basis where the problem is diagonalized. So the ideal simulation will be a population between the $|L,+\rangle$ and $|R,-\rangle$ as a function of time. This is identical to the values of $\mathcal{O}^-_R(t)$ and $\mathcal{O}^+_L(t)$, respectively. The transportation is of course not done spontaneously, but with a help of detuning $\Omega_R$ at times $t>0$ in such a way that the energy levels from left and right come in resonance. This is done by setting $\Omega_R=E^0_R-E^0_L$.\\

\begin{figure}[t!]
\includegraphics[width=\linewidth]{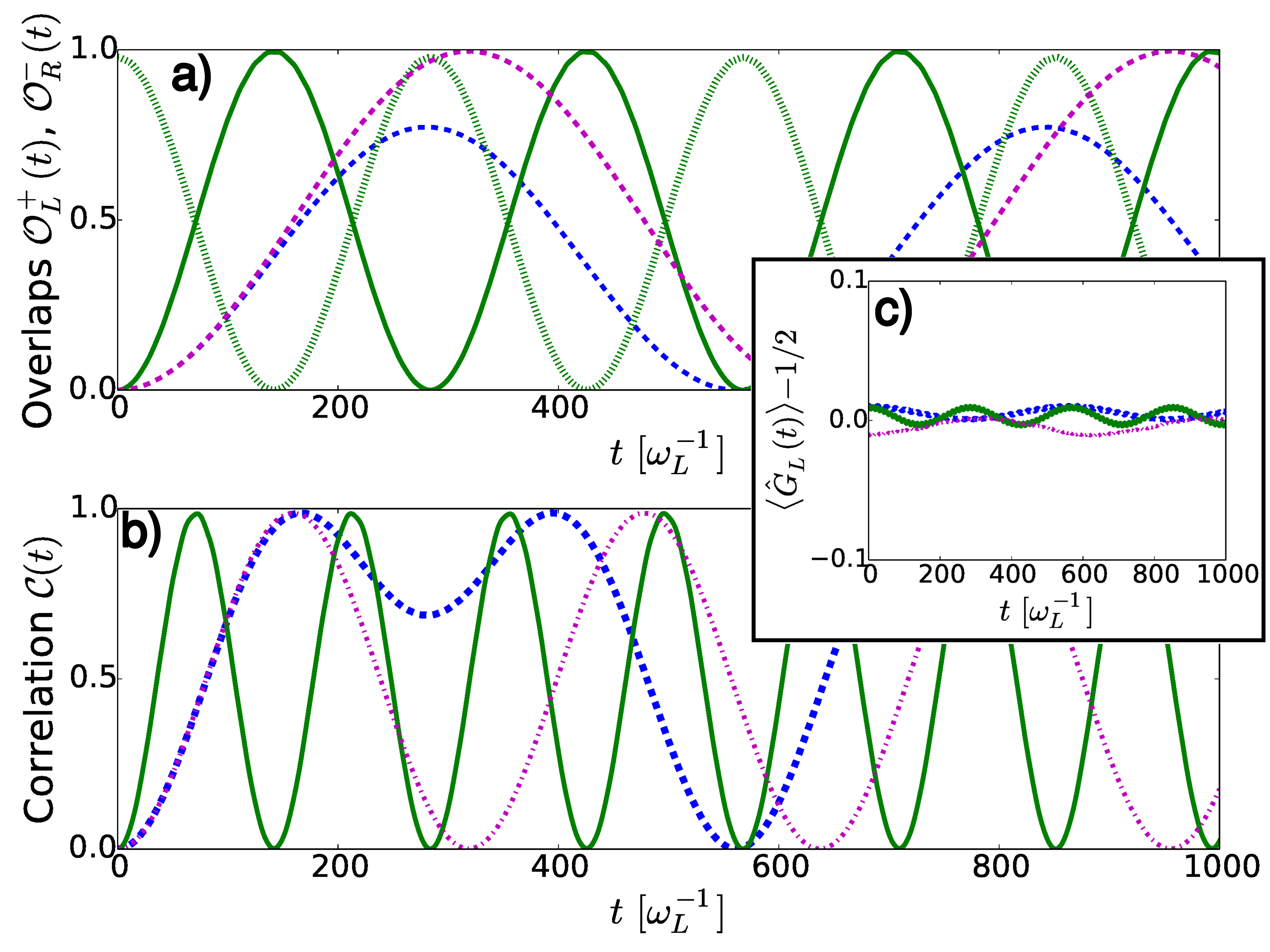}
\caption{a) Shows the evolution of $\mathcal{O}_{L}^{+}(t)$ for different cases. The double well is constructed by the following values: $d_R=2.0$, $\delta=0.5$ and $r=1.0$. Then the population, starting from $|R,-\rangle$ to $|L,+\rangle$ has been recorded. Green solid line shows the population of $|L,+\rangle$ for $g_{\mathrm{1D}}^{e,\uparrow}=-g_{\mathrm{1D}}^{e,\downarrow}=1$ and $g_{\mathrm{1D}}^{o,\uparrow}=-g_{\mathrm{1D}}^{o,\downarrow}=1$. The Green dashed line shows the evolution of $\mathcal{O}_{R}^{-}(t)$. Blue dashed line shows the evolution of $\mathcal{O}_{L}^{+}(t)$ for $g_{\mathrm{1D}}^{e,\uparrow}=g_{\mathrm{1D}}^{e,\downarrow}=1$ and $g_{\mathrm{1D}}^{o,\uparrow}=g_{\mathrm{1D}}^{o,\downarrow}=0.1$. This behavior has been improved by choosing a better detuning of $\Omega_R$ as explained in the text. b) Shows the correlated particle-impurity hopping $\mathcal{C}(t)$, while c) shows the expectation value of the operator $\hat G_L-1/2$. Figure is adapted from \cite{2017arXiv170400664D}.}
\label{gauge_supp_fig3}
\end{figure}

\noindent
An example of such a calculation is illustrated in Fig.~\ref{gauge_supp_fig3} a). $\mathcal{O}^+_L(t)$ is shown by solid green line and oscillates, starting from zero and $\mathcal{O}^-_R(t)$ is shown by green dashed line with the same behavior but starting from 1. Both graphs are done for a value of $g_{\mathrm{1D}}^{e,\uparrow}=-g_{\mathrm{1D}}^{e,\downarrow}=1$ and $g_{\mathrm{1D}}^{o,\uparrow}=-g_{\mathrm{1D}}^{o,\downarrow}=1$ with $d_R=2.0$, $\delta=0.5$ and $r=1.0$. These values are neither crucial nor unique in any way. They were chosen by doing a normally distributed random search between some intervals, where the fidelity above $95\%$ was set as the criterion. This is seen in panel a) (the green lines). Therefore other values are also possible. However, I have chosen to stick with these values throughout the discussion. Note that a well separation $d_L + d_R$ must be chosen in such a way that $J_{\uparrow,\downarrow}\ne 0$. Remember that $J$ is a measure of how much the $|L\rangle$ and $|R\rangle$ states couple to the impurity positioned in the middle of the double well. Otherwise the oscillation becomes exponentially damped and the impurity and the boson are no longer in contact. So with the given values of $d_R$, $\delta$ and $r$, the double well is prepared and the full solution is found. The question is then, how about the values of $g_{\mathrm{1D}}^{o,\uparrow}$ and $g_{\mathrm{1D}}^{e,\uparrow}$. It turns out that the choices of these parameters are many. In Fig.~\ref{gauge_supp_fig3} a) blue dashed line, the solution for the $g_{\mathrm{1D}}^{e,\uparrow}=g_{\mathrm{1D}}^{e,\downarrow}=1$ and $g_{\mathrm{1D}}^{o,\uparrow}=g_{\mathrm{1D}}^{o,\downarrow}=0.1$ is shown. As it is plotted these values decrease the optimal oscillation, in other words the oscillation gets damped. This is due to the fact that when the even and odd interactions are different, the choice of $\Omega_R$ is crucial. Remember that $\Omega_R$ was chosen to be the energy difference between the right and left energy levels. With the different values of interactions this frequency has to be changed correspondingly. A general choice of this value must therefore be set to: $\Omega_R=E^0_R-E^0_L+(J^{\downarrow}_{RR}+J^{\uparrow}_{RR}-J^{\downarrow}_{LL}-J^{\uparrow}_{LL})/2$. By this substitution, one can see in Fig.~\ref{gauge_supp_fig3} a) purple dashed line that the oscillation becomes full again.

The overall behavior is clear: the population of the four-level scheme follows a sine-like population, where the particle goes from $|R,+\rangle$ to $|L,-\rangle$ and vice versa. This full transfer from one state to the other and flipping the spin of the impurity is not dependent on the interaction strength, as discussed above, making the procedure quite general. Other values of $d_R$, $\delta$ and $r$ have also been tested and here the results are again consistent. Only the period of the oscillation changes when one plays with other values, but the overall transportation from right to left can be made 100\%.

Furthermore, two observables are interesting to investigate in more details: i) The correlated hopping defined as $\mathcal{C}_{L,R}(t) = \langle ( \hat n_L - \hat n_R) \hat\sigma_{L}^{x} \rangle - \langle\hat n_L - \hat n_{R}\rangle \langle \hat\sigma_{L}^{x} \rangle$ and shown in Fig.~\ref{gauge_supp_fig3} b), where $\hat n$ is the number operator in the second quantization formalism and $\sigma^x$ is the Pauli matrix \index{Pauli Matrices} with eigenstates $\{|+\rangle,|-\rangle\}$. ii) Conservation of the total number of particles defined as: $\hat G_{L} = \hat n_{L} - \frac{\hat\sigma^{x}_{L}}{2}$ (and similarly for the right well: $\hat G_{R}=\hat n_{R} + \frac{\hat\sigma^{x}_{L}}{2}$) and shown in Fig.~\ref{gauge_supp_fig3} c).

While the conservation of the number of particles has to be constant in order to keep the number of particles constant: $N=n_\downarrow+n_\uparrow$, the correlated hopping is assumed to follow a similar sinusoidal graph. This quantity reveals the correlated particle-impurity hopping rather than the independent hopping of the particle and non-correlated flip of the impurity. In fact, this is the case shown in Fig.~\ref{gauge_supp_fig3} b)-c)

Up until now, the desired simulation has been captured here, but the question is what happens when one allows more states to get populated. In this case, the initial state is the same, but one could imagine that the lowest states could be coupled to higher excited states. Fig.~\ref{gauge_fig2} shows a similar graph as Fig.~\ref{gauge_supp_fig3} but for a much higher basis cut ($>30$ states). The solid cyan line shows how the overlap evolves for the stationary impurity. It is shown that after a time $t_{\max}\approx 104 /\omega_L$ the desired target state is reached with a population of $94\%$. The same for the correlation hopping in panel b) is shown that the hopping of the boson is correlated with the spin flip of the impurity. In panel c) the value of the $\hat G_{L}$ is fairly kept constant.

All these results were done for a stationary impurity, where $\omega_{\text{I}} = 10^3\omega_L$, however in real experiments this assumption might be problematic to achieve. Therefore setting $\omega_{\text{I}} = 10\omega_L$ where the impurity has a broader interval on the axis, has also been investigated. The procedure of simulating such situation can be done by first separating the wells from each other at times $t<0$ and then move them together for $t>0$. The results for such a configuration is shown in Fig.~\ref{gauge_fig2} with the red dashed line. Even though the graph gains some additional wiggles the overall behavior is still clearly captured.

\subsection{Micro-motion\index{Micro Motion}}\label{ch:latticegauge:subsec:micromotion}
The previous derived method is actually so general that one can even replace the impurity in the middle with an ion, which has been investigated before in \cite{PhysRevA.89.063621}. The ion introduces a so-called micro-motion effect that might have an impact on the dynamics. The micro-motion occurs when the ion is trapped in a static and time-dependent electric field, the so-called Paul trap\index{Paul Trap} \cite{Leibfried2003}. In this case the ion gets an oscillatory motion, but the Hamiltonian can be written as \cite{Cook1985}:
\begin{align}
\hat H_{\text i}(t) = -\frac{\hbar^2}{2m_{\text i}}\frac{\partial^2}{\partial x_{\text i}^2}+\frac{m_{\text i}\omega_{\text i}^2}{2}x_{\text i}^2+\frac{m_{\text i}\omega_{\text i}^2}{2}x_{\text i}^2 \left[\frac{\Omega(t)^2}{4\omega_{\text i}^2}-1\right],
\end{align}
where $\Omega(t)^2=\Omega_{\text{rf}}^2 \cdot [a+2q\cos(\Omega_{\text{rf}} t)]$ with $\Omega_{\rm rf}$ the driving frequency, and $\omega_{\text i}=\frac{\Omega_{\text{rf}}}{2}\sqrt{a+q^2/2}$. In addition, $|a|\ll |q|<1$ are some constants that are geometry dependent but experimental values are typically $|a| \ll |q|\approx0.2$ for linear Paul traps.

\begin{figure}[t!]
\includegraphics*[width=1\linewidth]{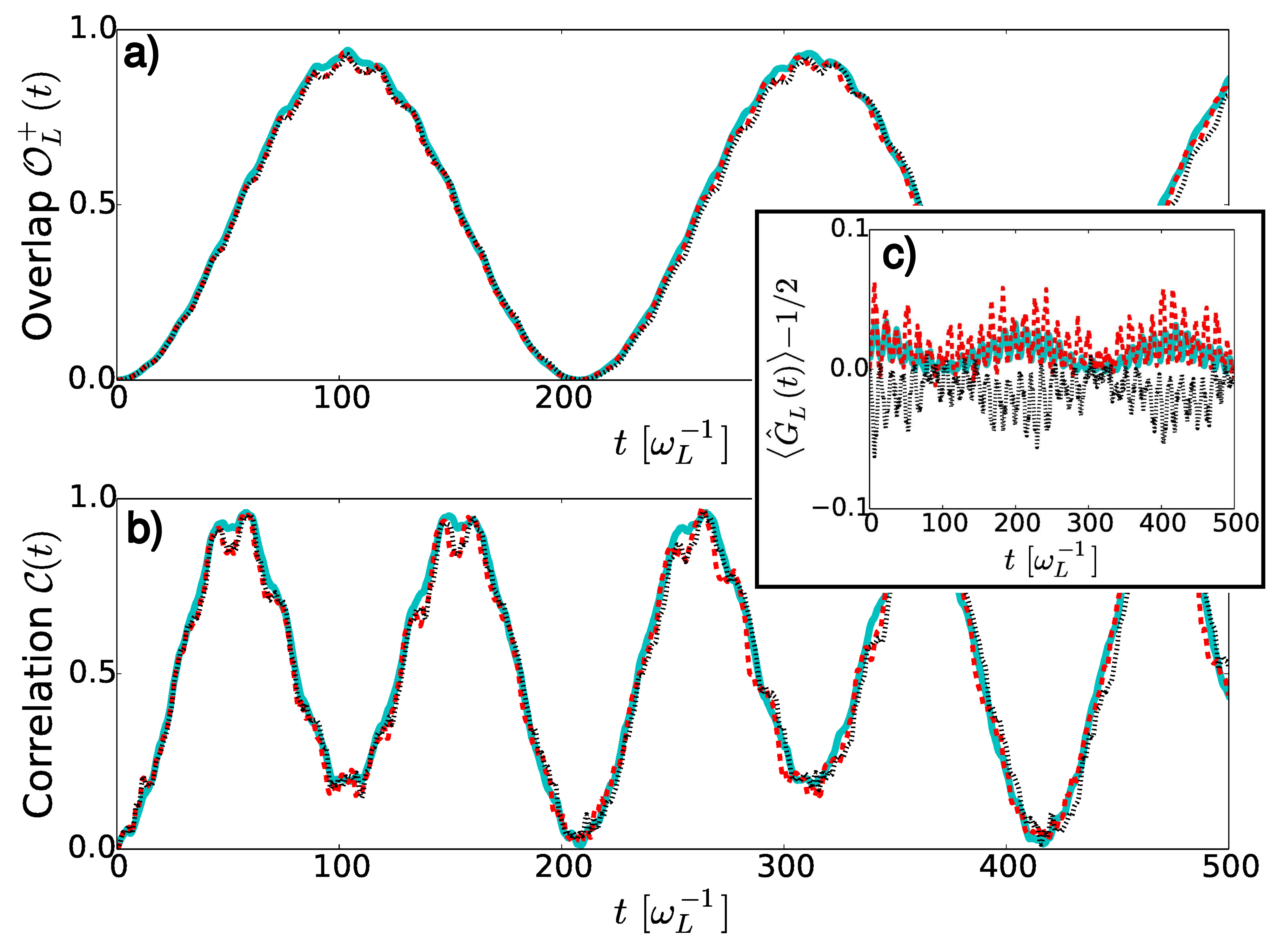}
\caption[]{Correlated Hopping.\index{Correlated Hopping} a) Shows the evolution of $\mathcal{O}_{L}^{+}(t)$ for different cases. The static impurity is shown with cyan solid line, while the moving impurity with a quenching start is shown with red dashed line. Finally, the micro-motion is shown with black dotted line. b) shows the correlation function as described in the text and c) is the expectation value of $\hat G_L - 1/2$. In all cases:$m_{\text p}/m_{\text I} = 1$, $g_{\mathrm{1D}}^{e,\uparrow} = -g_{\mathrm{1D}}^{e,\downarrow} = 1$, $g_{\mathrm{1D}}^{o,\uparrow} = -g_{\mathrm{1D}}^{o,\downarrow} = 0.1$. Furthermore, $r=1$, $d_L=\sqrt{5}$, $d_R = 2$ for a barrier height of $4\,\hbar\omega_L$, whereas in the quenched case $r= 1.4$, $d_L \simeq 2.97$, $d_R = 2$ for a barrier height of $5.92\,\hbar\omega_L$. In addition, $\omega_{\text{i}} = 10\omega_L$ for non-stationary impurity, and $\omega_{\text{i}} = 100\omega_L$ for the ion with micro-motion frequency $\Omega_{\text{rf}}/\omega_L = 2500$. Figure is adapted from \cite{2017arXiv170400664D}.}
\label{gauge_fig2}
\end{figure}

I still assume a contact interaction between the ion and the boson in the double well, which is reasonable if the separation between the particles is larger than the atom-ion interaction range. Therefore, the previous method can easily be applied and the results are shown in Fig.~\ref{gauge_fig2} (black dotted lines). The driving frequency is set to $\Omega_{\text{rf}}/\omega_L = 2500$ in this case. The graph shows a clear small local oscillatory pattern in the overlap, which is due to the movement of the ion. However, the overall oscillation is still there, which still confirms a transportation of the boson from one well to the other at the cost of flipping the ion's internal state.

\section{Lattice Gauge\index{Lattice Gauge} Quantum Simulation}\label{ch:latticegauge:sec:latticegaugequantumsimulation}
In the previous sections I spent many pages on discussing how one can setup a double well and move a boson from one well to the other by changing the internal quantum state of the impurity or an ion. In this section, I will discuss how this procedure can be used in lattice gauge quantum simulations. Generally, the lattice gauge theory is a study of gauge theories in discretized lattices. Usually the matter fields \index{Matter Fields} are described by the creation $\hat{b}_{k}^\dagger$ and annihilation $\hat{b}_{k}$ operators at the $k$th site, which could represent a boson or a fermion, while gauge fields  \index{Gauge Fields}are described by $\hat{U}_{k,k+1}$, which could be described by an internal quantum state. Under local gauge transformations it is required by definition that the dynamics must be invariant. Therefore, any simulation must reproduce a term of locally invariant term as $\hat{b}^{\dag}_{k} \hat{U}_{k,k+1} \hat{b}_{k+1}+h.c.$ \cite{2017arXiv170400664D}. This term indicates a matter-field hopping from one site to the other at the expense of exciting the gauge field. The gauge field is on a ``link'' of the lattice and that is the index of it. The goal is to simulate such a Hamiltonian in a quantum configuration.

\begin{figure}[t!]
\includegraphics[width=\linewidth]{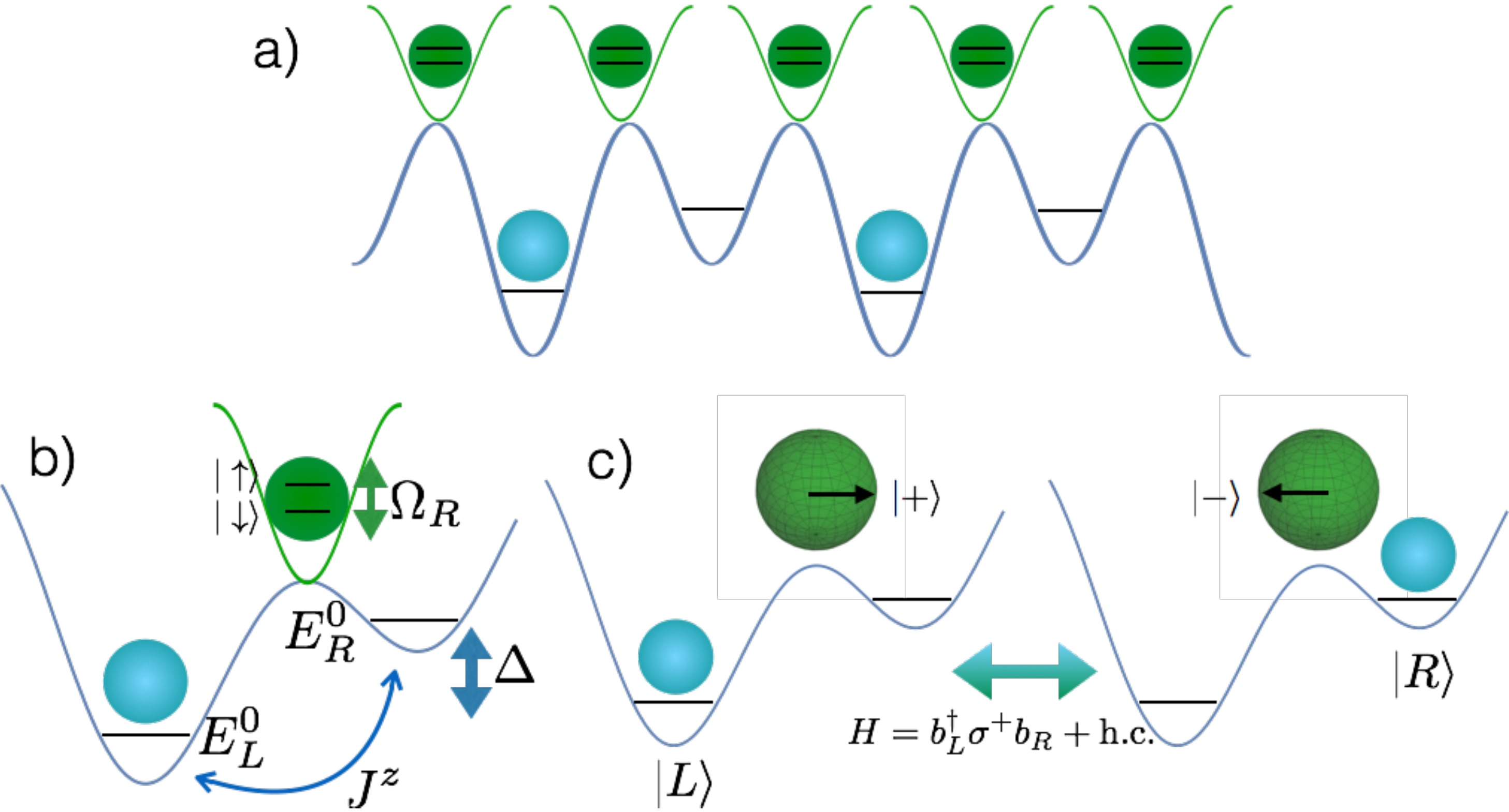}
\caption[]{a) Illustration of the overall setup where a boson is trapped in a irregular superlattice (blue lines) and the impurity is trapped in a periodic lattice (green line) centered in between the superlattice. b) Shows a close picture of the superlattice where only two wells are considered. The generalized tilted double harmonic trap can model this. The impurity has two quantum internal states which can be populated by the frequency $\Omega_R$. c) Shows the desired correlated particle-impurity hopping, that is a boson is created in the next well at the cost of flipping the impurity state from $|+\rangle$ to $|-\rangle$. Figure is adapted from \cite{2017arXiv170400664D}.}
\label{gauge_fig1}
\end{figure}

Such a system is illustrated in Fig.~\ref{gauge_fig1}. Here, a superlattice traps a boson at different irregular sites and the impurity is trapped in a periodic regular lattice. In Fig.~\ref{gauge_fig1} this is illustrated in panel a) with the blue line for the bosons and green line for the impurity. Notice that the figure can be split into small parts each with a double well and an impurity in the middle, see Fig.~\ref{gauge_fig1} b). This situation is the one that I have been analyzing so far in this chapter. However, remember that I started in a $|+\rangle$ state and flipped it into a $|-\rangle$ state when going from $|R\rangle$ to $|L\rangle$ and vice versa. Even though the impurity has two quantum internal states ${\uparrow},\, {\downarrow}$, the desired situation to simulate is illustrated in Fig.~\ref{gauge_fig1} c).

In modern experiments it is possible to setup a lattice environment like the one in Fig.~\ref{gauge_fig1} a), such that trapped particles can flow from one well to the other by going through a barrier. This flow can be controlled by the internal quantum state of an impurity \cite{PhysRevLett.93.140408, PhysRevA.93.063602, PhysRevLett.109.080402}, which interacts with the particles. The interaction between the boson and the impurity is of the form,

\begin{align}
\label{eq:Vx}
\!\!\!\!V_s(x) = \sum_k \upsilon_{\mathrm{1D}}^{e,s}(x - x_k) + \sum_k \upsilon_{\mathrm{1D}}^{o,s}(x - x_k)\,\,\,\,\,s=\uparrow,\,\downarrow,
\end{align} 
with even $\upsilon_{\mathrm{1D}}^{e,s}(x) = g_{\mathrm{1D}}^{e,s} \hat\delta_{\pm}(x)$ and odd $\upsilon_{\mathrm{1D}}^{o,s}(x) = g_{\mathrm{1D}}^{o,s} \delta^{\prime}(x)\hat\partial_{\pm}$ terms \cite{GirardeauPRA2004} as treated earlier in this chapter. It is shown in \cite{PhysRevB.90.155426} that in such a situation the following Hamiltonian can be written:

\begin{align}
\label{eq:BHsigma}
\hat H =\sum_k \hat J_ k\hat b^{\dag}_{k+1}\hat b_k + \text{H.c.} +\sum_k \frac{\hat U_k}{2}\hat n_k(\hat n_k -1),
\end{align}
where the last term is the bosonic many-body self-interaction term at each site. The above equation is for $\hat N = \sum_k\hat n_k = \sum_k\hat b^{\dag}_k\hat b_k$ particles with state-dependent hopping $\hat{J}_k = J_{\downarrow,k} |{\downarrow_k} \rangle \langle \downarrow_k |+J_{\uparrow,k}| \uparrow_k\rangle \langle \uparrow_k |$. $x_k$ is the position of the impurity. After some calculations, it is shown in \cite{2017arXiv170400664D} that the first term in Eq.~(\ref{eq:BHsigma}) can be transformed into $\propto \sum_k \hat b^\dag_k \tilde\sigma^+_k \hat b_{k+1} + h.c.$. Note that $\tilde\sigma^+=\tilde\sigma_x+i\tilde\sigma_y$, where $\sigma_x$ and $\sigma_y$ are the Pauli matrices \index{Pauli Matrices}. The eigenstates of $\tilde\sigma^+$ are the $\vert \pm\rangle = (|\uparrow\rangle\pm |\downarrow\rangle)/\sqrt{2}$ states. What this really means is this; a particle tunnels from one well to the other by flipping the impurity quantum state, which is described in the $|\pm\rangle$-basis. This situation is illustrated in Fig.~\ref{gauge_fig1} c) and this is also the reason why I initially started in the $|+\rangle$ state in my calculations. In addition, notice that the $\hat b^\dag_k \tilde\sigma^+_k \hat b_{k+1}$ term is of the $\hat{b}^{\dag}_{k} \hat{U}_{k,k+1} \hat{b}_{k+1}$ type, where $\tilde\sigma^+_k$ plays the role of the vector field $\hat{U}$.

Now, in order to show that the desired scheme has been successful, one must investigate i) the correlated hopping of the boson and impurity and ii) the conservation of the number of particles. These conditions were indeed calculated and plotted in Fig.~\ref{gauge_supp_fig3} and Fig.~\ref{gauge_fig2} panel b) and c). As already discussed, the conditions were fully satisfied under the scheme of hopping, which confirms that one can simulate such a lattice gauge by setting up a potential as the tilted double harmonic trap.

\newpage % Create empty back of side
\thispagestyle{empty}

%-----% NUMERICAL METHODS
% !TeX root = ../Main_publish.tex

\chapter{Numerical Methods}\label{ch:numerics}
\epigraph{\it “The complete is more than the sum of its pieces.”}{\rm ---Aristotle}

Analytical methods, whenever possible, are the most powerful tools that one can use in order to study quantum systems. These tools can be used to describe the systems in a very detailed fashion, and even use this knowledge to come up with some new ideas about other quantum systems. However, these methods are not always available and easily applied on all kind of systems. Therefore one has to get help from numerical as well. During my PhD studies, I have contributed and worked with three major codes in order to solve the 1D two-component interacting systems. These codes are:
\begin{itemize}
\item Effective Exact Diagonalizing Method (EEDM)
\item Correlated Gaussian Method (CGM)
\item Density Matrix Renormalization Group (DMRG)
\end{itemize}
In this chapter, I will go through the above list one by one and explain in more details about how the methods work in practice. In addition, I will summarize and discuss some results obtained by the codes.

\section{Effective Exact Diagonalizing Method\index{Effective Exact Diagonalizing Method}}\label{ch:numerics:sec:EEDM}
This method relies on the exact diagonalization of the many-body Schr{\"o}dinger equation. However, it is not a brute-force diagonalization, since the convergence of this kind of methods is very slow especially in the strong interacting regime. Nevertheless, the full Hamiltonian is projected onto a finite basis of harmonic oscillator single-particle states, $|n\rangle$ and each many-body basis state is then written as a tensor product of a symmetrized harmonic oscillator state of the $N_A$ type-A bosons and the $N_B$ type-B bosons, $|n_1\dots n_{N_A}\rangle \otimes |n_1\dots n_{N_B}\rangle$. The matrix is diagonalized in this basis. However, instead of bare zero-range interaction between the particles, the method makes use of an effective two-body short-range interaction, which speeds up the convergence in the calculations. This method was first used in \cite{LindgrenNJoP2014} for two-compoenent fermionic many-body states. I have contributed and extended it to also be applicable for {\it bosonic} two-component systems.

The $N_A+N_B$ systems are considered in the following, where $N_A$ corresponds to the number of particles in one component, called type-A, and $N_B$ corresponds to type-B. In the calculations the same mass, $m$, and trapping frequency, $\omega$, is assumed. In addition, the intra-species interactions are neglected. However, these interactions can be implemented in the code for future studies. For now, the Hamiltonian is given as,
\begin{equation}
\mathcal{H}=\sum_{i=1}^{N_A}\frac{1}{2}\left(p_{A,i}^2+q_{A,i}^2\right)+\sum_{i=1}^{N_B}\frac{1}{2}\left(p_{B,i}^2+q_{B,i}^2\right)+\sum_{i_A=0}^{N_A}\sum_{i_B=0}^{N_B}V_{i_A,i_B},
\label{numericsmanyhamiltonian}
\end{equation}
where $V_{i_A,i_B}$'s are the interaction terms between the particles. The first two sums in the above equation correspond to the non-interacting harmonic oscillator terms, which are well known. $p_{k,i}$ and $q_{k,i}$ are the momentum and coordinate operators for particle $i$ in subsystem $k\in\{A,B\}$, respectively. They each operate in their own subspace, so $p_{A,i}=p_{i}\otimes I$ and $p_{B,i}=I\otimes p_{i}$.

The total wave function of the full system consists of a tensor product of two subsystems. Each subsystem consists of another (symmetric) tensor product of few-body single-particle states. The total wave function is a bosonic state, and therefore the few-body states are totally symmetric under the exchange of any two particles within each subsystem. Each subsystem can therefore be written as a tensor product as,

\begin{equation}
|(m_1 m_2 \cdots m_N)\rangle \equiv \frac{1}{\sqrt{{N!~\prod_j n_j!}}} \sum_{p\in \text{perm(N)}} |m_{p(1)}\rangle |m_{p(2)}\rangle \cdots |m_{p(N)}\rangle,
\end{equation}
where by convention, a specific order of the labeling has been made, where $m_1\geq m_2\geq \dots \geq m_N$. On the other hand, the sum is over the all-possible permutations of $N$ elements. The factor in the front takes care of any repetition of the single-particle states, $|m\rangle$'s, which represent harmonic oscillator eigenstates with eigenvalue $m$. The second quantization formalism is also adapted here, where the number operator, $a_i^\dagger a_i$, reveals the number of particles in a specific eigenstate. The total wave function is written as:
\begin{equation}
|\Psi\rangle=|(m_1 m_2 \cdots m_{N_A})\rangle\otimes|(k_1 k_2 \cdots k_{N_B})\rangle.
\end{equation}
In the non-interacting case, the single-particle harmonic oscillator states are the full eigenstates of the system with the corresponding eigenvalue (in units of $\hbar\omega$) given as:
\begin{equation}
E=(\frac{N_A+N_B}{2}+m_1+\dots+m_{N_A}+k_1+\dots+k_{N_B}).
\end{equation}
This choice of basis is also convenient because only the interacting part of the total Hamiltonian couples the different eigenstates. Therefore the problem reduces to only finding the interaction matrix-elements.

\subsection{Matrix Elements}
{\it This section contains some updated parts from my qualifying exam report.}

In order to diagonalize the Hamiltonian, the matrix representation of the Hamiltonian is needed. In other words, one needs to calculate the matrix elements of each term of the Hamiltonian. Conveniently enough, one only needs to represent the interaction term of the Hamiltonian, due to the choice of the basis, as the kinetic and harmonic terms are trivial and well known. In this basis the $V_{i_A,i_B}$ operator couples two particles from two different species. Even though the subspaces consist of identical particles and therefore not separable in terms of which two particles interact with each other, the symmetrized few-body states are constructed from basis states of separable particles. And this is what is taken into account when calculating the interaction term. All the terms in the double sum of the interaction part in Eq.~(\ref{numericsmanyhamiltonian}) contribute equally as the first term, so the matrix element gives:
\begin{align*}
\scriptstyle M=&\scriptstyle N_A~N_B~\langle(m_1 m_2 \cdots m_{N_A})|\langle(k_1 k_2 \cdots k_{N_B})| V_{00}|(m_1 m_2 \cdots m_{N_A})\rangle|(k_1 k_2 \cdots k_{N_B})\rangle\\
&\scriptstyle =\frac{ ({ \prod_j n_j! \cdot\prod_i h_i! \cdot\prod_{j'} n'_{j'}! \cdot\prod_{i'} h'_{i'}! })^{-1/2}} {(N_A-1)!(N_B-1)!} \cdot \sum_{\substack{\sigma,\sigma'\in \text{perm($N_A$)}\\ \tau,\tau\in\text{perm($N_B$)}}} V_{m_{\sigma(1)},k_{\tau(1)},m'_{\sigma'(1)},k'_{\tau'(1)}} \times\\ 
&\scriptstyle \quad\quad\quad\times\delta_{m_{\sigma(2)},m'_{\sigma'(2)}}\dots\delta_{m_{\sigma(N_A)},m'_{\sigma'(N_A)}}\cdots \delta_{k_{\tau(2)},k'_{\tau'(2)}}\dots\delta_{k_{\tau(N_B)},k'_{\tau'(N_B)}},
\end{align*}
where $V_{a,b,c,d}\equiv\langle a,b|V|c,d\rangle$ is the two-body subspace matrix element. The delta-functions ensure that only the two states that differ at most with one single particle quantum number contribute to the matrix elements. Hence, in order to the matrix element not to vanish, only the $m_{\sigma(1)},k_{\tau(1)},m'_{\sigma'(1)},k'_{\tau'(1)}$ can be different and the others must be identical: $m_{\sigma(i)}=m'_{\sigma'(i)}$ and $k_{\tau(i)}=k'_{\tau'(i)}$ for $i>1$. This restricts the permutations, and hence for a given permutation $\sigma$ one can only fix $\sigma'$ in $(N_A-1)!$ ways such that the delta functions do not vanish. In case of bosons, there is of course the option, that some of the particles can also occupy the same state. If these states are not among the $m_{\sigma(1)},k_{\tau(1)},m'_{\sigma'(1)},k'_{\tau'(1)}$, then they will cancel out exactly the pre-factor in the above equation. On the other hand if they are, then they will also contribute with an additional factor of $(n_{\sigma(1)}! h_{\tau(1)}! n'_{\sigma'(1)}!h'_{\tau'(1)}!)$. This simplifies the above equations to:
\begin{align}
M=\sqrt{n_{\sigma}!\cdot h_{\tau}! \cdot n'_{\sigma'}!\cdot h'_{\tau'}!}\cdot V_{n_{\sigma(1)},h_{\tau(1)},n'_{\sigma'(1)},h'_{\tau'(1)}},
\label{eq:matrixelementsreduced_bosons}
\end{align}
where $l_\gamma$ is the number of times the specific single particle state $l_{\gamma(1)}$ repeats itself in the few-body state for all $l\in\{m,k,m',k'\}$ and $\gamma\in\{\sigma,\tau,\sigma',\tau'\}$. For fermionic systems the matrix element is instead given as:
\begin{align}
M=sgn[\sigma\tau\sigma'\tau'] V_{n_{\sigma(1)},h_{\tau(1)},n'_{\sigma'(1)},h'_{\tau'(1)}},
\label{eq:matrixelementsreduced_fermions}
\end{align}
where the $sgn$ adds the correct sign in front of the wave function, depending on how many permutations have been made.

Looking back at the bosonic case, one should also remember to add the diagonal elements where either one or both of the few-body states are equal. In order to illustrate the above discussion, a simple example is considered for the matrix elements between $\langle d'|\langle (a'b'c')|~V_{11}+V_{12}+V_{13}~|d\rangle|(abc)\rangle$. First note that the potentials contribute equally, hence a factor of 3 can replace the three terms. The state $|(abc)\rangle$ is given as: $|(abc)\rangle=N_{abc}\cdot(|abc\rangle\pm|acb\rangle+|bca\rangle\pm|bac\rangle+|cab\rangle\pm|cba\rangle)$ where the normalization is given as: $N_{abc}=(6\cdot(1+\delta_{bc}+\delta_{ab}+\delta_{ca}+
2\delta_{bc}\delta_{ac}\delta_{ab})^{-1/2}$ and $\pm$ is for bosons or fermions, respectively. Finally, the matrix element is:
\begin{align*}
M=&N_{abc}~N_{a'b'c'}~3\cdot 2! \cdot\\
&\Big\{V_{d'a',da}(\delta_{b'b,c'c}\pm\delta_{b'c,c'b})+\dots+V_{d'c',dc}(\delta_{a'a,b'b}\pm\delta_{a'b,b'a})\Big\},
\end{align*}
where the curly parentheses contain 9 terms for all the combinations one can make with 3 distinct particles. The factor of $2!$ comes from the all permutation one can make when $\sigma(1)$ and $\sigma'(1)$ are fixed. However, many terms vanish due to the delta-functions. For example $\langle 0|\langle (100)|~V~|0\rangle|(200)\rangle=\frac{1}{2}V_{01,02}(\delta_{00,00}+\delta_{00,00})=V_{01,02}$, as the factors cancel out due to the delta-functions, but if $\langle 0|\langle (110)|~V~|0\rangle|(210)\rangle=\sqrt{\frac{1}{2}}V_{01,02}(\delta_{11,00})\cdot 2=\sqrt{2}V_{01,02}$, because 1 and 2 are the only different numbers in the two few-body states, but 1 appears twice in the few-body state, so there will be 2! ways to get $V_{01,02}$.

\subsection{Center of Mass Excitations and Jacobi Coordinates}\label{ch:numerics:subsec:centerofmass}
{\it This section contains some updated parts from my qualifying exam report.}

Since only the intrinsic dynamics of the system are interesting, Lawson projection term \cite{GLOECKNER1974313} is used to push away the many-body solutions corresponding to excitations of the center of mass. This is done by shifting the Hamiltonian as: $\mathcal{H}\rightarrow\mathcal{H}+\lambda N-\frac{1}{2}$, where $N$ is the number operator for the center of mass excitations. The relative motion states are unaffected by this transformation and the extra $\frac{1}{2}$ is just to remove the ground state energy of the center of mass oscillator. The number operator can be expanded in the single-particle coordinates and the final Hamiltonian becomes:
\begin{equation}
\mathcal{H}=\mathcal{H}+\lambda H_{osc}+\lambda V_{osc} -\frac{1}{2}(\lambda+1)\hbar\omega,
\end{equation}
where $H_{osc}=\sum_i\frac{1}{2}(p_i^2+x_i^2)$ is like an external trap that must be added, and $V_{osc}=-\frac{1}{N_A+N_B}\sum_{i,j}\frac{1}{2}({|p_i-p_j|^2}+|x_i-x_j|^2)$ is like an interaction.\\

\noindent
Transformation of the relative coordinates into single-particle coordinates is done by normalized Jacobi coordinates, with center-of-mass coordinate, $X=\frac{x_1+x_2}{\sqrt{2}}$, and relative coordinate $x=\frac{x_2-x_1}{\sqrt{2}}$, where the same is true with the momentum: $P=\frac{p_1+p_2}{\sqrt{2}}$, $p=\frac{p_2-p_1}{\sqrt{2}}$. The two single-particle Hamiltonian can then be rewritten as:
\begin{equation*}
H_{single}=\sum_i^2(\frac{p_i^2}{2m}+\frac{1}{2}m\omega^2x_i^2)\rightarrow H_{jacobi}=\frac{1}{2m}(P^2+p^2)+\frac{1}{2}m\omega^2(x^2+X^2).
\end{equation*}
The change of basis is then done by: $|N;n\rangle=\sum_{n_1,n_2}|n_1,n_2\rangle\langle n_1,n_2|N;n\rangle$, where $|N\rangle$ and $|n\rangle$ are the eigenstates to the center of mass part and the relative part of $H_{Jacobi}$, respectively.

\subsection{Effective Interaction\index{Effective Exact Diagonalizing Method!Effective Interaction}}\label{ch:numerics:subsec:effectiveinteraction}

In order to speed up the convergence of the calculations for a given truncated basis, an effective two-body interaction is considered. This interaction is faster than the bare zero-range interaction and has recently been applied in connection with cold atoms in \cite{LindgrenNJoP2014,PhysRevA.79.012707,refId0}. The model is based on a two-body space, $P$, where two-body interacting solutions are designed using the Busch model\index{Busch Model} in \cite{busch1998} or Eq.~(\ref{HO2_energy}). The radial quantum number of the relative harmonic oscillator states are assumed to be much smaller than a cutoff, $n_{max}$, resulting in converged solutions. The effective Hamiltonian is built of a unitary transformed two-body Hamiltonian:
\begin{equation}
H_p^{eff}=\frac{U_{PP}^\dagger}{\sqrt{(U_{PP}^\dagger U_{PP})}}E_{PP}^{(2)}\frac{U_{PP}^\dagger}{\sqrt{(U_{PP}^\dagger U_{PP})}},
\end{equation}
where $E_{PP}^{(2)}$ and $U_{PP}$ contain the eigenvalues and eigenvectors from the $P$-space. $P$-space is everything up to $n_{max}$. However, when $n_{max}\rightarrow\infty$, the exact bare Hamiltonian results is reconstructed. In the implemented code, calculations are converged relatively fast. For example for a $2+2$ systems a $n_{max}=20$ in enough to obtain energies with an accuracy of three point decimals. The latter calculation takes approximately 60-70 seconds to calculate.

\begin{figure}
\centering
\includegraphics[width=\textwidth]{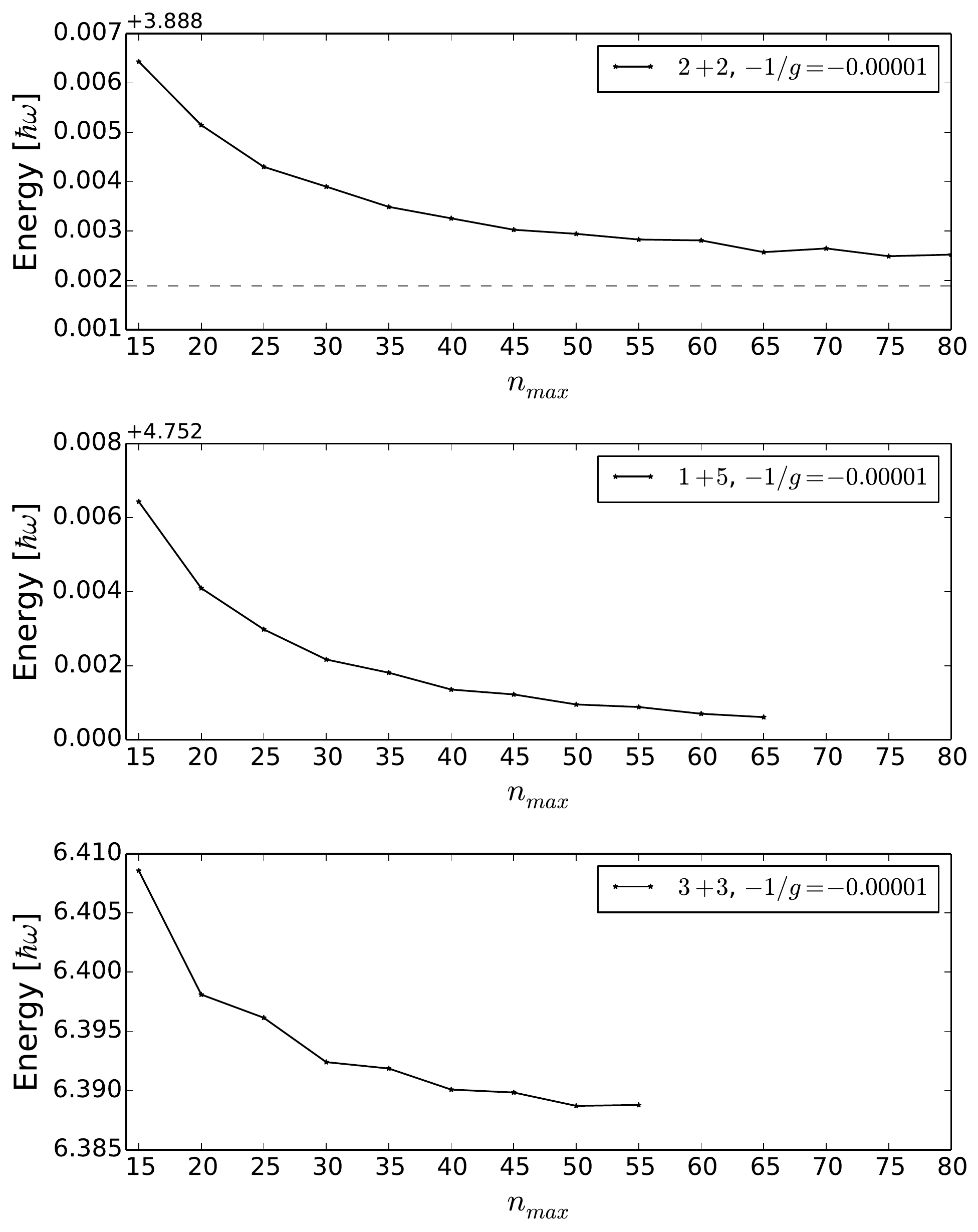}
\caption{Numerical energy calculations as a function of basis cut at $g=100~000$. The calculations are done for $2+2$ (upper panel), $1+5$ (middle panel) and $3+3$ (lower panel). The dashed horizontal line on the upper panel indicates the (semi)-analytical result for the ground state at $g\rightarrow\infty$ as discussed in Chapter 3. As the number of particles grows it takes longer time and occupies bigger space in RAM. The above calculations were made over a 7 days calculation time with a 16Gb RAM computer. The method can easily be extended to higher number of basis with better computers.}
\label{numerical_convergence}
\end{figure}

In order to show the convergence for the many-body bosons, in Fig.~\ref{numerical_convergence} I have plotted the obtained ground state energy for the $2+2$, $1+5$ and $3+3$ bosonic systems as a function of $n_{max}$. The fermionic convergences are discussed in \cite{LindgrenNJoP2014}. As is clear, Fig.~\ref{numerical_convergence}, shows a relatively quick convergence for all the systems. Notice that the calculations are done for an interaction strength of $g=100~000$. The dashed horizontal line on the upper panel of Fig.~\ref{numerical_convergence} indicates the (semi)-analytical result ($E=3.88988$) for the ground state at $g\rightarrow\infty$ as discussed in Chapter 3. This convergence is slower for bigger systems, but the overall calculations are very satisfying. With my experience, I know many brute-force diagonalization methods have problems for values from $g=5$ to $10$ in order to reproduce the correct ground state energy to just one decimal point.

\section{Correlated Gaussian Method}\label{ch:numerics:sec:CGM}

The Correlated Gaussian Method\index{Correlated Gaussian Method} builds on the idea that one can make a variationally upper limit for the bound energy state using Gaussian functions. Allow me to explain how it works in more details: The Hamiltonian in harmonic oscillator units, $b=\sqrt{\hbar/(m\omega)}$, is written as usual like this:
\begin{equation}
\mathcal{H}=\sum_{i=1}^{N_{tot}}\frac{1}{2}\left(p_i^2+q_i^2\right)+\sum_{i,j}V(q_i,q_j),
\label{gcm:eq:manyhamiltonian}
\end{equation}
where $p_i$ and $q_i$ are the $i$'th particles momentum and coordinate, respectively. Notice that even the external trap is assumed to be a harmonic oscillator trap, the method can easily be extended to other kind of traps. Finally, the last term, $V(q_i,q_j)$ is the 2-body interaction term given as $V(q_i,q_j)=g\delta(q_i-q_j)$, that is short-range with strength $g$. The variational theory is fundamentally based on the following functional:
\begin{equation}
E_{upper}[f]=\frac{\langle f|H|f\rangle}{\langle f|f\rangle}.
\end{equation}
This equation assumes a normalizable and differentiable ansatz, $|f\rangle$, which can be varied in such a way that minimizes the energy as much as possible. The ansatz is chosen to be built as a linear combination of states from a complete basis, $\{\phi_k\}$, with a truncation cutoff, $l$:
\begin{equation}
|f\rangle=\sum_{k=1}^l c_k\phi_k.
\end{equation}
The larger $l$ is, the better is the precision and therefore also longer computation time is required. The complete basis, $\{\phi_k\}$ are chosen as general Gaussian functions\index{General Gaussian Functions}:
\begin{equation}
\phi_k=e^{-(q_j-s_j^k)A_{jf}^k(q_f-s_f^k)}\equiv e^{-(\boldsymbol{q}-\boldsymbol{s})^T\mathbf{A}(\boldsymbol{q}-\boldsymbol{s'})},
\end{equation}
where $\{q_j\}$ are the coordinates of the system that are shifted by an amount of $\{s_j\}$. The $A_{jf}^k$ are some numbers that can be randomly chosen and reproduce a Gaussian function. However, $A_{jf}^k$ must be symmetric and positive-definite, so the final ansatz can be normalizable. One of the big advantages of using the Gaussian functions as basis, is that one can easily find an analytical solution to matrix elements like $\langle \phi_k|\phi_k\rangle$ or $\langle \phi_k|\frac{\partial^2}{\partial x^2}|\phi_k\rangle$. This speeds up the numerical calculations and one can therefore use larger truncated basis. In addition, if one is looking for symmetric eigenstates, then the ansatz can be made symmetric very easily and Jacobi coordinates\index{Jacobi Coordinates} can also be implemented in the basis.

After defining the Hamiltonian and the ansatz, one way to continue is to setup a plan on how to choose $l$ elements from a complete basis. There are many ways to do that. The method used here, chooses $A_{jl}^k$ and $s_j^k$ randomly from an exponential distribution and generates the ansatz with a truncated cutoff $k<l$. The ansatz can then be used to calculate the energy. The procedure can be repeated several times, say $x$ times, and then the best ansatz with the lowest energy among the $x$ trials can be chosen to be a candidate to continue with. In the next step, one chooses a randomly created basis-element and adds it to the chosen candidate. If the new extended ansatz, with $k+1\leq l$ lowers the energy even further, then the candidate gets substituted by this extended ansatz, otherwise it gets dismissed. This step can be repeated over $y$ times. Finally, by adding and extending the ansatz one by one, the truncated limit will eventually be reached and the best candidate will produce an upper limit for the lowest ground energy. Obviously, if $k$, $x$ and $y$ are a huge number it will take longer time to find the energy with better precision. It should be noted that when the Hamiltonian involves delta-functions it is harder to approximate the ansatz to vanish at contact points when the interaction is infinitely strong. This is due to the Gaussian properties that have no sharp zero points. In the calculations for the results in Chapter 2, a precision up to 4 point decimals was obtained with $x$ and $y$ set to $500$ each, which had an approximate calculation time of one hour on a decent computer with 4 cores. On the other hand, the method can implement different masses and still have analytically expressed matrix elements, and therefore making it very fast to calculate the upper-bound solutions even for mass-imbalanced systems.

In order to give an example, below I summarize the results for the $1+2$ systems. First the Hamiltonian in Eq.~(\ref{gcm:eq:manyhamiltonian}) is transformed into Jacobi coordinates and separated from the center-of-mass motion, $\mathcal{H}=H_{rel}+H_{CM}$, with:
\begin{equation}
H_{rel}=\frac{1}{2}(p_x^2+p_y^2)+\frac{1}{2}(x^2+y^2)+V,
\end{equation}
where
\begin{align*}
V=g\left[\frac{\mu_{12}}{\sqrt{\mu}}\delta(x)+\frac{\sqrt{\mu}}{\mu_{12}}\delta(\frac{\mu}{m_1}x+y)+\frac{\sqrt{\mu}}{\mu_{12}}\delta(-\frac{\mu}{m_2}x+y)\right],
\end{align*}
where the variables are defined back in Chapter 2 in Eq.~(\ref{ch2:eq:jacobicoordinates_diffmass}) and (\ref{ch2:eq:jacobihamiltonian_diffmass}). Accordingly, one can show the following analytical quantities, that are necessary in order to make a numerically fast upper limit estimate of the energy $\langle f|H|f\rangle$:

\begin{align*}
\scriptstyle N_{ij}\equiv\langle\phi_i|\phi_j\rangle&\scriptstyle= e^{-(\boldsymbol{s}^T\boldsymbol{A}\boldsymbol{s}
+\boldsymbol{s}'^T\boldsymbol{A}'\boldsymbol{s}')+\frac{\boldsymbol{v}^T
(\boldsymbol{B}^{-1})^T\boldsymbol{v}}{4}}\frac{\pi^{3/2}}{\sqrt{det(\boldsymbol{B})}}\\
\scriptstyle T_{rel}\equiv\frac{1}{2}\langle\phi_i|\boldsymbol{p}^T\boldsymbol{p}|\phi_j\rangle&\scriptstyle =\frac{1}{2}N_{ij}\cdot \left(Tr(\boldsymbol{A}')-Tr(\boldsymbol{A}'^2\boldsymbol{B}^{-1})-2\boldsymbol{\alpha}^T\boldsymbol{A}'^2
\boldsymbol{\alpha}+4\boldsymbol{s}'^T\boldsymbol{A}'^2\boldsymbol{\alpha}-2
\boldsymbol{s}'^T\boldsymbol{A}'^2\boldsymbol{s}'\right)\\
\scriptstyle V_{ext,rel}\equiv\frac{1}{2}\langle\phi_i|\boldsymbol{r}^T\boldsymbol{r}|\phi_j\rangle&\scriptstyle =\frac{1}{2}N_{ij}\cdot \left(\frac{1}{2}Tr(\boldsymbol{B}^{-1})+\frac{1}{4}\boldsymbol{v}^T (\boldsymbol{B}^{-1})^2 \boldsymbol{v}\right)\\
\scriptstyle \langle\phi_i|\delta(x)|\phi_j\rangle&\scriptstyle =\frac{\sqrt{det(\boldsymbol{B})}}{\pi^{3/2}}N_{ij}e^{-det(\boldsymbol{B})\cdot \alpha_1^2/B_{22} }\\
\scriptstyle \langle\phi_i|\delta(\pm\frac{\mu}{m}x+y)|\phi_j\rangle&\scriptstyle =\frac{\sqrt{det(\boldsymbol{B})}}{\pi^{3/2}}N_{ij}\frac{m}{\mu}\frac{\sqrt{\pi}}{\sqrt{B_{22}+\frac{m^2}{\mu^2}B_{11}\mp\frac{m}{\mu}2B_{12}}}e^{-det(\boldsymbol{B})\cdot \frac{(\alpha_1\pm\frac{m}{\mu}\alpha_2)^2}{B_{22}+\frac{m^2}{\mu^2}B_{11}\mp\frac{m}{\mu}2B_{12}} },
\end{align*}
where $\boldsymbol{B}=\boldsymbol{A}+\boldsymbol{A'}$ and $\boldsymbol{v}=2(\boldsymbol{A}\boldsymbol{s}+\boldsymbol{A'}\boldsymbol{s'})$ and $\boldsymbol{\alpha}=\frac{1}{2}\boldsymbol{B}^{-1}\boldsymbol{v}$ and $m_1=m_2=m$. The results for this code is shown in Chapter 2 Fig.~\ref{three-particles-energy-spectrum-unequal-masses}.

\section{Density Matrix Renormalization Group}\label{ch:numerics:sec:DMRG}

The Density Matrix Renormalization Group\index{Density Matrix Renormalization Group (DMRG)} (DMRG) method is an iterative variational technique that is one of the most successful methods applied on 1D systems. It was first introduced in \cite{whitePRL1992} and then closely linked to matrix product states, which opened up for a variety of applications \cite{UlrichRMP2005,SchollwockAP2011, FuehringerAdP2008, WallNJoP2012}. In recent years, it has even been pushed to the strongly interacting regime of continuous systems \cite{FangPRA2011,HuNJoP2016a}.

In the following, I will give a brief introduction to the method, but then mostly focus on how one can use one of the many libraries that have implemented the DMRG method into their core in order to investigate 1D problems. As the continuous models are of my main focus in this thesis, I will focus on how one can use the method to simulate 1D two-component particles in a harmonic oscillator potential.

\subsection{DMRG - The Basics}\label{ch:numerics:subsec:DMRGthebasics}
Imagine a 1D quantum chain/system with $L$ lattice sites and $d$ local states \{$\sigma_i$\} on each site. One example could be a chain of spin-1/2 fermions with local states: \{$\uparrow,\downarrow$\}. The main problem in numerical simulations is that in the quantum many-body physics the Hilbert space grows exponentially with $L$. Approximations such as mean-field approximation suppress the quantum entanglement nature of the systems, meaning that one loses information about the system. The DMRG method makes use of Matrix Product States \index{Matrix Product States}(MPS), where any quantum state can be represented by such states, although not unique. The representation is done one site at a time by using Singular Value Decomposition (SVD)\index{Singular Value Decomposition}. Indeed, the MPS representation decomposes the state into a {\it left} and {\it right} set of orthonormal basis by the Schmidt decomposition. Nevertheless, representing the wave function as a MPS does not reduce the complexity in the simulations. However, the SVD also reveals, which components are the most important ones in the representation of the wave function. Sometimes the wave function might contain only very few elements that are important and by reducing the size of such matrices and cutting the unimportant elements one is able to approximate the state with fewer information. The ground state energy is found by doing a variational ansatz, where $\langle\psi|H|\psi\rangle/\langle\psi|\psi\rangle$ is calculated iteratively and minimized. Starting by a random initial ansatz, one splits the system into a left and right orthonormal basis set. By projecting the ansatz into the subspaces, one calculates the energy and improves it iteratively at each step of projecting. Finally a converged energy and corresponding wave function is reached.

\subsection{From Discrete to Continuous Hamiltonian}\label{ch:numerics:subsec:fromdiscretetocontinuous}

Originally the DMRG method was designed to solve discrete systems such as the Hubbard and Bose-Hubbard Model \cite{PhysRevB.57.10324, PhysRevB.61.12474}, but it is also pushed to simulate continuous systems. There are a lot of libraries out there that perform a DMRG calculation, \cite{WallNJoP2012, Wall2009,itensor}, however, I chose to use the \cite{itensor} library implemented in C++. It should be noted that the following results and mapping are not library-dependent and there is a quite good consistency between the codes that my co-author, F.~F.~Bellotti, and I have experimented with \cite{BellottiPJD2017}. In addition, for the iTensor liberary there are a lot of instructions and examples on the source \cite{itensor}, which are available for free use. The library that I have used, the iTensor library, is still under development and the continuous case has to be implemented manually. In the following I have chosen to present the results and implementation for the Bose-Fermi four-body mixtures trapped in a 1D harmonic oscillator for any interaction strength. Notice that the formalism can easily be changed to pure bosonic or fermionic systems. Furthermore, a thorough investigation on pure systems has also been done in \cite{BellottiPJD2017}.

Let me first define our discrete system for two-component 1D mixtures. Imagine a system with $N_b$ bosons and $N_f$ fermions, so in total $N=N_b+N_f$ particles with the same mass, $m$. All the particles are trapped in the same 1D harmonic oscillator potential with the same trapping frequency, $\omega$. As the previous chapters, the interaction is repulsive and assumed to be short-range modeled by the Dirac delta-function. The total (continuous) Hamiltonian, $\mathcal{H}_c$, of the system is therefore written as (in units of harmonic oscillator, $b=\sqrt{\frac{\hbar}{m\omega}}$),
\begin{equation} 
\mathcal{H}_c=\sum_{i=1}^{N} \left( -\frac{1}{2} \frac{\partial^2}{\partial x_i^2} + \frac{1}{2}x_i^2 \right) + g_{ij}\sum_{i}^{N_f}\sum_{j}^{N_b} \delta(x_i-x_j) + g_{BB}\sum_{i<j}^{N_b} \delta(x_i-x_j),
\label{hamil}
\end{equation}
with the $g_{ij}=g>0$ and $g_{BB}>0$. Notice that due to the Pauli principle there is no self-interaction term among the fermions. In order to use the DMRG method to simulate the latter continuous equation, a discrete lattice model is needed. Fortunately, this is easy to construct when the bosonic and fermionic creation and annihilation operators are known. The following equation is the discrete lattice version of the above Hamiltonian:
\begin{align} 
\begin{split}
\mathcal{H}_d = -t \sum_{j=1}^{L-1} \left( b_{j}^{\dagger} b_{j+1} + b_{j+1}^{\dagger} b_{j} \right) 
-t \sum_{j=1}^{L-1} \left( f_{j}^{\dagger} f_{j+1} + f_{j+1}^{\dagger} f_{j} \right) 
+ U_{bf} \sum_{j=1}^{L} n_{b,j} n_{f,j} \\
+ \frac{U_{bb}}{2} \sum_{j=1}^{L} n_{b,j} \left(n_{b,j}-1 \right) 
+ V_h \sum_{j=1}^{L} (j-L/2)^2 \left(n_{b,j}+n_{f,j} \right),
\label{hub}
\end{split}
\end{align} 
where $b_j$ and $f_j$ are the bosonic and fermionic field operators acting on a site $j$, respectively. The number operators are correspondingly defined as $n_{b,j}=b_{j}^{\dagger} b_{j}$ and $n_{f,j}=f_{j}^{\dagger} f_{j}$. The first sum in the Hamiltonian is called the tunneling term with the tunneling constant, $t$. This term is responsible for the creation of a particle in the neighboring site at the expense of annihilating one in the original site. This is therefore related to the kinetic term in the continuous case. The same is true for the second sum but with fermionic particles. The 3rd sum is the on-site interaction term between the fermions and bosons with strength $U_{bf}$. The same is true for the 4th sum, $U_{bb}$, but this is the intra-species interactions. Notice that this term is over $n(n-1)/2$ due to the double counting of interactions. Finally, the last sum is the harmonic oscillator potential with strength, $V_h$. The last term is dependent on where the particles are on the lattice similar to the quadratic form in the harmonic oscillator potential. The following calculations are done for $U_{bb}=U_{bf}=U$, that is the same strength of interaction between all types of particles.

By implementing the DMRG code with these terms and applying the fermionic,
\begin{align*}
\{f_i,f_j\}&=\{f_i^\dagger,f_j^\dagger\}=0\\
\{f_i^\dagger,f_j\}&=\delta_{ij},
\end{align*}
and bosonic operators:
\begin{align*}
[b_i,b_j]&=[b_i^\dagger,b_j^\dagger]=0\\
[b_i,b_j^\dagger]&=\delta_{ij},
\end{align*}
the iTensor \cite{itensor} library can produce the ground state and energy of the Hamiltonian. However, the produced results are not in the correct units and therefore a mapping and calibration of the code is needed. For example the value of $U$ in the discrete case is not necessarily the same as $g$ in the continuous case. In fact, further analysis shows that the relation is: 
\begin{equation}
U=0.10291\cdot g.
\label{Uvsg}
\end{equation}
The relation is not that hard to find: If one knows the relation between the energy, $E_d$, calculated from the discrete Hamiltonian in Eq.~\eqref{hub}, and the energy, $E_c$ from the continuous Hamiltonian in Eq.~\eqref{hamil}, then one can relate the interaction strengths in both models together. Furthermore, it is settled that the energy units in the continuous case are in $\hbar\omega$, while the units in the discrete model are unspecified. By calculating the discrete one particle system in a harmonic trap for the ground and 1st excited states, the energy difference between these two states will correspond to exactly one $\hbar \omega_d$. This difference settles the units of the DMRG method. Therefore the energy calculated in the discrete model and the continuous model are related as follows,
\begin{equation}
\frac{E_c}{\hbar \omega}-N \frac{1}{2}=\frac{E_{d}}{\hbar \omega_{d}} -N\frac{ E_{d_{\mathrm{1p}}}}{\hbar \omega_{d}},
\label{eshift}
\end{equation} 
where $E_{d_\mathrm{1p}}$ is the discrete energy of the one particle system in a harmonic trap. The second term on both sides of the equal sign is to make sure to obtain the energy gained due to the interaction. From this equation the relation between $g$ and $U$ in Eq.~(\ref{Uvsg}) is found. Knowing what value of $U$ is needed to simulate the exact same model in the continuous case, makes the DMRG method applicable for any two-component 1D systems.

\subsection{Results: $\mathbf{2b+2f}$}\label{ch:numerics:subsec:DMRGresults}
Along with the exact analytical methods, the DMRG method can be put to a test. In the strongly interacting regime the exact wave function can be constructed using the methods in Chapter 3, \cite{DehkharghaniJoPBAMaOP2016}, or the other methods in \cite{VolosnievNC2014, VolosnievFS2014, DecampAe2016}, while for the intermediate region the interpolatory method derived in \cite{AndersenSR2016} can be used to test the results.

In the following I will present the results for the $2b+2f$ systems with $g_{BB}=g$. This is somewhat different than the results presented in Chapter 3, where $g_{BB}=0$, but the methods in that chapter can be used to obtain the results for the strong interacting regime $g_{BB}=g\rightarrow\infty$. Other systems with more than four particle are investigated in \cite{2017arXiv170301836D}.

\begin{figure}[t]%
\includegraphics[width=\linewidth]{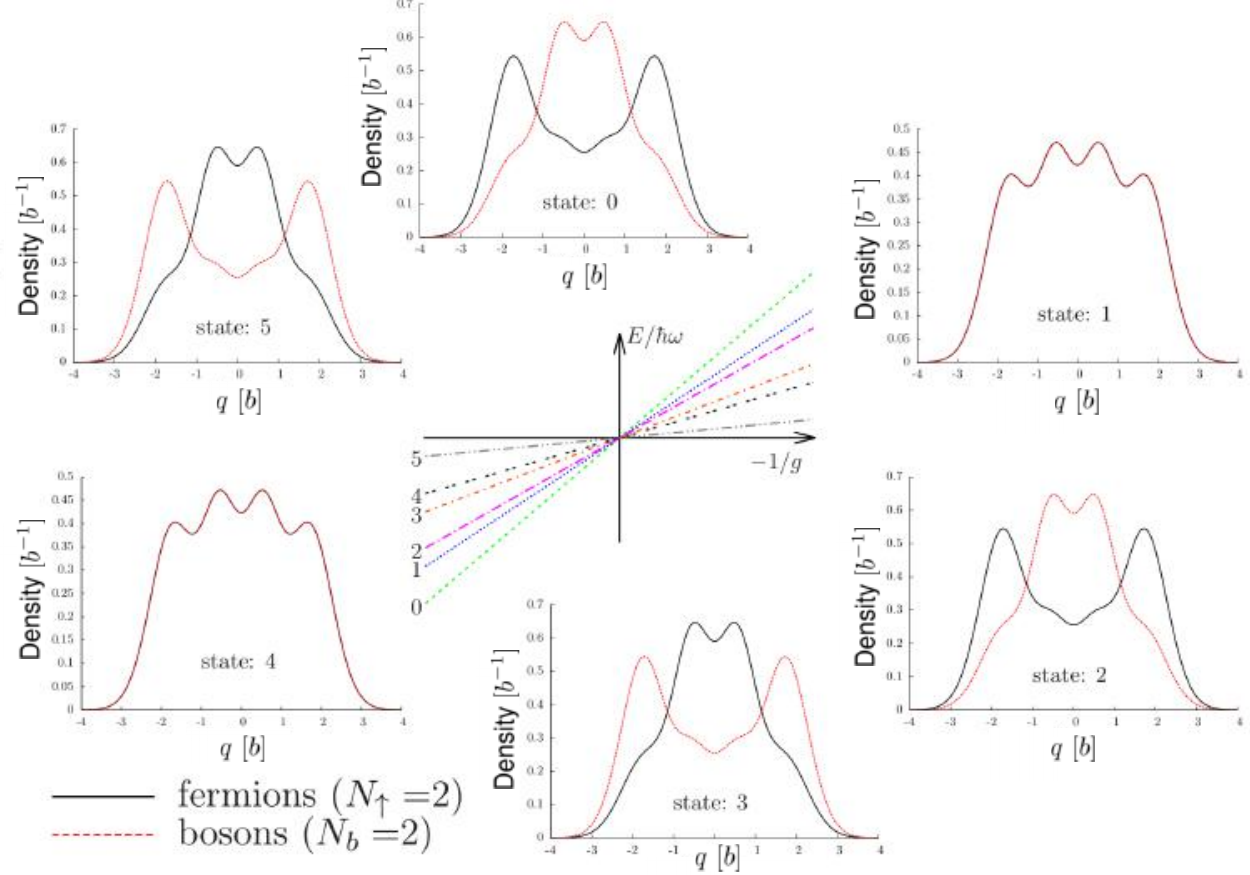}%
\caption{Middle panel shows the energy slopes as a function of $1/g=0$. The panels around the middle panel show the density profiles for the first six-fold degenerated states. Figure is adapted from \cite{2017arXiv170301836D}.}%
\label{DMRGfigure7}%
\end{figure}

Using the techniques of Ref.~\cite{VolosnievNC2014} and Ref.~\cite{LoftCPC2016} one can show that at the infinite interacting regime the system is $4!/(2!2!)=6$-fold degenerated with $\{a_1, a_2, a_3, a_4, a_5, a_6\}$ coefficients being the non-trivial linear combination coefficients of all the possible distinguishable configurations:
\begin{equation}
\psi(x_1,...,x_N)=
\begin{cases}
a_1 \Psi_A & \text{for $q_{f_1}<q_{f_2}<q_{b_1}<q_{b_2}$ $(\uparrow\uparrow BB)$}\\
a_2 \Psi_A & \text{for $q_{f_1}<q_{b_1}<q_{f_2}<q_{b_2}$ $(\uparrow B\uparrow B)$}\\
a_3 \Psi_A & \text{for $q_{f_1}<q_{b_1}<q_{b_2}<q_{f_2}$ $(\uparrow BB \uparrow)$}\\
a_4 \Psi_A & \text{for $q_{b_1}<q_{f_1}<q_{f_2}<q_{b_2}$ $(B\uparrow\uparrow B)$}\\
a_5 \Psi_A & \text{for $q_{b_1}<q_{f_1}<q_{b_2}<q_{f_2}$ $(B\uparrow B\uparrow)$}\\
a_6 \Psi_A & \text{for $q_{b_1}<q_{b_2}<q_{f_1}<q_{f_2}$ $(BB\uparrow\uparrow)$}
\end{cases},
\label{coefficientsch7}
\end{equation}
where $\Psi_A$ is the antisymmetric product of the first $N$ eigenstates of the non-interacting harmonic oscillator and $q_n$ is the coordinate of the $n^{th}$ particle. Fig.~\ref{DMRGfigure7} shows the energy slope plot and the density distribution of such a $2b+2f$ same mass system. The ground state is denoted by state: $0$ and the excited states as 1, 2, \dots, 5. As expected the slope of the ground state is steeper than the others and the value of the slope has been used to calculate the coefficients. For example the coefficients for the ground state has been shown in \cite{2017arXiv170301836D} to be ,
\begin{equation*}
(a_1,a_2,a_3,a_4,a_5,a_6) =(-0.222, 0.448, -0.669, -0.226, 0.448, -0.222),
\end{equation*}
while for the 1st excited state it is, 
\begin{align*}
(a_1,a_2,a_3,a_4,a_5,a_6)=(0.5, -0.5, 0.0, 0.0, 0.5, -0.5),
\end{align*}
The non-trivial combination is now clear and the states are in no means written as a trivial $(1,1,1,1,1,1)$ combination. Furthermore, using these coefficients one can plot the density plot of each state as illustrated in Fig.~\ref{DMRGfigure7}. Solid black line is for fermions whereas dashed red line is for bosons. One of the striking observations from this analytics is that each state populates somewhat a unique spatial configuration in the ordering of the particles on a line. This means that if one can find an easy way to populate the excited states, then some properties of quantum magnetism can be addressed here.
%%%%%%%%%%%%%%%
%%%%%%%%%%%%%%%

\begin{figure}[t]%
\includegraphics[width=\linewidth]{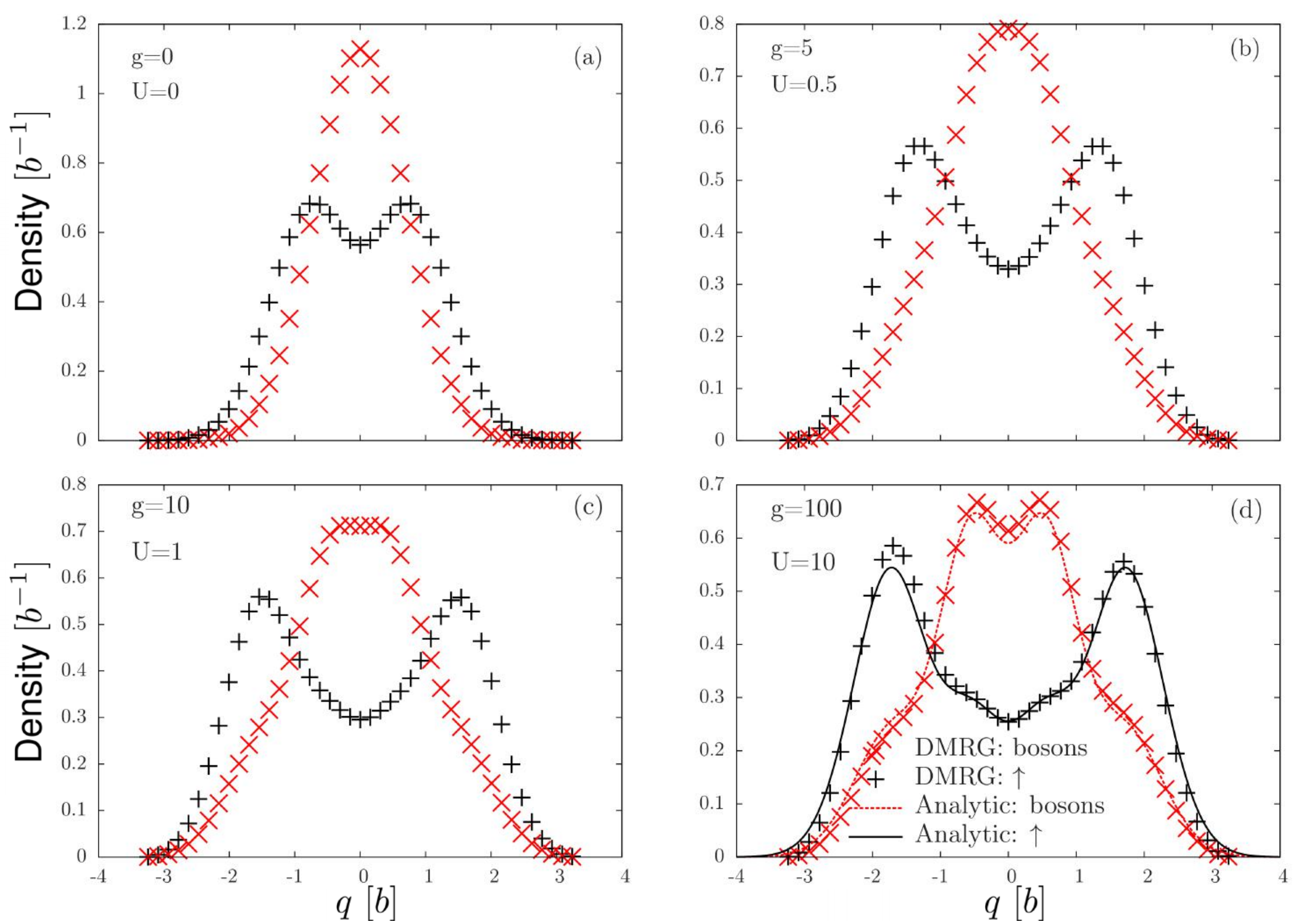}%
\caption{DMRG calculations for the ground state in the $2b+2f$ system. The panels show the evolution of the density of the fermions (black plus markers) and the bosons (red cross markers) as the interacting strength is varied $g$ ($U=0.10291 g$). Lower right panel compares the DMRG results with the analytically calculated results (black and dotted lines). The values used in the parameters for this calculation are $V_h/t=7\cdot 10^{-6}$, $t=1$, $L=128$, and $U_{bf}=U_{bb}=U$. Figure is adapted from \cite{2017arXiv170301836D}.}%
\label{DMRGfigure8}%
\end{figure}

After setting up the analytical results, I turn back to the question of how well the DMRG manages to capture these kind of results. Fig.~\ref{DMRGfigure8} shows the density profiles of the ground state of each component as a function of the interaction strength. In the non-interacting case, which is plotted in Fig.~\ref{DMRGfigure8} upper left panel it is clear that the bosons are in a nice Gaussian shape, while fermions have their characteristic two-peaked profile that is due to the Pauli principle. As the interaction increases, one can see that in the strongly interacting regime, Fig.~\ref{DMRGfigure8} lower right panel the results are a bit deviating from the analytically exact results. Although the DMRG method seems to have survived the first test, further investigation in the very strong regime, $g>100$, is therefore necessary.
%%%%%%%%%%%%%%
%%%%%%%%%%%%%%%

\begin{figure}[t]%
\includegraphics[width=\linewidth]{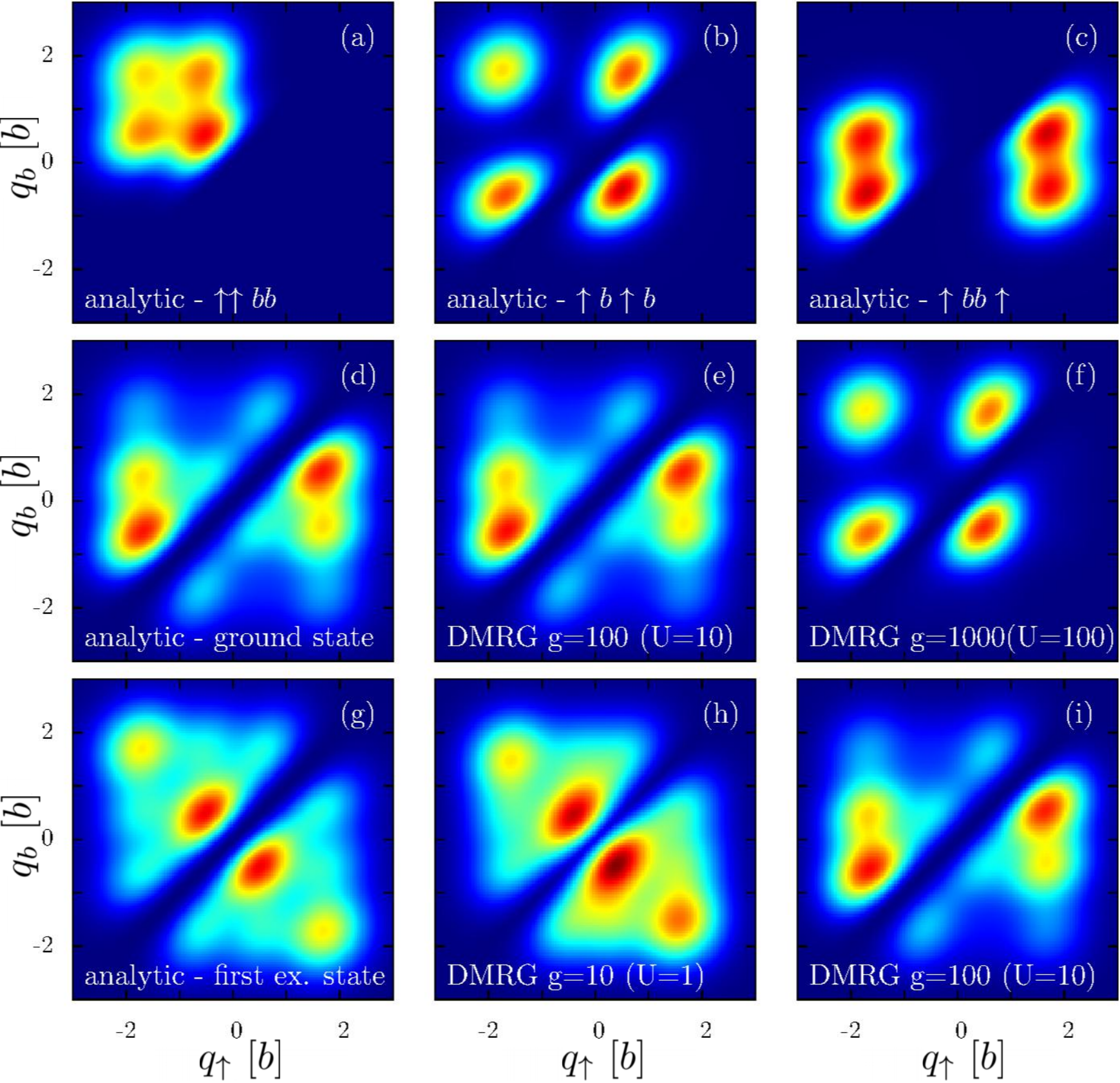}%
\caption{Pair correlation plot of the $2b+2f$ system. Panel (a)-(c) show the analytical results for only one type of configuration: (a): $\uparrow \uparrow BB$, (b): $\uparrow B \uparrow B$ and (c): $\uparrow BB \uparrow$. Panel (d) shows the analytical ground state, which is a non-trivial linear combination of the configurations in (a)-(c). Panel (g) shows the analytical 1st excited state. Correspondingly, (e) and (h) show the same plots calculated with the DMRG method. However, these results seem to be not convergent as the interaction strength is increased. Figure is adapted from \cite{2017arXiv170301836D}.}%
\label{DMRGfigure9}%
\end{figure}

Further analysis is done by the pair-correlation plot, which was also used in previous chapters, (see fx. Fig.~\ref{polaron_paircorrelation}). Fig.~\ref{DMRGfigure9} shows such a plot for the $2b+2f$ system. I will start by the DMRG results: Panel (e) and (h) show the results for the ground and 1st excited state, respectively. Accordingly, panel (f) and (i) show the same plots but this time for stronger interactions. As is apparent from these plots, the DMRG starts to fail in terms of convergence. The corresponding analytical results for the infinite interacting regime are shown in (d) and (g). It is clear that the DMRG has managed to capture these patterns in (e) and (h), but at the same time, it is clear that it runs into problems when the interaction becomes strong in (f) and (i). It seems that the method no longer gives the same results as before when the interaction strength is increased. For the sake of argument, the analytical results for one type of configurations is plotted in Fig.~\ref{DMRGfigure9} with (a): $\uparrow \uparrow BB$, (b): $\uparrow B \uparrow B$ and (c): $\uparrow BB \uparrow$. These panels show each configuration, which can be associated with their weighted coefficients as in Eq.~(\ref{coefficientsch7}). The ground and 1st excited states are built by a weighted linear combination of these single states. Now, looking back at (f) and comparing it with (b) it seems that the DMRG method has converged into this particular configuration and stayed there. The same is true for panel (i). 

Further analysis has shown that if one changes the DMRG parameters just a little bit, the DMRG result might converge into another spatial configuration in (a)-(c). This reveals that in the very strong interaction regime, the DMRG method can no longer distinguish between the degenerated states and stays in only one spatial configuration. This error starts to happen even earlier for the excited states than the ground state. Therefore the use of DMRG method has to be done with great care and for higher states, this should be even more carefully executed.
%%%%%%%%%%%%%%
%%%%%%%%%%%%%%%

Once again, as discussed in Chapter 2, the availability of the analytical results is very necessary in order to check these types of errors. In addition, it is very important to note that in general one can simply not construct the eigenstates of the strongly interacting regime by a simple and trivial linear combination of the non-interacting antisymmetric particles. DMRG is a very effective tool to calculate the 1D discrete quantum problems, and sometimes even the continuous systems, but in the latter case, one should be very careful in the strongly interacting regime.

\newpage % Create empty back of side
\thispagestyle{empty}

%-----% FINAL REMARKS
% !TeX root = ../Main_publish.tex

\chapter{Conclusion and Final Remarks}\label{sec:Conclusion}
\epigraph{\it “One, remember to look up at the stars and not down at your feet. Two, never give up work. Work gives you meaning and purpose and life is empty without it. Three, if you are lucky enough to find love, remember it is there and don't throw it away.”}{\rm ---Stephen Hawking}

In this final chapter, I will briefly summarize some of the important outcomes that were concluded in each previous chapter. However, this final chapter is not meant to be independent and for the full details of each chapter and a thorough derivation and discussion of the results, the reader is encouraged to go through the corresponding chapters. Finally, at the end of this chapter, I will give final remarks on what to expect from this field in the near future.

\section{Chapter 1: Introduction}\label{ch:conclusion:sec:1}
One-dimensional cold atoms are now easily prepared and controlled in many experiments. This allows one to verify some of the few well-known exact solutions in one-dimensional quantum systems. At the same time, the experiments allow for the theoreticians to develop some new models in some regimes that have not been explored before. During my studies, I have developed models both numerically and analytically to describe the one-dimensional few- and many-component interacting quantum systems trapped in harmonic oscillator potentials.

\section{Chapter 2: Three Particles in a 1D Harmonic Trap}\label{ch:conclusion:sec:2}
It turned out that when the interaction between the particles is very strong, then one is able to solve the three-particle 1D system analytically in a harmonic trap. In the same mass case, the fermionic systems turned out to be triply degenerated, while the bosons were doubly degenerated. For the $1+2$ fermionic systems the ground state was found to be in a linear combination of the $\uparrow\downarrow\uparrow$ and $\downarrow\uparrow\uparrow$ configurations, while for the $1+2$ bosonic case, the ground state was only found in the $AAB$ configuration. In addition, it was shown that the degeneracy breaks immediately as soon as the particles have slightly different mass ratios. Another striking conclusion was that the solutions at infinite interactions were non-trivial and one must be careful in using a Tonks-Girardeau state to construct the ground state. When it came to the intermediate case, a newly developed method based on an interpolation of the energies turned out to be a very good tool to produce these states. This method was built on a simple linear combination of the non- and strongly interacting solutions that is able to recreate any intermediate state within a few percent of error.

\section{Chapter 3: Four Particles in a 1D Harmonic Trap}\label{ch:conclusion:sec:3}
The ideas of the three-particle systems were passed on to the four-particle case. However, it turned out that the solution could not be done fully analytical and some numerical calculations were needed in order to solve the problem in full details. Despite the numerical approximations, a converged semi-analytical solution was obtained in the strong interacting regime for a two-component four-particle system. Here it was shown that the $2b+2b$ system was purely in a $AABB$ combination for any mass ratio $\beta>1$, where $\beta=m(B)/m(A)$. This observation was also supported by the developed Jacobi sphere, which clearly showed a bigger volume for this ordering. For the $2f+2f$ system the $\uparrow\downarrow\downarrow\uparrow$ combination was the favorite ordering in the ground state for $\beta>1$, where $\beta=m(\downarrow)/m(\uparrow)$. The same observation was made for the $2f+2b$, which had the $\uparrow BB \uparrow$ ordering for $\beta>1$, where $\beta=m(B)/m(\uparrow)$. In another mixture of bosons and fermions, the $2b+2f$ system, it turned out that the $\downarrow AA \downarrow$ was the favorite ground state configuration for $1<\beta<1.3$ and $\downarrow \downarrow AA$ for $\beta>1.3$, where $\beta=m(\downarrow)/m(A)$. In other words there was a transition in the ground state between the orderings. This was because of the mass term and Pauli principle where each have an important role in minimizing the energy. The interpolatory ansatz from chapter 3 was also applied here for the intermediate case for any mass ratio and further investigation again showed that one could indeed reproduce many of the intermediate properties by just taking a linear combination of the non- and strong interacting solutions. Finally, the four-component four-particle $1+1+1+1$ system was solved by almost the same method as the two-component system. Here, it turned out that the four-component system produces some chaotic and integrable solutions depending on what the mass ratio and ordering are between the particles. This can be compared to triangular quantum billiards, which also behave chaotic for some variables. Specifically, it turned out that for some specific mass and correct ordering of the particles in 1D systems, one could relate the system to a set of symmetries in three dimensions similar to the Platonic solids that have three families of solvable masses.

\section{Chapter 4: Many Particles: $\mathbf{N_A+N_B}$ Systems in a 1D Harmonic Trap}\label{ch:conclusion:sec:4}
In this chapter the many-body two-component $\mathrm{N_A+N_B}$ bosonic systems were discussed. Some of the important results here for the many-body balanced systems were that in the strong interacting limit the system clearly transforms into a ferromagnetic ground state, where the condensate separates into two distinct condensates. The distinction becomes even clearer as the number of the particles in each component grows. In the imbalanced case, it turned out that the impurity was pushed to the side contrary to what is seen in fermionic systems where the impurity is found in the middle. It was also shown that the systems both in balanced and imbalanced cases were doubly degenerated in the strong regime where one could find a $AA\dots A\pm BB\dots B$ and $A\pm BB\dots B$ configuration, respectively. Finally, A full energy spectrum for a $\mathrm{N_A+N_B}$ bosonic systems with $\{\mathrm{N_A},\mathrm{N_B}\}\in\{1,2,3,4,5\}$ and a momentum distribution were created.

\section{Chapter 5: Quantum Impurities: $\mathbf{1+N}$ and $\mathbf{2+N}$ Systems}\label{ch:conclusion:sec:5}
In this chapter the $1+N$ bosonic systems were investigated by a newly developed method. The method produces and solves an effective potential for the impurity by integrating out the bosonic particles that are obtained from a 1D harmonic trap with an impurity. The method is close to the Born-Oppenheimer approximation, but the physics is different. Despite the approximation, it turned out that the method does very good for the $1+8$ system and gets even better as the number of bosons grows. This is a huge advantage because most experiments usually work with many particles and also because many methods get worse in precision as the number of bosons grows. Furthermore, it meant that the method could give an accurate extrapolation between the few-body and the many-body systems. The method also works in any trapping potential with or without displaced traps and different masses, and even different trapping frequencies. Last but not least, the method could even incorporate the Gross-Pitaevskii equation (GPE) for the bosons. Entanglement entropy between the impurity and the bosons was also investigated. Here, as expected, it was shown that the entropy was increasing as a function of $g$, meaning that the interaction increases the correlation between the bosons and the impurity. At the end of this chapter, the ideas behind the $1+N$ system were applied on $2+N$ systems with great success. The $2+N$ system turned out to be a three-body problem between the two impurities and the condensate as a whole.

\section{Chapter 6: Lattice Gauge Simulation}\label{ch:conclusion:sec:6}
Here, the solutions for the generalized tilted double harmonic oscillator potential were derived. From these general solutions one could obtain the well-known single harmonic solutions. Later, these solutions were used to setup a quantum system, which could be used as a building block in simulating a lattice gauge model. By putting a boson in one of the wells in the double well, it was shown that one could move the boson to the opposite well by letting it interact with a fermionic particle, which is placed in the middle of the double well. By a specific scheme, the boson was able to tunnel to the other well by flipping the internal state of the fermion. It was also shown that the particle in the middle could also be an ion interacting with boson and still get the same transition.

\section{Chapter 7: Numerical Methods}\label{ch:conclusion:sec:7}
Three numerical methods were developed and presented in this chapter: Effective Exact Diagonalization Method (EEDM), Correlated Gaussian Method (CGM) and Density Matrix Renormalization Group (DMRG). EEDM was a very effective and fast method to solve the $\mathrm{N_A+N_B}$ systems for any $g$. The convergence results for the $2+2$, $1+5$ and $3+3$ systems were shown for $g=100~000$. The fast convergence was due to the effective two-body interaction that was recently developed and used here. However, the method was only developed to calculate systems with same masses and trapping frequencies. For different masses the CGM was developed, which builds on an idea of using Gaussian functions as a variationally upper limit ansatz for the system. The strength of this method was that one could calculate matrix elements of the Hamiltonian analytically, which made the numerical calculations much faster. The DMRG was used to investigate the continuous Hamiltonians in 1D systems. Here it was shown that the DMRG method was not able to reconstruct the wave functions in the strong interacting limit ($g>100$). Despite the wide use of DMRG, it was shown here that one should use the DMRG calculations for the continuous systems with care.

\section{Final Remarks}\label{ch:conclusion:sec:7}
One-dimensional few- and many-body quantum systems have been a very exciting and interesting topic to follow and investigate in details during my studies. Many developed models and results were new and surprising. I also really liked the fact that I could start from the bottom and add one particle at a time in the harmonic trap and analyze the system step by step both analytically and numerically. The models have indeed attracted a lot of attention from the cold atom community resulting in much collaboration across the different groups. Many of the results have been time independent, which is mostly because we wanted to build a solid base for future dynamical quantum calculations. Therefore, I am quite sure that experimentalists can use these results as a starting framework in their setups and build other exotic quantum systems. The good thing about one dimension is also that it can be used as a building block in building higher dimensions line-by-line and layer-by-layer. I also believe that in the near future we will see more dynamical manipulation towards quantum applications, as we understand the quantum properties better. One thing is for certain: Humans have always wanted to figure out how the Universe and everything in it, from tiny atoms to huge black holes, actually work and exploit this information for our own benefit.

\newpage % Create empty back of side
\thispagestyle{empty}

% REFERENCES / BIBLIOGRAPHY
\cleardoublepage
\addcontentsline{toc}{chapter}{Bibliography}
\bibliography{referencias} 

{\hypersetup{linkcolor=black}
% List of figures (add to table of contents)
\addcontentsline{toc}{chapter}{\listfigurename} \listoffigures
% List of tables (add to table of contents)
%\addcontentsline{toc}{chapter}{\listtablename}  \listoftables
}

\addcontentsline{toc}{chapter}{Index}
\small\printindex

\end{document}